\documentclass[twocolumn, prl, superscriptaddress,notitlepage]{revtex4-2}
\usepackage{graphicx}
\usepackage{dcolumn}
\usepackage{bm}
\usepackage[usenames,dvipsnames]{color}
\usepackage[most]{tcolorbox}
\usepackage{multirow}
\usepackage{gensymb}
\usepackage[normalem]{ulem}
\usepackage{CJK}
\usepackage{comment}
\usepackage{hyperref}
\hypersetup{colorlinks,allcolors=blue}
\usepackage{amssymb}
\usepackage{pifont}
\usepackage{physics}
\usepackage{natbib}

\usepackage{color,soul}

\begin{document}
\begin{CJK*}{UTF8}{}
\title{Phonon engineering of atomic-scale defects in superconducting quantum circuits}

\author{Mo Chen \CJKfamily{gbsn}(陈墨)}
\affiliation{Thomas J. Watson, Sr., Laboratory of Applied Physics, California Institute of Technology, Pasadena, CA 91125, USA}
\affiliation{Institute for Quantum Information and Matter, California Institute of Technology, Pasadena, CA 91125, USA}
\affiliation{Kavli Nanoscience Institute, California Institute of Technology, Pasadena, CA 91125, USA}

\author{John Clai Owens}\thanks{Present address: AWS Center for Quantum Computing, Pasadena, CA 91125, USA}
\affiliation{Thomas J. Watson, Sr., Laboratory of Applied Physics, California Institute of Technology, Pasadena, CA 91125, USA}
\affiliation{Institute for Quantum Information and Matter, California Institute of Technology, Pasadena, CA 91125, USA}
\affiliation{Kavli Nanoscience Institute, California Institute of Technology, Pasadena, CA 91125, USA}

\author{Harald Putterman}
\affiliation{AWS Center for Quantum Computing, Pasadena, CA 91125, USA}

\author{Max Sch{\"a}fer}\thanks{Present address: Department of Applied Physics and Physics, Yale University, New Haven, Connecticut 06520, USA}
\affiliation{Thomas J. Watson, Sr., Laboratory of Applied Physics, California Institute of Technology, Pasadena, CA 91125, USA}
\affiliation{Institute for Quantum Information and Matter, California Institute of Technology, Pasadena, CA 91125, USA}
\affiliation{Kavli Nanoscience Institute, California Institute of Technology, Pasadena, CA 91125, USA}

\author{Oskar Painter}
\email{opainter@caltech.edu}
\homepage{https://painterlab.caltech.edu}
\affiliation{Thomas J. Watson, Sr., Laboratory of Applied Physics, California Institute of Technology, Pasadena, CA 91125, USA}
\affiliation{Institute for Quantum Information and Matter, California Institute of Technology, Pasadena, CA 91125, USA}
\affiliation{Kavli Nanoscience Institute, California Institute of Technology, Pasadena, CA 91125, USA}
\affiliation{AWS Center for Quantum Computing, Pasadena, CA 91125, USA}

\date{\today}

\begin{abstract}
Noise within solid-state systems at low temperatures, where many of the degrees of freedom of the host material are frozen out, can typically be traced back to material defects that support low-energy excitations. These defects can take a wide variety of microscopic forms, and for amorphous materials are broadly described using generic models such as the tunneling two-level systems (TLS) model~\cite{Phillips72,Anderson72}. Although the details of TLS, and their impact on the low-temperature behavior of materials have been studied since the 1970s, these states have recently taken on further relevance in the field of quantum computing~\cite{Arute19,Wu21,Kjaergaard20,Krantz19}, where the limits to the coherence of superconducting microwave quantum circuits are dominated by TLS~\cite{Wang15a,Gambetta17}. Efforts to mitigate the impact of TLS have thus far focused on circuit design, material selection, and material surface treatment~\cite{Paik11,Bruno15,Barends13,Kjaergaard20,Lisenfeld23}. In this work, we take a new approach that seeks to directly modify the properties of TLS through nanoscale-engineering~\cite{Agarwal13, Behunin16, Rosen19}. This is achieved by periodically structuring the host material~\cite{Chan12,MacCabe20}, forming an acoustic bandgap that suppresses all microwave-frequency phonons in a GHz-wide frequency band around the operating frequency of a transmon qubit superconducting quantum circuit~\cite{Keller17}. For embedded TLS that are strongly coupled to the electric qubit, we measure a pronounced increase in relaxation time by two orders of magnitude when the TLS transition frequency lies within the acoustic bandgap, with the longest $T_1$ time exceeding $5$~milliseconds. Our work paves the way for in-depth investigation and coherent control of TLS, which is essential for deepening our understanding of noise in amorphous materials and advancing solid-state quantum devices.
\end{abstract}

\maketitle
\end{CJK*}	

Glassy materials exhibit abnormal thermal transport behaviors at low temperatures, $T<1$~K. These anomalies include specific heat and thermal conductivity that deviate from predictions of the Debye model. This is counter-intuitive, because the wavelengths of relevant phonons at low temperatures are too long to distinguish between structurally amorphous and crystalline solids. To address this mystery, Phillips~\cite{Phillips72} and Anderson, \textit{et al.}~\cite{Anderson72}, independently proposed the ubiquitous existence of microscopic two-level systems (TLS), which are defect states tunneling between two nearly equivalent local potential wells. These TLS defects are known to distribute nearly uniformly over a broad frequency range, and they have both elastic and electric dipoles that allow them to couple to strain and electric fields~\cite{Muller19}. The TLS model successfully explains the aforementioned thermal anomalies of glassy materials, as well as their acoustic and dielectric behaviors at low temperatures. Additionally, due to the omnipresence of TLS in amorphous materials, their wide frequency distribution, and their ability to couple through both phonons and photons, TLS have been associated with noise in various solid-state quantum systems, including superconducting (SC) quantum circuits~\cite{Gao08,Paladino14,Wang15a,Gambetta17,Klimov18}, nanomechanical resonators~\cite{Aspelmeyer08,Wollack21,Cleland23}, and optomechanical cavities~\cite{Riviere11,MacCabe20}.

\begin{figure*}[tbp]
    \centering
    \includegraphics[width=1.0\textwidth]{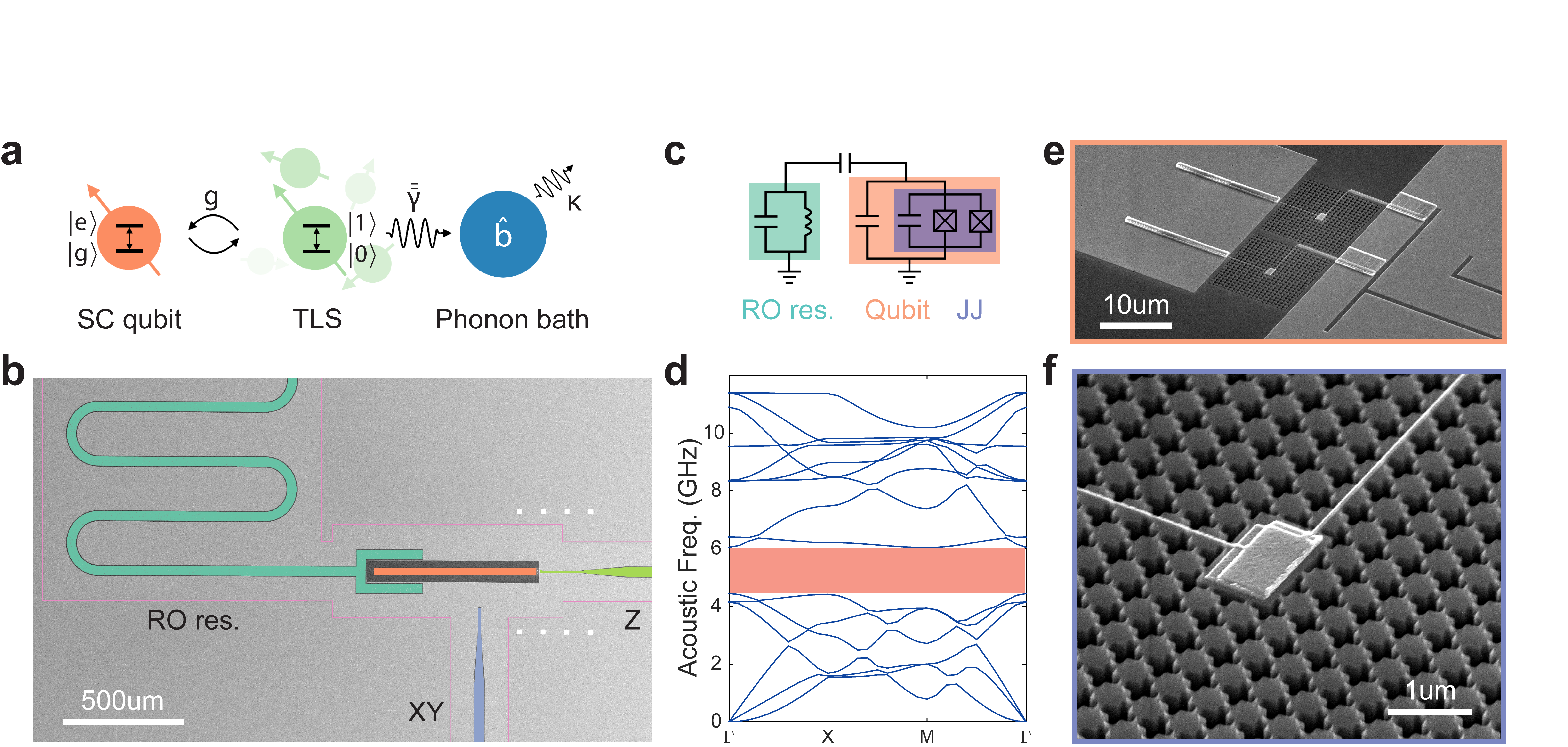}
    \caption{\textbf{A hybrid platform for phonon engineering in superconducting quantum circuits.} 
    \textbf{a,} Schematic illustrating the two-step energy dissipation process in a SC qubit. First, energy decays from the SC qubit to a bath of near-resonant TLS, with coupling rate $g$ set by the interaction between the electric field generated by the SC qubit and the electric dipole of the TLS. Next, energy further dissipates into the local environment of TLS. This environment is presumed to be dominated by a bath of phonon modes $\hat{b}$, relaxing at a rate $\kappa$, and interacting with the elastic dipole $\bar{\bar{\gamma}}$ of the TLS. \textbf{b,} SEM image (false colored) of the fabricated hybrid transmon qubit device. Each qubit couples to its dedicated $\lambda/4$ readout (RO) resonator (turquoise), Z-control line (green), and XY-control line (blue). The entire device area, as outlined by the pink lines, is suspended on the $220$~nm thick Si device layer, which is released from the underlying $3$~$\mu$m thick oxide BOX layer of the SOI chip.
    \textbf{c,} Circuit diagram of the transmon qubit and readout resonator. Approximately $40\%$ of the transmon capacitance comes from the shunt capacitor (orange), with the remaining $60\%$ from the JJs. 
    \textbf{d,} The simulated acoustic band structure of a unit cell of the Si cross-shield acoustic structure, with the acoustic bandgap centered around $5.1$~GHz shaded in pink. 
    \textbf{e,} A zoomed-in view of the region surrounding the SQUID loop of the transmon qubit device. The SQUID loop is formed with two JJs in parallel between the shunt capacitor and ground. Each JJ is fabricated on top of a micron-scale Si platform, which is tethered to the SOI substrate by a cross-shield acoustic bandgap structure. \textbf{f,} A detailed SEM image of one of the JJs of a qubit device, showing the cross-shield patterning of the acoustic bandgap structure. The contacts from the shunt capacitor and ground to the top and bottom electrodes of the JJ, respectively, are visible as narrow Al leads that run across the connected cross-shield lattice.  
    }
    \label{fig1:device}
\end{figure*}

In the context of SC quantum circuits, TLS have been identified as a primary limitation to the energy lifetime, coherence, and overall stability of the physical qubits being explored for scalable quantum computing architectures~\cite{Barends13, Paladino14, Wang15a, Gambetta17, Klimov18, Muller19, Burnett19, Schlor19, Kjaergaard20}. TLS are thought to reside primarily at the amorphous material interfaces that make up the physical qubit device, and cause dielectric loss through the interaction between their electric dipoles and the electric field of the qubit. The interaction leads to a two-step energy dissipation process, where energy first transfers from the SC qubit to resonant TLS, and subsequently dissipates into the local environment of the TLS~\cite{Muller09,Agarwal13,Behunin16,Rosen19}. Despite awareness of the two-step dissipation process for SC qubit decay, past research efforts have focused on investigating and mitigating the first step, namely, the energy decay from the SC qubit to TLS. This choice is in part because of the challenge in accessing and controlling atomic-scale TLS defects~\cite{Lisenfeld19,Spiecker23}, and therefore, in one's ability to modify the second step of the dissipation process. 
As a result, TLS have long been viewed as an intrinsic material defect to be avoided~\cite{Paik11,Bruno15,Barends13,Lisenfeld23}. In the pursuit of better SC qubits, material investigations have focused on finding superconductors with a surface oxide layer that has low TLS density~\cite{Oh06,Chang13,Place21}. Similarly, circuit designs of SC qubits aim to minimize the electric-field strength of the electromagnetic field produced by the qubit at material interfaces to reduce the interaction between the SC qubit and TLS~\cite{Paik11,Bruno15,Barends13}. 
These efforts have led to microwave-frequency SC qubits with energy relaxation times that extend over hundreds of microseconds~\cite{Kjaergaard20}. 

In the hopes of further understanding TLS and improving SC qubits, in this work we take direct aim at modifying the second step of the dissipation process of SC qubits, namely, the interaction of TLS with the reservoir of phonons of the material host as illustrated in Fig.~\ref{fig1:device}a. The phonon bath is targeted due to the roughly five-orders-of-magnitude difference in the speed of sound and the speed of light in materials, and the correspondingly much larger density of states (DOS) at microwave frequencies for phonons versus photons. For typical Debye-level electric dipole moments and eV-level deformation potentials of TLS, this makes the dominant bath that of phonons. We design and fabricate a frequency-tunable transmon qubit~\cite{Koch07,Schreier08} with its Josephson junctions (JJs) embedded in an engineered acoustic structure that features a GHz-wide acoustic bandgap, as shown in Fig.~\ref{fig1:device}b--f. TLS within the junctions and with transition frequency within the acoustic bandgap range, experience a suppressed two-dimensional (2D) phonon DOS, which effectively shields them from resonant decay into the phonon bath. Using the transmon qubit as a quantum sensor, we are able to individually address and characterize the coherence properties of strongly-coupled TLS within the JJs of the electric qubit. Our experimental results show that the $T_1$ times for TLS with resonant frequency lying inside the acoustic bandgap are, on average, extended by two orders of magnitude compared to when the TLS frequency lies outside the acoustic bandgap. The coherence and the temperature dependence of the $T_1$ relaxation process of the acoustically-shielded TLS are also studied, indicating that coherence is limited to the microsecond-level due to low-frequency noise, and thermally-activated relaxation channels open up above $50$~mK.  

\subsection{Transmon Qubit with Acoustically-Shielded Junctions: Design and Fabrication}
We fabricate our transmon qubit device on a silicon-on-insulator (SOI) substrate~\cite{Keller17}, which allows for nanoscale fabrication of high-quality acoustic bandgap structures in the microwave frequency range (see Sec.~I of the Supplementary Information (SI) for design and fabrication details). Figs.~\ref{fig1:device}b, ~\ref{fig1:device}e, and ~\ref{fig1:device}f show scanning electron micrographs (SEM) of various parts of the device. The transmon qubit has a shunt capacitor, which is used to couple to a coplanar-waveguide resonator for readout of the qubit state. It also has two aluminum-aluminum oxide-aluminum (Al-AlO$_x$-Al) JJs forming a magnetic flux sensitive SQUID loop for tuning of the electric qubit state via a current-carrying Z-control line. An XY-control line is added for direct charge excitation of the qubit. The transmon qubit is unremarkable in its design, except for the fact that each JJ in the SQUID loop (Fig.~\ref{fig1:device}f) is located on top of a suspended Si platform formed by the release of the $220$~nm thick Si device layer from the SOI substrate, and tethered to the SOI substrate by nine periods of an acoustic bandgap structure (Fig.~\ref{fig1:device}f). The acoustic bandgap structure of this work is based on a cross-shield design~\cite{Chan12,MacCabe20}, which possesses a $1.372$~GHz-wide acoustic bandgap centered at $5.128$~GHz~\cite{SOM}. In theory, this effectively isolates the JJs of the transmon qubit, and any TLS defects that may be within the amorphous oxide layer of each junction, from acoustic modes of the SOI substrate.

The TLS within the acoustically-isolated JJs are distinguished from other TLS in different regions of the circuit by their signature strong coupling to the transmon qubit due to the strong electric field of the qubit mode in the atomically-thin AlO$_x$ barrier layer of the JJs. To increase the occurrence of these TLS of interest, the JJs of our device are chosen to have a relatively large area of $0.83$~$\mu$m$^2$ each. The AlO$_x$ barrier layer is also grown slightly thicker to maintain the transmon operating frequency close to $5$~GHz (see Sec.~I.B of the SI for details). As a result, the JJs have a significant junction capacitance in addition to the nonlinear inductance, which is similar to the merged-element transmon qubit~\cite{Zhao20,Mamin21}. 
In the current design, the JJs account for approximately $60$~fF, or $60\%$ of the total transmon capacitance. 

\subsection{Transmon Qubit and TLS Characterization}
Characterization of transmon qubit devices, and any coupled TLS, are performed in a dilution refrigerator, where the chip-scale sample containing the devices is mounted to the mixing chamber plate of the fridge. The fridge reaches a base temperature of $7$~mK, which cools down both the SC qubit and TLS close to their respective ground states. The transmon qubit is first characterized in the time domain with pulsed excitation and dispersive readout. The transmon qubits of this work are measured to have excited-to-ground state relaxation times of $T_1 \sim 3$~$\mu$s, which is comparable to the best reported SOI qubit~\cite{Keller17}. We attribute the qubit $T_1$ relaxation to both dielectric loss at the shunt capacitor and Purcell decay to the readout resonator. Further details of the qubit characterization and the qubit parameters for all seven qubit devices studied in this work can be found in Sec. I.A of the SI.

\begin{figure}[!htbp]
    \centering
    \includegraphics[width=0.45\textwidth]{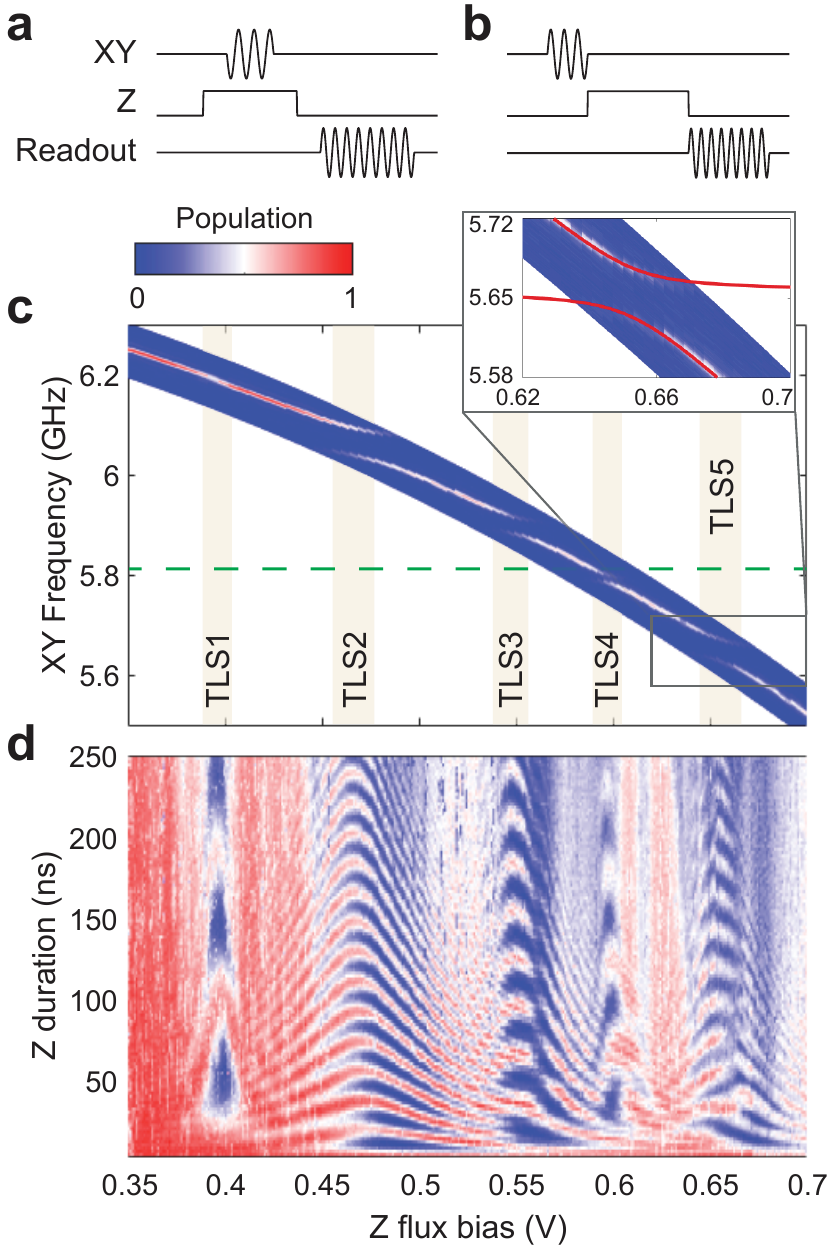}
    \caption{\textbf{Characterization of a hybrid transmon-TLS system.} \textbf{a,} Pulse sequence used for microwave spectroscopy of the qubit, and \textbf{c,} corresponding measured transmon qubit spectrum for Q$_1$ of Chip-A. In this measurement protocol, a Z-pulse flux biases the transmon qubit away from its flux-insensitive sweet spot. An overlapping XY-pulse probes the excitation of the transmon qubit at different flux biases. When the transmon is in resonance with a TLS, their hybridization results in avoided crossings. For Q$_1$ of Chip-A we measure five distinct TLS, labelled TLS1 through TLS5. Notably, the avoided crossings of TLS4 and TLS5 are situated within the simulated acoustic bandgap, with the upper frequency bandedge indicated by the green dashed line.
    The inset provides a magnified view of the avoided crossing of TLS5. The red solid line in the inset is a fitting curve with $\omega_{\mathrm{TLS5}}/2\pi = 5.6563$~GHz and $g/2\pi = 21.7$~MHz. \textbf{b,} Pulse sequence of the transmon-TLS SWAP gate spectroscopy, and \textbf{d,} corresponding measured transmon-TLS SWAP spectrum for device Q$_1$ of Chip-A. In this measurement protocol, the transmon is first excited by an XY $\pi$-pulse, then tuned by a Z-pulse with varying amplitude and duration, and finally the transmon qubit population is dispersively read out upon tuning back to its starting frequency. The resulting chevron patterns correspond to vacuum Rabi oscillations between the transmon qubit and the strongly coupled TLS1--5.
    }
    \label{fig2:qubit}
\end{figure}

Once characterized, the transmon qubit is used as a quantum sensor to identify individual TLS in the JJs that experience a structured acoustic environment. Using the pulse sequence depicted in Fig.~\ref{fig2:qubit}a, we perform pulsed microwave spectroscopy to explore the electrically-active transitions of a given transmon qubit device. The measured microwave spectrum of one such transmon qubit device (Chip-A, Q$_1$) is shown in Fig.~\ref{fig2:qubit}c, with qubit frequency tuned between $5.5$~GHz and $6.3$~GHz using a flux bias pulse via the Z-control line. Strong couplings of the transmon to five TLS (labelled TLS1 through TLS5) manifest as avoided crossings in the spectrum, from which we extract the TLS frequencies and coupling strengths $g$ to the transmon qubit. In these experiments, the microwave power on the XY-control line is chosen to yield approximately $100$~ns transmon $\pi$-pulses, which allows us to resolve TLS with a coupling strength of $g\gtrsim 5$~MHz. The strong couplings of TLS1--5 are a signature that these TLS are physically located inside the JJs of Q$_1$ on Chip-A, and hence, inside the acoustic bandgap structures. 

Next, in order to probe further the properties of the strongly coupled TLS, we calibrate coherent SWAP operations between transmon qubit and TLS states using SWAP spectroscopy~\cite{Martinis05,Neeley08}, as illustrated in Fig.~\ref{fig2:qubit}b. Figure~\ref{fig2:qubit}d shows a representative measurement of SWAP spectroscopy performed on device Q$_1$ of Chip-A, where five pronounced vacuum Rabi oscillation patterns appear at the same pulsed flux-bias amplitudes as the previously measured anticrossings for TLS1--5. This confirms that these fringes arise from the resonant exchange interactions between the transmon qubit and individual strongly coupled TLS, and signifies the coherent nature of these TLS. 
Based on these vacuum Rabi oscillations, we identify optimal SWAP gates for each TLS. Utilizing the SWAP gate, we are able to selectively prepare any one of the five TLS in their first excited-state. Moreover, sequential application of this technique allows us to look for the presence of second (and higher-order) excited-states of the TLS. The absence of observable higher-order excited states (see Sec.~II.F of the SI), indiccates that the strongly coupled TLS to the transmon qubits measured here are highly anharmonic. Importantly, this eliminates any concerns that the TLS-like behavior observed in this study originates from high-$Q$ harmonic acoustic modes of the acoustic bandgap structure.

\begin{figure*}[!htbp]
    \centering
    \includegraphics[width = 0.9\textwidth]{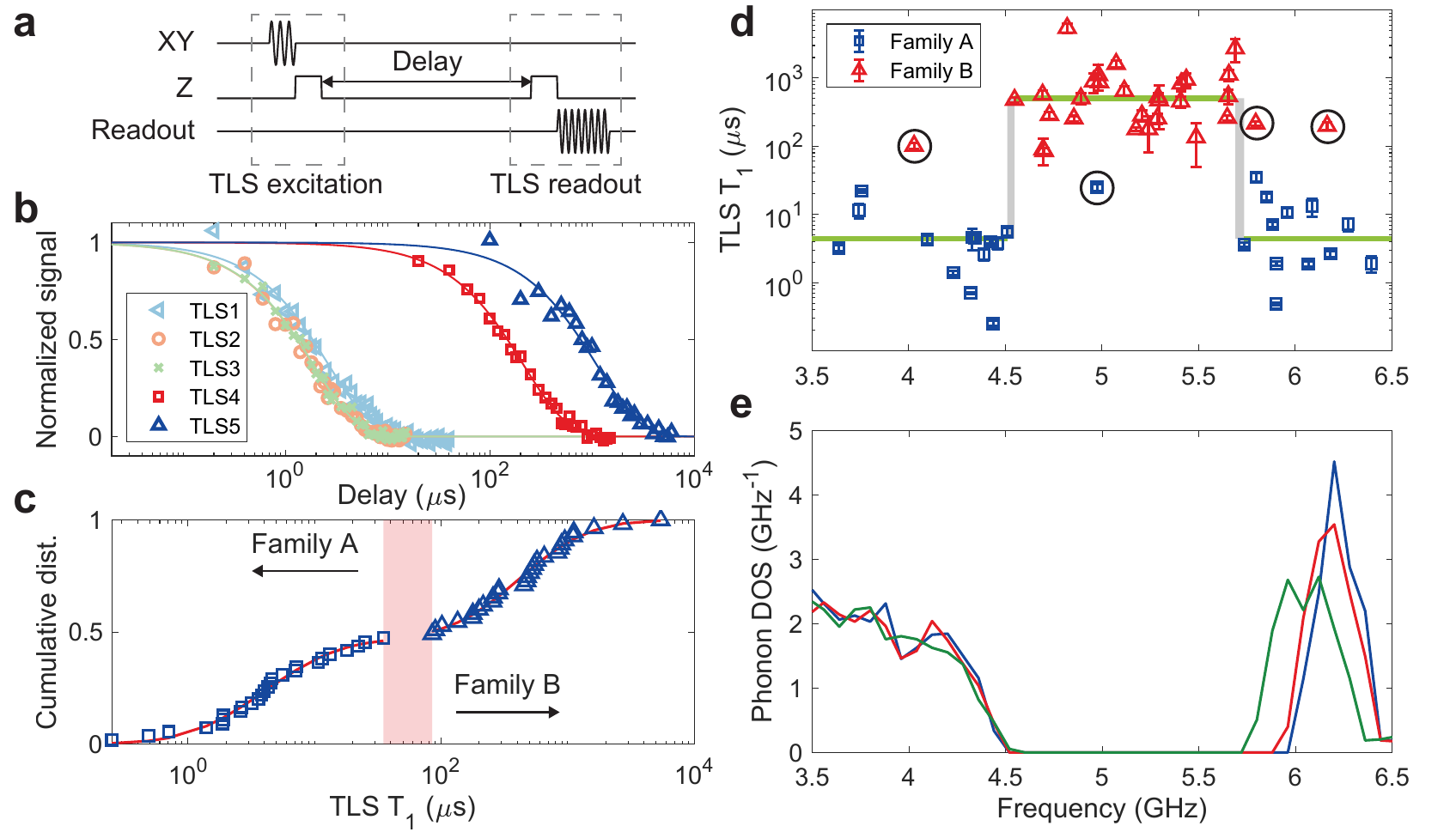}
    \caption{\textbf{TLS $T_1$ relaxations in an engineered acoustic bath.} \textbf{a,} Pulse sequence to measure the $T_1$ energy relaxation time of TLS. The qubit is first driven to its excited state via an XY $\pi$-pulse. This excitation is then transferred to the TLS using a Z SWAP pulse. After a variable delay, the TLS state is swapped back to the transmon qubit and the qubit popultation is subsequently read out. 
    \textbf{b,} Relaxation curves for each of TLS1--5 of Q$_1$ on Chip-A identified in the microwave spectroscopy of Fig.~\ref{fig2:qubit}. The measured signals have been normalized and fitted to a decaying exponential. 
    \textbf{c,} Cumulative distribution of TLS $T_1$ relaxation times from 55 TLS measured across two chips, seven devices, and two cooldowns. A gap between $35$--$85~\mu$s (pink shade) divides the cumulative distribution into two families, A and B. Each family is fitted to a log-normal distribution, represented by the red solid lines. 
    \textbf{d,} $T_1$ relaxation times of the TLS plotted against their frequencies. TLS in family A are represented by blue squares, and TLS in family B by red triangles.
    The two TLS families exhibit a strong correlation with TLS frequency. The shaded vertical gray regions represent the edges of a central frequency band where almost all of the TLS in family B reside, and outside of which almost all of the TLS in family A reside. The solid green line is a guide for the eye, representing the median $T_1$ values for TLS with transition frequency inside and outside the central frequency region. Four outlier TLS from family A and B, whose $T_1$ values are not correlated with frequency in the same way as the rest, are marked by black circles. 
    \textbf{e,} Simulated phonon DOS of an infinitely-periodic cross-shield acoustic bandgap structure. The blue, red, and green curves are derived from slightly different acoustic structure unit cells, representing variations in the Al leads from the JJs (see Sec.~I.A of the SI).
    }
    \label{fig3:TLS_T1}
\end{figure*}

\subsection{T1 Lifetime of Acoustically-Shielded TLS}
To characterize the lifetime of TLS, we begin by preparing a TLS in its excited-state using the SWAP gate, and then let the TLS relax for a variable amount of time. 
Finally, we map the TLS state back to the transmon using a second SWAP gate, and measure the final transmon state using its dispersive readout circuit (Fig.~\ref{fig3:TLS_T1}a). During the TLS relaxation time, the transmon qubit is tuned to its uppermost frequency, far from resonance with the TLS to avoid relaxation of the TLS through hybridization with the transmon qubit. The resulting $T_1$ relaxation curves for the five characterized TLS of Q$_1$ on Chip-A are shown in Fig.~\ref{fig3:TLS_T1}b, with TLS1--3 having $T_1\sim 2$~$\mu$s, while TLS4 and TLS5 exhibit two to three orders-of-magnitude longer relaxation times of $T_{1,\mathrm{TLS4}}=215 \pm 15$~$\mu$s and $T_{1,\mathrm{TLS5}}=1100 \pm 200$~$\mu$s, respectively (here the one standard deviation uncertainty in the $T_1$ is quoted). Notably, both TLS4 and TLS5 have transition frequencies that lie within the expected acoustic bandgap of the crosss-shield structure based on numerical finite-element simulations, whereas TLS1--3 all have frequencies above the simulated bandgap. 

To gather further statistical data on the correlation between the $T_1$ of strongly coupled TLS and their transition frequency, we examined TLS across seven transmon qubit devices (nominally, all of the same by design), on two different fabricated chips. Additionally, we thermally cycled the devices up to room temperature and back down to milliKelvin temperatures to redistribute the TLS frequencies~\cite{Shalibo10}. In total, $56$ TLS (or more accurately TLS with unique frequencies) have been characterized. These TLS span frequencies from $3.7421$~GHz to $6.3935$~GHz, and their $T_1$ values range from $0.25 \pm 0.02$~$\mu$s to $5400\pm 800$~$\mu$s. For a full list of measured TLS parameters, see Sec.~II.A of the SI. Here, we focus on the statistical properties of the TLS, and show compelling evidence that the extended TLS $T_1$ relaxation times, as observed in TLS4 and TLS5, originate from the acoustic bandgap. 

We begin our analysis by plotting in Fig.~\ref{fig3:TLS_T1}c the cumulative distribution of TLS versus their measured $T_1$ relaxation time (here we have excluded one TLS data point whose $T_1$ value we estimate to be limited by decay through the transmon qubit). From this plot, we identify a gap in measured $T_1$ times, between $35 \textrm{--} 85~\mu$s (pink shaded region). This gap in $T_1$ divides the TLS naturally into two distinct families, referred to as family A (blue squares) and family B (blue triangles). Each family is fitted to a log-normal distribution, represented by the red solid lines. The fitted parameters yield median $T_1$ values of $4.1\pm 0.2~\mu$s and $414\pm 17~\mu$s for family A and B, respectively (we use the median as opposed to the mean due to the large skewness of $1.8$ and $3.3$ for the TLS $T_1$ distributions of the two families).

Next, we plot in Fig.~\ref{fig3:TLS_T1}d the measured TLS $T_1$ relaxation times against their frequencies, marking those TLS in family A with blue squares and those in family B with red triangles. As is clearly visible, the two TLS families, categorized solely by their distinct $T_1$ values, exhibit a strong correlation with frequency. Specifically, TLS in family A predominantly occupy frequencies outside a frequency band centered around $5.1$~GHz, whereas those in family B mostly reside within this frequency band. We can define this central frequency band quantitatively by using the following cost function,
\begin{equation}\label{eq:cost_fun}
    \mathcal{C}(f_1,f_2) = \log[1 - F_A(f_1,f_2)\times F_B(f_1,f_2)],
\end{equation}
where the central frequency band is defined between lower and upper bandedges $f_1$ and $f_2$, respectively. $F_A(f_1,f_2)$ denotes the fraction of TLS in family A whose frequencies lie outside this frequency band, while $F_B(f_1,f_2)$ represents the fraction of TLS in family B that fall within this frequency band. Upon minimizing the cost function, we obtain $\mathcal{C}_\mathrm{min} = -1.98$, with the lower bandedge $f_1$ lying between $4.510$--$4.547$~GHz, and the upper bandedge $f_2$ lying between $5.690$--$5.735$~GHz. These empirically defined bandedges are marked as vertical gray shaded regions in Fig.~\ref{fig3:TLS_T1}d. This can be compared to the expected frequency bandgap region of the cross-shield acoustic structure that the JJs of the transmon qubits are embedded within. In Fig.~\ref{fig3:TLS_T1}e, we present the numerically-simulated phonon DOS for three slightly different unit cells of the acoustic cross-shield, taking into account variations in the Al leads that connect the JJs to the rest of the circuit (see Sec.~I.A, I.D of the SI). Even for the simulation with the heaviest loading by the Al leads (green curve in Fig.~\ref{fig3:TLS_T1}e), we find excellent correspondence between the frequency band defined by high TLS $T_1$ values and the acoustic bandgap with zero phonon DOS of the simulated cross-shield structure.  

The above correlations between TLS $T_1$, TLS frequency, and the designed acoustic bandgap frequency serve as compelling evidence that the observed several orders-of-magnitude increase in TLS $T_1$ relaxation time, from a median of $M_{\mathrm{out, 2D}}(T_1) = 4.4$~$\mu$s outside the central frequency band region, to a median of $M_{\mathrm{in, 2D}}(T_1) = 506$~$\mu$s inside the central frequency band region, originates from the presence of an acoustic bandgap in the acoustic bath seen by the strongly coupled TLS within the JJs. This result also indicates that for TLS that couple to the electric field of microwave-frequency SC qubits, the dominant relaxation channel is spontaneous phonon emission into the acoustic bath, in agreement with the two-step dissipation chain shown in Fig.~\ref{fig1:device}a. There are, however, several outliers in the measured TLS data. These are marked by black circles in Fig.~\ref{fig3:TLS_T1}d. The TLS from family B with low $T_1$ at a frequency of $5$~GHz lying within the central frequency band, is from a device (Q$_2$, Chip-A) that seems to have a frequency bandgap which is shifted to higher frequencies (see Sec.~II.D in the SI) due to fabrication variation from device to device. Similarly, the TLS of family B with high $T_1$ lying just outside the central frequency band at $5.8$~GHz, is from the Q$_1$ Chip-A device, which has an upper bandedge frequency slightly higher than the average (see Sec.~II.D in the SI). Despite these outliers, the consistency of the inferred bandgap region from all seven devices indicates that the fabrication process, although not perfect, is relatively accurate on the scale of a percent standard deviation. 

The two TLS of family B with high $T_1$ values that are far from the central frequency band, at approximately $4$~GHz and $6.2$~GHz, represent a different type of outlier. We believe that these TLS, although lying outside the cross-shield bandgap, have either acoustic dipole orientations that are orthogonal to the polarization of the acoustic modes of the cross-shield structure, or, are decoupled from the acoustic bulk phonon modes of the SOI substrate due to the finite extent of the cross-shield and acoustic reflections at its perimeter. These effects might also explain a portion of the large variation seen in the $T_1$ values for TLS outside of the acoustic bandgap, although varying acoustic dipole strength between TLS would also contribute to the observed $T_1$ variance. This discussion draws attention to the fact that all of the measured TLS in this work live within a 2D Si membrane, with an effective 2D phonon DOS. Comparing our results to previous studies of TLS in the JJs of a phase qubit fabricated on a high resistivity Si substrate, where the median $T_1$ values were of order $M_{\mathrm{3D}}(T_1)\sim 200$~ns~\cite{Shalibo10}, highlights that the observed increase in TLS $T_1$ within the acoustic bandgap of our structures is three orders-of-magnitude above that for TLS in a three-dimensional bulk material.      

\subsection{TLS Coherence and Temperature-Dependence of $T_1$ Relaxation}
The significantly extended TLS $T_1$ time naturally raises questions about their coherence time. There is also the question of what limits the TLS $T_1$ values once the direct resonant coupling to a phonon bath is removed. Here we perform further studies on TLS5 of device $Q_1$ on Chip-A, which displays a long $T_1 = 1100\pm200$~$\mu$s, making it a sensitive probe of these effects. Coherent control of the TLS can be achieved by sending a strong microwave pulse resonant with the TLS down the XY-control line of the transmon qubit~\cite{Lisenfeld10}. This pulse is able to directly control the TLS due to the hybridization between the transmon qubit and the TLS, as described in Sec.~II.I of the SI. 
We calibrate control pulses for TLS5 of Q$_1$ on Chip-A, and perform Ramsey spectroscopy to characterize its $T_2^*$. 

\begin{figure*}
    \centering
    \includegraphics[width = 0.9\textwidth]{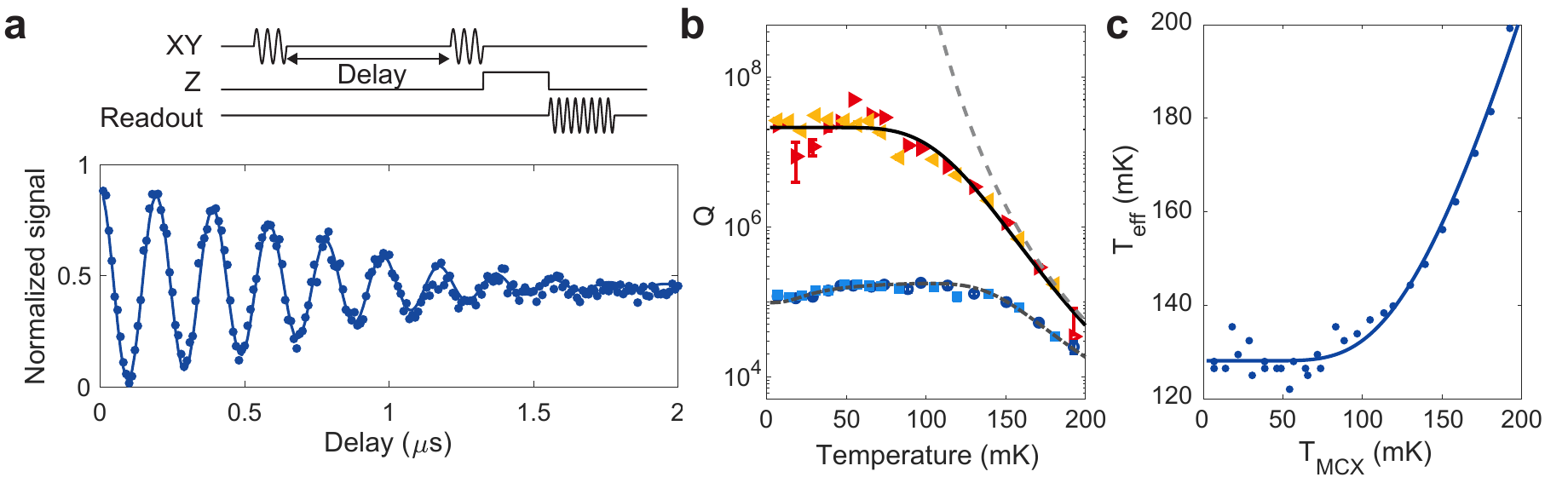}
    \caption{\textbf{Coherence and temperature-dependent $T_1$ relaxation of TLS5.} 
    \textbf{a,} (top) Pulse sequence to measure the TLS Ramsey spectroscopy. The sequence consists of two TLS $\pi/2$-pulses separated by a variable delay, followed by a TLS-transmon SWAP gate and subsequent transmon readout. (bottom) Measured Ramsey curve for TLS5. Fitting the data to an oscillatory decaying curve yields $T_2^* = 0.91\pm0.05~\mu$s.
    Blue dot markers represent experimental data, and the blue solid line represents the fit to the data. The XY-control line driving power for the TLS is $17$~dB stronger than what is typically used for controlling the transmon qubit. This increased driving power allows for fast TLS control with $t_\pi = 160$~ns.
    \textbf{b,}
    Plot of the $Q$-factor of both TLS5 and the transmon qubit as a function of the mixing plate temperature. 
    The red (dark blue) markers denote measurements of the TLS (transmon qubit) during device warm-up (WU), and the yellow (light blue) markers during the device cooldown (CD). The gray dashed line corresponds to a phenomenological model of QP damping of TLS using the mixing plate temperature.
    The black solid line represents a correction to the gray dashed line, when using the effective temperature from \textbf{c}. The gray dash-dotted line is a fit to the transmon curve with a model including thermal saturation of weakly coupled TLS and damping from thermally-activated QPs.
    \textbf{c,}
    Plot of the effective temperature against the mixing plate temperature. The effective temperature is deduced by assuming a single TLS energy relaxation channel of QPs. The empirical fit assumes the functional form $T_\mathrm{eff}(T_\mathrm{MXC}) = A\sqrt{1+B\tanh(C/T_\mathrm{MXC})}/\tanh(C/T_\mathrm{MXC})$.
    }
    \label{fig4:TLS_coherence}
\end{figure*}

Using the Ramsey sequence illustrated in Fig.~\ref{fig4:TLS_coherence}a, we plot the excited state population of the transmon qubit mapped from the TLS as a function of the free precession time of the TLS in Fig.~\ref{fig4:TLS_coherence}b. Fitting the resulting pattern to an oscillatory decaying curve, $A\cos(\omega t +\phi_0)\exp[-(t/T_2^*)^n] + B$, yields $T_2^* = 0.91\pm0.05~\mu$s, with $n=1.8\pm 0.3$. The obtained TLS coherence time does not show a significant improvement over TLS without the engineered acoustic environment~\cite{Neeley08,Lisenfeld10}. This is perhaps to be expected, as the decoherence of TLS is thought to be dominated by the bath of thermally-activated TLS through an effective $ZZ$ interaction, and the properties of these low-frequency TLS, with frequencies $\hbar\omega\lesssim k_B T$, are not fundamentally altered by the presence of the microwave-frequency acoustic bandgap of the structures in this work. This model is consistent with the exponent of the TLS coherence decay: being close to quadratic it is associated with coherent-like, low-frequency noise on the TLS transition frequency. The extended $T_1$ time of the acoustically-shielded TLS, along with dynamical decoupling sequences, presents an opportunity to use the microwave-frequency TLS as a highly sensitive nanoscale sensor~\cite{Taminiau12,Degen17,Bylander11}, capable of revealing the structure of its low-frequency bath. A detailed investigation, highlighting the properties of low-frequency and thermally-activated TLS, will be the subject of a future work.

Finally, in Fig.~\ref{fig4:TLS_coherence}b we present the temperature-dependence of the $T_1$ relaxation rate of TLS5 as we warm up the mixing plate of the fridge from a base temperature of $7$~mK to $193$~mK, and cool down back to base temperature. No hysteresis is observed between the warm-up (red markers) and cooldown (yellow markers) paths. Here we plot the quality factor, defined as $Q=\omega T_1$, to compare the relaxation of both TLS5 at transition frequency $\omega_{\mathrm{TLS5}}/2\pi=5.65$~GHz and the transmon qubit at transition frequency $\omega_{\mathrm{q}}/2\pi=6.48$~GHz. The variation of the transmon $Q$-factor with temperature agrees with a widely adopted model that considers effects from thermal saturation of weakly-coupled resonant TLS and the thermal excitation of quasiparticles (QPs) in the superconducting Al layers of the transmon qubit~\cite{Catelani11,Crowley23,Gao08,SOM}. A fit of this model (see Sec.~II.G of the SI for details) to the transmon data is shown as a gray dash-dotted line. 

The $Q$-factor of the TLS, on the other hand, stays roughly constant at $Q\sim 2.5\times10^7$ for temperatures below $75$~mK, then drops by three orders of magnitude to $Q<5\times10^4$ at $193$~mK. For temperatures above $150$~mK, the drop in TLS $Q$-factor looks to follow that of the transmon qubit. Fitting a similar QP loss model to the TLS data in this region yields the dashed curved in Fig.~\ref{fig4:TLS_coherence}b. As can be seen, the plateau in the TLS $Q$-factor at the lowest temperatures is not captured by this simple model. Possible explanations for the limited TLS $Q$-factor below $75$~mK include temperature-independent phenomena such as TLS coupling to heavily damped grain-boundary motion in the polycrystalline Al films~\cite{Ke47,Wollack21}, or coupling to non-equilibrium QPs induced by high-energy particle events~\cite{Vepsalainen20}. Using the high temperature fit of the QP model to the TLS $Q$-factor data, and then fitting for an effective temperature $T_{\mathrm{eff}}$ for the QPs that tracks the measured TLS $Q$-factor data at lower mixing plate temperatures ($T_{\mathrm{MXC}}$), yields the temperature curve shown in Fig.~\ref{fig4:TLS_coherence}c. The corresponding fitting curve for the TLS $Q$-factor assuming $T_{\mathrm{eff}}(T_{\mathrm{MXC}})$ is shown as a solid black line in Fig.~\ref{fig4:TLS_coherence}b (the transmon fit using $T_{\mathrm{eff}}$ remains largely unaffected). If QPs were to explain the plateau in TLS $Q$-factor, this analysis predicts a QP saturation temperature of approximately $130$~mK. This is consistent with recent studies~\cite{Vepsalainen20,Gustavsson16,Serniak18} of non-equilibrium QP population in Al superconducting circuits, which infer a QP population with effective temperature of $120$-$150$~mK. Details of the QP loss model and further discussion of possible sources of thermally-activated TLS relaxation channels are given in Sec.~II.G and II.H of the SI. We emphasize that TLS-QP coupling is simply one possible explanation for the observed temperature-dependent TLS relaxation behavior. This behavior deviates from predictions of the standard tunneling model of TLS~\cite{Behunin16}, and reveals previously unexplored TLS physics that requires further investigation. 

\subsection{Outlook}
Despite the success of using TLS as a phenomenological model, in-depth knowledge of the nature and origin of TLS remains elusive in many specific situations~\cite{Muller19,deGraaf22}. The long-lived coherent TLS realized in this work through phonon engineering of the host material should be able to shed new light on previous TLS studies. These TLS can serve as nanoscale sensors, providing valuable information about the local environment of TLS~\cite{Faoro15,Meibner18}.
Of particular interest are low-frequency TLS that can be thermally activated. These TLS exhibit fluctuations due to interactions with thermal phonons, and have been linked to parameter fluctuations of SC qubits over extended timescales~\cite{Klimov18,Burnett19, Schlor19}, as well as $1/f$ noise in conductors~\cite{Dutta81}. In each of these cases, a better understanding of the TLS could potentially lead to strategies for their elimination through materials selection, growth, and processing~\cite{Siddiqi21,deLeon21,Muller19,Murray21}. Alternatively, long-lived TLS may find applications as qubits or quantum memories themselves~\cite{Zagoskin06,Neeley08}, with coherent control provided by superconducting devices such as the transmon qubit in this work. Finally, there is low-temperature flux noise found in SQUID loops, which limits the coherence of frequency-tunable SC qubits~\cite{Yoshihara06, Kakuyanagi07, Bialczak07, Rower23}. This noise is hypothesized to arise from surface defects in the vicinity of the SQUID loop, that carry spin degrees of freedom~\cite{Paladino14}. This notion draws interesting connections between the TLS of amorphous solids, and the color center defects of crystalline hosts such as diamond, which also experience a major source of noise from surface spins~\cite{deLeon21, Dwyer22}.
Similarly, phonon engineering could mitigate unwanted phonon damping for acoustically-active transitions of the defect center, such as found in the groundstate manifold of SiV~\cite{Klotz22}. 

\vspace{2mm}

\noindent\textbf{Acknowledgement}\\ 
We thank P. Cappellaro, J. Gao, L. Jiang, S. Lloyd and G. Rafael for helpful discussions, A. Butler, L. De Rose, V. Ferreira, E. Kim, G. Kim, S. Meesala, and X. Zhang for various help, B. Baker, M. McCoy, and B. Larrowe for experimental support, the MIT Lincoln Laboratories for the provision of Josephson traveling wave parametric amplifiers, and the AWS Center for Quantum Computing for help with the fridge setup.
This work is supported by Amazon Web Services, the Army Research Office and Laboratory for Physical Sciences (grant W911NF-18-1-0103), and the Moore Foundation (grant 7435).\\

\noindent\textbf{Author contributions}\\ 
O.P. and M.C. conceived the concept. M.C. designed and fabricated the device. M.C. performed the experiments and analyzed the data. J.C.O. and H.P. designed, fabricated and tested early versions of the devices studied here, consisting of merged-element qubits on a Si substrate. M.S. assisted in device simulation. M.C. wrote the manuscript with feedback from all authors. 
O.P. supervised the project.\\

\bibliography{Biblio}

\begin{thebibliography}{66}%
\makeatletter
\providecommand \@ifxundefined [1]{%
 \@ifx{#1\undefined}
}%
\providecommand \@ifnum [1]{%
 \ifnum #1\expandafter \@firstoftwo
 \else \expandafter \@secondoftwo
 \fi
}%
\providecommand \@ifx [1]{%
 \ifx #1\expandafter \@firstoftwo
 \else \expandafter \@secondoftwo
 \fi
}%
\providecommand \natexlab [1]{#1}%
\providecommand \enquote  [1]{``#1''}%
\providecommand \bibnamefont  [1]{#1}%
\providecommand \bibfnamefont [1]{#1}%
\providecommand \citenamefont [1]{#1}%
\providecommand \href@noop [0]{\@secondoftwo}%
\providecommand \href [0]{\begingroup \@sanitize@url \@href}%
\providecommand \@href[1]{\@@startlink{#1}\@@href}%
\providecommand \@@href[1]{\endgroup#1\@@endlink}%
\providecommand \@sanitize@url [0]{\catcode `\\12\catcode `\$12\catcode
  `\&12\catcode `\#12\catcode `\^12\catcode `\_12\catcode `\%12\relax}%
\providecommand \@@startlink[1]{}%
\providecommand \@@endlink[0]{}%
\providecommand \url  [0]{\begingroup\@sanitize@url \@url }%
\providecommand \@url [1]{\endgroup\@href {#1}{\urlprefix }}%
\providecommand \urlprefix  [0]{URL }%
\providecommand \Eprint [0]{\href }%
\providecommand \doibase [0]{https://doi.org/}%
\providecommand \selectlanguage [0]{\@gobble}%
\providecommand \bibinfo  [0]{\@secondoftwo}%
\providecommand \bibfield  [0]{\@secondoftwo}%
\providecommand \translation [1]{[#1]}%
\providecommand \BibitemOpen [0]{}%
\providecommand \bibitemStop [0]{}%
\providecommand \bibitemNoStop [0]{.\EOS\space}%
\providecommand \EOS [0]{\spacefactor3000\relax}%
\providecommand \BibitemShut  [1]{\csname bibitem#1\endcsname}%
\let\auto@bib@innerbib\@empty
\bibitem [{\citenamefont {Phillips}(1972)}]{Phillips72}%
  \BibitemOpen
  \bibfield  {author} {\bibinfo {author} {\bibfnamefont {W.~A.}\ \bibnamefont
  {Phillips}},\ }\bibfield  {title} {\bibinfo {title} {Tunneling states in
  amorphous solids},\ }\href {https://doi.org/10.1007/BF00660072} {\bibfield
  {journal} {\bibinfo  {journal} {Journal of Low Temperature Physics}\ }\textbf
  {\bibinfo {volume} {7}},\ \bibinfo {pages} {351} (\bibinfo {year}
  {1972})}\BibitemShut {NoStop}%
\bibitem [{\citenamefont {Anderson}\ \emph {et~al.}(1972)\citenamefont
  {Anderson}, \citenamefont {Halperin},\ and\ \citenamefont
  {Varma}}]{Anderson72}%
  \BibitemOpen
  \bibfield  {author} {\bibinfo {author} {\bibfnamefont {P.~W.}\ \bibnamefont
  {Anderson}}, \bibinfo {author} {\bibfnamefont {B.~I.}\ \bibnamefont
  {Halperin}},\ and\ \bibinfo {author} {\bibfnamefont {C.~M.}\ \bibnamefont
  {Varma}},\ }\bibfield  {title} {\bibinfo {title} {Anomalous low-temperature
  thermal properties of glasses and spin glasses},\ }\href
  {https://doi.org/10.1080/14786437208229210} {\bibfield  {journal} {\bibinfo
  {journal} {The Philosophical Magazine: A Journal of Theoretical Experimental
  and Applied Physics}\ }\textbf {\bibinfo {volume} {25}},\ \bibinfo {pages}
  {1} (\bibinfo {year} {1972})}\BibitemShut {NoStop}%
\bibitem [{\citenamefont {Arute}\ \emph {et~al.}(2019)\citenamefont {Arute},
  \citenamefont {Arya}, \citenamefont {Babbush}, \citenamefont {Bacon},
  \citenamefont {Bardin}, \citenamefont {Barends}, \citenamefont {Biswas},
  \citenamefont {Boixo}, \citenamefont {Brandao}, \citenamefont {Buell},
  \citenamefont {Burkett}, \citenamefont {Chen}, \citenamefont {Chen},
  \citenamefont {Chiaro}, \citenamefont {Collins}, \citenamefont {Courtney},
  \citenamefont {Dunsworth}, \citenamefont {Farhi}, \citenamefont {Foxen},
  \citenamefont {Fowler}, \citenamefont {Gidney}, \citenamefont {Giustina},
  \citenamefont {Graff}, \citenamefont {Guerin}, \citenamefont {Habegger},
  \citenamefont {Harrigan}, \citenamefont {Hartmann}, \citenamefont {Ho},
  \citenamefont {Hoffmann}, \citenamefont {Huang}, \citenamefont {Humble},
  \citenamefont {Isakov}, \citenamefont {Jeffrey}, \citenamefont {Jiang},
  \citenamefont {Kafri}, \citenamefont {Kechedzhi}, \citenamefont {Kelly},
  \citenamefont {Klimov}, \citenamefont {Knysh}, \citenamefont {Korotkov},
  \citenamefont {Kostritsa}, \citenamefont {Landhuis}, \citenamefont
  {Lindmark}, \citenamefont {Lucero}, \citenamefont {Lyakh}, \citenamefont
  {Mandr{\`a}}, \citenamefont {McClean}, \citenamefont {McEwen}, \citenamefont
  {Megrant}, \citenamefont {Mi}, \citenamefont {Michielsen}, \citenamefont
  {Mohseni}, \citenamefont {Mutus}, \citenamefont {Naaman}, \citenamefont
  {Neeley}, \citenamefont {Neill}, \citenamefont {Niu}, \citenamefont {Ostby},
  \citenamefont {Petukhov}, \citenamefont {Platt}, \citenamefont {Quintana},
  \citenamefont {Rieffel}, \citenamefont {Roushan}, \citenamefont {Rubin},
  \citenamefont {Sank}, \citenamefont {Satzinger}, \citenamefont {Smelyanskiy},
  \citenamefont {Sung}, \citenamefont {Trevithick}, \citenamefont
  {Vainsencher}, \citenamefont {Villalonga}, \citenamefont {White},
  \citenamefont {Yao}, \citenamefont {Yeh}, \citenamefont {Zalcman},
  \citenamefont {Neven},\ and\ \citenamefont {Martinis}}]{Arute19}%
  \BibitemOpen
  \bibfield  {author} {\bibinfo {author} {\bibfnamefont {F.}~\bibnamefont
  {Arute}}, \bibinfo {author} {\bibfnamefont {K.}~\bibnamefont {Arya}},
  \bibinfo {author} {\bibfnamefont {R.}~\bibnamefont {Babbush}}, \bibinfo
  {author} {\bibfnamefont {D.}~\bibnamefont {Bacon}}, \bibinfo {author}
  {\bibfnamefont {J.~C.}\ \bibnamefont {Bardin}}, \bibinfo {author}
  {\bibfnamefont {R.}~\bibnamefont {Barends}}, \bibinfo {author} {\bibfnamefont
  {R.}~\bibnamefont {Biswas}}, \bibinfo {author} {\bibfnamefont
  {S.}~\bibnamefont {Boixo}}, \bibinfo {author} {\bibfnamefont {F.~G. S.~L.}\
  \bibnamefont {Brandao}}, \bibinfo {author} {\bibfnamefont {D.~A.}\
  \bibnamefont {Buell}}, \bibinfo {author} {\bibfnamefont {B.}~\bibnamefont
  {Burkett}}, \bibinfo {author} {\bibfnamefont {Y.}~\bibnamefont {Chen}},
  \bibinfo {author} {\bibfnamefont {Z.}~\bibnamefont {Chen}}, \bibinfo {author}
  {\bibfnamefont {B.}~\bibnamefont {Chiaro}}, \bibinfo {author} {\bibfnamefont
  {R.}~\bibnamefont {Collins}}, \bibinfo {author} {\bibfnamefont
  {W.}~\bibnamefont {Courtney}}, \bibinfo {author} {\bibfnamefont
  {A.}~\bibnamefont {Dunsworth}}, \bibinfo {author} {\bibfnamefont
  {E.}~\bibnamefont {Farhi}}, \bibinfo {author} {\bibfnamefont
  {B.}~\bibnamefont {Foxen}}, \bibinfo {author} {\bibfnamefont
  {A.}~\bibnamefont {Fowler}}, \bibinfo {author} {\bibfnamefont
  {C.}~\bibnamefont {Gidney}}, \bibinfo {author} {\bibfnamefont
  {M.}~\bibnamefont {Giustina}}, \bibinfo {author} {\bibfnamefont
  {R.}~\bibnamefont {Graff}}, \bibinfo {author} {\bibfnamefont
  {K.}~\bibnamefont {Guerin}}, \bibinfo {author} {\bibfnamefont
  {S.}~\bibnamefont {Habegger}}, \bibinfo {author} {\bibfnamefont {M.~P.}\
  \bibnamefont {Harrigan}}, \bibinfo {author} {\bibfnamefont {M.~J.}\
  \bibnamefont {Hartmann}}, \bibinfo {author} {\bibfnamefont {A.}~\bibnamefont
  {Ho}}, \bibinfo {author} {\bibfnamefont {M.}~\bibnamefont {Hoffmann}},
  \bibinfo {author} {\bibfnamefont {T.}~\bibnamefont {Huang}}, \bibinfo
  {author} {\bibfnamefont {T.~S.}\ \bibnamefont {Humble}}, \bibinfo {author}
  {\bibfnamefont {S.~V.}\ \bibnamefont {Isakov}}, \bibinfo {author}
  {\bibfnamefont {E.}~\bibnamefont {Jeffrey}}, \bibinfo {author} {\bibfnamefont
  {Z.}~\bibnamefont {Jiang}}, \bibinfo {author} {\bibfnamefont
  {D.}~\bibnamefont {Kafri}}, \bibinfo {author} {\bibfnamefont
  {K.}~\bibnamefont {Kechedzhi}}, \bibinfo {author} {\bibfnamefont
  {J.}~\bibnamefont {Kelly}}, \bibinfo {author} {\bibfnamefont {P.~V.}\
  \bibnamefont {Klimov}}, \bibinfo {author} {\bibfnamefont {S.}~\bibnamefont
  {Knysh}}, \bibinfo {author} {\bibfnamefont {A.}~\bibnamefont {Korotkov}},
  \bibinfo {author} {\bibfnamefont {F.}~\bibnamefont {Kostritsa}}, \bibinfo
  {author} {\bibfnamefont {D.}~\bibnamefont {Landhuis}}, \bibinfo {author}
  {\bibfnamefont {M.}~\bibnamefont {Lindmark}}, \bibinfo {author}
  {\bibfnamefont {E.}~\bibnamefont {Lucero}}, \bibinfo {author} {\bibfnamefont
  {D.}~\bibnamefont {Lyakh}}, \bibinfo {author} {\bibfnamefont
  {S.}~\bibnamefont {Mandr{\`a}}}, \bibinfo {author} {\bibfnamefont {J.~R.}\
  \bibnamefont {McClean}}, \bibinfo {author} {\bibfnamefont {M.}~\bibnamefont
  {McEwen}}, \bibinfo {author} {\bibfnamefont {A.}~\bibnamefont {Megrant}},
  \bibinfo {author} {\bibfnamefont {X.}~\bibnamefont {Mi}}, \bibinfo {author}
  {\bibfnamefont {K.}~\bibnamefont {Michielsen}}, \bibinfo {author}
  {\bibfnamefont {M.}~\bibnamefont {Mohseni}}, \bibinfo {author} {\bibfnamefont
  {J.}~\bibnamefont {Mutus}}, \bibinfo {author} {\bibfnamefont
  {O.}~\bibnamefont {Naaman}}, \bibinfo {author} {\bibfnamefont
  {M.}~\bibnamefont {Neeley}}, \bibinfo {author} {\bibfnamefont
  {C.}~\bibnamefont {Neill}}, \bibinfo {author} {\bibfnamefont {M.~Y.}\
  \bibnamefont {Niu}}, \bibinfo {author} {\bibfnamefont {E.}~\bibnamefont
  {Ostby}}, \bibinfo {author} {\bibfnamefont {A.}~\bibnamefont {Petukhov}},
  \bibinfo {author} {\bibfnamefont {J.~C.}\ \bibnamefont {Platt}}, \bibinfo
  {author} {\bibfnamefont {C.}~\bibnamefont {Quintana}}, \bibinfo {author}
  {\bibfnamefont {E.~G.}\ \bibnamefont {Rieffel}}, \bibinfo {author}
  {\bibfnamefont {P.}~\bibnamefont {Roushan}}, \bibinfo {author} {\bibfnamefont
  {N.~C.}\ \bibnamefont {Rubin}}, \bibinfo {author} {\bibfnamefont
  {D.}~\bibnamefont {Sank}}, \bibinfo {author} {\bibfnamefont {K.~J.}\
  \bibnamefont {Satzinger}}, \bibinfo {author} {\bibfnamefont {V.}~\bibnamefont
  {Smelyanskiy}}, \bibinfo {author} {\bibfnamefont {K.~J.}\ \bibnamefont
  {Sung}}, \bibinfo {author} {\bibfnamefont {M.~D.}\ \bibnamefont
  {Trevithick}}, \bibinfo {author} {\bibfnamefont {A.}~\bibnamefont
  {Vainsencher}}, \bibinfo {author} {\bibfnamefont {B.}~\bibnamefont
  {Villalonga}}, \bibinfo {author} {\bibfnamefont {T.}~\bibnamefont {White}},
  \bibinfo {author} {\bibfnamefont {Z.~J.}\ \bibnamefont {Yao}}, \bibinfo
  {author} {\bibfnamefont {P.}~\bibnamefont {Yeh}}, \bibinfo {author}
  {\bibfnamefont {A.}~\bibnamefont {Zalcman}}, \bibinfo {author} {\bibfnamefont
  {H.}~\bibnamefont {Neven}},\ and\ \bibinfo {author} {\bibfnamefont {J.~M.}\
  \bibnamefont {Martinis}},\ }\bibfield  {title} {\bibinfo {title} {Quantum
  supremacy using a programmable superconducting processor},\ }\href
  {https://doi.org/10.1038/s41586-019-1666-5} {\bibfield  {journal} {\bibinfo
  {journal} {Nature}\ }\textbf {\bibinfo {volume} {574}},\ \bibinfo {pages}
  {505} (\bibinfo {year} {2019})}\BibitemShut {NoStop}%
\bibitem [{\citenamefont {Wu}\ \emph {et~al.}(2021)\citenamefont {Wu},
  \citenamefont {Bao}, \citenamefont {Cao}, \citenamefont {Chen}, \citenamefont
  {Chen}, \citenamefont {Chen}, \citenamefont {Chung}, \citenamefont {Deng},
  \citenamefont {Du}, \citenamefont {Fan}, \citenamefont {Gong}, \citenamefont
  {Guo}, \citenamefont {Guo}, \citenamefont {Guo}, \citenamefont {Han},
  \citenamefont {Hong}, \citenamefont {Huang}, \citenamefont {Huo},
  \citenamefont {Li}, \citenamefont {Li}, \citenamefont {Li}, \citenamefont
  {Li}, \citenamefont {Liang}, \citenamefont {Lin}, \citenamefont {Lin},
  \citenamefont {Qian}, \citenamefont {Qiao}, \citenamefont {Rong},
  \citenamefont {Su}, \citenamefont {Sun}, \citenamefont {Wang}, \citenamefont
  {Wang}, \citenamefont {Wu}, \citenamefont {Xu}, \citenamefont {Yan},
  \citenamefont {Yang}, \citenamefont {Yang}, \citenamefont {Ye}, \citenamefont
  {Yin}, \citenamefont {Ying}, \citenamefont {Yu}, \citenamefont {Zha},
  \citenamefont {Zhang}, \citenamefont {Zhang}, \citenamefont {Zhang},
  \citenamefont {Zhang}, \citenamefont {Zhao}, \citenamefont {Zhao},
  \citenamefont {Zhou}, \citenamefont {Zhu}, \citenamefont {Lu}, \citenamefont
  {Peng}, \citenamefont {Zhu},\ and\ \citenamefont {Pan}}]{Wu21}%
  \BibitemOpen
  \bibfield  {author} {\bibinfo {author} {\bibfnamefont {Y.}~\bibnamefont
  {Wu}}, \bibinfo {author} {\bibfnamefont {W.-S.}\ \bibnamefont {Bao}},
  \bibinfo {author} {\bibfnamefont {S.}~\bibnamefont {Cao}}, \bibinfo {author}
  {\bibfnamefont {F.}~\bibnamefont {Chen}}, \bibinfo {author} {\bibfnamefont
  {M.-C.}\ \bibnamefont {Chen}}, \bibinfo {author} {\bibfnamefont
  {X.}~\bibnamefont {Chen}}, \bibinfo {author} {\bibfnamefont {T.-H.}\
  \bibnamefont {Chung}}, \bibinfo {author} {\bibfnamefont {H.}~\bibnamefont
  {Deng}}, \bibinfo {author} {\bibfnamefont {Y.}~\bibnamefont {Du}}, \bibinfo
  {author} {\bibfnamefont {D.}~\bibnamefont {Fan}}, \bibinfo {author}
  {\bibfnamefont {M.}~\bibnamefont {Gong}}, \bibinfo {author} {\bibfnamefont
  {C.}~\bibnamefont {Guo}}, \bibinfo {author} {\bibfnamefont {C.}~\bibnamefont
  {Guo}}, \bibinfo {author} {\bibfnamefont {S.}~\bibnamefont {Guo}}, \bibinfo
  {author} {\bibfnamefont {L.}~\bibnamefont {Han}}, \bibinfo {author}
  {\bibfnamefont {L.}~\bibnamefont {Hong}}, \bibinfo {author} {\bibfnamefont
  {H.-L.}\ \bibnamefont {Huang}}, \bibinfo {author} {\bibfnamefont {Y.-H.}\
  \bibnamefont {Huo}}, \bibinfo {author} {\bibfnamefont {L.}~\bibnamefont
  {Li}}, \bibinfo {author} {\bibfnamefont {N.}~\bibnamefont {Li}}, \bibinfo
  {author} {\bibfnamefont {S.}~\bibnamefont {Li}}, \bibinfo {author}
  {\bibfnamefont {Y.}~\bibnamefont {Li}}, \bibinfo {author} {\bibfnamefont
  {F.}~\bibnamefont {Liang}}, \bibinfo {author} {\bibfnamefont
  {C.}~\bibnamefont {Lin}}, \bibinfo {author} {\bibfnamefont {J.}~\bibnamefont
  {Lin}}, \bibinfo {author} {\bibfnamefont {H.}~\bibnamefont {Qian}}, \bibinfo
  {author} {\bibfnamefont {D.}~\bibnamefont {Qiao}}, \bibinfo {author}
  {\bibfnamefont {H.}~\bibnamefont {Rong}}, \bibinfo {author} {\bibfnamefont
  {H.}~\bibnamefont {Su}}, \bibinfo {author} {\bibfnamefont {L.}~\bibnamefont
  {Sun}}, \bibinfo {author} {\bibfnamefont {L.}~\bibnamefont {Wang}}, \bibinfo
  {author} {\bibfnamefont {S.}~\bibnamefont {Wang}}, \bibinfo {author}
  {\bibfnamefont {D.}~\bibnamefont {Wu}}, \bibinfo {author} {\bibfnamefont
  {Y.}~\bibnamefont {Xu}}, \bibinfo {author} {\bibfnamefont {K.}~\bibnamefont
  {Yan}}, \bibinfo {author} {\bibfnamefont {W.}~\bibnamefont {Yang}}, \bibinfo
  {author} {\bibfnamefont {Y.}~\bibnamefont {Yang}}, \bibinfo {author}
  {\bibfnamefont {Y.}~\bibnamefont {Ye}}, \bibinfo {author} {\bibfnamefont
  {J.}~\bibnamefont {Yin}}, \bibinfo {author} {\bibfnamefont {C.}~\bibnamefont
  {Ying}}, \bibinfo {author} {\bibfnamefont {J.}~\bibnamefont {Yu}}, \bibinfo
  {author} {\bibfnamefont {C.}~\bibnamefont {Zha}}, \bibinfo {author}
  {\bibfnamefont {C.}~\bibnamefont {Zhang}}, \bibinfo {author} {\bibfnamefont
  {H.}~\bibnamefont {Zhang}}, \bibinfo {author} {\bibfnamefont
  {K.}~\bibnamefont {Zhang}}, \bibinfo {author} {\bibfnamefont
  {Y.}~\bibnamefont {Zhang}}, \bibinfo {author} {\bibfnamefont
  {H.}~\bibnamefont {Zhao}}, \bibinfo {author} {\bibfnamefont {Y.}~\bibnamefont
  {Zhao}}, \bibinfo {author} {\bibfnamefont {L.}~\bibnamefont {Zhou}}, \bibinfo
  {author} {\bibfnamefont {Q.}~\bibnamefont {Zhu}}, \bibinfo {author}
  {\bibfnamefont {C.-Y.}\ \bibnamefont {Lu}}, \bibinfo {author} {\bibfnamefont
  {C.-Z.}\ \bibnamefont {Peng}}, \bibinfo {author} {\bibfnamefont
  {X.}~\bibnamefont {Zhu}},\ and\ \bibinfo {author} {\bibfnamefont {J.-W.}\
  \bibnamefont {Pan}},\ }\bibfield  {title} {\bibinfo {title} {Strong quantum
  computational advantage using a superconducting quantum processor},\ }\href
  {https://doi.org/10.1103/PhysRevLett.127.180501} {\bibfield  {journal}
  {\bibinfo  {journal} {Phys. Rev. Lett.}\ }\textbf {\bibinfo {volume} {127}},\
  \bibinfo {pages} {180501} (\bibinfo {year} {2021})}\BibitemShut {NoStop}%
\bibitem [{\citenamefont {Kjaergaard}\ \emph {et~al.}(2020)\citenamefont
  {Kjaergaard}, \citenamefont {Schwartz}, \citenamefont {Braum\"{u}ller},
  \citenamefont {Krantz}, \citenamefont {Wang}, \citenamefont {Gustavsson},\
  and\ \citenamefont {Oliver}}]{Kjaergaard20}%
  \BibitemOpen
  \bibfield  {author} {\bibinfo {author} {\bibfnamefont {M.}~\bibnamefont
  {Kjaergaard}}, \bibinfo {author} {\bibfnamefont {M.~E.}\ \bibnamefont
  {Schwartz}}, \bibinfo {author} {\bibfnamefont {J.}~\bibnamefont
  {Braum\"{u}ller}}, \bibinfo {author} {\bibfnamefont {P.}~\bibnamefont
  {Krantz}}, \bibinfo {author} {\bibfnamefont {J.~I.-J.}\ \bibnamefont {Wang}},
  \bibinfo {author} {\bibfnamefont {S.}~\bibnamefont {Gustavsson}},\ and\
  \bibinfo {author} {\bibfnamefont {W.~D.}\ \bibnamefont {Oliver}},\ }\bibfield
   {title} {\bibinfo {title} {Superconducting qubits: Current state of play},\
  }\href {https://doi.org/10.1146/annurev-conmatphys-031119-050605} {\bibfield
  {journal} {\bibinfo  {journal} {Annual Review of Condensed Matter Physics}\
  }\textbf {\bibinfo {volume} {11}},\ \bibinfo {pages} {369} (\bibinfo {year}
  {2020})}\BibitemShut {NoStop}%
\bibitem [{\citenamefont {Krantz}\ \emph {et~al.}(2019)\citenamefont {Krantz},
  \citenamefont {Kjaergaard}, \citenamefont {Yan}, \citenamefont {Orlando},
  \citenamefont {Gustavsson},\ and\ \citenamefont {Oliver}}]{Krantz19}%
  \BibitemOpen
  \bibfield  {author} {\bibinfo {author} {\bibfnamefont {P.}~\bibnamefont
  {Krantz}}, \bibinfo {author} {\bibfnamefont {M.}~\bibnamefont {Kjaergaard}},
  \bibinfo {author} {\bibfnamefont {F.}~\bibnamefont {Yan}}, \bibinfo {author}
  {\bibfnamefont {T.~P.}\ \bibnamefont {Orlando}}, \bibinfo {author}
  {\bibfnamefont {S.}~\bibnamefont {Gustavsson}},\ and\ \bibinfo {author}
  {\bibfnamefont {W.~D.}\ \bibnamefont {Oliver}},\ }\bibfield  {title}
  {\bibinfo {title} {{A quantum engineer's guide to superconducting qubits}},\
  }\href {https://doi.org/10.1063/1.5089550} {\bibfield  {journal} {\bibinfo
  {journal} {Applied Physics Reviews}\ }\textbf {\bibinfo {volume} {6}},\
  \bibinfo {pages} {021318} (\bibinfo {year} {2019})}\BibitemShut {NoStop}%
\bibitem [{\citenamefont {Wang}\ \emph {et~al.}(2015)\citenamefont {Wang},
  \citenamefont {Axline}, \citenamefont {Gao}, \citenamefont {Brecht},
  \citenamefont {Chu}, \citenamefont {Frunzio}, \citenamefont {Devoret},\ and\
  \citenamefont {Schoelkopf}}]{Wang15a}%
  \BibitemOpen
  \bibfield  {author} {\bibinfo {author} {\bibfnamefont {C.}~\bibnamefont
  {Wang}}, \bibinfo {author} {\bibfnamefont {C.}~\bibnamefont {Axline}},
  \bibinfo {author} {\bibfnamefont {Y.~Y.}\ \bibnamefont {Gao}}, \bibinfo
  {author} {\bibfnamefont {T.}~\bibnamefont {Brecht}}, \bibinfo {author}
  {\bibfnamefont {Y.}~\bibnamefont {Chu}}, \bibinfo {author} {\bibfnamefont
  {L.}~\bibnamefont {Frunzio}}, \bibinfo {author} {\bibfnamefont {M.~H.}\
  \bibnamefont {Devoret}},\ and\ \bibinfo {author} {\bibfnamefont {R.~J.}\
  \bibnamefont {Schoelkopf}},\ }\bibfield  {title} {\bibinfo {title} {Surface
  participation and dielectric loss in superconducting qubits},\ }\href
  {https://doi.org/10.1063/1.4934486} {\bibfield  {journal} {\bibinfo
  {journal} {apl}\ }\textbf {\bibinfo {volume} {107}},\ \bibinfo {pages}
  {162601} (\bibinfo {year} {2015})}\BibitemShut {NoStop}%
\bibitem [{\citenamefont {Gambetta}\ \emph {et~al.}(2017)\citenamefont
  {Gambetta}, \citenamefont {Murray}, \citenamefont {Fung}, \citenamefont
  {McClure}, \citenamefont {Dial}, \citenamefont {Shanks}, \citenamefont
  {Sleight},\ and\ \citenamefont {Steffen}}]{Gambetta17}%
  \BibitemOpen
  \bibfield  {author} {\bibinfo {author} {\bibfnamefont {J.~M.}\ \bibnamefont
  {Gambetta}}, \bibinfo {author} {\bibfnamefont {C.~E.}\ \bibnamefont
  {Murray}}, \bibinfo {author} {\bibfnamefont {Y.-K.-K.}\ \bibnamefont {Fung}},
  \bibinfo {author} {\bibfnamefont {D.~T.}\ \bibnamefont {McClure}}, \bibinfo
  {author} {\bibfnamefont {O.}~\bibnamefont {Dial}}, \bibinfo {author}
  {\bibfnamefont {W.}~\bibnamefont {Shanks}}, \bibinfo {author} {\bibfnamefont
  {J.~W.}\ \bibnamefont {Sleight}},\ and\ \bibinfo {author} {\bibfnamefont
  {M.}~\bibnamefont {Steffen}},\ }\bibfield  {title} {\bibinfo {title}
  {Investigating surface loss effects in superconducting transmon qubits},\
  }\href {https://doi.org/10.1109/TASC.2016.2629670} {\bibfield  {journal}
  {\bibinfo  {journal} {IEEE Transactions on Applied Superconductivity}\
  }\textbf {\bibinfo {volume} {27}},\ \bibinfo {pages} {1} (\bibinfo {year}
  {2017})}\BibitemShut {NoStop}%
\bibitem [{\citenamefont {Paik}\ \emph {et~al.}(2011)\citenamefont {Paik},
  \citenamefont {Schuster}, \citenamefont {Bishop}, \citenamefont {Kirchmair},
  \citenamefont {Catelani}, \citenamefont {Sears}, \citenamefont {Johnson},
  \citenamefont {Reagor}, \citenamefont {Frunzio}, \citenamefont {Glazman},
  \citenamefont {Girvin}, \citenamefont {Devoret},\ and\ \citenamefont
  {Schoelkopf}}]{Paik11}%
  \BibitemOpen
  \bibfield  {author} {\bibinfo {author} {\bibfnamefont {H.}~\bibnamefont
  {Paik}}, \bibinfo {author} {\bibfnamefont {D.~I.}\ \bibnamefont {Schuster}},
  \bibinfo {author} {\bibfnamefont {L.~S.}\ \bibnamefont {Bishop}}, \bibinfo
  {author} {\bibfnamefont {G.}~\bibnamefont {Kirchmair}}, \bibinfo {author}
  {\bibfnamefont {G.}~\bibnamefont {Catelani}}, \bibinfo {author}
  {\bibfnamefont {A.~P.}\ \bibnamefont {Sears}}, \bibinfo {author}
  {\bibfnamefont {B.~R.}\ \bibnamefont {Johnson}}, \bibinfo {author}
  {\bibfnamefont {M.~J.}\ \bibnamefont {Reagor}}, \bibinfo {author}
  {\bibfnamefont {L.}~\bibnamefont {Frunzio}}, \bibinfo {author} {\bibfnamefont
  {L.~I.}\ \bibnamefont {Glazman}}, \bibinfo {author} {\bibfnamefont {S.~M.}\
  \bibnamefont {Girvin}}, \bibinfo {author} {\bibfnamefont {M.~H.}\
  \bibnamefont {Devoret}},\ and\ \bibinfo {author} {\bibfnamefont {R.~J.}\
  \bibnamefont {Schoelkopf}},\ }\bibfield  {title} {\bibinfo {title}
  {Observation of high coherence in josephson junction qubits measured in a
  three-dimensional circuit qed architecture},\ }\href
  {https://doi.org/10.1103/PhysRevLett.107.240501} {\bibfield  {journal}
  {\bibinfo  {journal} {Phys. Rev. Lett.}\ }\textbf {\bibinfo {volume} {107}},\
  \bibinfo {pages} {240501} (\bibinfo {year} {2011})}\BibitemShut {NoStop}%
\bibitem [{\citenamefont {Bruno}\ \emph {et~al.}(2015)\citenamefont {Bruno},
  \citenamefont {de~Lange}, \citenamefont {Asaad}, \citenamefont {van~der
  Enden}, \citenamefont {Langford},\ and\ \citenamefont {DiCarlo}}]{Bruno15}%
  \BibitemOpen
  \bibfield  {author} {\bibinfo {author} {\bibfnamefont {A.}~\bibnamefont
  {Bruno}}, \bibinfo {author} {\bibfnamefont {G.}~\bibnamefont {de~Lange}},
  \bibinfo {author} {\bibfnamefont {S.}~\bibnamefont {Asaad}}, \bibinfo
  {author} {\bibfnamefont {K.~L.}\ \bibnamefont {van~der Enden}}, \bibinfo
  {author} {\bibfnamefont {N.~K.}\ \bibnamefont {Langford}},\ and\ \bibinfo
  {author} {\bibfnamefont {L.}~\bibnamefont {DiCarlo}},\ }\bibfield  {title}
  {\bibinfo {title} {{Reducing intrinsic loss in superconducting resonators by
  surface treatment and deep etching of silicon substrates}},\ }\href
  {https://doi.org/10.1063/1.4919761} {\bibfield  {journal} {\bibinfo
  {journal} {Applied Physics Letters}\ }\textbf {\bibinfo {volume} {106}},\
  \bibinfo {pages} {182601} (\bibinfo {year} {2015})}\BibitemShut {NoStop}%
\bibitem [{\citenamefont {Barends}\ \emph {et~al.}(2013)\citenamefont
  {Barends}, \citenamefont {Kelly}, \citenamefont {Megrant}, \citenamefont
  {Sank}, \citenamefont {Jeffrey}, \citenamefont {Chen}, \citenamefont {Yin},
  \citenamefont {Chiaro}, \citenamefont {Mutus}, \citenamefont {Neill},
  \citenamefont {O'Malley}, \citenamefont {Roushan}, \citenamefont {Wenner},
  \citenamefont {White}, \citenamefont {Cleland},\ and\ \citenamefont
  {Martinis}}]{Barends13}%
  \BibitemOpen
  \bibfield  {author} {\bibinfo {author} {\bibfnamefont {R.}~\bibnamefont
  {Barends}}, \bibinfo {author} {\bibfnamefont {J.}~\bibnamefont {Kelly}},
  \bibinfo {author} {\bibfnamefont {A.}~\bibnamefont {Megrant}}, \bibinfo
  {author} {\bibfnamefont {D.}~\bibnamefont {Sank}}, \bibinfo {author}
  {\bibfnamefont {E.}~\bibnamefont {Jeffrey}}, \bibinfo {author} {\bibfnamefont
  {Y.}~\bibnamefont {Chen}}, \bibinfo {author} {\bibfnamefont {Y.}~\bibnamefont
  {Yin}}, \bibinfo {author} {\bibfnamefont {B.}~\bibnamefont {Chiaro}},
  \bibinfo {author} {\bibfnamefont {J.}~\bibnamefont {Mutus}}, \bibinfo
  {author} {\bibfnamefont {C.}~\bibnamefont {Neill}}, \bibinfo {author}
  {\bibfnamefont {P.}~\bibnamefont {O'Malley}}, \bibinfo {author}
  {\bibfnamefont {P.}~\bibnamefont {Roushan}}, \bibinfo {author} {\bibfnamefont
  {J.}~\bibnamefont {Wenner}}, \bibinfo {author} {\bibfnamefont {T.~C.}\
  \bibnamefont {White}}, \bibinfo {author} {\bibfnamefont {A.~N.}\ \bibnamefont
  {Cleland}},\ and\ \bibinfo {author} {\bibfnamefont {J.~M.}\ \bibnamefont
  {Martinis}},\ }\bibfield  {title} {\bibinfo {title} {Coherent josephson qubit
  suitable for scalable quantum integrated circuits},\ }\href
  {https://doi.org/10.1103/PhysRevLett.111.080502} {\bibfield  {journal}
  {\bibinfo  {journal} {Phys. Rev. Lett.}\ }\textbf {\bibinfo {volume} {111}},\
  \bibinfo {pages} {080502} (\bibinfo {year} {2013})}\BibitemShut {NoStop}%
\bibitem [{\citenamefont {Lisenfeld}\ \emph {et~al.}(2023)\citenamefont
  {Lisenfeld}, \citenamefont {Bilmes},\ and\ \citenamefont
  {Ustinov}}]{Lisenfeld23}%
  \BibitemOpen
  \bibfield  {author} {\bibinfo {author} {\bibfnamefont {J.}~\bibnamefont
  {Lisenfeld}}, \bibinfo {author} {\bibfnamefont {A.}~\bibnamefont {Bilmes}},\
  and\ \bibinfo {author} {\bibfnamefont {A.~V.}\ \bibnamefont {Ustinov}},\
  }\bibfield  {title} {\bibinfo {title} {Enhancing the coherence of
  superconducting quantum bits with electric fields},\ }\href
  {https://doi.org/10.1038/s41534-023-00678-9} {\bibfield  {journal} {\bibinfo
  {journal} {npj Quantum Information}\ }\textbf {\bibinfo {volume} {9}},\
  \bibinfo {pages} {8} (\bibinfo {year} {2023})}\BibitemShut {NoStop}%
\bibitem [{\citenamefont {Agarwal}\ \emph {et~al.}(2013)\citenamefont
  {Agarwal}, \citenamefont {Martin}, \citenamefont {Lukin},\ and\ \citenamefont
  {Demler}}]{Agarwal13}%
  \BibitemOpen
  \bibfield  {author} {\bibinfo {author} {\bibfnamefont {K.}~\bibnamefont
  {Agarwal}}, \bibinfo {author} {\bibfnamefont {I.}~\bibnamefont {Martin}},
  \bibinfo {author} {\bibfnamefont {M.~D.}\ \bibnamefont {Lukin}},\ and\
  \bibinfo {author} {\bibfnamefont {E.}~\bibnamefont {Demler}},\ }\bibfield
  {title} {\bibinfo {title} {Polaronic model of two-level systems in amorphous
  solids},\ }\href {https://doi.org/10.1103/PhysRevB.87.144201} {\bibfield
  {journal} {\bibinfo  {journal} {Phys. Rev. B}\ }\textbf {\bibinfo {volume}
  {87}},\ \bibinfo {pages} {144201} (\bibinfo {year} {2013})}\BibitemShut
  {NoStop}%
\bibitem [{\citenamefont {Behunin}\ \emph {et~al.}(2016)\citenamefont
  {Behunin}, \citenamefont {Intravaia},\ and\ \citenamefont
  {Rakich}}]{Behunin16}%
  \BibitemOpen
  \bibfield  {author} {\bibinfo {author} {\bibfnamefont {R.~O.}\ \bibnamefont
  {Behunin}}, \bibinfo {author} {\bibfnamefont {F.}~\bibnamefont {Intravaia}},\
  and\ \bibinfo {author} {\bibfnamefont {P.~T.}\ \bibnamefont {Rakich}},\
  }\bibfield  {title} {\bibinfo {title} {Dimensional transformation of
  defect-induced noise, dissipation, and nonlinearity},\ }\href
  {https://doi.org/10.1103/PhysRevB.93.224110} {\bibfield  {journal} {\bibinfo
  {journal} {Phys. Rev. B}\ }\textbf {\bibinfo {volume} {93}},\ \bibinfo
  {pages} {224110} (\bibinfo {year} {2016})}\BibitemShut {NoStop}%
\bibitem [{\citenamefont {Rosen}\ \emph {et~al.}(2019)\citenamefont {Rosen},
  \citenamefont {Horsley}, \citenamefont {Harrison}, \citenamefont {Holland},
  \citenamefont {Chang}, \citenamefont {Bond},\ and\ \citenamefont
  {DuBois}}]{Rosen19}%
  \BibitemOpen
  \bibfield  {author} {\bibinfo {author} {\bibfnamefont {Y.~J.}\ \bibnamefont
  {Rosen}}, \bibinfo {author} {\bibfnamefont {M.~A.}\ \bibnamefont {Horsley}},
  \bibinfo {author} {\bibfnamefont {S.~E.}\ \bibnamefont {Harrison}}, \bibinfo
  {author} {\bibfnamefont {E.~T.}\ \bibnamefont {Holland}}, \bibinfo {author}
  {\bibfnamefont {A.~S.}\ \bibnamefont {Chang}}, \bibinfo {author}
  {\bibfnamefont {T.}~\bibnamefont {Bond}},\ and\ \bibinfo {author}
  {\bibfnamefont {J.~L.}\ \bibnamefont {DuBois}},\ }\bibfield  {title}
  {\bibinfo {title} {{Protecting superconducting qubits from phonon mediated
  decay}},\ }\bibfield  {journal} {\bibinfo  {journal} {Applied Physics
  Letters}\ }\textbf {\bibinfo {volume} {114}},\ \href
  {https://doi.org/10.1063/1.5096182} {10.1063/1.5096182} (\bibinfo {year}
  {2019}),\ \bibinfo {note} {202601}\BibitemShut {NoStop}%
\bibitem [{\citenamefont {Chan}\ \emph {et~al.}(2012)\citenamefont {Chan},
  \citenamefont {Safavi-Naeini}, \citenamefont {Hill}, \citenamefont
  {Meenehan},\ and\ \citenamefont {Painter}}]{Chan12}%
  \BibitemOpen
  \bibfield  {author} {\bibinfo {author} {\bibfnamefont {J.}~\bibnamefont
  {Chan}}, \bibinfo {author} {\bibfnamefont {A.~H.}\ \bibnamefont
  {Safavi-Naeini}}, \bibinfo {author} {\bibfnamefont {J.~T.}\ \bibnamefont
  {Hill}}, \bibinfo {author} {\bibfnamefont {S.}~\bibnamefont {Meenehan}},\
  and\ \bibinfo {author} {\bibfnamefont {O.}~\bibnamefont {Painter}},\
  }\bibfield  {title} {\bibinfo {title} {{Optimized optomechanical crystal
  cavity with acoustic radiation shield}},\ }\bibfield  {journal} {\bibinfo
  {journal} {Applied Physics Letters}\ }\textbf {\bibinfo {volume} {101}},\
  \href {https://doi.org/10.1063/1.4747726} {10.1063/1.4747726} (\bibinfo
  {year} {2012}),\ \bibinfo {note} {081115}\BibitemShut {NoStop}%
\bibitem [{\citenamefont {MacCabe}\ \emph {et~al.}(2020)\citenamefont
  {MacCabe}, \citenamefont {Ren}, \citenamefont {Luo}, \citenamefont {Cohen},
  \citenamefont {Zhou}, \citenamefont {Sipahigil}, \citenamefont
  {Mirhosseini},\ and\ \citenamefont {Painter}}]{MacCabe20}%
  \BibitemOpen
  \bibfield  {author} {\bibinfo {author} {\bibfnamefont {G.~S.}\ \bibnamefont
  {MacCabe}}, \bibinfo {author} {\bibfnamefont {H.}~\bibnamefont {Ren}},
  \bibinfo {author} {\bibfnamefont {J.}~\bibnamefont {Luo}}, \bibinfo {author}
  {\bibfnamefont {J.~D.}\ \bibnamefont {Cohen}}, \bibinfo {author}
  {\bibfnamefont {H.}~\bibnamefont {Zhou}}, \bibinfo {author} {\bibfnamefont
  {A.}~\bibnamefont {Sipahigil}}, \bibinfo {author} {\bibfnamefont
  {M.}~\bibnamefont {Mirhosseini}},\ and\ \bibinfo {author} {\bibfnamefont
  {O.}~\bibnamefont {Painter}},\ }\bibfield  {title} {\bibinfo {title}
  {Nano-acoustic resonator with ultralong phonon lifetime},\ }\href
  {https://doi.org/10.1126/science.abc7312} {\bibfield  {journal} {\bibinfo
  {journal} {Science}\ }\textbf {\bibinfo {volume} {370}},\ \bibinfo {pages}
  {840} (\bibinfo {year} {2020})}\BibitemShut {NoStop}%
\bibitem [{\citenamefont {Keller}\ \emph {et~al.}(2017)\citenamefont {Keller},
  \citenamefont {Dieterle}, \citenamefont {Fang}, \citenamefont {Berger},
  \citenamefont {Fink},\ and\ \citenamefont {Painter}}]{Keller17}%
  \BibitemOpen
  \bibfield  {author} {\bibinfo {author} {\bibfnamefont {A.~J.}\ \bibnamefont
  {Keller}}, \bibinfo {author} {\bibfnamefont {P.~B.}\ \bibnamefont
  {Dieterle}}, \bibinfo {author} {\bibfnamefont {M.}~\bibnamefont {Fang}},
  \bibinfo {author} {\bibfnamefont {B.}~\bibnamefont {Berger}}, \bibinfo
  {author} {\bibfnamefont {J.~M.}\ \bibnamefont {Fink}},\ and\ \bibinfo
  {author} {\bibfnamefont {O.}~\bibnamefont {Painter}},\ }\bibfield  {title}
  {\bibinfo {title} {Al transmon qubits on silicon-on-insulator for quantum
  device integration},\ }\href {https://doi.org/10.1063/1.4994661} {\bibfield
  {journal} {\bibinfo  {journal} {apl}\ }\textbf {\bibinfo {volume} {111}},\
  \bibinfo {pages} {042603} (\bibinfo {year} {2017})}\BibitemShut {NoStop}%
\bibitem [{\citenamefont {M\"{u}ller}\ \emph {et~al.}(2019)\citenamefont
  {M\"{u}ller}, \citenamefont {Cole},\ and\ \citenamefont
  {Lisenfeld}}]{Muller19}%
  \BibitemOpen
  \bibfield  {author} {\bibinfo {author} {\bibfnamefont {C.}~\bibnamefont
  {M\"{u}ller}}, \bibinfo {author} {\bibfnamefont {J.~H.}\ \bibnamefont
  {Cole}},\ and\ \bibinfo {author} {\bibfnamefont {J.}~\bibnamefont
  {Lisenfeld}},\ }\bibfield  {title} {\bibinfo {title} {Towards understanding
  two-level-systems in amorphous solids: insights from quantum circuits},\
  }\href {https://doi.org/10.1088/1361-6633/ab3a7e} {\bibfield  {journal}
  {\bibinfo  {journal} {Reports on Progress in Physics}\ }\textbf {\bibinfo
  {volume} {82}},\ \bibinfo {pages} {124501} (\bibinfo {year}
  {2019})}\BibitemShut {NoStop}%
\bibitem [{\citenamefont {Gao}(2008)}]{Gao08}%
  \BibitemOpen
  \bibfield  {author} {\bibinfo {author} {\bibfnamefont {J.}~\bibnamefont
  {Gao}},\ }\emph {\bibinfo {title} {The Physics of Superconducting Microwave
  Resonators}},\ \href
  {https://thesis.library.caltech.edu/2530/1/thesismain_0610.pdf} {Ph.D.
  thesis},\ \bibinfo  {school} {California Institute of Technology} (\bibinfo
  {year} {2008})\BibitemShut {NoStop}%
\bibitem [{\citenamefont {Paladino}\ \emph {et~al.}(2014)\citenamefont
  {Paladino}, \citenamefont {Galperin}, \citenamefont {Falci},\ and\
  \citenamefont {Altshuler}}]{Paladino14}%
  \BibitemOpen
  \bibfield  {author} {\bibinfo {author} {\bibfnamefont {E.}~\bibnamefont
  {Paladino}}, \bibinfo {author} {\bibfnamefont {Y.~M.}\ \bibnamefont
  {Galperin}}, \bibinfo {author} {\bibfnamefont {G.}~\bibnamefont {Falci}},\
  and\ \bibinfo {author} {\bibfnamefont {B.~L.}\ \bibnamefont {Altshuler}},\
  }\bibfield  {title} {\bibinfo {title} {1/f noise: Implications for
  solid-state quantum information},\ }\href
  {https://doi.org/10.1103/RevModPhys.86.361} {\bibfield  {journal} {\bibinfo
  {journal} {Rev. Mod. Phys.}\ }\textbf {\bibinfo {volume} {86}},\ \bibinfo
  {pages} {361} (\bibinfo {year} {2014})}\BibitemShut {NoStop}%
\bibitem [{\citenamefont {Klimov}\ \emph {et~al.}(2018)\citenamefont {Klimov},
  \citenamefont {Kelly}, \citenamefont {Chen}, \citenamefont {Neeley},
  \citenamefont {Megrant}, \citenamefont {Burkett}, \citenamefont {Barends},
  \citenamefont {Arya}, \citenamefont {Chiaro}, \citenamefont {Chen},
  \citenamefont {Dunsworth}, \citenamefont {Fowler}, \citenamefont {Foxen},
  \citenamefont {Gidney}, \citenamefont {Giustina}, \citenamefont {Graff},
  \citenamefont {Huang}, \citenamefont {Jeffrey}, \citenamefont {Lucero},
  \citenamefont {Mutus}, \citenamefont {Naaman}, \citenamefont {Neill},
  \citenamefont {Quintana}, \citenamefont {Roushan}, \citenamefont {Sank},
  \citenamefont {Vainsencher}, \citenamefont {Wenner}, \citenamefont {White},
  \citenamefont {Boixo}, \citenamefont {Babbush}, \citenamefont {Smelyanskiy},
  \citenamefont {Neven},\ and\ \citenamefont {Martinis}}]{Klimov18}%
  \BibitemOpen
  \bibfield  {author} {\bibinfo {author} {\bibfnamefont {P.~V.}\ \bibnamefont
  {Klimov}}, \bibinfo {author} {\bibfnamefont {J.}~\bibnamefont {Kelly}},
  \bibinfo {author} {\bibfnamefont {Z.}~\bibnamefont {Chen}}, \bibinfo {author}
  {\bibfnamefont {M.}~\bibnamefont {Neeley}}, \bibinfo {author} {\bibfnamefont
  {A.}~\bibnamefont {Megrant}}, \bibinfo {author} {\bibfnamefont
  {B.}~\bibnamefont {Burkett}}, \bibinfo {author} {\bibfnamefont
  {R.}~\bibnamefont {Barends}}, \bibinfo {author} {\bibfnamefont
  {K.}~\bibnamefont {Arya}}, \bibinfo {author} {\bibfnamefont {B.}~\bibnamefont
  {Chiaro}}, \bibinfo {author} {\bibfnamefont {Y.}~\bibnamefont {Chen}},
  \bibinfo {author} {\bibfnamefont {A.}~\bibnamefont {Dunsworth}}, \bibinfo
  {author} {\bibfnamefont {A.}~\bibnamefont {Fowler}}, \bibinfo {author}
  {\bibfnamefont {B.}~\bibnamefont {Foxen}}, \bibinfo {author} {\bibfnamefont
  {C.}~\bibnamefont {Gidney}}, \bibinfo {author} {\bibfnamefont
  {M.}~\bibnamefont {Giustina}}, \bibinfo {author} {\bibfnamefont
  {R.}~\bibnamefont {Graff}}, \bibinfo {author} {\bibfnamefont
  {T.}~\bibnamefont {Huang}}, \bibinfo {author} {\bibfnamefont
  {E.}~\bibnamefont {Jeffrey}}, \bibinfo {author} {\bibfnamefont
  {E.}~\bibnamefont {Lucero}}, \bibinfo {author} {\bibfnamefont {J.~Y.}\
  \bibnamefont {Mutus}}, \bibinfo {author} {\bibfnamefont {O.}~\bibnamefont
  {Naaman}}, \bibinfo {author} {\bibfnamefont {C.}~\bibnamefont {Neill}},
  \bibinfo {author} {\bibfnamefont {C.}~\bibnamefont {Quintana}}, \bibinfo
  {author} {\bibfnamefont {P.}~\bibnamefont {Roushan}}, \bibinfo {author}
  {\bibfnamefont {D.}~\bibnamefont {Sank}}, \bibinfo {author} {\bibfnamefont
  {A.}~\bibnamefont {Vainsencher}}, \bibinfo {author} {\bibfnamefont
  {J.}~\bibnamefont {Wenner}}, \bibinfo {author} {\bibfnamefont {T.~C.}\
  \bibnamefont {White}}, \bibinfo {author} {\bibfnamefont {S.}~\bibnamefont
  {Boixo}}, \bibinfo {author} {\bibfnamefont {R.}~\bibnamefont {Babbush}},
  \bibinfo {author} {\bibfnamefont {V.~N.}\ \bibnamefont {Smelyanskiy}},
  \bibinfo {author} {\bibfnamefont {H.}~\bibnamefont {Neven}},\ and\ \bibinfo
  {author} {\bibfnamefont {J.~M.}\ \bibnamefont {Martinis}},\ }\bibfield
  {title} {\bibinfo {title} {Fluctuations of energy-relaxation times in
  superconducting qubits},\ }\href
  {https://doi.org/10.1103/PhysRevLett.121.090502} {\bibfield  {journal}
  {\bibinfo  {journal} {Phys. Rev. Lett.}\ }\textbf {\bibinfo {volume} {121}},\
  \bibinfo {pages} {090502} (\bibinfo {year} {2018})}\BibitemShut {NoStop}%
\bibitem [{\citenamefont {Aspelmeyer}\ and\ \citenamefont
  {Schwab}(2008)}]{Aspelmeyer08}%
  \BibitemOpen
  \bibfield  {author} {\bibinfo {author} {\bibfnamefont {M.}~\bibnamefont
  {Aspelmeyer}}\ and\ \bibinfo {author} {\bibfnamefont {K.}~\bibnamefont
  {Schwab}},\ }\bibfield  {title} {\bibinfo {title} {Focus on mechanical
  systems at the quantum limit},\ }\href
  {https://doi.org/10.1088/1367-2630/10/9/095001} {\bibfield  {journal}
  {\bibinfo  {journal} {New Journal of Physics}\ }\textbf {\bibinfo {volume}
  {10}},\ \bibinfo {pages} {095001} (\bibinfo {year} {2008})}\BibitemShut
  {NoStop}%
\bibitem [{\citenamefont {Wollack}\ \emph {et~al.}(2021)\citenamefont
  {Wollack}, \citenamefont {Cleland}, \citenamefont {Arrangoiz-Arriola},
  \citenamefont {McKenna}, \citenamefont {Gruenke}, \citenamefont {Patel},
  \citenamefont {Jiang}, \citenamefont {Sarabalis},\ and\ \citenamefont
  {Safavi-Naeini}}]{Wollack21}%
  \BibitemOpen
  \bibfield  {author} {\bibinfo {author} {\bibfnamefont {E.~A.}\ \bibnamefont
  {Wollack}}, \bibinfo {author} {\bibfnamefont {A.~Y.}\ \bibnamefont
  {Cleland}}, \bibinfo {author} {\bibfnamefont {P.}~\bibnamefont
  {Arrangoiz-Arriola}}, \bibinfo {author} {\bibfnamefont {T.~P.}\ \bibnamefont
  {McKenna}}, \bibinfo {author} {\bibfnamefont {R.~G.}\ \bibnamefont
  {Gruenke}}, \bibinfo {author} {\bibfnamefont {R.~N.}\ \bibnamefont {Patel}},
  \bibinfo {author} {\bibfnamefont {W.}~\bibnamefont {Jiang}}, \bibinfo
  {author} {\bibfnamefont {C.~J.}\ \bibnamefont {Sarabalis}},\ and\ \bibinfo
  {author} {\bibfnamefont {A.~H.}\ \bibnamefont {Safavi-Naeini}},\ }\bibfield
  {title} {\bibinfo {title} {{Loss channels affecting lithium niobate phononic
  crystal resonators at cryogenic temperature}},\ }\href
  {https://doi.org/10.1063/5.0034909} {\bibfield  {journal} {\bibinfo
  {journal} {Applied Physics Letters}\ }\textbf {\bibinfo {volume} {118}},\
  \bibinfo {pages} {123501} (\bibinfo {year} {2021})}\BibitemShut {NoStop}%
\bibitem [{\citenamefont {Cleland}\ \emph {et~al.}(2023)\citenamefont
  {Cleland}, \citenamefont {Wollack},\ and\ \citenamefont
  {Safavi-Naeini}}]{Cleland23}%
  \BibitemOpen
  \bibfield  {author} {\bibinfo {author} {\bibfnamefont {A.~Y.}\ \bibnamefont
  {Cleland}}, \bibinfo {author} {\bibfnamefont {E.~A.}\ \bibnamefont
  {Wollack}},\ and\ \bibinfo {author} {\bibfnamefont {A.~H.}\ \bibnamefont
  {Safavi-Naeini}},\ }\href@noop {} {\bibinfo {title} {Studying phonon
  coherence with a quantum sensor}} (\bibinfo {year} {2023}),\ \Eprint
  {https://arxiv.org/abs/2302.00221} {arXiv:2302.00221 [quant-ph]} \BibitemShut
  {NoStop}%
\bibitem [{\citenamefont {Rivi\`ere}\ \emph {et~al.}(2011)\citenamefont
  {Rivi\`ere}, \citenamefont {Del\'eglise}, \citenamefont {Weis}, \citenamefont
  {Gavartin}, \citenamefont {Arcizet}, \citenamefont {Schliesser},\ and\
  \citenamefont {Kippenberg}}]{Riviere11}%
  \BibitemOpen
  \bibfield  {author} {\bibinfo {author} {\bibfnamefont {R.}~\bibnamefont
  {Rivi\`ere}}, \bibinfo {author} {\bibfnamefont {S.}~\bibnamefont
  {Del\'eglise}}, \bibinfo {author} {\bibfnamefont {S.}~\bibnamefont {Weis}},
  \bibinfo {author} {\bibfnamefont {E.}~\bibnamefont {Gavartin}}, \bibinfo
  {author} {\bibfnamefont {O.}~\bibnamefont {Arcizet}}, \bibinfo {author}
  {\bibfnamefont {A.}~\bibnamefont {Schliesser}},\ and\ \bibinfo {author}
  {\bibfnamefont {T.~J.}\ \bibnamefont {Kippenberg}},\ }\bibfield  {title}
  {\bibinfo {title} {Optomechanical sideband cooling of a micromechanical
  oscillator close to the quantum ground state},\ }\href
  {https://doi.org/10.1103/PhysRevA.83.063835} {\bibfield  {journal} {\bibinfo
  {journal} {Phys. Rev. A}\ }\textbf {\bibinfo {volume} {83}},\ \bibinfo
  {pages} {063835} (\bibinfo {year} {2011})}\BibitemShut {NoStop}%
\bibitem [{\citenamefont {Burnett}\ \emph {et~al.}(2019)\citenamefont
  {Burnett}, \citenamefont {Bengtsson}, \citenamefont {Scigliuzzo},
  \citenamefont {Niepce}, \citenamefont {Kudra}, \citenamefont {Delsing},\ and\
  \citenamefont {Bylander}}]{Burnett19}%
  \BibitemOpen
  \bibfield  {author} {\bibinfo {author} {\bibfnamefont {J.~J.}\ \bibnamefont
  {Burnett}}, \bibinfo {author} {\bibfnamefont {A.}~\bibnamefont {Bengtsson}},
  \bibinfo {author} {\bibfnamefont {M.}~\bibnamefont {Scigliuzzo}}, \bibinfo
  {author} {\bibfnamefont {D.}~\bibnamefont {Niepce}}, \bibinfo {author}
  {\bibfnamefont {M.}~\bibnamefont {Kudra}}, \bibinfo {author} {\bibfnamefont
  {P.}~\bibnamefont {Delsing}},\ and\ \bibinfo {author} {\bibfnamefont
  {J.}~\bibnamefont {Bylander}},\ }\bibfield  {title} {\bibinfo {title}
  {Decoherence benchmarking of superconducting qubits},\ }\href
  {https://doi.org/10.1038/s41534-019-0168-5} {\bibfield  {journal} {\bibinfo
  {journal} {npj Quantum Information}\ }\textbf {\bibinfo {volume} {5}},\
  \bibinfo {pages} {54} (\bibinfo {year} {2019})}\BibitemShut {NoStop}%
\bibitem [{\citenamefont {Schl\"or}\ \emph {et~al.}(2019)\citenamefont
  {Schl\"or}, \citenamefont {Lisenfeld}, \citenamefont {M\"uller},
  \citenamefont {Bilmes}, \citenamefont {Schneider}, \citenamefont {Pappas},
  \citenamefont {Ustinov},\ and\ \citenamefont {Weides}}]{Schlor19}%
  \BibitemOpen
  \bibfield  {author} {\bibinfo {author} {\bibfnamefont {S.}~\bibnamefont
  {Schl\"or}}, \bibinfo {author} {\bibfnamefont {J.}~\bibnamefont {Lisenfeld}},
  \bibinfo {author} {\bibfnamefont {C.}~\bibnamefont {M\"uller}}, \bibinfo
  {author} {\bibfnamefont {A.}~\bibnamefont {Bilmes}}, \bibinfo {author}
  {\bibfnamefont {A.}~\bibnamefont {Schneider}}, \bibinfo {author}
  {\bibfnamefont {D.~P.}\ \bibnamefont {Pappas}}, \bibinfo {author}
  {\bibfnamefont {A.~V.}\ \bibnamefont {Ustinov}},\ and\ \bibinfo {author}
  {\bibfnamefont {M.}~\bibnamefont {Weides}},\ }\bibfield  {title} {\bibinfo
  {title} {Correlating decoherence in transmon qubits: Low frequency noise by
  single fluctuators},\ }\href {https://doi.org/10.1103/PhysRevLett.123.190502}
  {\bibfield  {journal} {\bibinfo  {journal} {Phys. Rev. Lett.}\ }\textbf
  {\bibinfo {volume} {123}},\ \bibinfo {pages} {190502} (\bibinfo {year}
  {2019})}\BibitemShut {NoStop}%
\bibitem [{\citenamefont {M\"uller}\ \emph {et~al.}(2009)\citenamefont
  {M\"uller}, \citenamefont {Shnirman},\ and\ \citenamefont
  {Makhlin}}]{Muller09}%
  \BibitemOpen
  \bibfield  {author} {\bibinfo {author} {\bibfnamefont {C.}~\bibnamefont
  {M\"uller}}, \bibinfo {author} {\bibfnamefont {A.}~\bibnamefont {Shnirman}},\
  and\ \bibinfo {author} {\bibfnamefont {Y.}~\bibnamefont {Makhlin}},\
  }\bibfield  {title} {\bibinfo {title} {Relaxation of josephson qubits due to
  strong coupling to two-level systems},\ }\href
  {https://doi.org/10.1103/PhysRevB.80.134517} {\bibfield  {journal} {\bibinfo
  {journal} {Phys. Rev. B}\ }\textbf {\bibinfo {volume} {80}},\ \bibinfo
  {pages} {134517} (\bibinfo {year} {2009})}\BibitemShut {NoStop}%
\bibitem [{\citenamefont {Lisenfeld}\ \emph {et~al.}(2019)\citenamefont
  {Lisenfeld}, \citenamefont {Bilmes}, \citenamefont {Megrant}, \citenamefont
  {Barends}, \citenamefont {Kelly}, \citenamefont {Klimov}, \citenamefont
  {Weiss}, \citenamefont {Martinis},\ and\ \citenamefont
  {Ustinov}}]{Lisenfeld19}%
  \BibitemOpen
  \bibfield  {author} {\bibinfo {author} {\bibfnamefont {J.}~\bibnamefont
  {Lisenfeld}}, \bibinfo {author} {\bibfnamefont {A.}~\bibnamefont {Bilmes}},
  \bibinfo {author} {\bibfnamefont {A.}~\bibnamefont {Megrant}}, \bibinfo
  {author} {\bibfnamefont {R.}~\bibnamefont {Barends}}, \bibinfo {author}
  {\bibfnamefont {J.}~\bibnamefont {Kelly}}, \bibinfo {author} {\bibfnamefont
  {P.}~\bibnamefont {Klimov}}, \bibinfo {author} {\bibfnamefont
  {G.}~\bibnamefont {Weiss}}, \bibinfo {author} {\bibfnamefont {J.~M.}\
  \bibnamefont {Martinis}},\ and\ \bibinfo {author} {\bibfnamefont {A.~V.}\
  \bibnamefont {Ustinov}},\ }\bibfield  {title} {\bibinfo {title} {Electric
  field spectroscopy of material defects in transmon qubits},\ }\href
  {https://doi.org/10.1038/s41534-019-0224-1} {\bibfield  {journal} {\bibinfo
  {journal} {npj Quantum Information}\ }\textbf {\bibinfo {volume} {5}},\
  \bibinfo {pages} {105} (\bibinfo {year} {2019})}\BibitemShut {NoStop}%
\bibitem [{\citenamefont {Spiecker}\ \emph {et~al.}(2023)\citenamefont
  {Spiecker}, \citenamefont {Paluch}, \citenamefont {Gosling}, \citenamefont
  {Drucker}, \citenamefont {Matityahu}, \citenamefont {Gusenkova},
  \citenamefont {G{\"u}nzler}, \citenamefont {Rieger}, \citenamefont
  {Takmakov}, \citenamefont {Valenti}, \citenamefont {Winkel}, \citenamefont
  {Gebauer}, \citenamefont {Sander}, \citenamefont {Catelani}, \citenamefont
  {Shnirman}, \citenamefont {Ustinov}, \citenamefont {Wernsdorfer},
  \citenamefont {Cohen},\ and\ \citenamefont {Pop}}]{Spiecker23}%
  \BibitemOpen
  \bibfield  {author} {\bibinfo {author} {\bibfnamefont {M.}~\bibnamefont
  {Spiecker}}, \bibinfo {author} {\bibfnamefont {P.}~\bibnamefont {Paluch}},
  \bibinfo {author} {\bibfnamefont {N.}~\bibnamefont {Gosling}}, \bibinfo
  {author} {\bibfnamefont {N.}~\bibnamefont {Drucker}}, \bibinfo {author}
  {\bibfnamefont {S.}~\bibnamefont {Matityahu}}, \bibinfo {author}
  {\bibfnamefont {D.}~\bibnamefont {Gusenkova}}, \bibinfo {author}
  {\bibfnamefont {S.}~\bibnamefont {G{\"u}nzler}}, \bibinfo {author}
  {\bibfnamefont {D.}~\bibnamefont {Rieger}}, \bibinfo {author} {\bibfnamefont
  {I.}~\bibnamefont {Takmakov}}, \bibinfo {author} {\bibfnamefont
  {F.}~\bibnamefont {Valenti}}, \bibinfo {author} {\bibfnamefont
  {P.}~\bibnamefont {Winkel}}, \bibinfo {author} {\bibfnamefont
  {R.}~\bibnamefont {Gebauer}}, \bibinfo {author} {\bibfnamefont
  {O.}~\bibnamefont {Sander}}, \bibinfo {author} {\bibfnamefont
  {G.}~\bibnamefont {Catelani}}, \bibinfo {author} {\bibfnamefont
  {A.}~\bibnamefont {Shnirman}}, \bibinfo {author} {\bibfnamefont {A.~V.}\
  \bibnamefont {Ustinov}}, \bibinfo {author} {\bibfnamefont {W.}~\bibnamefont
  {Wernsdorfer}}, \bibinfo {author} {\bibfnamefont {Y.}~\bibnamefont {Cohen}},\
  and\ \bibinfo {author} {\bibfnamefont {I.~M.}\ \bibnamefont {Pop}},\
  }\bibfield  {title} {\bibinfo {title} {Two-level system hyperpolarization
  using a quantum szilard engine},\ }\href
  {https://doi.org/10.1038/s41567-023-02082-8} {\bibfield  {journal} {\bibinfo
  {journal} {Nature Physics}\ }\textbf {\bibinfo {volume} {19}},\ \bibinfo
  {pages} {1320} (\bibinfo {year} {2023})}\BibitemShut {NoStop}%
\bibitem [{\citenamefont {Oh}\ \emph {et~al.}(2006)\citenamefont {Oh},
  \citenamefont {Cicak}, \citenamefont {Kline}, \citenamefont {Sillanp\"a\"a},
  \citenamefont {Osborn}, \citenamefont {Whittaker}, \citenamefont {Simmonds},\
  and\ \citenamefont {Pappas}}]{Oh06}%
  \BibitemOpen
  \bibfield  {author} {\bibinfo {author} {\bibfnamefont {S.}~\bibnamefont
  {Oh}}, \bibinfo {author} {\bibfnamefont {K.}~\bibnamefont {Cicak}}, \bibinfo
  {author} {\bibfnamefont {J.~S.}\ \bibnamefont {Kline}}, \bibinfo {author}
  {\bibfnamefont {M.~A.}\ \bibnamefont {Sillanp\"a\"a}}, \bibinfo {author}
  {\bibfnamefont {K.~D.}\ \bibnamefont {Osborn}}, \bibinfo {author}
  {\bibfnamefont {J.~D.}\ \bibnamefont {Whittaker}}, \bibinfo {author}
  {\bibfnamefont {R.~W.}\ \bibnamefont {Simmonds}},\ and\ \bibinfo {author}
  {\bibfnamefont {D.~P.}\ \bibnamefont {Pappas}},\ }\bibfield  {title}
  {\bibinfo {title} {Elimination of two level fluctuators in superconducting
  quantum bits by an epitaxial tunnel barrier},\ }\href
  {https://doi.org/10.1103/PhysRevB.74.100502} {\bibfield  {journal} {\bibinfo
  {journal} {Phys. Rev. B}\ }\textbf {\bibinfo {volume} {74}},\ \bibinfo
  {pages} {100502} (\bibinfo {year} {2006})}\BibitemShut {NoStop}%
\bibitem [{\citenamefont {Chang}\ \emph {et~al.}(2013)\citenamefont {Chang},
  \citenamefont {Vissers}, \citenamefont {C\'{o}rcoles}, \citenamefont
  {Sandberg}, \citenamefont {Gao}, \citenamefont {Abraham}, \citenamefont
  {Chow}, \citenamefont {Gambetta}, \citenamefont {Beth~Rothwell},
  \citenamefont {Keefe}, \citenamefont {Steffen},\ and\ \citenamefont
  {Pappas}}]{Chang13}%
  \BibitemOpen
  \bibfield  {author} {\bibinfo {author} {\bibfnamefont {J.~B.}\ \bibnamefont
  {Chang}}, \bibinfo {author} {\bibfnamefont {M.~R.}\ \bibnamefont {Vissers}},
  \bibinfo {author} {\bibfnamefont {A.~D.}\ \bibnamefont {C\'{o}rcoles}},
  \bibinfo {author} {\bibfnamefont {M.}~\bibnamefont {Sandberg}}, \bibinfo
  {author} {\bibfnamefont {J.}~\bibnamefont {Gao}}, \bibinfo {author}
  {\bibfnamefont {D.~W.}\ \bibnamefont {Abraham}}, \bibinfo {author}
  {\bibfnamefont {J.~M.}\ \bibnamefont {Chow}}, \bibinfo {author}
  {\bibfnamefont {J.~M.}\ \bibnamefont {Gambetta}}, \bibinfo {author}
  {\bibfnamefont {M.}~\bibnamefont {Beth~Rothwell}}, \bibinfo {author}
  {\bibfnamefont {G.~A.}\ \bibnamefont {Keefe}}, \bibinfo {author}
  {\bibfnamefont {M.}~\bibnamefont {Steffen}},\ and\ \bibinfo {author}
  {\bibfnamefont {D.~P.}\ \bibnamefont {Pappas}},\ }\bibfield  {title}
  {\bibinfo {title} {Improved superconducting qubit coherence using titanium
  nitride},\ }\bibfield  {journal} {\bibinfo  {journal} {Applied Physics
  Letters}\ }\textbf {\bibinfo {volume} {103}},\ \href
  {https://doi.org/10.1063/1.4813269} {10.1063/1.4813269} (\bibinfo {year}
  {2013}),\ \bibinfo {note} {012602}\BibitemShut {NoStop}%
\bibitem [{\citenamefont {Place}\ \emph {et~al.}(2021)\citenamefont {Place},
  \citenamefont {Rodgers}, \citenamefont {Mundada}, \citenamefont {Smitham},
  \citenamefont {Fitzpatrick}, \citenamefont {Leng}, \citenamefont {Premkumar},
  \citenamefont {Bryon}, \citenamefont {Vrajitoarea}, \citenamefont {Sussman},
  \citenamefont {Cheng}, \citenamefont {Madhavan}, \citenamefont {Babla},
  \citenamefont {Le}, \citenamefont {Gang}, \citenamefont {J{\"a}ck},
  \citenamefont {Gyenis}, \citenamefont {Yao}, \citenamefont {Cava},
  \citenamefont {de~Leon},\ and\ \citenamefont {Houck}}]{Place21}%
  \BibitemOpen
  \bibfield  {author} {\bibinfo {author} {\bibfnamefont {A.~P.~M.}\
  \bibnamefont {Place}}, \bibinfo {author} {\bibfnamefont {L.~V.~H.}\
  \bibnamefont {Rodgers}}, \bibinfo {author} {\bibfnamefont {P.}~\bibnamefont
  {Mundada}}, \bibinfo {author} {\bibfnamefont {B.~M.}\ \bibnamefont
  {Smitham}}, \bibinfo {author} {\bibfnamefont {M.}~\bibnamefont
  {Fitzpatrick}}, \bibinfo {author} {\bibfnamefont {Z.}~\bibnamefont {Leng}},
  \bibinfo {author} {\bibfnamefont {A.}~\bibnamefont {Premkumar}}, \bibinfo
  {author} {\bibfnamefont {J.}~\bibnamefont {Bryon}}, \bibinfo {author}
  {\bibfnamefont {A.}~\bibnamefont {Vrajitoarea}}, \bibinfo {author}
  {\bibfnamefont {S.}~\bibnamefont {Sussman}}, \bibinfo {author} {\bibfnamefont
  {G.}~\bibnamefont {Cheng}}, \bibinfo {author} {\bibfnamefont
  {T.}~\bibnamefont {Madhavan}}, \bibinfo {author} {\bibfnamefont {H.~K.}\
  \bibnamefont {Babla}}, \bibinfo {author} {\bibfnamefont {X.~H.}\ \bibnamefont
  {Le}}, \bibinfo {author} {\bibfnamefont {Y.}~\bibnamefont {Gang}}, \bibinfo
  {author} {\bibfnamefont {B.}~\bibnamefont {J{\"a}ck}}, \bibinfo {author}
  {\bibfnamefont {A.}~\bibnamefont {Gyenis}}, \bibinfo {author} {\bibfnamefont
  {N.}~\bibnamefont {Yao}}, \bibinfo {author} {\bibfnamefont {R.~J.}\
  \bibnamefont {Cava}}, \bibinfo {author} {\bibfnamefont {N.~P.}\ \bibnamefont
  {de~Leon}},\ and\ \bibinfo {author} {\bibfnamefont {A.~A.}\ \bibnamefont
  {Houck}},\ }\bibfield  {title} {\bibinfo {title} {New material platform for
  superconducting transmon qubits with coherence times exceeding 0.3
  milliseconds},\ }\href {https://doi.org/10.1038/s41467-021-22030-5}
  {\bibfield  {journal} {\bibinfo  {journal} {Nature Communications}\ }\textbf
  {\bibinfo {volume} {12}},\ \bibinfo {pages} {1779} (\bibinfo {year}
  {2021})}\BibitemShut {NoStop}%
\bibitem [{\citenamefont {Koch}\ \emph {et~al.}(2007)\citenamefont {Koch},
  \citenamefont {Yu}, \citenamefont {Gambetta}, \citenamefont {Houck},
  \citenamefont {Schuster}, \citenamefont {Majer}, \citenamefont {Blais},
  \citenamefont {Devoret}, \citenamefont {Girvin},\ and\ \citenamefont
  {Schoelkopf}}]{Koch07}%
  \BibitemOpen
  \bibfield  {author} {\bibinfo {author} {\bibfnamefont {J.}~\bibnamefont
  {Koch}}, \bibinfo {author} {\bibfnamefont {T.~M.}\ \bibnamefont {Yu}},
  \bibinfo {author} {\bibfnamefont {J.}~\bibnamefont {Gambetta}}, \bibinfo
  {author} {\bibfnamefont {A.~A.}\ \bibnamefont {Houck}}, \bibinfo {author}
  {\bibfnamefont {D.~I.}\ \bibnamefont {Schuster}}, \bibinfo {author}
  {\bibfnamefont {J.}~\bibnamefont {Majer}}, \bibinfo {author} {\bibfnamefont
  {A.}~\bibnamefont {Blais}}, \bibinfo {author} {\bibfnamefont {M.~H.}\
  \bibnamefont {Devoret}}, \bibinfo {author} {\bibfnamefont {S.~M.}\
  \bibnamefont {Girvin}},\ and\ \bibinfo {author} {\bibfnamefont {R.~J.}\
  \bibnamefont {Schoelkopf}},\ }\bibfield  {title} {\bibinfo {title}
  {Charge-insensitive qubit design derived from the cooper pair box},\ }\href
  {https://doi.org/10.1103/PhysRevA.76.042319} {\bibfield  {journal} {\bibinfo
  {journal} {Phys. Rev. A}\ }\textbf {\bibinfo {volume} {76}},\ \bibinfo
  {pages} {042319} (\bibinfo {year} {2007})}\BibitemShut {NoStop}%
\bibitem [{\citenamefont {Schreier}\ \emph {et~al.}(2008)\citenamefont
  {Schreier}, \citenamefont {Houck}, \citenamefont {Koch}, \citenamefont
  {Schuster}, \citenamefont {Johnson}, \citenamefont {Chow}, \citenamefont
  {Gambetta}, \citenamefont {Majer}, \citenamefont {Frunzio}, \citenamefont
  {Devoret}, \citenamefont {Girvin},\ and\ \citenamefont
  {Schoelkopf}}]{Schreier08}%
  \BibitemOpen
  \bibfield  {author} {\bibinfo {author} {\bibfnamefont {J.~A.}\ \bibnamefont
  {Schreier}}, \bibinfo {author} {\bibfnamefont {A.~A.}\ \bibnamefont {Houck}},
  \bibinfo {author} {\bibfnamefont {J.}~\bibnamefont {Koch}}, \bibinfo {author}
  {\bibfnamefont {D.~I.}\ \bibnamefont {Schuster}}, \bibinfo {author}
  {\bibfnamefont {B.~R.}\ \bibnamefont {Johnson}}, \bibinfo {author}
  {\bibfnamefont {J.~M.}\ \bibnamefont {Chow}}, \bibinfo {author}
  {\bibfnamefont {J.~M.}\ \bibnamefont {Gambetta}}, \bibinfo {author}
  {\bibfnamefont {J.}~\bibnamefont {Majer}}, \bibinfo {author} {\bibfnamefont
  {L.}~\bibnamefont {Frunzio}}, \bibinfo {author} {\bibfnamefont {M.~H.}\
  \bibnamefont {Devoret}}, \bibinfo {author} {\bibfnamefont {S.~M.}\
  \bibnamefont {Girvin}},\ and\ \bibinfo {author} {\bibfnamefont {R.~J.}\
  \bibnamefont {Schoelkopf}},\ }\bibfield  {title} {\bibinfo {title}
  {Suppressing charge noise decoherence in superconducting charge qubits},\
  }\href {https://doi.org/10.1103/PhysRevB.77.180502} {\bibfield  {journal}
  {\bibinfo  {journal} {Phys. Rev. B}\ }\textbf {\bibinfo {volume} {77}},\
  \bibinfo {pages} {180502} (\bibinfo {year} {2008})}\BibitemShut {NoStop}%
\bibitem [{SOM()}]{SOM}%
  \BibitemOpen
  \href@noop {} {}\bibinfo {howpublished} {See supplementary online
  material.}\BibitemShut {Stop}%
\bibitem [{\citenamefont {Zhao}\ \emph {et~al.}(2020)\citenamefont {Zhao},
  \citenamefont {Park}, \citenamefont {Zhao}, \citenamefont {Bal},
  \citenamefont {McRae}, \citenamefont {Long},\ and\ \citenamefont
  {Pappas}}]{Zhao20}%
  \BibitemOpen
  \bibfield  {author} {\bibinfo {author} {\bibfnamefont {R.}~\bibnamefont
  {Zhao}}, \bibinfo {author} {\bibfnamefont {S.}~\bibnamefont {Park}}, \bibinfo
  {author} {\bibfnamefont {T.}~\bibnamefont {Zhao}}, \bibinfo {author}
  {\bibfnamefont {M.}~\bibnamefont {Bal}}, \bibinfo {author} {\bibfnamefont
  {C.}~\bibnamefont {McRae}}, \bibinfo {author} {\bibfnamefont
  {J.}~\bibnamefont {Long}},\ and\ \bibinfo {author} {\bibfnamefont
  {D.}~\bibnamefont {Pappas}},\ }\bibfield  {title} {\bibinfo {title}
  {Merged-element transmon},\ }\href
  {https://doi.org/10.1103/PhysRevApplied.14.064006} {\bibfield  {journal}
  {\bibinfo  {journal} {Phys. Rev. Appl.}\ }\textbf {\bibinfo {volume} {14}},\
  \bibinfo {pages} {064006} (\bibinfo {year} {2020})}\BibitemShut {NoStop}%
\bibitem [{\citenamefont {Mamin}\ \emph {et~al.}(2021)\citenamefont {Mamin},
  \citenamefont {Huang}, \citenamefont {Carnevale}, \citenamefont {Rettner},
  \citenamefont {Arellano}, \citenamefont {Sherwood}, \citenamefont {Kurter},
  \citenamefont {Trimm}, \citenamefont {Sandberg}, \citenamefont {Shelby},
  \citenamefont {Mueed}, \citenamefont {Madon}, \citenamefont {Pushp},
  \citenamefont {Steffen},\ and\ \citenamefont {Rugar}}]{Mamin21}%
  \BibitemOpen
  \bibfield  {author} {\bibinfo {author} {\bibfnamefont {H.}~\bibnamefont
  {Mamin}}, \bibinfo {author} {\bibfnamefont {E.}~\bibnamefont {Huang}},
  \bibinfo {author} {\bibfnamefont {S.}~\bibnamefont {Carnevale}}, \bibinfo
  {author} {\bibfnamefont {C.}~\bibnamefont {Rettner}}, \bibinfo {author}
  {\bibfnamefont {N.}~\bibnamefont {Arellano}}, \bibinfo {author}
  {\bibfnamefont {M.}~\bibnamefont {Sherwood}}, \bibinfo {author}
  {\bibfnamefont {C.}~\bibnamefont {Kurter}}, \bibinfo {author} {\bibfnamefont
  {B.}~\bibnamefont {Trimm}}, \bibinfo {author} {\bibfnamefont
  {M.}~\bibnamefont {Sandberg}}, \bibinfo {author} {\bibfnamefont
  {R.}~\bibnamefont {Shelby}}, \bibinfo {author} {\bibfnamefont
  {M.}~\bibnamefont {Mueed}}, \bibinfo {author} {\bibfnamefont
  {B.}~\bibnamefont {Madon}}, \bibinfo {author} {\bibfnamefont
  {A.}~\bibnamefont {Pushp}}, \bibinfo {author} {\bibfnamefont
  {M.}~\bibnamefont {Steffen}},\ and\ \bibinfo {author} {\bibfnamefont
  {D.}~\bibnamefont {Rugar}},\ }\bibfield  {title} {\bibinfo {title}
  {Merged-element transmons: Design and qubit performance},\ }\href
  {https://doi.org/10.1103/PhysRevApplied.16.024023} {\bibfield  {journal}
  {\bibinfo  {journal} {Phys. Rev. Applied}\ }\textbf {\bibinfo {volume}
  {16}},\ \bibinfo {pages} {024023} (\bibinfo {year} {2021})}\BibitemShut
  {NoStop}%
\bibitem [{\citenamefont {Martinis}\ \emph {et~al.}(2005)\citenamefont
  {Martinis}, \citenamefont {Cooper}, \citenamefont {McDermott}, \citenamefont
  {Steffen}, \citenamefont {Ansmann}, \citenamefont {Osborn}, \citenamefont
  {Cicak}, \citenamefont {Oh}, \citenamefont {Pappas}, \citenamefont
  {Simmonds},\ and\ \citenamefont {Yu}}]{Martinis05}%
  \BibitemOpen
  \bibfield  {author} {\bibinfo {author} {\bibfnamefont {J.~M.}\ \bibnamefont
  {Martinis}}, \bibinfo {author} {\bibfnamefont {K.~B.}\ \bibnamefont
  {Cooper}}, \bibinfo {author} {\bibfnamefont {R.}~\bibnamefont {McDermott}},
  \bibinfo {author} {\bibfnamefont {M.}~\bibnamefont {Steffen}}, \bibinfo
  {author} {\bibfnamefont {M.}~\bibnamefont {Ansmann}}, \bibinfo {author}
  {\bibfnamefont {K.~D.}\ \bibnamefont {Osborn}}, \bibinfo {author}
  {\bibfnamefont {K.}~\bibnamefont {Cicak}}, \bibinfo {author} {\bibfnamefont
  {S.}~\bibnamefont {Oh}}, \bibinfo {author} {\bibfnamefont {D.~P.}\
  \bibnamefont {Pappas}}, \bibinfo {author} {\bibfnamefont {R.~W.}\
  \bibnamefont {Simmonds}},\ and\ \bibinfo {author} {\bibfnamefont {C.~C.}\
  \bibnamefont {Yu}},\ }\bibfield  {title} {\bibinfo {title} {Decoherence in
  josephson qubits from dielectric loss},\ }\href
  {https://doi.org/10.1103/PhysRevLett.95.210503} {\bibfield  {journal}
  {\bibinfo  {journal} {Phys. Rev. Lett.}\ }\textbf {\bibinfo {volume} {95}},\
  \bibinfo {pages} {210503} (\bibinfo {year} {2005})}\BibitemShut {NoStop}%
\bibitem [{\citenamefont {Neeley}\ \emph {et~al.}(2008)\citenamefont {Neeley},
  \citenamefont {Ansmann}, \citenamefont {Bialczak}, \citenamefont {Hofheinz},
  \citenamefont {Katz1}, \citenamefont {Lucero}, \citenamefont {O'Connell},
  \citenamefont {Wang}, \citenamefont {Cleland},\ and\ \citenamefont
  {Martinis}}]{Neeley08}%
  \BibitemOpen
  \bibfield  {author} {\bibinfo {author} {\bibfnamefont {M.}~\bibnamefont
  {Neeley}}, \bibinfo {author} {\bibfnamefont {M.}~\bibnamefont {Ansmann}},
  \bibinfo {author} {\bibfnamefont {R.~C.}\ \bibnamefont {Bialczak}}, \bibinfo
  {author} {\bibfnamefont {M.}~\bibnamefont {Hofheinz}}, \bibinfo {author}
  {\bibfnamefont {N.}~\bibnamefont {Katz1}}, \bibinfo {author} {\bibfnamefont
  {E.}~\bibnamefont {Lucero}}, \bibinfo {author} {\bibfnamefont
  {A.}~\bibnamefont {O'Connell}}, \bibinfo {author} {\bibfnamefont
  {H.}~\bibnamefont {Wang}}, \bibinfo {author} {\bibfnamefont {A.~N.}\
  \bibnamefont {Cleland}},\ and\ \bibinfo {author} {\bibfnamefont {J.~M.}\
  \bibnamefont {Martinis}},\ }\bibfield  {title} {\bibinfo {title} {Process
  tomography of quantum memory in a josephson-phase qubit coupled to a
  two-level state},\ }\href {https://doi.org/doi:10.1038/nphys972} {\bibfield
  {journal} {\bibinfo  {journal} {Nat. Phys.}\ }\textbf {\bibinfo {volume}
  {4}},\ \bibinfo {pages} {523 } (\bibinfo {year} {2008})}\BibitemShut
  {NoStop}%
\bibitem [{\citenamefont {Shalibo}\ \emph {et~al.}(2010)\citenamefont
  {Shalibo}, \citenamefont {Rofe}, \citenamefont {Shwa}, \citenamefont
  {Zeides}, \citenamefont {Neeley}, \citenamefont {Martinis},\ and\
  \citenamefont {Katz}}]{Shalibo10}%
  \BibitemOpen
  \bibfield  {author} {\bibinfo {author} {\bibfnamefont {Y.}~\bibnamefont
  {Shalibo}}, \bibinfo {author} {\bibfnamefont {Y.}~\bibnamefont {Rofe}},
  \bibinfo {author} {\bibfnamefont {D.}~\bibnamefont {Shwa}}, \bibinfo {author}
  {\bibfnamefont {F.}~\bibnamefont {Zeides}}, \bibinfo {author} {\bibfnamefont
  {M.}~\bibnamefont {Neeley}}, \bibinfo {author} {\bibfnamefont {J.~M.}\
  \bibnamefont {Martinis}},\ and\ \bibinfo {author} {\bibfnamefont
  {N.}~\bibnamefont {Katz}},\ }\bibfield  {title} {\bibinfo {title} {Lifetime
  and coherence of two-level defects in a josephson junction},\ }\href
  {https://doi.org/10.1103/PhysRevLett.105.177001} {\bibfield  {journal}
  {\bibinfo  {journal} {Phys. Rev. Lett.}\ }\textbf {\bibinfo {volume} {105}},\
  \bibinfo {pages} {177001} (\bibinfo {year} {2010})}\BibitemShut {NoStop}%
\bibitem [{\citenamefont {Lisenfeld}\ \emph {et~al.}(2010)\citenamefont
  {Lisenfeld}, \citenamefont {M\"uller}, \citenamefont {Cole}, \citenamefont
  {Bushev}, \citenamefont {Lukashenko}, \citenamefont {Shnirman},\ and\
  \citenamefont {Ustinov}}]{Lisenfeld10}%
  \BibitemOpen
  \bibfield  {author} {\bibinfo {author} {\bibfnamefont {J.}~\bibnamefont
  {Lisenfeld}}, \bibinfo {author} {\bibfnamefont {C.}~\bibnamefont {M\"uller}},
  \bibinfo {author} {\bibfnamefont {J.~H.}\ \bibnamefont {Cole}}, \bibinfo
  {author} {\bibfnamefont {P.}~\bibnamefont {Bushev}}, \bibinfo {author}
  {\bibfnamefont {A.}~\bibnamefont {Lukashenko}}, \bibinfo {author}
  {\bibfnamefont {A.}~\bibnamefont {Shnirman}},\ and\ \bibinfo {author}
  {\bibfnamefont {A.~V.}\ \bibnamefont {Ustinov}},\ }\bibfield  {title}
  {\bibinfo {title} {Measuring the temperature dependence of individual
  two-level systems by direct coherent control},\ }\href
  {https://doi.org/10.1103/PhysRevLett.105.230504} {\bibfield  {journal}
  {\bibinfo  {journal} {Phys. Rev. Lett.}\ }\textbf {\bibinfo {volume} {105}},\
  \bibinfo {pages} {230504} (\bibinfo {year} {2010})}\BibitemShut {NoStop}%
\bibitem [{\citenamefont {Taminiau}\ \emph {et~al.}(2012)\citenamefont
  {Taminiau}, \citenamefont {Wagenaar}, \citenamefont {van~der Sar},
  \citenamefont {Jelezko}, \citenamefont {Dobrovitski},\ and\ \citenamefont
  {Hanson}}]{Taminiau12}%
  \BibitemOpen
  \bibfield  {author} {\bibinfo {author} {\bibfnamefont {T.~H.}\ \bibnamefont
  {Taminiau}}, \bibinfo {author} {\bibfnamefont {J.~J.~T.}\ \bibnamefont
  {Wagenaar}}, \bibinfo {author} {\bibfnamefont {T.}~\bibnamefont {van~der
  Sar}}, \bibinfo {author} {\bibfnamefont {F.}~\bibnamefont {Jelezko}},
  \bibinfo {author} {\bibfnamefont {V.~V.}\ \bibnamefont {Dobrovitski}},\ and\
  \bibinfo {author} {\bibfnamefont {R.}~\bibnamefont {Hanson}},\ }\bibfield
  {title} {\bibinfo {title} {Detection and control of individual nuclear spins
  using a weakly coupled electron spin},\ }\href
  {https://doi.org/10.1103/PhysRevLett.109.137602} {\bibfield  {journal}
  {\bibinfo  {journal} {Phys. Rev. Lett.}\ }\textbf {\bibinfo {volume} {109}},\
  \bibinfo {pages} {137602} (\bibinfo {year} {2012})}\BibitemShut {NoStop}%
\bibitem [{\citenamefont {Degen}\ \emph {et~al.}(2017)\citenamefont {Degen},
  \citenamefont {Reinhard},\ and\ \citenamefont {Cappellaro}}]{Degen17}%
  \BibitemOpen
  \bibfield  {author} {\bibinfo {author} {\bibfnamefont {C.~L.}\ \bibnamefont
  {Degen}}, \bibinfo {author} {\bibfnamefont {F.}~\bibnamefont {Reinhard}},\
  and\ \bibinfo {author} {\bibfnamefont {P.}~\bibnamefont {Cappellaro}},\
  }\bibfield  {title} {\bibinfo {title} {Quantum sensing},\ }\href
  {https://doi.org/10.1103/RevModPhys.89.035002} {\bibfield  {journal}
  {\bibinfo  {journal} {Rev. Mod. Phys.}\ }\textbf {\bibinfo {volume} {89}},\
  \bibinfo {pages} {035002} (\bibinfo {year} {2017})}\BibitemShut {NoStop}%
\bibitem [{\citenamefont {Bylander}\ \emph {et~al.}(2011)\citenamefont
  {Bylander}, \citenamefont {Gustavsson}, \citenamefont {Yan}, \citenamefont
  {Yoshihara}, \citenamefont {Harrabi}, \citenamefont {Fitch}, \citenamefont
  {Cory},\ and\ \citenamefont {Oliver}}]{Bylander11}%
  \BibitemOpen
  \bibfield  {author} {\bibinfo {author} {\bibfnamefont {J.}~\bibnamefont
  {Bylander}}, \bibinfo {author} {\bibfnamefont {S.}~\bibnamefont
  {Gustavsson}}, \bibinfo {author} {\bibfnamefont {F.}~\bibnamefont {Yan}},
  \bibinfo {author} {\bibfnamefont {F.}~\bibnamefont {Yoshihara}}, \bibinfo
  {author} {\bibfnamefont {K.}~\bibnamefont {Harrabi}}, \bibinfo {author}
  {\bibfnamefont {G.}~\bibnamefont {Fitch}}, \bibinfo {author} {\bibfnamefont
  {D.~G.}\ \bibnamefont {Cory}},\ and\ \bibinfo {author} {\bibfnamefont
  {W.~D.}\ \bibnamefont {Oliver}},\ }\bibfield  {title} {\bibinfo {title}
  {Noise spectroscopy through dynamical decoupling with a superconducting flux
  qubit},\ }\href {https://doi.org/10.1038/nphys1994} {\bibfield  {journal}
  {\bibinfo  {journal} {Nat. Phys.}\ }\textbf {\bibinfo {volume} {7}},\
  \bibinfo {pages} {565} (\bibinfo {year} {2011})}\BibitemShut {NoStop}%
\bibitem [{\citenamefont {Catelani}\ \emph {et~al.}(2011)\citenamefont
  {Catelani}, \citenamefont {Koch}, \citenamefont {Frunzio}, \citenamefont
  {Schoelkopf}, \citenamefont {Devoret},\ and\ \citenamefont
  {Glazman}}]{Catelani11}%
  \BibitemOpen
  \bibfield  {author} {\bibinfo {author} {\bibfnamefont {G.}~\bibnamefont
  {Catelani}}, \bibinfo {author} {\bibfnamefont {J.}~\bibnamefont {Koch}},
  \bibinfo {author} {\bibfnamefont {L.}~\bibnamefont {Frunzio}}, \bibinfo
  {author} {\bibfnamefont {R.~J.}\ \bibnamefont {Schoelkopf}}, \bibinfo
  {author} {\bibfnamefont {M.~H.}\ \bibnamefont {Devoret}},\ and\ \bibinfo
  {author} {\bibfnamefont {L.~I.}\ \bibnamefont {Glazman}},\ }\bibfield
  {title} {\bibinfo {title} {Quasiparticle relaxation of superconducting qubits
  in the presence of flux},\ }\href
  {https://doi.org/10.1103/PhysRevLett.106.077002} {\bibfield  {journal}
  {\bibinfo  {journal} {Phys. Rev. Lett.}\ }\textbf {\bibinfo {volume} {106}},\
  \bibinfo {pages} {077002} (\bibinfo {year} {2011})}\BibitemShut {NoStop}%
\bibitem [{\citenamefont {Crowley}\ \emph {et~al.}(2023)\citenamefont
  {Crowley}, \citenamefont {McLellan}, \citenamefont {Dutta}, \citenamefont
  {Shumiya}, \citenamefont {Place}, \citenamefont {Le}, \citenamefont {Gang},
  \citenamefont {Madhavan}, \citenamefont {Khedkar}, \citenamefont {Feng},
  \citenamefont {Umbarkar}, \citenamefont {Gui}, \citenamefont {Rodgers},
  \citenamefont {Jia}, \citenamefont {Feldman}, \citenamefont {Lyon},
  \citenamefont {Liu}, \citenamefont {Cava}, \citenamefont {Houck},\ and\
  \citenamefont {de~Leon}}]{Crowley23}%
  \BibitemOpen
  \bibfield  {author} {\bibinfo {author} {\bibfnamefont {K.~D.}\ \bibnamefont
  {Crowley}}, \bibinfo {author} {\bibfnamefont {R.~A.}\ \bibnamefont
  {McLellan}}, \bibinfo {author} {\bibfnamefont {A.}~\bibnamefont {Dutta}},
  \bibinfo {author} {\bibfnamefont {N.}~\bibnamefont {Shumiya}}, \bibinfo
  {author} {\bibfnamefont {A.~P.~M.}\ \bibnamefont {Place}}, \bibinfo {author}
  {\bibfnamefont {X.~H.}\ \bibnamefont {Le}}, \bibinfo {author} {\bibfnamefont
  {Y.}~\bibnamefont {Gang}}, \bibinfo {author} {\bibfnamefont {T.}~\bibnamefont
  {Madhavan}}, \bibinfo {author} {\bibfnamefont {N.}~\bibnamefont {Khedkar}},
  \bibinfo {author} {\bibfnamefont {Y.~C.}\ \bibnamefont {Feng}}, \bibinfo
  {author} {\bibfnamefont {E.~A.}\ \bibnamefont {Umbarkar}}, \bibinfo {author}
  {\bibfnamefont {X.}~\bibnamefont {Gui}}, \bibinfo {author} {\bibfnamefont
  {L.~V.~H.}\ \bibnamefont {Rodgers}}, \bibinfo {author} {\bibfnamefont
  {Y.}~\bibnamefont {Jia}}, \bibinfo {author} {\bibfnamefont {M.~M.}\
  \bibnamefont {Feldman}}, \bibinfo {author} {\bibfnamefont {S.~A.}\
  \bibnamefont {Lyon}}, \bibinfo {author} {\bibfnamefont {M.}~\bibnamefont
  {Liu}}, \bibinfo {author} {\bibfnamefont {R.~J.}\ \bibnamefont {Cava}},
  \bibinfo {author} {\bibfnamefont {A.~A.}\ \bibnamefont {Houck}},\ and\
  \bibinfo {author} {\bibfnamefont {N.~P.}\ \bibnamefont {de~Leon}},\
  }\href@noop {} {\bibinfo {title} {Disentangling losses in tantalum
  superconducting circuits}} (\bibinfo {year} {2023}),\ \Eprint
  {https://arxiv.org/abs/2301.07848} {arXiv:2301.07848 [quant-ph]} \BibitemShut
  {NoStop}%
\bibitem [{\citenamefont {K\^e}(1947)}]{Ke47}%
  \BibitemOpen
  \bibfield  {author} {\bibinfo {author} {\bibfnamefont {T.-S.}\ \bibnamefont
  {K\^e}},\ }\bibfield  {title} {\bibinfo {title} {Experimental evidence of the
  viscous behavior of grain boundaries in metals},\ }\href
  {https://doi.org/10.1103/PhysRev.71.533} {\bibfield  {journal} {\bibinfo
  {journal} {Phys. Rev.}\ }\textbf {\bibinfo {volume} {71}},\ \bibinfo {pages}
  {533} (\bibinfo {year} {1947})}\BibitemShut {NoStop}%
\bibitem [{\citenamefont {Veps{\"a}l{\"a}inen}\ \emph
  {et~al.}(2020)\citenamefont {Veps{\"a}l{\"a}inen}, \citenamefont {Karamlou},
  \citenamefont {Orrell}, \citenamefont {Dogra}, \citenamefont {Loer},
  \citenamefont {Vasconcelos}, \citenamefont {Kim}, \citenamefont {Melville},
  \citenamefont {Niedzielski}, \citenamefont {Yoder}, \citenamefont
  {Gustavsson}, \citenamefont {Formaggio}, \citenamefont {VanDevender},\ and\
  \citenamefont {Oliver}}]{Vepsalainen20}%
  \BibitemOpen
  \bibfield  {author} {\bibinfo {author} {\bibfnamefont {A.~P.}\ \bibnamefont
  {Veps{\"a}l{\"a}inen}}, \bibinfo {author} {\bibfnamefont {A.~H.}\
  \bibnamefont {Karamlou}}, \bibinfo {author} {\bibfnamefont {J.~L.}\
  \bibnamefont {Orrell}}, \bibinfo {author} {\bibfnamefont {A.~S.}\
  \bibnamefont {Dogra}}, \bibinfo {author} {\bibfnamefont {B.}~\bibnamefont
  {Loer}}, \bibinfo {author} {\bibfnamefont {F.}~\bibnamefont {Vasconcelos}},
  \bibinfo {author} {\bibfnamefont {D.~K.}\ \bibnamefont {Kim}}, \bibinfo
  {author} {\bibfnamefont {A.~J.}\ \bibnamefont {Melville}}, \bibinfo {author}
  {\bibfnamefont {B.~M.}\ \bibnamefont {Niedzielski}}, \bibinfo {author}
  {\bibfnamefont {J.~L.}\ \bibnamefont {Yoder}}, \bibinfo {author}
  {\bibfnamefont {S.}~\bibnamefont {Gustavsson}}, \bibinfo {author}
  {\bibfnamefont {J.~A.}\ \bibnamefont {Formaggio}}, \bibinfo {author}
  {\bibfnamefont {B.~A.}\ \bibnamefont {VanDevender}},\ and\ \bibinfo {author}
  {\bibfnamefont {W.~D.}\ \bibnamefont {Oliver}},\ }\bibfield  {title}
  {\bibinfo {title} {Impact of ionizing radiation on superconducting qubit
  coherence},\ }\href {https://doi.org/10.1038/s41586-020-2619-8} {\bibfield
  {journal} {\bibinfo  {journal} {Nature}\ }\textbf {\bibinfo {volume} {584}},\
  \bibinfo {pages} {551} (\bibinfo {year} {2020})}\BibitemShut {NoStop}%
\bibitem [{\citenamefont {Gustavsson}\ \emph {et~al.}(2016)\citenamefont
  {Gustavsson}, \citenamefont {Yan}, \citenamefont {Catelani}, \citenamefont
  {Bylander}, \citenamefont {Kamal}, \citenamefont {Birenbaum}, \citenamefont
  {Hover}, \citenamefont {Rosenberg}, \citenamefont {Samach}, \citenamefont
  {Sears}, \citenamefont {Weber}, \citenamefont {Yoder}, \citenamefont
  {Clarke}, \citenamefont {Kerman}, \citenamefont {Yoshihara}, \citenamefont
  {Nakamura}, \citenamefont {Orlando},\ and\ \citenamefont
  {Oliver}}]{Gustavsson16}%
  \BibitemOpen
  \bibfield  {author} {\bibinfo {author} {\bibfnamefont {S.}~\bibnamefont
  {Gustavsson}}, \bibinfo {author} {\bibfnamefont {F.}~\bibnamefont {Yan}},
  \bibinfo {author} {\bibfnamefont {G.}~\bibnamefont {Catelani}}, \bibinfo
  {author} {\bibfnamefont {J.}~\bibnamefont {Bylander}}, \bibinfo {author}
  {\bibfnamefont {A.}~\bibnamefont {Kamal}}, \bibinfo {author} {\bibfnamefont
  {J.}~\bibnamefont {Birenbaum}}, \bibinfo {author} {\bibfnamefont
  {D.}~\bibnamefont {Hover}}, \bibinfo {author} {\bibfnamefont
  {D.}~\bibnamefont {Rosenberg}}, \bibinfo {author} {\bibfnamefont
  {G.}~\bibnamefont {Samach}}, \bibinfo {author} {\bibfnamefont {A.~P.}\
  \bibnamefont {Sears}}, \bibinfo {author} {\bibfnamefont {S.~J.}\ \bibnamefont
  {Weber}}, \bibinfo {author} {\bibfnamefont {J.~L.}\ \bibnamefont {Yoder}},
  \bibinfo {author} {\bibfnamefont {J.}~\bibnamefont {Clarke}}, \bibinfo
  {author} {\bibfnamefont {A.~J.}\ \bibnamefont {Kerman}}, \bibinfo {author}
  {\bibfnamefont {F.}~\bibnamefont {Yoshihara}}, \bibinfo {author}
  {\bibfnamefont {Y.}~\bibnamefont {Nakamura}}, \bibinfo {author}
  {\bibfnamefont {T.~P.}\ \bibnamefont {Orlando}},\ and\ \bibinfo {author}
  {\bibfnamefont {W.~D.}\ \bibnamefont {Oliver}},\ }\bibfield  {title}
  {\bibinfo {title} {Suppressing relaxation in superconducting qubits by
  quasiparticle pumping},\ }\href {https://doi.org/10.1126/science.aah5844}
  {\bibfield  {journal} {\bibinfo  {journal} {Science}\ }\textbf {\bibinfo
  {volume} {354}},\ \bibinfo {pages} {1573} (\bibinfo {year}
  {2016})}\BibitemShut {NoStop}%
\bibitem [{\citenamefont {Serniak}\ \emph {et~al.}(2018)\citenamefont
  {Serniak}, \citenamefont {Hays}, \citenamefont {de~Lange}, \citenamefont
  {Diamond}, \citenamefont {Shankar}, \citenamefont {Burkhart}, \citenamefont
  {Frunzio}, \citenamefont {Houzet},\ and\ \citenamefont
  {Devoret}}]{Serniak18}%
  \BibitemOpen
  \bibfield  {author} {\bibinfo {author} {\bibfnamefont {K.}~\bibnamefont
  {Serniak}}, \bibinfo {author} {\bibfnamefont {M.}~\bibnamefont {Hays}},
  \bibinfo {author} {\bibfnamefont {G.}~\bibnamefont {de~Lange}}, \bibinfo
  {author} {\bibfnamefont {S.}~\bibnamefont {Diamond}}, \bibinfo {author}
  {\bibfnamefont {S.}~\bibnamefont {Shankar}}, \bibinfo {author} {\bibfnamefont
  {L.~D.}\ \bibnamefont {Burkhart}}, \bibinfo {author} {\bibfnamefont
  {L.}~\bibnamefont {Frunzio}}, \bibinfo {author} {\bibfnamefont
  {M.}~\bibnamefont {Houzet}},\ and\ \bibinfo {author} {\bibfnamefont {M.~H.}\
  \bibnamefont {Devoret}},\ }\bibfield  {title} {\bibinfo {title} {Hot
  nonequilibrium quasiparticles in transmon qubits},\ }\href
  {https://doi.org/10.1103/PhysRevLett.121.157701} {\bibfield  {journal}
  {\bibinfo  {journal} {Phys. Rev. Lett.}\ }\textbf {\bibinfo {volume} {121}},\
  \bibinfo {pages} {157701} (\bibinfo {year} {2018})}\BibitemShut {NoStop}%
\bibitem [{\citenamefont {de~Graaf}\ \emph {et~al.}(2022)\citenamefont
  {de~Graaf}, \citenamefont {Un}, \citenamefont {Shard},\ and\ \citenamefont
  {Lindstr\"{o}m}}]{deGraaf22}%
  \BibitemOpen
  \bibfield  {author} {\bibinfo {author} {\bibfnamefont {S.~E.}\ \bibnamefont
  {de~Graaf}}, \bibinfo {author} {\bibfnamefont {S.}~\bibnamefont {Un}},
  \bibinfo {author} {\bibfnamefont {A.~G.}\ \bibnamefont {Shard}},\ and\
  \bibinfo {author} {\bibfnamefont {T.}~\bibnamefont {Lindstr\"{o}m}},\
  }\bibfield  {title} {\bibinfo {title} {Chemical and structural identification
  of material defects in superconducting quantum circuits},\ }\href
  {https://doi.org/10.1088/2633-4356/ac78ba} {\bibfield  {journal} {\bibinfo
  {journal} {Materials for Quantum Technology}\ }\textbf {\bibinfo {volume}
  {2}},\ \bibinfo {pages} {032001} (\bibinfo {year} {2022})}\BibitemShut
  {NoStop}%
\bibitem [{\citenamefont {Faoro}\ and\ \citenamefont {Ioffe}(2015)}]{Faoro15}%
  \BibitemOpen
  \bibfield  {author} {\bibinfo {author} {\bibfnamefont {L.}~\bibnamefont
  {Faoro}}\ and\ \bibinfo {author} {\bibfnamefont {L.~B.}\ \bibnamefont
  {Ioffe}},\ }\bibfield  {title} {\bibinfo {title} {Interacting tunneling model
  for two-level systems in amorphous materials and its predictions for their
  dephasing and noise in superconducting microresonators},\ }\href
  {https://doi.org/10.1103/PhysRevB.91.014201} {\bibfield  {journal} {\bibinfo
  {journal} {Phys. Rev. B}\ }\textbf {\bibinfo {volume} {91}},\ \bibinfo
  {pages} {014201} (\bibinfo {year} {2015})}\BibitemShut {NoStop}%
\bibitem [{\citenamefont {Mei\ss{}ner}\ \emph {et~al.}(2018)\citenamefont
  {Mei\ss{}ner}, \citenamefont {Seiler}, \citenamefont {Lisenfeld},
  \citenamefont {Ustinov},\ and\ \citenamefont {Weiss}}]{Meibner18}%
  \BibitemOpen
  \bibfield  {author} {\bibinfo {author} {\bibfnamefont {S.~M.}\ \bibnamefont
  {Mei\ss{}ner}}, \bibinfo {author} {\bibfnamefont {A.}~\bibnamefont {Seiler}},
  \bibinfo {author} {\bibfnamefont {J.}~\bibnamefont {Lisenfeld}}, \bibinfo
  {author} {\bibfnamefont {A.~V.}\ \bibnamefont {Ustinov}},\ and\ \bibinfo
  {author} {\bibfnamefont {G.}~\bibnamefont {Weiss}},\ }\bibfield  {title}
  {\bibinfo {title} {Probing individual tunneling fluctuators with coherently
  controlled tunneling systems},\ }\href
  {https://doi.org/10.1103/PhysRevB.97.180505} {\bibfield  {journal} {\bibinfo
  {journal} {Phys. Rev. B}\ }\textbf {\bibinfo {volume} {97}},\ \bibinfo
  {pages} {180505} (\bibinfo {year} {2018})}\BibitemShut {NoStop}%
\bibitem [{\citenamefont {Dutta}\ and\ \citenamefont {Horn}(1981)}]{Dutta81}%
  \BibitemOpen
  \bibfield  {author} {\bibinfo {author} {\bibfnamefont {P.}~\bibnamefont
  {Dutta}}\ and\ \bibinfo {author} {\bibfnamefont {P.~M.}\ \bibnamefont
  {Horn}},\ }\bibfield  {title} {\bibinfo {title} {Low-frequency fluctuations
  in solids: 1/f noise},\ }\href {https://doi.org/10.1103/RevModPhys.53.497}
  {\bibfield  {journal} {\bibinfo  {journal} {Rev. Mod. Phys.}\ }\textbf
  {\bibinfo {volume} {53}},\ \bibinfo {pages} {497} (\bibinfo {year}
  {1981})}\BibitemShut {NoStop}%
\bibitem [{\citenamefont {Siddiqi}(2021)}]{Siddiqi21}%
  \BibitemOpen
  \bibfield  {author} {\bibinfo {author} {\bibfnamefont {I.}~\bibnamefont
  {Siddiqi}},\ }\bibfield  {title} {\bibinfo {title} {Engineering
  high-coherence superconducting qubits},\ }\href
  {https://doi.org/10.1038/s41578-021-00370-4} {\bibfield  {journal} {\bibinfo
  {journal} {Nature Reviews Materials}\ }\textbf {\bibinfo {volume} {6}},\
  \bibinfo {pages} {875} (\bibinfo {year} {2021})}\BibitemShut {NoStop}%
\bibitem [{\citenamefont {de~Leon}\ \emph {et~al.}(2021)\citenamefont
  {de~Leon}, \citenamefont {Itoh}, \citenamefont {Kim}, \citenamefont {Mehta},
  \citenamefont {Northup}, \citenamefont {Paik}, \citenamefont {Palmer},
  \citenamefont {Samarth}, \citenamefont {Sangtawesin},\ and\ \citenamefont
  {Steuerman}}]{deLeon21}%
  \BibitemOpen
  \bibfield  {author} {\bibinfo {author} {\bibfnamefont {N.~P.}\ \bibnamefont
  {de~Leon}}, \bibinfo {author} {\bibfnamefont {K.~M.}\ \bibnamefont {Itoh}},
  \bibinfo {author} {\bibfnamefont {D.}~\bibnamefont {Kim}}, \bibinfo {author}
  {\bibfnamefont {K.~K.}\ \bibnamefont {Mehta}}, \bibinfo {author}
  {\bibfnamefont {T.~E.}\ \bibnamefont {Northup}}, \bibinfo {author}
  {\bibfnamefont {H.}~\bibnamefont {Paik}}, \bibinfo {author} {\bibfnamefont
  {B.~S.}\ \bibnamefont {Palmer}}, \bibinfo {author} {\bibfnamefont
  {N.}~\bibnamefont {Samarth}}, \bibinfo {author} {\bibfnamefont
  {S.}~\bibnamefont {Sangtawesin}},\ and\ \bibinfo {author} {\bibfnamefont
  {D.~W.}\ \bibnamefont {Steuerman}},\ }\bibfield  {title} {\bibinfo {title}
  {Materials challenges and opportunities for quantum computing hardware},\
  }\href {https://doi.org/10.1126/science.abb2823} {\bibfield  {journal}
  {\bibinfo  {journal} {Science}\ }\textbf {\bibinfo {volume} {372}},\ \bibinfo
  {pages} {eabb2823} (\bibinfo {year} {2021})}\BibitemShut {NoStop}%
\bibitem [{\citenamefont {Murray}(2021)}]{Murray21}%
  \BibitemOpen
  \bibfield  {author} {\bibinfo {author} {\bibfnamefont {C.~E.}\ \bibnamefont
  {Murray}},\ }\bibfield  {title} {\bibinfo {title} {Material matters in
  superconducting qubits},\ }\href
  {https://doi.org/https://doi.org/10.1016/j.mser.2021.100646} {\bibfield
  {journal} {\bibinfo  {journal} {Materials Science and Engineering: R:
  Reports}\ }\textbf {\bibinfo {volume} {146}},\ \bibinfo {pages} {100646}
  (\bibinfo {year} {2021})}\BibitemShut {NoStop}%
\bibitem [{\citenamefont {Zagoskin}\ \emph {et~al.}(2006)\citenamefont
  {Zagoskin}, \citenamefont {Ashhab}, \citenamefont {Johansson},\ and\
  \citenamefont {Nori}}]{Zagoskin06}%
  \BibitemOpen
  \bibfield  {author} {\bibinfo {author} {\bibfnamefont {A.~M.}\ \bibnamefont
  {Zagoskin}}, \bibinfo {author} {\bibfnamefont {S.}~\bibnamefont {Ashhab}},
  \bibinfo {author} {\bibfnamefont {J.~R.}\ \bibnamefont {Johansson}},\ and\
  \bibinfo {author} {\bibfnamefont {F.}~\bibnamefont {Nori}},\ }\bibfield
  {title} {\bibinfo {title} {Quantum two-level systems in josephson junctions
  as naturally formed qubits},\ }\href
  {https://doi.org/10.1103/PhysRevLett.97.077001} {\bibfield  {journal}
  {\bibinfo  {journal} {Phys. Rev. Lett.}\ }\textbf {\bibinfo {volume} {97}},\
  \bibinfo {pages} {077001} (\bibinfo {year} {2006})}\BibitemShut {NoStop}%
\bibitem [{\citenamefont {Yoshihara}\ \emph {et~al.}(2006)\citenamefont
  {Yoshihara}, \citenamefont {Harrabi}, \citenamefont {Niskanen}, \citenamefont
  {Nakamura},\ and\ \citenamefont {Tsai}}]{Yoshihara06}%
  \BibitemOpen
  \bibfield  {author} {\bibinfo {author} {\bibfnamefont {F.}~\bibnamefont
  {Yoshihara}}, \bibinfo {author} {\bibfnamefont {K.}~\bibnamefont {Harrabi}},
  \bibinfo {author} {\bibfnamefont {A.~O.}\ \bibnamefont {Niskanen}}, \bibinfo
  {author} {\bibfnamefont {Y.}~\bibnamefont {Nakamura}},\ and\ \bibinfo
  {author} {\bibfnamefont {J.~S.}\ \bibnamefont {Tsai}},\ }\bibfield  {title}
  {\bibinfo {title} {Decoherence of flux qubits due to $1/f$ flux noise},\
  }\href {https://doi.org/10.1103/PhysRevLett.97.167001} {\bibfield  {journal}
  {\bibinfo  {journal} {Phys. Rev. Lett.}\ }\textbf {\bibinfo {volume} {97}},\
  \bibinfo {pages} {167001} (\bibinfo {year} {2006})}\BibitemShut {NoStop}%
\bibitem [{\citenamefont {{Kakuyanagi}}\ \emph {et~al.}(2007)\citenamefont
  {{Kakuyanagi}}, \citenamefont {{Meno}}, \citenamefont {{Saito}},
  \citenamefont {{Nakano}}, \citenamefont {{Semba}}, \citenamefont
  {{Takayanagi}}, \citenamefont {{Deppe}},\ and\ \citenamefont
  {{Shnirman}}}]{Kakuyanagi07}%
  \BibitemOpen
  \bibfield  {author} {\bibinfo {author} {\bibfnamefont {K.}~\bibnamefont
  {{Kakuyanagi}}}, \bibinfo {author} {\bibfnamefont {T.}~\bibnamefont
  {{Meno}}}, \bibinfo {author} {\bibfnamefont {S.}~\bibnamefont {{Saito}}},
  \bibinfo {author} {\bibfnamefont {H.}~\bibnamefont {{Nakano}}}, \bibinfo
  {author} {\bibfnamefont {K.}~\bibnamefont {{Semba}}}, \bibinfo {author}
  {\bibfnamefont {H.}~\bibnamefont {{Takayanagi}}}, \bibinfo {author}
  {\bibfnamefont {F.}~\bibnamefont {{Deppe}}},\ and\ \bibinfo {author}
  {\bibfnamefont {A.}~\bibnamefont {{Shnirman}}},\ }\bibfield  {title}
  {\bibinfo {title} {Dephasing of a superconducting flux qubit},\ }\href
  {https://doi.org/10.1103/PhysRevLett.98.047004} {\bibfield  {journal}
  {\bibinfo  {journal} {Phys. Rev. Lett.}\ }\textbf {\bibinfo {volume} {98}},\
  \bibinfo {eid} {047004} (\bibinfo {year} {2007})}\BibitemShut {NoStop}%
\bibitem [{\citenamefont {Bialczak}\ \emph {et~al.}(2007)\citenamefont
  {Bialczak}, \citenamefont {McDermott}, \citenamefont {Ansmann}, \citenamefont
  {Hofheinz}, \citenamefont {Katz}, \citenamefont {Lucero}, \citenamefont
  {Neeley}, \citenamefont {O'Connell}, \citenamefont {Wang}, \citenamefont
  {Cleland},\ and\ \citenamefont {Martinis}}]{Bialczak07}%
  \BibitemOpen
  \bibfield  {author} {\bibinfo {author} {\bibfnamefont {R.~C.}\ \bibnamefont
  {Bialczak}}, \bibinfo {author} {\bibfnamefont {R.}~\bibnamefont {McDermott}},
  \bibinfo {author} {\bibfnamefont {M.}~\bibnamefont {Ansmann}}, \bibinfo
  {author} {\bibfnamefont {M.}~\bibnamefont {Hofheinz}}, \bibinfo {author}
  {\bibfnamefont {N.}~\bibnamefont {Katz}}, \bibinfo {author} {\bibfnamefont
  {E.}~\bibnamefont {Lucero}}, \bibinfo {author} {\bibfnamefont
  {M.}~\bibnamefont {Neeley}}, \bibinfo {author} {\bibfnamefont {A.~D.}\
  \bibnamefont {O'Connell}}, \bibinfo {author} {\bibfnamefont {H.}~\bibnamefont
  {Wang}}, \bibinfo {author} {\bibfnamefont {A.~N.}\ \bibnamefont {Cleland}},\
  and\ \bibinfo {author} {\bibfnamefont {J.~M.}\ \bibnamefont {Martinis}},\
  }\bibfield  {title} {\bibinfo {title} {$1/f$ flux noise in josephson phase
  qubits},\ }\href {https://doi.org/10.1103/PhysRevLett.99.187006} {\bibfield
  {journal} {\bibinfo  {journal} {Phys. Rev. Lett.}\ }\textbf {\bibinfo
  {volume} {99}},\ \bibinfo {pages} {187006} (\bibinfo {year}
  {2007})}\BibitemShut {NoStop}%
\bibitem [{\citenamefont {Rower}\ \emph {et~al.}(2023)\citenamefont {Rower},
  \citenamefont {Ateshian}, \citenamefont {Li}, \citenamefont {Hays},
  \citenamefont {Bluvstein}, \citenamefont {Ding}, \citenamefont {Kannan},
  \citenamefont {Almanakly}, \citenamefont {Braum\"uller}, \citenamefont {Kim},
  \citenamefont {Melville}, \citenamefont {Niedzielski}, \citenamefont
  {Schwartz}, \citenamefont {Yoder}, \citenamefont {Orlando}, \citenamefont
  {Wang}, \citenamefont {Gustavsson}, \citenamefont {Grover}, \citenamefont
  {Serniak}, \citenamefont {Comin},\ and\ \citenamefont {Oliver}}]{Rower23}%
  \BibitemOpen
  \bibfield  {author} {\bibinfo {author} {\bibfnamefont {D.~A.}\ \bibnamefont
  {Rower}}, \bibinfo {author} {\bibfnamefont {L.}~\bibnamefont {Ateshian}},
  \bibinfo {author} {\bibfnamefont {L.~H.}\ \bibnamefont {Li}}, \bibinfo
  {author} {\bibfnamefont {M.}~\bibnamefont {Hays}}, \bibinfo {author}
  {\bibfnamefont {D.}~\bibnamefont {Bluvstein}}, \bibinfo {author}
  {\bibfnamefont {L.}~\bibnamefont {Ding}}, \bibinfo {author} {\bibfnamefont
  {B.}~\bibnamefont {Kannan}}, \bibinfo {author} {\bibfnamefont
  {A.}~\bibnamefont {Almanakly}}, \bibinfo {author} {\bibfnamefont
  {J.}~\bibnamefont {Braum\"uller}}, \bibinfo {author} {\bibfnamefont {D.~K.}\
  \bibnamefont {Kim}}, \bibinfo {author} {\bibfnamefont {A.}~\bibnamefont
  {Melville}}, \bibinfo {author} {\bibfnamefont {B.~M.}\ \bibnamefont
  {Niedzielski}}, \bibinfo {author} {\bibfnamefont {M.~E.}\ \bibnamefont
  {Schwartz}}, \bibinfo {author} {\bibfnamefont {J.~L.}\ \bibnamefont {Yoder}},
  \bibinfo {author} {\bibfnamefont {T.~P.}\ \bibnamefont {Orlando}}, \bibinfo
  {author} {\bibfnamefont {J.~I.-J.}\ \bibnamefont {Wang}}, \bibinfo {author}
  {\bibfnamefont {S.}~\bibnamefont {Gustavsson}}, \bibinfo {author}
  {\bibfnamefont {J.~A.}\ \bibnamefont {Grover}}, \bibinfo {author}
  {\bibfnamefont {K.}~\bibnamefont {Serniak}}, \bibinfo {author} {\bibfnamefont
  {R.}~\bibnamefont {Comin}},\ and\ \bibinfo {author} {\bibfnamefont {W.~D.}\
  \bibnamefont {Oliver}},\ }\bibfield  {title} {\bibinfo {title} {Evolution of
  $1/f$ flux noise in superconducting qubits with weak magnetic fields},\
  }\href {https://doi.org/10.1103/PhysRevLett.130.220602} {\bibfield  {journal}
  {\bibinfo  {journal} {Phys. Rev. Lett.}\ }\textbf {\bibinfo {volume} {130}},\
  \bibinfo {pages} {220602} (\bibinfo {year} {2023})}\BibitemShut {NoStop}%
\bibitem [{\citenamefont {Dwyer}\ \emph {et~al.}(2022)\citenamefont {Dwyer},
  \citenamefont {Rodgers}, \citenamefont {Urbach}, \citenamefont {Bluvstein},
  \citenamefont {Sangtawesin}, \citenamefont {Zhou}, \citenamefont {Nassab},
  \citenamefont {Fitzpatrick}, \citenamefont {Yuan}, \citenamefont {De~Greve},
  \citenamefont {Peterson}, \citenamefont {Knowles}, \citenamefont {Sumarac},
  \citenamefont {Chou}, \citenamefont {Gali}, \citenamefont {Dobrovitski},
  \citenamefont {Lukin},\ and\ \citenamefont {de~Leon}}]{Dwyer22}%
  \BibitemOpen
  \bibfield  {author} {\bibinfo {author} {\bibfnamefont {B.~L.}\ \bibnamefont
  {Dwyer}}, \bibinfo {author} {\bibfnamefont {L.~V.}\ \bibnamefont {Rodgers}},
  \bibinfo {author} {\bibfnamefont {E.~K.}\ \bibnamefont {Urbach}}, \bibinfo
  {author} {\bibfnamefont {D.}~\bibnamefont {Bluvstein}}, \bibinfo {author}
  {\bibfnamefont {S.}~\bibnamefont {Sangtawesin}}, \bibinfo {author}
  {\bibfnamefont {H.}~\bibnamefont {Zhou}}, \bibinfo {author} {\bibfnamefont
  {Y.}~\bibnamefont {Nassab}}, \bibinfo {author} {\bibfnamefont
  {M.}~\bibnamefont {Fitzpatrick}}, \bibinfo {author} {\bibfnamefont
  {Z.}~\bibnamefont {Yuan}}, \bibinfo {author} {\bibfnamefont {K.}~\bibnamefont
  {De~Greve}}, \bibinfo {author} {\bibfnamefont {E.~L.}\ \bibnamefont
  {Peterson}}, \bibinfo {author} {\bibfnamefont {H.}~\bibnamefont {Knowles}},
  \bibinfo {author} {\bibfnamefont {T.}~\bibnamefont {Sumarac}}, \bibinfo
  {author} {\bibfnamefont {J.-P.}\ \bibnamefont {Chou}}, \bibinfo {author}
  {\bibfnamefont {A.}~\bibnamefont {Gali}}, \bibinfo {author} {\bibfnamefont
  {V.}~\bibnamefont {Dobrovitski}}, \bibinfo {author} {\bibfnamefont {M.~D.}\
  \bibnamefont {Lukin}},\ and\ \bibinfo {author} {\bibfnamefont {N.~P.}\
  \bibnamefont {de~Leon}},\ }\bibfield  {title} {\bibinfo {title} {Probing spin
  dynamics on diamond surfaces using a single quantum sensor},\ }\href
  {https://doi.org/10.1103/PRXQuantum.3.040328} {\bibfield  {journal} {\bibinfo
   {journal} {PRX Quantum}\ }\textbf {\bibinfo {volume} {3}},\ \bibinfo {pages}
  {040328} (\bibinfo {year} {2022})}\BibitemShut {NoStop}%
\bibitem [{\citenamefont {Klotz}\ \emph {et~al.}(2022)\citenamefont {Klotz},
  \citenamefont {Fehler}, \citenamefont {Waltrich}, \citenamefont {Steiger},
  \citenamefont {H\"au\ss{}ler}, \citenamefont {Reddy}, \citenamefont
  {Kulikova}, \citenamefont {Davydov}, \citenamefont {Agafonov}, \citenamefont
  {Doherty},\ and\ \citenamefont {Kubanek}}]{Klotz22}%
  \BibitemOpen
  \bibfield  {author} {\bibinfo {author} {\bibfnamefont {M.}~\bibnamefont
  {Klotz}}, \bibinfo {author} {\bibfnamefont {K.~G.}\ \bibnamefont {Fehler}},
  \bibinfo {author} {\bibfnamefont {R.}~\bibnamefont {Waltrich}}, \bibinfo
  {author} {\bibfnamefont {E.~S.}\ \bibnamefont {Steiger}}, \bibinfo {author}
  {\bibfnamefont {S.}~\bibnamefont {H\"au\ss{}ler}}, \bibinfo {author}
  {\bibfnamefont {P.}~\bibnamefont {Reddy}}, \bibinfo {author} {\bibfnamefont
  {L.~F.}\ \bibnamefont {Kulikova}}, \bibinfo {author} {\bibfnamefont {V.~A.}\
  \bibnamefont {Davydov}}, \bibinfo {author} {\bibfnamefont {V.~N.}\
  \bibnamefont {Agafonov}}, \bibinfo {author} {\bibfnamefont {M.~W.}\
  \bibnamefont {Doherty}},\ and\ \bibinfo {author} {\bibfnamefont
  {A.}~\bibnamefont {Kubanek}},\ }\bibfield  {title} {\bibinfo {title}
  {Prolonged orbital relaxation by locally modified phonon density of states
  for the ${\mathrm{si}v}^{\ensuremath{-}}$ center in nanodiamonds},\ }\href
  {https://doi.org/10.1103/PhysRevLett.128.153602} {\bibfield  {journal}
  {\bibinfo  {journal} {Phys. Rev. Lett.}\ }\textbf {\bibinfo {volume} {128}},\
  \bibinfo {pages} {153602} (\bibinfo {year} {2022})}\BibitemShut {NoStop}%
\end{thebibliography}%


\begin{thebibliography}{46}%
\makeatletter
\providecommand \@ifxundefined [1]{%
 \@ifx{#1\undefined}
}%
\providecommand \@ifnum [1]{%
 \ifnum #1\expandafter \@firstoftwo
 \else \expandafter \@secondoftwo
 \fi
}%
\providecommand \@ifx [1]{%
 \ifx #1\expandafter \@firstoftwo
 \else \expandafter \@secondoftwo
 \fi
}%
\providecommand \natexlab [1]{#1}%
\providecommand \enquote  [1]{``#1''}%
\providecommand \bibnamefont  [1]{#1}%
\providecommand \bibfnamefont [1]{#1}%
\providecommand \citenamefont [1]{#1}%
\providecommand \href@noop [0]{\@secondoftwo}%
\providecommand \href [0]{\begingroup \@sanitize@url \@href}%
\providecommand \@href[1]{\@@startlink{#1}\@@href}%
\providecommand \@@href[1]{\endgroup#1\@@endlink}%
\providecommand \@sanitize@url [0]{\catcode `\\12\catcode `\$12\catcode
  `\&12\catcode `\#12\catcode `\^12\catcode `\_12\catcode `\%12\relax}%
\providecommand \@@startlink[1]{}%
\providecommand \@@endlink[0]{}%
\providecommand \url  [0]{\begingroup\@sanitize@url \@url }%
\providecommand \@url [1]{\endgroup\@href {#1}{\urlprefix }}%
\providecommand \urlprefix  [0]{URL }%
\providecommand \Eprint [0]{\href }%
\providecommand \doibase [0]{https://doi.org/}%
\providecommand \selectlanguage [0]{\@gobble}%
\providecommand \bibinfo  [0]{\@secondoftwo}%
\providecommand \bibfield  [0]{\@secondoftwo}%
\providecommand \translation [1]{[#1]}%
\providecommand \BibitemOpen [0]{}%
\providecommand \bibitemStop [0]{}%
\providecommand \bibitemNoStop [0]{.\EOS\space}%
\providecommand \EOS [0]{\spacefactor3000\relax}%
\providecommand \BibitemShut  [1]{\csname bibitem#1\endcsname}%
\let\auto@bib@innerbib\@empty
\bibitem [{\citenamefont {Martinis}\ \emph {et~al.}(2005)\citenamefont
  {Martinis}, \citenamefont {Cooper}, \citenamefont {McDermott}, \citenamefont
  {Steffen}, \citenamefont {Ansmann}, \citenamefont {Osborn}, \citenamefont
  {Cicak}, \citenamefont {Oh}, \citenamefont {Pappas}, \citenamefont
  {Simmonds},\ and\ \citenamefont {Yu}}]{Martinis05}%
  \BibitemOpen
  \bibfield  {author} {\bibinfo {author} {\bibfnamefont {J.~M.}\ \bibnamefont
  {Martinis}}, \bibinfo {author} {\bibfnamefont {K.~B.}\ \bibnamefont
  {Cooper}}, \bibinfo {author} {\bibfnamefont {R.}~\bibnamefont {McDermott}},
  \bibinfo {author} {\bibfnamefont {M.}~\bibnamefont {Steffen}}, \bibinfo
  {author} {\bibfnamefont {M.}~\bibnamefont {Ansmann}}, \bibinfo {author}
  {\bibfnamefont {K.~D.}\ \bibnamefont {Osborn}}, \bibinfo {author}
  {\bibfnamefont {K.}~\bibnamefont {Cicak}}, \bibinfo {author} {\bibfnamefont
  {S.}~\bibnamefont {Oh}}, \bibinfo {author} {\bibfnamefont {D.~P.}\
  \bibnamefont {Pappas}}, \bibinfo {author} {\bibfnamefont {R.~W.}\
  \bibnamefont {Simmonds}},\ and\ \bibinfo {author} {\bibfnamefont {C.~C.}\
  \bibnamefont {Yu}},\ }\bibfield  {title} {\bibinfo {title} {Decoherence in
  josephson qubits from dielectric loss},\ }\href
  {https://doi.org/10.1103/PhysRevLett.95.210503} {\bibfield  {journal}
  {\bibinfo  {journal} {Phys. Rev. Lett.}\ }\textbf {\bibinfo {volume} {95}},\
  \bibinfo {pages} {210503} (\bibinfo {year} {2005})}\BibitemShut {NoStop}%
\bibitem [{\citenamefont {Neeley}\ \emph {et~al.}(2008)\citenamefont {Neeley},
  \citenamefont {Ansmann}, \citenamefont {Bialczak}, \citenamefont {Hofheinz},
  \citenamefont {Katz1}, \citenamefont {Lucero}, \citenamefont {O'Connell},
  \citenamefont {Wang}, \citenamefont {Cleland},\ and\ \citenamefont
  {Martinis}}]{Neeley08}%
  \BibitemOpen
  \bibfield  {author} {\bibinfo {author} {\bibfnamefont {M.}~\bibnamefont
  {Neeley}}, \bibinfo {author} {\bibfnamefont {M.}~\bibnamefont {Ansmann}},
  \bibinfo {author} {\bibfnamefont {R.~C.}\ \bibnamefont {Bialczak}}, \bibinfo
  {author} {\bibfnamefont {M.}~\bibnamefont {Hofheinz}}, \bibinfo {author}
  {\bibfnamefont {N.}~\bibnamefont {Katz1}}, \bibinfo {author} {\bibfnamefont
  {E.}~\bibnamefont {Lucero}}, \bibinfo {author} {\bibfnamefont
  {A.}~\bibnamefont {O'Connell}}, \bibinfo {author} {\bibfnamefont
  {H.}~\bibnamefont {Wang}}, \bibinfo {author} {\bibfnamefont {A.~N.}\
  \bibnamefont {Cleland}},\ and\ \bibinfo {author} {\bibfnamefont {J.~M.}\
  \bibnamefont {Martinis}},\ }\bibfield  {title} {\bibinfo {title} {Process
  tomography of quantum memory in a josephson-phase qubit coupled to a
  two-level state},\ }\href {https://doi.org/doi:10.1038/nphys972} {\bibfield
  {journal} {\bibinfo  {journal} {Nat. Phys.}\ }\textbf {\bibinfo {volume}
  {4}},\ \bibinfo {pages} {523 } (\bibinfo {year} {2008})}\BibitemShut
  {NoStop}%
\bibitem [{\citenamefont {Zhao}\ \emph {et~al.}(2020)\citenamefont {Zhao},
  \citenamefont {Park}, \citenamefont {Zhao}, \citenamefont {Bal},
  \citenamefont {McRae}, \citenamefont {Long},\ and\ \citenamefont
  {Pappas}}]{Zhao20}%
  \BibitemOpen
  \bibfield  {author} {\bibinfo {author} {\bibfnamefont {R.}~\bibnamefont
  {Zhao}}, \bibinfo {author} {\bibfnamefont {S.}~\bibnamefont {Park}}, \bibinfo
  {author} {\bibfnamefont {T.}~\bibnamefont {Zhao}}, \bibinfo {author}
  {\bibfnamefont {M.}~\bibnamefont {Bal}}, \bibinfo {author} {\bibfnamefont
  {C.}~\bibnamefont {McRae}}, \bibinfo {author} {\bibfnamefont
  {J.}~\bibnamefont {Long}},\ and\ \bibinfo {author} {\bibfnamefont
  {D.}~\bibnamefont {Pappas}},\ }\bibfield  {title} {\bibinfo {title}
  {Merged-element transmon},\ }\href
  {https://doi.org/10.1103/PhysRevApplied.14.064006} {\bibfield  {journal}
  {\bibinfo  {journal} {Phys. Rev. Appl.}\ }\textbf {\bibinfo {volume} {14}},\
  \bibinfo {pages} {064006} (\bibinfo {year} {2020})}\BibitemShut {NoStop}%
\bibitem [{\citenamefont {Mamin}\ \emph {et~al.}(2021)\citenamefont {Mamin},
  \citenamefont {Huang}, \citenamefont {Carnevale}, \citenamefont {Rettner},
  \citenamefont {Arellano}, \citenamefont {Sherwood}, \citenamefont {Kurter},
  \citenamefont {Trimm}, \citenamefont {Sandberg}, \citenamefont {Shelby},
  \citenamefont {Mueed}, \citenamefont {Madon}, \citenamefont {Pushp},
  \citenamefont {Steffen},\ and\ \citenamefont {Rugar}}]{Mamin21}%
  \BibitemOpen
  \bibfield  {author} {\bibinfo {author} {\bibfnamefont {H.}~\bibnamefont
  {Mamin}}, \bibinfo {author} {\bibfnamefont {E.}~\bibnamefont {Huang}},
  \bibinfo {author} {\bibfnamefont {S.}~\bibnamefont {Carnevale}}, \bibinfo
  {author} {\bibfnamefont {C.}~\bibnamefont {Rettner}}, \bibinfo {author}
  {\bibfnamefont {N.}~\bibnamefont {Arellano}}, \bibinfo {author}
  {\bibfnamefont {M.}~\bibnamefont {Sherwood}}, \bibinfo {author}
  {\bibfnamefont {C.}~\bibnamefont {Kurter}}, \bibinfo {author} {\bibfnamefont
  {B.}~\bibnamefont {Trimm}}, \bibinfo {author} {\bibfnamefont
  {M.}~\bibnamefont {Sandberg}}, \bibinfo {author} {\bibfnamefont
  {R.}~\bibnamefont {Shelby}}, \bibinfo {author} {\bibfnamefont
  {M.}~\bibnamefont {Mueed}}, \bibinfo {author} {\bibfnamefont
  {B.}~\bibnamefont {Madon}}, \bibinfo {author} {\bibfnamefont
  {A.}~\bibnamefont {Pushp}}, \bibinfo {author} {\bibfnamefont
  {M.}~\bibnamefont {Steffen}},\ and\ \bibinfo {author} {\bibfnamefont
  {D.}~\bibnamefont {Rugar}},\ }\bibfield  {title} {\bibinfo {title}
  {Merged-element transmons: Design and qubit performance},\ }\href
  {https://doi.org/10.1103/PhysRevApplied.16.024023} {\bibfield  {journal}
  {\bibinfo  {journal} {Phys. Rev. Applied}\ }\textbf {\bibinfo {volume}
  {16}},\ \bibinfo {pages} {024023} (\bibinfo {year} {2021})}\BibitemShut
  {NoStop}%
\bibitem [{\citenamefont {Keller}\ \emph {et~al.}(2017)\citenamefont {Keller},
  \citenamefont {Dieterle}, \citenamefont {Fang}, \citenamefont {Berger},
  \citenamefont {Fink},\ and\ \citenamefont {Painter}}]{Keller17}%
  \BibitemOpen
  \bibfield  {author} {\bibinfo {author} {\bibfnamefont {A.~J.}\ \bibnamefont
  {Keller}}, \bibinfo {author} {\bibfnamefont {P.~B.}\ \bibnamefont
  {Dieterle}}, \bibinfo {author} {\bibfnamefont {M.}~\bibnamefont {Fang}},
  \bibinfo {author} {\bibfnamefont {B.}~\bibnamefont {Berger}}, \bibinfo
  {author} {\bibfnamefont {J.~M.}\ \bibnamefont {Fink}},\ and\ \bibinfo
  {author} {\bibfnamefont {O.}~\bibnamefont {Painter}},\ }\bibfield  {title}
  {\bibinfo {title} {Al transmon qubits on silicon-on-insulator for quantum
  device integration},\ }\href {https://doi.org/10.1063/1.4994661} {\bibfield
  {journal} {\bibinfo  {journal} {apl}\ }\textbf {\bibinfo {volume} {111}},\
  \bibinfo {pages} {042603} (\bibinfo {year} {2017})}\BibitemShut {NoStop}%
\bibitem [{\citenamefont {Zhang}\ \emph {et~al.}(2017)\citenamefont {Zhang},
  \citenamefont {Li}, \citenamefont {Liu}, \citenamefont {Yu},\ and\
  \citenamefont {Yu}}]{Zhang17sr}%
  \BibitemOpen
  \bibfield  {author} {\bibinfo {author} {\bibfnamefont {K.}~\bibnamefont
  {Zhang}}, \bibinfo {author} {\bibfnamefont {M.-M.}\ \bibnamefont {Li}},
  \bibinfo {author} {\bibfnamefont {Q.}~\bibnamefont {Liu}}, \bibinfo {author}
  {\bibfnamefont {H.-F.}\ \bibnamefont {Yu}},\ and\ \bibinfo {author}
  {\bibfnamefont {Y.}~\bibnamefont {Yu}},\ }\bibfield  {title} {\bibinfo
  {title} {Bridge-free fabrication process for al/alox/al josephson
  junctions*},\ }\href {https://doi.org/10.1088/1674-1056/26/7/078501}
  {\bibfield  {journal} {\bibinfo  {journal} {Chinese Physics B}\ }\textbf
  {\bibinfo {volume} {26}},\ \bibinfo {pages} {078501} (\bibinfo {year}
  {2017})}\BibitemShut {NoStop}%
\bibitem [{\citenamefont {Ferreira}\ \emph {et~al.}(2022)\citenamefont
  {Ferreira}, \citenamefont {Kim}, \citenamefont {Butler}, \citenamefont
  {Pichler},\ and\ \citenamefont {Painter}}]{Ferreira22}%
  \BibitemOpen
  \bibfield  {author} {\bibinfo {author} {\bibfnamefont {V.~S.}\ \bibnamefont
  {Ferreira}}, \bibinfo {author} {\bibfnamefont {G.}~\bibnamefont {Kim}},
  \bibinfo {author} {\bibfnamefont {A.}~\bibnamefont {Butler}}, \bibinfo
  {author} {\bibfnamefont {H.}~\bibnamefont {Pichler}},\ and\ \bibinfo {author}
  {\bibfnamefont {O.}~\bibnamefont {Painter}},\ }\href@noop {} {\bibinfo
  {title} {Deterministic generation of multidimensional photonic cluster states
  with a single quantum emitter}} (\bibinfo {year} {2022}),\ \Eprint
  {https://arxiv.org/abs/2206.10076} {arXiv:2206.10076 [quant-ph]} \BibitemShut
  {NoStop}%
\bibitem [{\citenamefont {Zhang}\ \emph {et~al.}(2023)\citenamefont {Zhang},
  \citenamefont {Kim}, \citenamefont {Mark}, \citenamefont {Choi},\ and\
  \citenamefont {Painter}}]{Zhang23}%
  \BibitemOpen
  \bibfield  {author} {\bibinfo {author} {\bibfnamefont {X.}~\bibnamefont
  {Zhang}}, \bibinfo {author} {\bibfnamefont {E.}~\bibnamefont {Kim}}, \bibinfo
  {author} {\bibfnamefont {D.~K.}\ \bibnamefont {Mark}}, \bibinfo {author}
  {\bibfnamefont {S.}~\bibnamefont {Choi}},\ and\ \bibinfo {author}
  {\bibfnamefont {O.}~\bibnamefont {Painter}},\ }\bibfield  {title} {\bibinfo
  {title} {A superconducting quantum simulator based on a photonic-bandgap
  metamaterial},\ }\href {https://doi.org/10.1126/science.ade7651} {\bibfield
  {journal} {\bibinfo  {journal} {Science}\ }\textbf {\bibinfo {volume}
  {379}},\ \bibinfo {pages} {278} (\bibinfo {year} {2023})}\BibitemShut
  {NoStop}%
\bibitem [{\citenamefont {Rol}\ \emph {et~al.}(2020)\citenamefont {Rol},
  \citenamefont {Ciorciaro}, \citenamefont {Malinowski}, \citenamefont
  {Tarasinski}, \citenamefont {Sagastizabal}, \citenamefont {Bultink},
  \citenamefont {Salathe}, \citenamefont {Haandbaek}, \citenamefont {Sedivy},\
  and\ \citenamefont {DiCarlo}}]{Rol20}%
  \BibitemOpen
  \bibfield  {author} {\bibinfo {author} {\bibfnamefont {M.~A.}\ \bibnamefont
  {Rol}}, \bibinfo {author} {\bibfnamefont {L.}~\bibnamefont {Ciorciaro}},
  \bibinfo {author} {\bibfnamefont {F.~K.}\ \bibnamefont {Malinowski}},
  \bibinfo {author} {\bibfnamefont {B.~M.}\ \bibnamefont {Tarasinski}},
  \bibinfo {author} {\bibfnamefont {R.~E.}\ \bibnamefont {Sagastizabal}},
  \bibinfo {author} {\bibfnamefont {C.~C.}\ \bibnamefont {Bultink}}, \bibinfo
  {author} {\bibfnamefont {Y.}~\bibnamefont {Salathe}}, \bibinfo {author}
  {\bibfnamefont {N.}~\bibnamefont {Haandbaek}}, \bibinfo {author}
  {\bibfnamefont {J.}~\bibnamefont {Sedivy}},\ and\ \bibinfo {author}
  {\bibfnamefont {L.}~\bibnamefont {DiCarlo}},\ }\bibfield  {title} {\bibinfo
  {title} {{Time-domain characterization and correction of on-chip distortion
  of control pulses in a quantum processor}},\ }\href
  {https://doi.org/10.1063/1.5133894} {\bibfield  {journal} {\bibinfo
  {journal} {Applied Physics Letters}\ }\textbf {\bibinfo {volume} {116}},\
  \bibinfo {pages} {054001} (\bibinfo {year} {2020})}\BibitemShut {NoStop}%
\bibitem [{\citenamefont {Shalibo}\ \emph {et~al.}(2010)\citenamefont
  {Shalibo}, \citenamefont {Rofe}, \citenamefont {Shwa}, \citenamefont
  {Zeides}, \citenamefont {Neeley}, \citenamefont {Martinis},\ and\
  \citenamefont {Katz}}]{Shalibo10}%
  \BibitemOpen
  \bibfield  {author} {\bibinfo {author} {\bibfnamefont {Y.}~\bibnamefont
  {Shalibo}}, \bibinfo {author} {\bibfnamefont {Y.}~\bibnamefont {Rofe}},
  \bibinfo {author} {\bibfnamefont {D.}~\bibnamefont {Shwa}}, \bibinfo {author}
  {\bibfnamefont {F.}~\bibnamefont {Zeides}}, \bibinfo {author} {\bibfnamefont
  {M.}~\bibnamefont {Neeley}}, \bibinfo {author} {\bibfnamefont {J.~M.}\
  \bibnamefont {Martinis}},\ and\ \bibinfo {author} {\bibfnamefont
  {N.}~\bibnamefont {Katz}},\ }\bibfield  {title} {\bibinfo {title} {Lifetime
  and coherence of two-level defects in a josephson junction},\ }\href
  {https://doi.org/10.1103/PhysRevLett.105.177001} {\bibfield  {journal}
  {\bibinfo  {journal} {Phys. Rev. Lett.}\ }\textbf {\bibinfo {volume} {105}},\
  \bibinfo {pages} {177001} (\bibinfo {year} {2010})}\BibitemShut {NoStop}%
\bibitem [{\citenamefont {Wang}\ \emph {et~al.}(2014)\citenamefont {Wang},
  \citenamefont {Gao}, \citenamefont {Pop}, \citenamefont {Vool}, \citenamefont
  {Axline}, \citenamefont {Brecht}, \citenamefont {Heeres}, \citenamefont
  {Frunzio}, \citenamefont {Devoret}, \citenamefont {Catelani}, \citenamefont
  {Glazman},\ and\ \citenamefont {Schoelkopf}}]{Wang14}%
  \BibitemOpen
  \bibfield  {author} {\bibinfo {author} {\bibfnamefont {C.}~\bibnamefont
  {Wang}}, \bibinfo {author} {\bibfnamefont {Y.~Y.}\ \bibnamefont {Gao}},
  \bibinfo {author} {\bibfnamefont {I.~M.}\ \bibnamefont {Pop}}, \bibinfo
  {author} {\bibfnamefont {U.}~\bibnamefont {Vool}}, \bibinfo {author}
  {\bibfnamefont {C.}~\bibnamefont {Axline}}, \bibinfo {author} {\bibfnamefont
  {T.}~\bibnamefont {Brecht}}, \bibinfo {author} {\bibfnamefont {R.~W.}\
  \bibnamefont {Heeres}}, \bibinfo {author} {\bibfnamefont {L.}~\bibnamefont
  {Frunzio}}, \bibinfo {author} {\bibfnamefont {M.~H.}\ \bibnamefont
  {Devoret}}, \bibinfo {author} {\bibfnamefont {G.}~\bibnamefont {Catelani}},
  \bibinfo {author} {\bibfnamefont {L.~I.}\ \bibnamefont {Glazman}},\ and\
  \bibinfo {author} {\bibfnamefont {R.~J.}\ \bibnamefont {Schoelkopf}},\
  }\bibfield  {title} {\bibinfo {title} {Measurement and control of
  quasiparticle dynamics in a superconducting qubit},\ }\href
  {https://doi.org/10.1038/ncomms6836} {\bibfield  {journal} {\bibinfo
  {journal} {Nature Communications}\ }\textbf {\bibinfo {volume} {5}},\
  \bibinfo {pages} {5836} (\bibinfo {year} {2014})}\BibitemShut {NoStop}%
\bibitem [{\citenamefont {Vool}\ \emph {et~al.}(2014)\citenamefont {Vool},
  \citenamefont {Pop}, \citenamefont {Sliwa}, \citenamefont {Abdo},
  \citenamefont {Wang}, \citenamefont {Brecht}, \citenamefont {Gao},
  \citenamefont {Shankar}, \citenamefont {Hatridge}, \citenamefont {Catelani},
  \citenamefont {Mirrahimi}, \citenamefont {Frunzio}, \citenamefont
  {Schoelkopf}, \citenamefont {Glazman},\ and\ \citenamefont
  {Devoret}}]{Vool14}%
  \BibitemOpen
  \bibfield  {author} {\bibinfo {author} {\bibfnamefont {U.}~\bibnamefont
  {Vool}}, \bibinfo {author} {\bibfnamefont {I.~M.}\ \bibnamefont {Pop}},
  \bibinfo {author} {\bibfnamefont {K.}~\bibnamefont {Sliwa}}, \bibinfo
  {author} {\bibfnamefont {B.}~\bibnamefont {Abdo}}, \bibinfo {author}
  {\bibfnamefont {C.}~\bibnamefont {Wang}}, \bibinfo {author} {\bibfnamefont
  {T.}~\bibnamefont {Brecht}}, \bibinfo {author} {\bibfnamefont {Y.~Y.}\
  \bibnamefont {Gao}}, \bibinfo {author} {\bibfnamefont {S.}~\bibnamefont
  {Shankar}}, \bibinfo {author} {\bibfnamefont {M.}~\bibnamefont {Hatridge}},
  \bibinfo {author} {\bibfnamefont {G.}~\bibnamefont {Catelani}}, \bibinfo
  {author} {\bibfnamefont {M.}~\bibnamefont {Mirrahimi}}, \bibinfo {author}
  {\bibfnamefont {L.}~\bibnamefont {Frunzio}}, \bibinfo {author} {\bibfnamefont
  {R.~J.}\ \bibnamefont {Schoelkopf}}, \bibinfo {author} {\bibfnamefont
  {L.~I.}\ \bibnamefont {Glazman}},\ and\ \bibinfo {author} {\bibfnamefont
  {M.~H.}\ \bibnamefont {Devoret}},\ }\bibfield  {title} {\bibinfo {title}
  {Non-poissonian quantum jumps of a fluxonium qubit due to quasiparticle
  excitations},\ }\href {https://doi.org/10.1103/PhysRevLett.113.247001}
  {\bibfield  {journal} {\bibinfo  {journal} {Phys. Rev. Lett.}\ }\textbf
  {\bibinfo {volume} {113}},\ \bibinfo {pages} {247001} (\bibinfo {year}
  {2014})}\BibitemShut {NoStop}%
\bibitem [{\citenamefont {Lisenfeld}\ \emph {et~al.}(2019)\citenamefont
  {Lisenfeld}, \citenamefont {Bilmes}, \citenamefont {Megrant}, \citenamefont
  {Barends}, \citenamefont {Kelly}, \citenamefont {Klimov}, \citenamefont
  {Weiss}, \citenamefont {Martinis},\ and\ \citenamefont
  {Ustinov}}]{Lisenfeld19}%
  \BibitemOpen
  \bibfield  {author} {\bibinfo {author} {\bibfnamefont {J.}~\bibnamefont
  {Lisenfeld}}, \bibinfo {author} {\bibfnamefont {A.}~\bibnamefont {Bilmes}},
  \bibinfo {author} {\bibfnamefont {A.}~\bibnamefont {Megrant}}, \bibinfo
  {author} {\bibfnamefont {R.}~\bibnamefont {Barends}}, \bibinfo {author}
  {\bibfnamefont {J.}~\bibnamefont {Kelly}}, \bibinfo {author} {\bibfnamefont
  {P.}~\bibnamefont {Klimov}}, \bibinfo {author} {\bibfnamefont
  {G.}~\bibnamefont {Weiss}}, \bibinfo {author} {\bibfnamefont {J.~M.}\
  \bibnamefont {Martinis}},\ and\ \bibinfo {author} {\bibfnamefont {A.~V.}\
  \bibnamefont {Ustinov}},\ }\bibfield  {title} {\bibinfo {title} {Electric
  field spectroscopy of material defects in transmon qubits},\ }\href
  {https://doi.org/10.1038/s41534-019-0224-1} {\bibfield  {journal} {\bibinfo
  {journal} {npj Quantum Information}\ }\textbf {\bibinfo {volume} {5}},\
  \bibinfo {pages} {105} (\bibinfo {year} {2019})}\BibitemShut {NoStop}%
\bibitem [{\citenamefont {Phillips}(1972)}]{Phillips72}%
  \BibitemOpen
  \bibfield  {author} {\bibinfo {author} {\bibfnamefont {W.~A.}\ \bibnamefont
  {Phillips}},\ }\bibfield  {title} {\bibinfo {title} {Tunneling states in
  amorphous solids},\ }\href {https://doi.org/10.1007/BF00660072} {\bibfield
  {journal} {\bibinfo  {journal} {Journal of Low Temperature Physics}\ }\textbf
  {\bibinfo {volume} {7}},\ \bibinfo {pages} {351} (\bibinfo {year}
  {1972})}\BibitemShut {NoStop}%
\bibitem [{\citenamefont {Anderson}\ \emph {et~al.}(1972)\citenamefont
  {Anderson}, \citenamefont {Halperin},\ and\ \citenamefont
  {Varma}}]{Anderson72}%
  \BibitemOpen
  \bibfield  {author} {\bibinfo {author} {\bibfnamefont {P.~W.}\ \bibnamefont
  {Anderson}}, \bibinfo {author} {\bibfnamefont {B.~I.}\ \bibnamefont
  {Halperin}},\ and\ \bibinfo {author} {\bibfnamefont {C.~M.}\ \bibnamefont
  {Varma}},\ }\bibfield  {title} {\bibinfo {title} {Anomalous low-temperature
  thermal properties of glasses and spin glasses},\ }\href
  {https://doi.org/10.1080/14786437208229210} {\bibfield  {journal} {\bibinfo
  {journal} {The Philosophical Magazine: A Journal of Theoretical Experimental
  and Applied Physics}\ }\textbf {\bibinfo {volume} {25}},\ \bibinfo {pages}
  {1} (\bibinfo {year} {1972})}\BibitemShut {NoStop}%
\bibitem [{\citenamefont {Behunin}\ \emph {et~al.}(2016)\citenamefont
  {Behunin}, \citenamefont {Intravaia},\ and\ \citenamefont
  {Rakich}}]{Behunin16}%
  \BibitemOpen
  \bibfield  {author} {\bibinfo {author} {\bibfnamefont {R.~O.}\ \bibnamefont
  {Behunin}}, \bibinfo {author} {\bibfnamefont {F.}~\bibnamefont {Intravaia}},\
  and\ \bibinfo {author} {\bibfnamefont {P.~T.}\ \bibnamefont {Rakich}},\
  }\bibfield  {title} {\bibinfo {title} {Dimensional transformation of
  defect-induced noise, dissipation, and nonlinearity},\ }\href
  {https://doi.org/10.1103/PhysRevB.93.224110} {\bibfield  {journal} {\bibinfo
  {journal} {Phys. Rev. B}\ }\textbf {\bibinfo {volume} {93}},\ \bibinfo
  {pages} {224110} (\bibinfo {year} {2016})}\BibitemShut {NoStop}%
\bibitem [{\citenamefont {MacCabe}\ \emph {et~al.}(2020)\citenamefont
  {MacCabe}, \citenamefont {Ren}, \citenamefont {Luo}, \citenamefont {Cohen},
  \citenamefont {Zhou}, \citenamefont {Sipahigil}, \citenamefont
  {Mirhosseini},\ and\ \citenamefont {Painter}}]{MacCabe20}%
  \BibitemOpen
  \bibfield  {author} {\bibinfo {author} {\bibfnamefont {G.~S.}\ \bibnamefont
  {MacCabe}}, \bibinfo {author} {\bibfnamefont {H.}~\bibnamefont {Ren}},
  \bibinfo {author} {\bibfnamefont {J.}~\bibnamefont {Luo}}, \bibinfo {author}
  {\bibfnamefont {J.~D.}\ \bibnamefont {Cohen}}, \bibinfo {author}
  {\bibfnamefont {H.}~\bibnamefont {Zhou}}, \bibinfo {author} {\bibfnamefont
  {A.}~\bibnamefont {Sipahigil}}, \bibinfo {author} {\bibfnamefont
  {M.}~\bibnamefont {Mirhosseini}},\ and\ \bibinfo {author} {\bibfnamefont
  {O.}~\bibnamefont {Painter}},\ }\bibfield  {title} {\bibinfo {title}
  {Nano-acoustic resonator with ultralong phonon lifetime},\ }\href
  {https://doi.org/10.1126/science.abc7312} {\bibfield  {journal} {\bibinfo
  {journal} {Science}\ }\textbf {\bibinfo {volume} {370}},\ \bibinfo {pages}
  {840} (\bibinfo {year} {2020})}\BibitemShut {NoStop}%
\bibitem [{\citenamefont {Wollack}\ \emph {et~al.}(2021)\citenamefont
  {Wollack}, \citenamefont {Cleland}, \citenamefont {Arrangoiz-Arriola},
  \citenamefont {McKenna}, \citenamefont {Gruenke}, \citenamefont {Patel},
  \citenamefont {Jiang}, \citenamefont {Sarabalis},\ and\ \citenamefont
  {Safavi-Naeini}}]{Wollack21}%
  \BibitemOpen
  \bibfield  {author} {\bibinfo {author} {\bibfnamefont {E.~A.}\ \bibnamefont
  {Wollack}}, \bibinfo {author} {\bibfnamefont {A.~Y.}\ \bibnamefont
  {Cleland}}, \bibinfo {author} {\bibfnamefont {P.}~\bibnamefont
  {Arrangoiz-Arriola}}, \bibinfo {author} {\bibfnamefont {T.~P.}\ \bibnamefont
  {McKenna}}, \bibinfo {author} {\bibfnamefont {R.~G.}\ \bibnamefont
  {Gruenke}}, \bibinfo {author} {\bibfnamefont {R.~N.}\ \bibnamefont {Patel}},
  \bibinfo {author} {\bibfnamefont {W.}~\bibnamefont {Jiang}}, \bibinfo
  {author} {\bibfnamefont {C.~J.}\ \bibnamefont {Sarabalis}},\ and\ \bibinfo
  {author} {\bibfnamefont {A.~H.}\ \bibnamefont {Safavi-Naeini}},\ }\bibfield
  {title} {\bibinfo {title} {{Loss channels affecting lithium niobate phononic
  crystal resonators at cryogenic temperature}},\ }\href
  {https://doi.org/10.1063/5.0034909} {\bibfield  {journal} {\bibinfo
  {journal} {Applied Physics Letters}\ }\textbf {\bibinfo {volume} {118}},\
  \bibinfo {pages} {123501} (\bibinfo {year} {2021})}\BibitemShut {NoStop}%
\bibitem [{\citenamefont {Lupa\ifmmode~\mbox{\c{s}}\else \c{s}\fi{}cu}\ \emph
  {et~al.}(2009)\citenamefont {Lupa\ifmmode~\mbox{\c{s}}\else \c{s}\fi{}cu},
  \citenamefont {Bertet}, \citenamefont {Driessen}, \citenamefont {Harmans},\
  and\ \citenamefont {Mooij}}]{Lupascu09}%
  \BibitemOpen
  \bibfield  {author} {\bibinfo {author} {\bibfnamefont {A.}~\bibnamefont
  {Lupa\ifmmode~\mbox{\c{s}}\else \c{s}\fi{}cu}}, \bibinfo {author}
  {\bibfnamefont {P.}~\bibnamefont {Bertet}}, \bibinfo {author} {\bibfnamefont
  {E.~F.~C.}\ \bibnamefont {Driessen}}, \bibinfo {author} {\bibfnamefont {C.~J.
  P.~M.}\ \bibnamefont {Harmans}},\ and\ \bibinfo {author} {\bibfnamefont
  {J.~E.}\ \bibnamefont {Mooij}},\ }\bibfield  {title} {\bibinfo {title} {One-
  and two-photon spectroscopy of a flux qubit coupled to a microscopic
  defect},\ }\href {https://doi.org/10.1103/PhysRevB.80.172506} {\bibfield
  {journal} {\bibinfo  {journal} {Phys. Rev. B}\ }\textbf {\bibinfo {volume}
  {80}},\ \bibinfo {pages} {172506} (\bibinfo {year} {2009})}\BibitemShut
  {NoStop}%
\bibitem [{\citenamefont {Bushev}\ \emph {et~al.}(2010)\citenamefont {Bushev},
  \citenamefont {M\"uller}, \citenamefont {Lisenfeld}, \citenamefont {Cole},
  \citenamefont {Lukashenko}, \citenamefont {Shnirman},\ and\ \citenamefont
  {Ustinov}}]{Bushev10}%
  \BibitemOpen
  \bibfield  {author} {\bibinfo {author} {\bibfnamefont {P.}~\bibnamefont
  {Bushev}}, \bibinfo {author} {\bibfnamefont {C.}~\bibnamefont {M\"uller}},
  \bibinfo {author} {\bibfnamefont {J.}~\bibnamefont {Lisenfeld}}, \bibinfo
  {author} {\bibfnamefont {J.~H.}\ \bibnamefont {Cole}}, \bibinfo {author}
  {\bibfnamefont {A.}~\bibnamefont {Lukashenko}}, \bibinfo {author}
  {\bibfnamefont {A.}~\bibnamefont {Shnirman}},\ and\ \bibinfo {author}
  {\bibfnamefont {A.~V.}\ \bibnamefont {Ustinov}},\ }\bibfield  {title}
  {\bibinfo {title} {Multiphoton spectroscopy of a hybrid quantum system},\
  }\href {https://doi.org/10.1103/PhysRevB.82.134530} {\bibfield  {journal}
  {\bibinfo  {journal} {Phys. Rev. B}\ }\textbf {\bibinfo {volume} {82}},\
  \bibinfo {pages} {134530} (\bibinfo {year} {2010})}\BibitemShut {NoStop}%
\bibitem [{\citenamefont {Cole}\ \emph {et~al.}(2010)\citenamefont {Cole},
  \citenamefont {M\"uller}, \citenamefont {Bushev}, \citenamefont {Grabovskij},
  \citenamefont {Lisenfeld}, \citenamefont {Lukashenko}, \citenamefont
  {Ustinov},\ and\ \citenamefont {Shnirman}}]{Cole10}%
  \BibitemOpen
  \bibfield  {author} {\bibinfo {author} {\bibfnamefont {J.~H.}\ \bibnamefont
  {Cole}}, \bibinfo {author} {\bibfnamefont {C.}~\bibnamefont {M\"uller}},
  \bibinfo {author} {\bibfnamefont {P.}~\bibnamefont {Bushev}}, \bibinfo
  {author} {\bibfnamefont {G.~J.}\ \bibnamefont {Grabovskij}}, \bibinfo
  {author} {\bibfnamefont {J.}~\bibnamefont {Lisenfeld}}, \bibinfo {author}
  {\bibfnamefont {A.}~\bibnamefont {Lukashenko}}, \bibinfo {author}
  {\bibfnamefont {A.~V.}\ \bibnamefont {Ustinov}},\ and\ \bibinfo {author}
  {\bibfnamefont {A.}~\bibnamefont {Shnirman}},\ }\bibfield  {title} {\bibinfo
  {title} {{Quantitative evaluation of defect-models in superconducting phase
  qubits}},\ }\href {https://doi.org/10.1063/1.3529457} {\bibfield  {journal}
  {\bibinfo  {journal} {Applied Physics Letters}\ }\textbf {\bibinfo {volume}
  {97}},\ \bibinfo {pages} {252501} (\bibinfo {year} {2010})}\BibitemShut
  {NoStop}%
\bibitem [{\citenamefont {M\"{u}ller}\ \emph {et~al.}(2019)\citenamefont
  {M\"{u}ller}, \citenamefont {Cole},\ and\ \citenamefont
  {Lisenfeld}}]{Muller19}%
  \BibitemOpen
  \bibfield  {author} {\bibinfo {author} {\bibfnamefont {C.}~\bibnamefont
  {M\"{u}ller}}, \bibinfo {author} {\bibfnamefont {J.~H.}\ \bibnamefont
  {Cole}},\ and\ \bibinfo {author} {\bibfnamefont {J.}~\bibnamefont
  {Lisenfeld}},\ }\bibfield  {title} {\bibinfo {title} {Towards understanding
  two-level-systems in amorphous solids: insights from quantum circuits},\
  }\href {https://doi.org/10.1088/1361-6633/ab3a7e} {\bibfield  {journal}
  {\bibinfo  {journal} {Reports on Progress in Physics}\ }\textbf {\bibinfo
  {volume} {82}},\ \bibinfo {pages} {124501} (\bibinfo {year}
  {2019})}\BibitemShut {NoStop}%
\bibitem [{\citenamefont {Klimov}\ \emph {et~al.}(2018)\citenamefont {Klimov},
  \citenamefont {Kelly}, \citenamefont {Chen}, \citenamefont {Neeley},
  \citenamefont {Megrant}, \citenamefont {Burkett}, \citenamefont {Barends},
  \citenamefont {Arya}, \citenamefont {Chiaro}, \citenamefont {Chen},
  \citenamefont {Dunsworth}, \citenamefont {Fowler}, \citenamefont {Foxen},
  \citenamefont {Gidney}, \citenamefont {Giustina}, \citenamefont {Graff},
  \citenamefont {Huang}, \citenamefont {Jeffrey}, \citenamefont {Lucero},
  \citenamefont {Mutus}, \citenamefont {Naaman}, \citenamefont {Neill},
  \citenamefont {Quintana}, \citenamefont {Roushan}, \citenamefont {Sank},
  \citenamefont {Vainsencher}, \citenamefont {Wenner}, \citenamefont {White},
  \citenamefont {Boixo}, \citenamefont {Babbush}, \citenamefont {Smelyanskiy},
  \citenamefont {Neven},\ and\ \citenamefont {Martinis}}]{Klimov18}%
  \BibitemOpen
  \bibfield  {author} {\bibinfo {author} {\bibfnamefont {P.~V.}\ \bibnamefont
  {Klimov}}, \bibinfo {author} {\bibfnamefont {J.}~\bibnamefont {Kelly}},
  \bibinfo {author} {\bibfnamefont {Z.}~\bibnamefont {Chen}}, \bibinfo {author}
  {\bibfnamefont {M.}~\bibnamefont {Neeley}}, \bibinfo {author} {\bibfnamefont
  {A.}~\bibnamefont {Megrant}}, \bibinfo {author} {\bibfnamefont
  {B.}~\bibnamefont {Burkett}}, \bibinfo {author} {\bibfnamefont
  {R.}~\bibnamefont {Barends}}, \bibinfo {author} {\bibfnamefont
  {K.}~\bibnamefont {Arya}}, \bibinfo {author} {\bibfnamefont {B.}~\bibnamefont
  {Chiaro}}, \bibinfo {author} {\bibfnamefont {Y.}~\bibnamefont {Chen}},
  \bibinfo {author} {\bibfnamefont {A.}~\bibnamefont {Dunsworth}}, \bibinfo
  {author} {\bibfnamefont {A.}~\bibnamefont {Fowler}}, \bibinfo {author}
  {\bibfnamefont {B.}~\bibnamefont {Foxen}}, \bibinfo {author} {\bibfnamefont
  {C.}~\bibnamefont {Gidney}}, \bibinfo {author} {\bibfnamefont
  {M.}~\bibnamefont {Giustina}}, \bibinfo {author} {\bibfnamefont
  {R.}~\bibnamefont {Graff}}, \bibinfo {author} {\bibfnamefont
  {T.}~\bibnamefont {Huang}}, \bibinfo {author} {\bibfnamefont
  {E.}~\bibnamefont {Jeffrey}}, \bibinfo {author} {\bibfnamefont
  {E.}~\bibnamefont {Lucero}}, \bibinfo {author} {\bibfnamefont {J.~Y.}\
  \bibnamefont {Mutus}}, \bibinfo {author} {\bibfnamefont {O.}~\bibnamefont
  {Naaman}}, \bibinfo {author} {\bibfnamefont {C.}~\bibnamefont {Neill}},
  \bibinfo {author} {\bibfnamefont {C.}~\bibnamefont {Quintana}}, \bibinfo
  {author} {\bibfnamefont {P.}~\bibnamefont {Roushan}}, \bibinfo {author}
  {\bibfnamefont {D.}~\bibnamefont {Sank}}, \bibinfo {author} {\bibfnamefont
  {A.}~\bibnamefont {Vainsencher}}, \bibinfo {author} {\bibfnamefont
  {J.}~\bibnamefont {Wenner}}, \bibinfo {author} {\bibfnamefont {T.~C.}\
  \bibnamefont {White}}, \bibinfo {author} {\bibfnamefont {S.}~\bibnamefont
  {Boixo}}, \bibinfo {author} {\bibfnamefont {R.}~\bibnamefont {Babbush}},
  \bibinfo {author} {\bibfnamefont {V.~N.}\ \bibnamefont {Smelyanskiy}},
  \bibinfo {author} {\bibfnamefont {H.}~\bibnamefont {Neven}},\ and\ \bibinfo
  {author} {\bibfnamefont {J.~M.}\ \bibnamefont {Martinis}},\ }\bibfield
  {title} {\bibinfo {title} {Fluctuations of energy-relaxation times in
  superconducting qubits},\ }\href
  {https://doi.org/10.1103/PhysRevLett.121.090502} {\bibfield  {journal}
  {\bibinfo  {journal} {Phys. Rev. Lett.}\ }\textbf {\bibinfo {volume} {121}},\
  \bibinfo {pages} {090502} (\bibinfo {year} {2018})}\BibitemShut {NoStop}%
\bibitem [{\citenamefont {Grabovskij}\ \emph {et~al.}(2011)\citenamefont
  {Grabovskij}, \citenamefont {Bushev}, \citenamefont {Cole}, \citenamefont
  {M\"uller}, \citenamefont {Lisenfeld}, \citenamefont {Lukashenko},\ and\
  \citenamefont {Ustinov}}]{Grabovskij11}%
  \BibitemOpen
  \bibfield  {author} {\bibinfo {author} {\bibfnamefont {G.~J.}\ \bibnamefont
  {Grabovskij}}, \bibinfo {author} {\bibfnamefont {P.}~\bibnamefont {Bushev}},
  \bibinfo {author} {\bibfnamefont {J.~H.}\ \bibnamefont {Cole}}, \bibinfo
  {author} {\bibfnamefont {C.}~\bibnamefont {M\"uller}}, \bibinfo {author}
  {\bibfnamefont {J.}~\bibnamefont {Lisenfeld}}, \bibinfo {author}
  {\bibfnamefont {A.}~\bibnamefont {Lukashenko}},\ and\ \bibinfo {author}
  {\bibfnamefont {A.~V.}\ \bibnamefont {Ustinov}},\ }\bibfield  {title}
  {\bibinfo {title} {Entangling microscopic defects via a macroscopic quantum
  shuttle},\ }\href {https://doi.org/10.1088/1367-2630/13/6/063015} {\bibfield
  {journal} {\bibinfo  {journal} {New Journal of Physics}\ }\textbf {\bibinfo
  {volume} {13}},\ \bibinfo {pages} {063015} (\bibinfo {year}
  {2011})}\BibitemShut {NoStop}%
\bibitem [{\citenamefont {Gao}(2008)}]{Gao08}%
  \BibitemOpen
  \bibfield  {author} {\bibinfo {author} {\bibfnamefont {J.}~\bibnamefont
  {Gao}},\ }\emph {\bibinfo {title} {The Physics of Superconducting Microwave
  Resonators}},\ \href
  {https://thesis.library.caltech.edu/2530/1/thesismain_0610.pdf} {Ph.D.
  thesis},\ \bibinfo  {school} {California Institute of Technology} (\bibinfo
  {year} {2008})\BibitemShut {NoStop}%
\bibitem [{\citenamefont {Catelani}\ \emph
  {et~al.}(2011{\natexlab{a}})\citenamefont {Catelani}, \citenamefont {Koch},
  \citenamefont {Frunzio}, \citenamefont {Schoelkopf}, \citenamefont
  {Devoret},\ and\ \citenamefont {Glazman}}]{Catelani11}%
  \BibitemOpen
  \bibfield  {author} {\bibinfo {author} {\bibfnamefont {G.}~\bibnamefont
  {Catelani}}, \bibinfo {author} {\bibfnamefont {J.}~\bibnamefont {Koch}},
  \bibinfo {author} {\bibfnamefont {L.}~\bibnamefont {Frunzio}}, \bibinfo
  {author} {\bibfnamefont {R.~J.}\ \bibnamefont {Schoelkopf}}, \bibinfo
  {author} {\bibfnamefont {M.~H.}\ \bibnamefont {Devoret}},\ and\ \bibinfo
  {author} {\bibfnamefont {L.~I.}\ \bibnamefont {Glazman}},\ }\bibfield
  {title} {\bibinfo {title} {Quasiparticle relaxation of superconducting qubits
  in the presence of flux},\ }\href
  {https://doi.org/10.1103/PhysRevLett.106.077002} {\bibfield  {journal}
  {\bibinfo  {journal} {Phys. Rev. Lett.}\ }\textbf {\bibinfo {volume} {106}},\
  \bibinfo {pages} {077002} (\bibinfo {year} {2011}{\natexlab{a}})}\BibitemShut
  {NoStop}%
\bibitem [{\citenamefont {Crowley}\ \emph {et~al.}(2023)\citenamefont
  {Crowley}, \citenamefont {McLellan}, \citenamefont {Dutta}, \citenamefont
  {Shumiya}, \citenamefont {Place}, \citenamefont {Le}, \citenamefont {Gang},
  \citenamefont {Madhavan}, \citenamefont {Khedkar}, \citenamefont {Feng},
  \citenamefont {Umbarkar}, \citenamefont {Gui}, \citenamefont {Rodgers},
  \citenamefont {Jia}, \citenamefont {Feldman}, \citenamefont {Lyon},
  \citenamefont {Liu}, \citenamefont {Cava}, \citenamefont {Houck},\ and\
  \citenamefont {de~Leon}}]{Crowley23}%
  \BibitemOpen
  \bibfield  {author} {\bibinfo {author} {\bibfnamefont {K.~D.}\ \bibnamefont
  {Crowley}}, \bibinfo {author} {\bibfnamefont {R.~A.}\ \bibnamefont
  {McLellan}}, \bibinfo {author} {\bibfnamefont {A.}~\bibnamefont {Dutta}},
  \bibinfo {author} {\bibfnamefont {N.}~\bibnamefont {Shumiya}}, \bibinfo
  {author} {\bibfnamefont {A.~P.~M.}\ \bibnamefont {Place}}, \bibinfo {author}
  {\bibfnamefont {X.~H.}\ \bibnamefont {Le}}, \bibinfo {author} {\bibfnamefont
  {Y.}~\bibnamefont {Gang}}, \bibinfo {author} {\bibfnamefont {T.}~\bibnamefont
  {Madhavan}}, \bibinfo {author} {\bibfnamefont {N.}~\bibnamefont {Khedkar}},
  \bibinfo {author} {\bibfnamefont {Y.~C.}\ \bibnamefont {Feng}}, \bibinfo
  {author} {\bibfnamefont {E.~A.}\ \bibnamefont {Umbarkar}}, \bibinfo {author}
  {\bibfnamefont {X.}~\bibnamefont {Gui}}, \bibinfo {author} {\bibfnamefont
  {L.~V.~H.}\ \bibnamefont {Rodgers}}, \bibinfo {author} {\bibfnamefont
  {Y.}~\bibnamefont {Jia}}, \bibinfo {author} {\bibfnamefont {M.~M.}\
  \bibnamefont {Feldman}}, \bibinfo {author} {\bibfnamefont {S.~A.}\
  \bibnamefont {Lyon}}, \bibinfo {author} {\bibfnamefont {M.}~\bibnamefont
  {Liu}}, \bibinfo {author} {\bibfnamefont {R.~J.}\ \bibnamefont {Cava}},
  \bibinfo {author} {\bibfnamefont {A.~A.}\ \bibnamefont {Houck}},\ and\
  \bibinfo {author} {\bibfnamefont {N.~P.}\ \bibnamefont {de~Leon}},\
  }\href@noop {} {\bibinfo {title} {Disentangling losses in tantalum
  superconducting circuits}} (\bibinfo {year} {2023}),\ \Eprint
  {https://arxiv.org/abs/2301.07848} {arXiv:2301.07848 [quant-ph]} \BibitemShut
  {NoStop}%
\bibitem [{\citenamefont {Lucas}\ \emph {et~al.}(2023)\citenamefont {Lucas},
  \citenamefont {Danilov}, \citenamefont {Levitin}, \citenamefont {Jayaraman},
  \citenamefont {Casey}, \citenamefont {Faoro}, \citenamefont {Tzalenchuk},
  \citenamefont {Kubatkin}, \citenamefont {Saunders},\ and\ \citenamefont
  {de~Graaf}}]{Lucas23}%
  \BibitemOpen
  \bibfield  {author} {\bibinfo {author} {\bibfnamefont {M.}~\bibnamefont
  {Lucas}}, \bibinfo {author} {\bibfnamefont {A.~V.}\ \bibnamefont {Danilov}},
  \bibinfo {author} {\bibfnamefont {L.~V.}\ \bibnamefont {Levitin}}, \bibinfo
  {author} {\bibfnamefont {A.}~\bibnamefont {Jayaraman}}, \bibinfo {author}
  {\bibfnamefont {A.~J.}\ \bibnamefont {Casey}}, \bibinfo {author}
  {\bibfnamefont {L.}~\bibnamefont {Faoro}}, \bibinfo {author} {\bibfnamefont
  {A.~Y.}\ \bibnamefont {Tzalenchuk}}, \bibinfo {author} {\bibfnamefont
  {S.~E.}\ \bibnamefont {Kubatkin}}, \bibinfo {author} {\bibfnamefont
  {J.}~\bibnamefont {Saunders}},\ and\ \bibinfo {author} {\bibfnamefont
  {S.~E.}\ \bibnamefont {de~Graaf}},\ }\bibfield  {title} {\bibinfo {title}
  {Quantum bath suppression in a superconducting circuit by immersion
  cooling},\ }\href {https://doi.org/10.1038/s41467-023-39249-z} {\bibfield
  {journal} {\bibinfo  {journal} {Nature Communications}\ }\textbf {\bibinfo
  {volume} {14}},\ \bibinfo {pages} {3522} (\bibinfo {year}
  {2023})}\BibitemShut {NoStop}%
\bibitem [{\citenamefont {Veps{\"a}l{\"a}inen}\ \emph
  {et~al.}(2020)\citenamefont {Veps{\"a}l{\"a}inen}, \citenamefont {Karamlou},
  \citenamefont {Orrell}, \citenamefont {Dogra}, \citenamefont {Loer},
  \citenamefont {Vasconcelos}, \citenamefont {Kim}, \citenamefont {Melville},
  \citenamefont {Niedzielski}, \citenamefont {Yoder}, \citenamefont
  {Gustavsson}, \citenamefont {Formaggio}, \citenamefont {VanDevender},\ and\
  \citenamefont {Oliver}}]{Vepsalainen20}%
  \BibitemOpen
  \bibfield  {author} {\bibinfo {author} {\bibfnamefont {A.~P.}\ \bibnamefont
  {Veps{\"a}l{\"a}inen}}, \bibinfo {author} {\bibfnamefont {A.~H.}\
  \bibnamefont {Karamlou}}, \bibinfo {author} {\bibfnamefont {J.~L.}\
  \bibnamefont {Orrell}}, \bibinfo {author} {\bibfnamefont {A.~S.}\
  \bibnamefont {Dogra}}, \bibinfo {author} {\bibfnamefont {B.}~\bibnamefont
  {Loer}}, \bibinfo {author} {\bibfnamefont {F.}~\bibnamefont {Vasconcelos}},
  \bibinfo {author} {\bibfnamefont {D.~K.}\ \bibnamefont {Kim}}, \bibinfo
  {author} {\bibfnamefont {A.~J.}\ \bibnamefont {Melville}}, \bibinfo {author}
  {\bibfnamefont {B.~M.}\ \bibnamefont {Niedzielski}}, \bibinfo {author}
  {\bibfnamefont {J.~L.}\ \bibnamefont {Yoder}}, \bibinfo {author}
  {\bibfnamefont {S.}~\bibnamefont {Gustavsson}}, \bibinfo {author}
  {\bibfnamefont {J.~A.}\ \bibnamefont {Formaggio}}, \bibinfo {author}
  {\bibfnamefont {B.~A.}\ \bibnamefont {VanDevender}},\ and\ \bibinfo {author}
  {\bibfnamefont {W.~D.}\ \bibnamefont {Oliver}},\ }\bibfield  {title}
  {\bibinfo {title} {Impact of ionizing radiation on superconducting qubit
  coherence},\ }\href {https://doi.org/10.1038/s41586-020-2619-8} {\bibfield
  {journal} {\bibinfo  {journal} {Nature}\ }\textbf {\bibinfo {volume} {584}},\
  \bibinfo {pages} {551} (\bibinfo {year} {2020})}\BibitemShut {NoStop}%
\bibitem [{SOM()}]{SOM}%
  \BibitemOpen
  \href@noop {} {}\bibinfo {howpublished} {See supplementary online
  material.}\BibitemShut {Stop}%
\bibitem [{\citenamefont {Gustavsson}\ \emph {et~al.}(2016)\citenamefont
  {Gustavsson}, \citenamefont {Yan}, \citenamefont {Catelani}, \citenamefont
  {Bylander}, \citenamefont {Kamal}, \citenamefont {Birenbaum}, \citenamefont
  {Hover}, \citenamefont {Rosenberg}, \citenamefont {Samach}, \citenamefont
  {Sears}, \citenamefont {Weber}, \citenamefont {Yoder}, \citenamefont
  {Clarke}, \citenamefont {Kerman}, \citenamefont {Yoshihara}, \citenamefont
  {Nakamura}, \citenamefont {Orlando},\ and\ \citenamefont
  {Oliver}}]{Gustavsson16}%
  \BibitemOpen
  \bibfield  {author} {\bibinfo {author} {\bibfnamefont {S.}~\bibnamefont
  {Gustavsson}}, \bibinfo {author} {\bibfnamefont {F.}~\bibnamefont {Yan}},
  \bibinfo {author} {\bibfnamefont {G.}~\bibnamefont {Catelani}}, \bibinfo
  {author} {\bibfnamefont {J.}~\bibnamefont {Bylander}}, \bibinfo {author}
  {\bibfnamefont {A.}~\bibnamefont {Kamal}}, \bibinfo {author} {\bibfnamefont
  {J.}~\bibnamefont {Birenbaum}}, \bibinfo {author} {\bibfnamefont
  {D.}~\bibnamefont {Hover}}, \bibinfo {author} {\bibfnamefont
  {D.}~\bibnamefont {Rosenberg}}, \bibinfo {author} {\bibfnamefont
  {G.}~\bibnamefont {Samach}}, \bibinfo {author} {\bibfnamefont {A.~P.}\
  \bibnamefont {Sears}}, \bibinfo {author} {\bibfnamefont {S.~J.}\ \bibnamefont
  {Weber}}, \bibinfo {author} {\bibfnamefont {J.~L.}\ \bibnamefont {Yoder}},
  \bibinfo {author} {\bibfnamefont {J.}~\bibnamefont {Clarke}}, \bibinfo
  {author} {\bibfnamefont {A.~J.}\ \bibnamefont {Kerman}}, \bibinfo {author}
  {\bibfnamefont {F.}~\bibnamefont {Yoshihara}}, \bibinfo {author}
  {\bibfnamefont {Y.}~\bibnamefont {Nakamura}}, \bibinfo {author}
  {\bibfnamefont {T.~P.}\ \bibnamefont {Orlando}},\ and\ \bibinfo {author}
  {\bibfnamefont {W.~D.}\ \bibnamefont {Oliver}},\ }\bibfield  {title}
  {\bibinfo {title} {Suppressing relaxation in superconducting qubits by
  quasiparticle pumping},\ }\href {https://doi.org/10.1126/science.aah5844}
  {\bibfield  {journal} {\bibinfo  {journal} {Science}\ }\textbf {\bibinfo
  {volume} {354}},\ \bibinfo {pages} {1573} (\bibinfo {year}
  {2016})}\BibitemShut {NoStop}%
\bibitem [{\citenamefont {Serniak}\ \emph {et~al.}(2018)\citenamefont
  {Serniak}, \citenamefont {Hays}, \citenamefont {de~Lange}, \citenamefont
  {Diamond}, \citenamefont {Shankar}, \citenamefont {Burkhart}, \citenamefont
  {Frunzio}, \citenamefont {Houzet},\ and\ \citenamefont
  {Devoret}}]{Serniak18}%
  \BibitemOpen
  \bibfield  {author} {\bibinfo {author} {\bibfnamefont {K.}~\bibnamefont
  {Serniak}}, \bibinfo {author} {\bibfnamefont {M.}~\bibnamefont {Hays}},
  \bibinfo {author} {\bibfnamefont {G.}~\bibnamefont {de~Lange}}, \bibinfo
  {author} {\bibfnamefont {S.}~\bibnamefont {Diamond}}, \bibinfo {author}
  {\bibfnamefont {S.}~\bibnamefont {Shankar}}, \bibinfo {author} {\bibfnamefont
  {L.~D.}\ \bibnamefont {Burkhart}}, \bibinfo {author} {\bibfnamefont
  {L.}~\bibnamefont {Frunzio}}, \bibinfo {author} {\bibfnamefont
  {M.}~\bibnamefont {Houzet}},\ and\ \bibinfo {author} {\bibfnamefont {M.~H.}\
  \bibnamefont {Devoret}},\ }\bibfield  {title} {\bibinfo {title} {Hot
  nonequilibrium quasiparticles in transmon qubits},\ }\href
  {https://doi.org/10.1103/PhysRevLett.121.157701} {\bibfield  {journal}
  {\bibinfo  {journal} {Phys. Rev. Lett.}\ }\textbf {\bibinfo {volume} {121}},\
  \bibinfo {pages} {157701} (\bibinfo {year} {2018})}\BibitemShut {NoStop}%
\bibitem [{\citenamefont {K\^e}(1947)}]{Ke47}%
  \BibitemOpen
  \bibfield  {author} {\bibinfo {author} {\bibfnamefont {T.-S.}\ \bibnamefont
  {K\^e}},\ }\bibfield  {title} {\bibinfo {title} {Experimental evidence of the
  viscous behavior of grain boundaries in metals},\ }\href
  {https://doi.org/10.1103/PhysRev.71.533} {\bibfield  {journal} {\bibinfo
  {journal} {Phys. Rev.}\ }\textbf {\bibinfo {volume} {71}},\ \bibinfo {pages}
  {533} (\bibinfo {year} {1947})}\BibitemShut {NoStop}%
\bibitem [{\citenamefont {Wilen}\ \emph {et~al.}(2021)\citenamefont {Wilen},
  \citenamefont {Abdullah}, \citenamefont {Kurinsky}, \citenamefont {Stanford},
  \citenamefont {Cardani}, \citenamefont {D'Imperio}, \citenamefont {Tomei},
  \citenamefont {Faoro}, \citenamefont {Ioffe}, \citenamefont {Liu},
  \citenamefont {Opremcak}, \citenamefont {Christensen}, \citenamefont
  {DuBois},\ and\ \citenamefont {McDermott}}]{Wilen21}%
  \BibitemOpen
  \bibfield  {author} {\bibinfo {author} {\bibfnamefont {C.~D.}\ \bibnamefont
  {Wilen}}, \bibinfo {author} {\bibfnamefont {S.}~\bibnamefont {Abdullah}},
  \bibinfo {author} {\bibfnamefont {N.~A.}\ \bibnamefont {Kurinsky}}, \bibinfo
  {author} {\bibfnamefont {C.}~\bibnamefont {Stanford}}, \bibinfo {author}
  {\bibfnamefont {L.}~\bibnamefont {Cardani}}, \bibinfo {author} {\bibfnamefont
  {G.}~\bibnamefont {D'Imperio}}, \bibinfo {author} {\bibfnamefont
  {C.}~\bibnamefont {Tomei}}, \bibinfo {author} {\bibfnamefont
  {L.}~\bibnamefont {Faoro}}, \bibinfo {author} {\bibfnamefont {L.~B.}\
  \bibnamefont {Ioffe}}, \bibinfo {author} {\bibfnamefont {C.~H.}\ \bibnamefont
  {Liu}}, \bibinfo {author} {\bibfnamefont {A.}~\bibnamefont {Opremcak}},
  \bibinfo {author} {\bibfnamefont {B.~G.}\ \bibnamefont {Christensen}},
  \bibinfo {author} {\bibfnamefont {J.~L.}\ \bibnamefont {DuBois}},\ and\
  \bibinfo {author} {\bibfnamefont {R.}~\bibnamefont {McDermott}},\ }\bibfield
  {title} {\bibinfo {title} {Correlated charge noise and relaxation errors in
  superconducting qubits},\ }\href {https://doi.org/10.1038/s41586-021-03557-5}
  {\bibfield  {journal} {\bibinfo  {journal} {Nature}\ }\textbf {\bibinfo
  {volume} {594}},\ \bibinfo {pages} {369} (\bibinfo {year}
  {2021})}\BibitemShut {NoStop}%
\bibitem [{\citenamefont {McEwen}\ \emph {et~al.}(2022)\citenamefont {McEwen},
  \citenamefont {Faoro}, \citenamefont {Arya}, \citenamefont {Dunsworth},
  \citenamefont {Huang}, \citenamefont {Kim}, \citenamefont {Burkett},
  \citenamefont {Fowler}, \citenamefont {Arute}, \citenamefont {Bardin},
  \citenamefont {Bengtsson}, \citenamefont {Bilmes}, \citenamefont {Buckley},
  \citenamefont {Bushnell}, \citenamefont {Chen}, \citenamefont {Collins},
  \citenamefont {Demura}, \citenamefont {Derk}, \citenamefont {Erickson},
  \citenamefont {Giustina}, \citenamefont {Harrington}, \citenamefont {Hong},
  \citenamefont {Jeffrey}, \citenamefont {Kelly}, \citenamefont {Klimov},
  \citenamefont {Kostritsa}, \citenamefont {Laptev}, \citenamefont {Locharla},
  \citenamefont {Mi}, \citenamefont {Miao}, \citenamefont {Montazeri},
  \citenamefont {Mutus}, \citenamefont {Naaman}, \citenamefont {Neeley},
  \citenamefont {Neill}, \citenamefont {Opremcak}, \citenamefont {Quintana},
  \citenamefont {Redd}, \citenamefont {Roushan}, \citenamefont {Sank},
  \citenamefont {Satzinger}, \citenamefont {Shvarts}, \citenamefont {White},
  \citenamefont {Yao}, \citenamefont {Yeh}, \citenamefont {Yoo}, \citenamefont
  {Chen}, \citenamefont {Smelyanskiy}, \citenamefont {Martinis}, \citenamefont
  {Neven}, \citenamefont {Megrant}, \citenamefont {Ioffe},\ and\ \citenamefont
  {Barends}}]{McEwen22}%
  \BibitemOpen
  \bibfield  {author} {\bibinfo {author} {\bibfnamefont {M.}~\bibnamefont
  {McEwen}}, \bibinfo {author} {\bibfnamefont {L.}~\bibnamefont {Faoro}},
  \bibinfo {author} {\bibfnamefont {K.}~\bibnamefont {Arya}}, \bibinfo {author}
  {\bibfnamefont {A.}~\bibnamefont {Dunsworth}}, \bibinfo {author}
  {\bibfnamefont {T.}~\bibnamefont {Huang}}, \bibinfo {author} {\bibfnamefont
  {S.}~\bibnamefont {Kim}}, \bibinfo {author} {\bibfnamefont {B.}~\bibnamefont
  {Burkett}}, \bibinfo {author} {\bibfnamefont {A.}~\bibnamefont {Fowler}},
  \bibinfo {author} {\bibfnamefont {F.}~\bibnamefont {Arute}}, \bibinfo
  {author} {\bibfnamefont {J.~C.}\ \bibnamefont {Bardin}}, \bibinfo {author}
  {\bibfnamefont {A.}~\bibnamefont {Bengtsson}}, \bibinfo {author}
  {\bibfnamefont {A.}~\bibnamefont {Bilmes}}, \bibinfo {author} {\bibfnamefont
  {B.~B.}\ \bibnamefont {Buckley}}, \bibinfo {author} {\bibfnamefont
  {N.}~\bibnamefont {Bushnell}}, \bibinfo {author} {\bibfnamefont
  {Z.}~\bibnamefont {Chen}}, \bibinfo {author} {\bibfnamefont {R.}~\bibnamefont
  {Collins}}, \bibinfo {author} {\bibfnamefont {S.}~\bibnamefont {Demura}},
  \bibinfo {author} {\bibfnamefont {A.~R.}\ \bibnamefont {Derk}}, \bibinfo
  {author} {\bibfnamefont {C.}~\bibnamefont {Erickson}}, \bibinfo {author}
  {\bibfnamefont {M.}~\bibnamefont {Giustina}}, \bibinfo {author}
  {\bibfnamefont {S.~D.}\ \bibnamefont {Harrington}}, \bibinfo {author}
  {\bibfnamefont {S.}~\bibnamefont {Hong}}, \bibinfo {author} {\bibfnamefont
  {E.}~\bibnamefont {Jeffrey}}, \bibinfo {author} {\bibfnamefont
  {J.}~\bibnamefont {Kelly}}, \bibinfo {author} {\bibfnamefont {P.~V.}\
  \bibnamefont {Klimov}}, \bibinfo {author} {\bibfnamefont {F.}~\bibnamefont
  {Kostritsa}}, \bibinfo {author} {\bibfnamefont {P.}~\bibnamefont {Laptev}},
  \bibinfo {author} {\bibfnamefont {A.}~\bibnamefont {Locharla}}, \bibinfo
  {author} {\bibfnamefont {X.}~\bibnamefont {Mi}}, \bibinfo {author}
  {\bibfnamefont {K.~C.}\ \bibnamefont {Miao}}, \bibinfo {author}
  {\bibfnamefont {S.}~\bibnamefont {Montazeri}}, \bibinfo {author}
  {\bibfnamefont {J.}~\bibnamefont {Mutus}}, \bibinfo {author} {\bibfnamefont
  {O.}~\bibnamefont {Naaman}}, \bibinfo {author} {\bibfnamefont
  {M.}~\bibnamefont {Neeley}}, \bibinfo {author} {\bibfnamefont
  {C.}~\bibnamefont {Neill}}, \bibinfo {author} {\bibfnamefont
  {A.}~\bibnamefont {Opremcak}}, \bibinfo {author} {\bibfnamefont
  {C.}~\bibnamefont {Quintana}}, \bibinfo {author} {\bibfnamefont
  {N.}~\bibnamefont {Redd}}, \bibinfo {author} {\bibfnamefont {P.}~\bibnamefont
  {Roushan}}, \bibinfo {author} {\bibfnamefont {D.}~\bibnamefont {Sank}},
  \bibinfo {author} {\bibfnamefont {K.~J.}\ \bibnamefont {Satzinger}}, \bibinfo
  {author} {\bibfnamefont {V.}~\bibnamefont {Shvarts}}, \bibinfo {author}
  {\bibfnamefont {T.}~\bibnamefont {White}}, \bibinfo {author} {\bibfnamefont
  {Z.~J.}\ \bibnamefont {Yao}}, \bibinfo {author} {\bibfnamefont
  {P.}~\bibnamefont {Yeh}}, \bibinfo {author} {\bibfnamefont {J.}~\bibnamefont
  {Yoo}}, \bibinfo {author} {\bibfnamefont {Y.}~\bibnamefont {Chen}}, \bibinfo
  {author} {\bibfnamefont {V.}~\bibnamefont {Smelyanskiy}}, \bibinfo {author}
  {\bibfnamefont {J.~M.}\ \bibnamefont {Martinis}}, \bibinfo {author}
  {\bibfnamefont {H.}~\bibnamefont {Neven}}, \bibinfo {author} {\bibfnamefont
  {A.}~\bibnamefont {Megrant}}, \bibinfo {author} {\bibfnamefont
  {L.}~\bibnamefont {Ioffe}},\ and\ \bibinfo {author} {\bibfnamefont
  {R.}~\bibnamefont {Barends}},\ }\bibfield  {title} {\bibinfo {title}
  {Resolving catastrophic error bursts from cosmic rays in large arrays of
  superconducting qubits},\ }\href {https://doi.org/10.1038/s41567-021-01432-8}
  {\bibfield  {journal} {\bibinfo  {journal} {Nature Physics}\ }\textbf
  {\bibinfo {volume} {18}},\ \bibinfo {pages} {107} (\bibinfo {year}
  {2022})}\BibitemShut {NoStop}%
\bibitem [{\citenamefont {Burnett}\ \emph {et~al.}(2014)\citenamefont
  {Burnett}, \citenamefont {Faoro}, \citenamefont {Wisby}, \citenamefont
  {Gurtovoi}, \citenamefont {Chernykh}, \citenamefont {Mikhailov},
  \citenamefont {Tulin}, \citenamefont {Shaikhaidarov}, \citenamefont
  {Antonov}, \citenamefont {Meeson}, \citenamefont {Tzalenchuk},\ and\
  \citenamefont {Lindstroem}}]{Burnett14}%
  \BibitemOpen
  \bibfield  {author} {\bibinfo {author} {\bibfnamefont {J.}~\bibnamefont
  {Burnett}}, \bibinfo {author} {\bibfnamefont {L.}~\bibnamefont {Faoro}},
  \bibinfo {author} {\bibfnamefont {I.}~\bibnamefont {Wisby}}, \bibinfo
  {author} {\bibfnamefont {V.~L.}\ \bibnamefont {Gurtovoi}}, \bibinfo {author}
  {\bibfnamefont {A.~V.}\ \bibnamefont {Chernykh}}, \bibinfo {author}
  {\bibfnamefont {G.~M.}\ \bibnamefont {Mikhailov}}, \bibinfo {author}
  {\bibfnamefont {V.~A.}\ \bibnamefont {Tulin}}, \bibinfo {author}
  {\bibfnamefont {R.}~\bibnamefont {Shaikhaidarov}}, \bibinfo {author}
  {\bibfnamefont {V.}~\bibnamefont {Antonov}}, \bibinfo {author} {\bibfnamefont
  {P.~J.}\ \bibnamefont {Meeson}}, \bibinfo {author} {\bibfnamefont {A.~Y.}\
  \bibnamefont {Tzalenchuk}},\ and\ \bibinfo {author} {\bibfnamefont
  {T.}~\bibnamefont {Lindstroem}},\ }\bibfield  {title} {\bibinfo {title}
  {Evidence for interacting two-level systems from the 1/f noise of a
  superconducting resonator},\ }\href {http://dx.doi.org/10.1038/ncomms5119}
  {\bibfield  {journal} {\bibinfo  {journal} {Nat. Commun.}\ }\textbf {\bibinfo
  {volume} {5}},\ \bibinfo {pages} {4119} (\bibinfo {year} {2014})}\BibitemShut
  {NoStop}%
\bibitem [{\citenamefont {Zolfagharkhani}\ \emph {et~al.}(2005)\citenamefont
  {Zolfagharkhani}, \citenamefont {Gaidarzhy}, \citenamefont {Shim},
  \citenamefont {Badzey},\ and\ \citenamefont {Mohanty}}]{Zolfagharkhani05}%
  \BibitemOpen
  \bibfield  {author} {\bibinfo {author} {\bibfnamefont {G.}~\bibnamefont
  {Zolfagharkhani}}, \bibinfo {author} {\bibfnamefont {A.}~\bibnamefont
  {Gaidarzhy}}, \bibinfo {author} {\bibfnamefont {S.-B.}\ \bibnamefont {Shim}},
  \bibinfo {author} {\bibfnamefont {R.~L.}\ \bibnamefont {Badzey}},\ and\
  \bibinfo {author} {\bibfnamefont {P.}~\bibnamefont {Mohanty}},\ }\bibfield
  {title} {\bibinfo {title} {Quantum friction in nanomechanical oscillators at
  millikelvin temperatures},\ }\href
  {https://doi.org/10.1103/PhysRevB.72.224101} {\bibfield  {journal} {\bibinfo
  {journal} {Phys. Rev. B}\ }\textbf {\bibinfo {volume} {72}},\ \bibinfo
  {pages} {224101} (\bibinfo {year} {2005})}\BibitemShut {NoStop}%
\bibitem [{\citenamefont {Orbach}\ and\ \citenamefont
  {Bleaney}(1961)}]{Orbach61}%
  \BibitemOpen
  \bibfield  {author} {\bibinfo {author} {\bibfnamefont {R.~.}\ \bibnamefont
  {Orbach}}\ and\ \bibinfo {author} {\bibfnamefont {B.}~\bibnamefont
  {Bleaney}},\ }\bibfield  {title} {\bibinfo {title} {Spin-lattice relaxation
  in rare-earth salts},\ }\href {https://doi.org/10.1098/rspa.1961.0211}
  {\bibfield  {journal} {\bibinfo  {journal} {Proceedings of the Royal Society
  of London. Series A. Mathematical and Physical Sciences}\ }\textbf {\bibinfo
  {volume} {264}},\ \bibinfo {pages} {458} (\bibinfo {year}
  {1961})}\BibitemShut {NoStop}%
\bibitem [{\citenamefont {Cambria}\ \emph {et~al.}(2021)\citenamefont
  {Cambria}, \citenamefont {Gardill}, \citenamefont {Li}, \citenamefont
  {Norambuena}, \citenamefont {Maze},\ and\ \citenamefont
  {Kolkowitz}}]{Cambria21}%
  \BibitemOpen
  \bibfield  {author} {\bibinfo {author} {\bibfnamefont {M.~C.}\ \bibnamefont
  {Cambria}}, \bibinfo {author} {\bibfnamefont {A.}~\bibnamefont {Gardill}},
  \bibinfo {author} {\bibfnamefont {Y.}~\bibnamefont {Li}}, \bibinfo {author}
  {\bibfnamefont {A.}~\bibnamefont {Norambuena}}, \bibinfo {author}
  {\bibfnamefont {J.~R.}\ \bibnamefont {Maze}},\ and\ \bibinfo {author}
  {\bibfnamefont {S.}~\bibnamefont {Kolkowitz}},\ }\bibfield  {title} {\bibinfo
  {title} {State-dependent phonon-limited spin relaxation of nitrogen-vacancy
  centers},\ }\href {https://doi.org/10.1103/PhysRevResearch.3.013123}
  {\bibfield  {journal} {\bibinfo  {journal} {Phys. Rev. Res.}\ }\textbf
  {\bibinfo {volume} {3}},\ \bibinfo {pages} {013123} (\bibinfo {year}
  {2021})}\BibitemShut {NoStop}%
\bibitem [{\citenamefont {Srivastava}(1990)}]{Srivastava90}%
  \BibitemOpen
  \bibfield  {author} {\bibinfo {author} {\bibfnamefont {G.~P.}\ \bibnamefont
  {Srivastava}},\ }\href
  {https://doi.org/https://doi.org/10.1201/9780203736241} {\emph {\bibinfo
  {title} {The Physics of Phonons}}}\ (\bibinfo  {publisher} {Routledge},\
  \bibinfo {year} {1990})\BibitemShut {NoStop}%
\bibitem [{\citenamefont {Catelani}\ \emph
  {et~al.}(2011{\natexlab{b}})\citenamefont {Catelani}, \citenamefont
  {Schoelkopf}, \citenamefont {Devoret},\ and\ \citenamefont
  {Glazman}}]{Catelani11b}%
  \BibitemOpen
  \bibfield  {author} {\bibinfo {author} {\bibfnamefont {G.}~\bibnamefont
  {Catelani}}, \bibinfo {author} {\bibfnamefont {R.~J.}\ \bibnamefont
  {Schoelkopf}}, \bibinfo {author} {\bibfnamefont {M.~H.}\ \bibnamefont
  {Devoret}},\ and\ \bibinfo {author} {\bibfnamefont {L.~I.}\ \bibnamefont
  {Glazman}},\ }\bibfield  {title} {\bibinfo {title} {Relaxation and frequency
  shifts induced by quasiparticles in superconducting qubits},\ }\href
  {https://doi.org/10.1103/PhysRevB.84.064517} {\bibfield  {journal} {\bibinfo
  {journal} {Phys. Rev. B}\ }\textbf {\bibinfo {volume} {84}},\ \bibinfo
  {pages} {064517} (\bibinfo {year} {2011}{\natexlab{b}})}\BibitemShut
  {NoStop}%
\bibitem [{\citenamefont {Lisenfeld}\ \emph {et~al.}(2010)\citenamefont
  {Lisenfeld}, \citenamefont {M\"uller}, \citenamefont {Cole}, \citenamefont
  {Bushev}, \citenamefont {Lukashenko}, \citenamefont {Shnirman},\ and\
  \citenamefont {Ustinov}}]{Lisenfeld10}%
  \BibitemOpen
  \bibfield  {author} {\bibinfo {author} {\bibfnamefont {J.}~\bibnamefont
  {Lisenfeld}}, \bibinfo {author} {\bibfnamefont {C.}~\bibnamefont {M\"uller}},
  \bibinfo {author} {\bibfnamefont {J.~H.}\ \bibnamefont {Cole}}, \bibinfo
  {author} {\bibfnamefont {P.}~\bibnamefont {Bushev}}, \bibinfo {author}
  {\bibfnamefont {A.}~\bibnamefont {Lukashenko}}, \bibinfo {author}
  {\bibfnamefont {A.}~\bibnamefont {Shnirman}},\ and\ \bibinfo {author}
  {\bibfnamefont {A.~V.}\ \bibnamefont {Ustinov}},\ }\bibfield  {title}
  {\bibinfo {title} {Measuring the temperature dependence of individual
  two-level systems by direct coherent control},\ }\href
  {https://doi.org/10.1103/PhysRevLett.105.230504} {\bibfield  {journal}
  {\bibinfo  {journal} {Phys. Rev. Lett.}\ }\textbf {\bibinfo {volume} {105}},\
  \bibinfo {pages} {230504} (\bibinfo {year} {2010})}\BibitemShut {NoStop}%
\bibitem [{\citenamefont {Paraoanu}(2006)}]{Paraoanu06}%
  \BibitemOpen
  \bibfield  {author} {\bibinfo {author} {\bibfnamefont {G.~S.}\ \bibnamefont
  {Paraoanu}},\ }\bibfield  {title} {\bibinfo {title} {Microwave-induced
  coupling of superconducting qubits},\ }\href
  {https://doi.org/10.1103/PhysRevB.74.140504} {\bibfield  {journal} {\bibinfo
  {journal} {Phys. Rev. B}\ }\textbf {\bibinfo {volume} {74}},\ \bibinfo
  {pages} {140504} (\bibinfo {year} {2006})}\BibitemShut {NoStop}%
\bibitem [{\citenamefont {Rigetti}\ and\ \citenamefont
  {Devoret}(2010)}]{Rigetti10}%
  \BibitemOpen
  \bibfield  {author} {\bibinfo {author} {\bibfnamefont {C.}~\bibnamefont
  {Rigetti}}\ and\ \bibinfo {author} {\bibfnamefont {M.}~\bibnamefont
  {Devoret}},\ }\bibfield  {title} {\bibinfo {title} {Fully microwave-tunable
  universal gates in superconducting qubits with linear couplings and fixed
  transition frequencies},\ }\href {https://doi.org/10.1103/PhysRevB.81.134507}
  {\bibfield  {journal} {\bibinfo  {journal} {Phys. Rev. B}\ }\textbf {\bibinfo
  {volume} {81}},\ \bibinfo {pages} {134507} (\bibinfo {year}
  {2010})}\BibitemShut {NoStop}%
\bibitem [{\citenamefont {Chen}\ \emph {et~al.}(2015)\citenamefont {Chen},
  \citenamefont {Hirose},\ and\ \citenamefont {Cappellaro}}]{Chen15}%
  \BibitemOpen
  \bibfield  {author} {\bibinfo {author} {\bibfnamefont {M.}~\bibnamefont
  {Chen}}, \bibinfo {author} {\bibfnamefont {M.}~\bibnamefont {Hirose}},\ and\
  \bibinfo {author} {\bibfnamefont {P.}~\bibnamefont {Cappellaro}},\ }\bibfield
   {title} {\bibinfo {title} {Measurement of transverse hyperfine interaction
  by forbidden transitions},\ }\href
  {https://doi.org/10.1103/PhysRevB.92.020101} {\bibfield  {journal} {\bibinfo
  {journal} {Phys. Rev. B}\ }\textbf {\bibinfo {volume} {92}},\ \bibinfo
  {pages} {020101} (\bibinfo {year} {2015})}\BibitemShut {NoStop}%
\bibitem [{\citenamefont {Pop}\ \emph {et~al.}(2014)\citenamefont {Pop},
  \citenamefont {Geerlings}, \citenamefont {Catelani}, \citenamefont
  {Schoelkopf}, \citenamefont {Glazman},\ and\ \citenamefont
  {Devoret}}]{Pop14}%
  \BibitemOpen
  \bibfield  {author} {\bibinfo {author} {\bibfnamefont {I.~M.}\ \bibnamefont
  {Pop}}, \bibinfo {author} {\bibfnamefont {K.}~\bibnamefont {Geerlings}},
  \bibinfo {author} {\bibfnamefont {G.}~\bibnamefont {Catelani}}, \bibinfo
  {author} {\bibfnamefont {R.~J.}\ \bibnamefont {Schoelkopf}}, \bibinfo
  {author} {\bibfnamefont {L.~I.}\ \bibnamefont {Glazman}},\ and\ \bibinfo
  {author} {\bibfnamefont {M.~H.}\ \bibnamefont {Devoret}},\ }\bibfield
  {title} {\bibinfo {title} {Coherent suppression of electromagnetic
  dissipation due to superconducting quasiparticles},\ }\href
  {https://doi.org/10.1038/nature13017} {\bibfield  {journal} {\bibinfo
  {journal} {Nature}\ }\textbf {\bibinfo {volume} {508}},\ \bibinfo {pages}
  {369} (\bibinfo {year} {2014})}\BibitemShut {NoStop}%
\end{thebibliography}%
\end{document}


\begin{CJK*}{UTF8}{}
\title{Supplementary Information: ``Phonon engineering of atomic-scale defects in superconducting quantum circuits''
}

\author{Mo Chen \CJKfamily{gbsn}(陈墨)}
\affiliation{Thomas J. Watson, Sr., Laboratory of Applied Physics, California Institute of Technology, Pasadena, CA 91125, USA}
\affiliation{Institute for Quantum Information and Matter, California Institute of Technology, Pasadena, CA 91125, USA}
\affiliation{Kavli Nanoscience Institute, California Institute of Technology, Pasadena, CA 91125, USA}

\author{John Clai Owens}\thanks{Present address: AWS Center for Quantum Computing, Pasadena, CA 91125, USA}
\affiliation{Thomas J. Watson, Sr., Laboratory of Applied Physics, California Institute of Technology, Pasadena, CA 91125, USA}
\affiliation{Institute for Quantum Information and Matter, California Institute of Technology, Pasadena, CA 91125, USA}
\affiliation{Kavli Nanoscience Institute, California Institute of Technology, Pasadena, CA 91125, USA}

\author{Harald Putterman}
\affiliation{AWS Center for Quantum Computing, Pasadena, CA 91125, USA}

\author{Max Sch{\"a}fer}\thanks{Present address: Department of Applied Physics and Physics, Yale University, New Haven, Connecticut 06520, USA}
\affiliation{Thomas J. Watson, Sr., Laboratory of Applied Physics, California Institute of Technology, Pasadena, CA 91125, USA}
\affiliation{Institute for Quantum Information and Matter, California Institute of Technology, Pasadena, CA 91125, USA}
\affiliation{Kavli Nanoscience Institute, California Institute of Technology, Pasadena, CA 91125, USA}

\author{Oskar~Painter}
\email{opainter@caltech.edu}
\homepage{https://painterlab.caltech.edu}
\affiliation{Thomas J. Watson, Sr., Laboratory of Applied Physics, California Institute of Technology, Pasadena, CA 91125, USA}
\affiliation{Institute for Quantum Information and Matter, California Institute of Technology, Pasadena, CA 91125, USA}
\affiliation{Kavli Nanoscience Institute, California Institute of Technology, Pasadena, CA 91125, USA}
\affiliation{AWS Center for Quantum Computing, Pasadena, CA 91125, USA}

\begin{abstract}

\end{abstract}

\maketitle
\end{CJK*}  

\tableofcontents
\newpage
\section{Methods}
In this section, we describe considerations that underlie the design of the hybrid transmon qubit device with Josephson junctions (JJs) embedded in acoustic bandgap structure. Our overarching goal is to strike a balance between the simplicity of the transmon qubit device and its effectiveness in demonstrating the phonon engineering of tunneling two-level systems (TLS) defects. This guiding principle is reflected in our decision of excluding Purcell filters, as well as the inclusion of a shunt capacitor for the transmon qubit. Further discussions on the device design will follow shortly. Along with the transmon qubit device design, we consider the acoustic metamaterials, as well as their integration into the transmon device. We will also discuss the device fabrication process, wherein a single resist layer Manhattan-style Josephson junction process plays a key role in the realization of our device. To conclude the Methods section, we provide brief descriptions of our experimental measurement setup, the calculation of phonon density of states using COMSOL, and a technique we used to generate new sets of TLS, known as thermal cycling.

\subsection{Device design}
The device serves two purposes: a) to identify individual TLS influenced by the engineered acoustic environment, and b) to characterize their relaxation behavior. To achieve this, we direct our attention towards TLS that are physically located inside the Josephson junction (JJ) tunnel barriers. This choice has three advantages. Firstly, their strong couplings to the transmon qubit, due to the intense electric field within the JJ, set them apart from TLS at circuit interfaces. Secondly, their physical confinement within a small area (the JJ) makes it convenient for phonon engineering. Lastly, individual addressing and characterization of TLS inside the JJ are well-established~\cite{Martinis05,Neeley08}. 

\subsubsection{Transmon qubit}
In the design of the transmon qubit, we decide to make our JJ an order-of-magnitude larger than typical JJs, with a size of approximately $\sim0.83$~$\mu$m$^2$, in order to increase occurrences of TLS inside the JJ. Such large JJ contributes to a substantial junction capacitance. A transmon qubit, characterized by its qubit capacitance consisting mainly of the junction capacitance, has been recently demonstrated in refs.~\cite{Zhao20,Mamin21}. In our design, the two JJs that form a symmetric SQUID (superconducting quantum interference device) loop collectively contribute a $\sim 60$~fF junction capacitance to the transmon qubit. It is noteworthy that the junction capacitance is large enough that the JJs alone can make up a transmon qubit, a configuration termed the `merged-element transmon'~\cite{Zhao20,Mamin21}. In this work, however, instead of implementing a full merged-element transmon, we introduce a shunt capacitor. The shunt capacitor conveniently facilitates coupling to control lines and readout resonators. The shunt capacitor accounts for $\sim 40$~fF, resulting in a total transmon capacitance of $\sim 100$~fF. It is important to emphasize that the shunt capacitor is not protected by the acoustic metamaterials, and its interaction with nearby resonant TLS is considered the major $T_1$ relaxation channel for the transmon qubit. Consequently, due to the influence of the shunt capacitor, we do not expect substantial effects of phonon engineering on the transmon qubit in the current design.

\begin{figure}
    \centering
    \includegraphics[width = 0.5\textwidth]{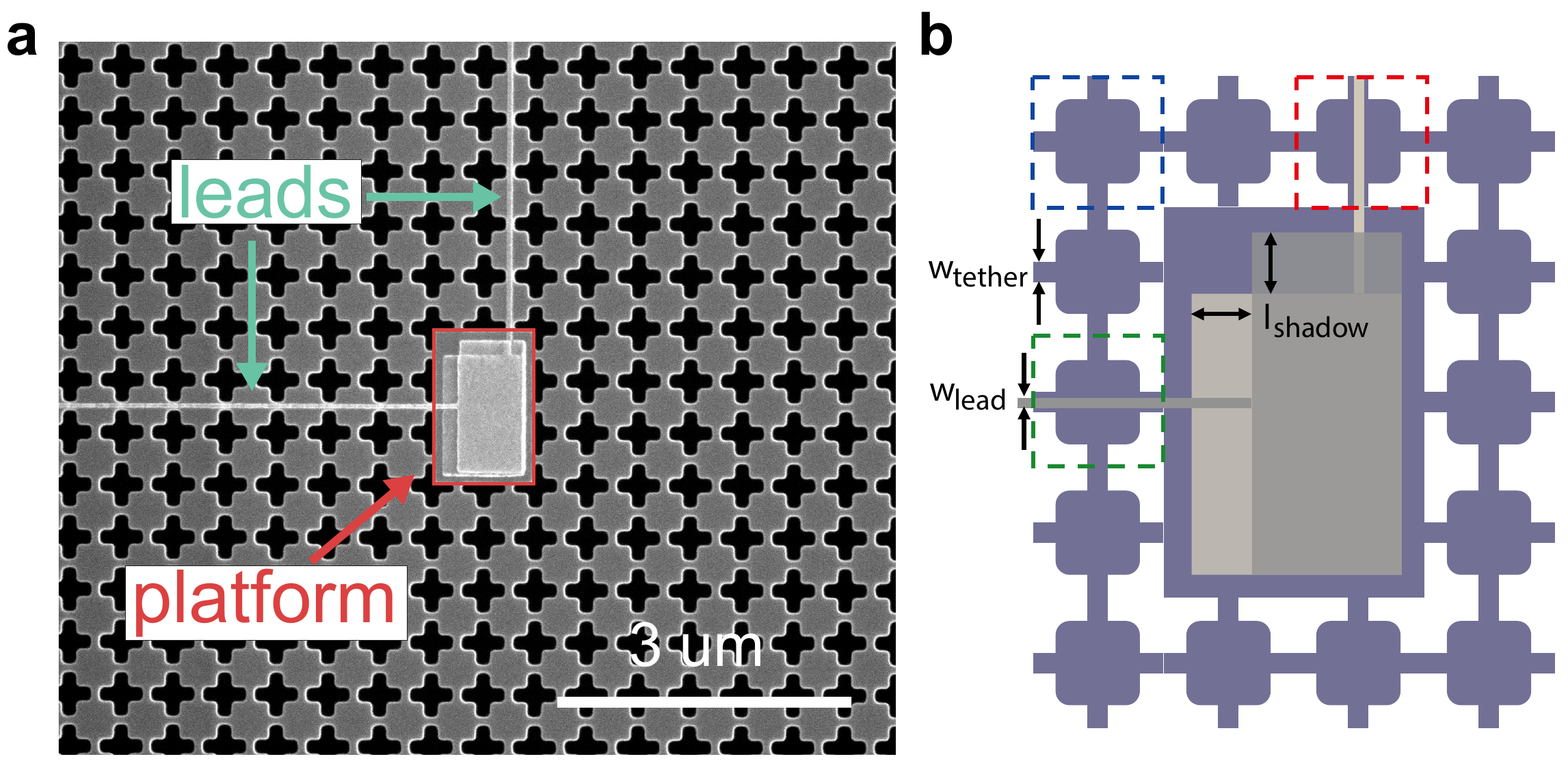}
    \caption{{JJ embedded in the acoustic metamaterial.} \textbf{a,} SEM showcasing the JJ, the rectangular platform, and the surrounding acoustic metamaterial. The JJ is positioned on top of an approximately $1\times1.6~\mu$m$^2$ rectangular Si platform, highlighted in red. The JJ is formed by two Al layers sandwiching a thin AlO$_x$ barrier layer. Each Al layer has a narrow and long lead, indicated by turquoise arrows. These JJ leads pass through the acoustic metamaterials and form a SQUID loop. 
    \textbf{b,} Schematic topview of \textbf{a}. For our fabricated devices, we have $w_\mathrm{lead}=45$~nm, $w_\mathrm{tether}=72$~nm, and $l_\mathrm{shadow}\approx150$~nm. The alignment between the JJ and the Si tether is better than $\sim 10$~nm. 
    Based on our device geometry, the unit cell of the acoustic structures has three different types, showcased in the blue, red, and green dashed boxes, representing the Si cross-shield, Si cross-shield with $30$~nm thick JJ lead, and Si cross-shield with $50$~nm thick JJ lead, respectively.}
    \label{fig:JJ_topview}
\end{figure}

To engineer the acoustic environment that the JJ and the TLS inside the JJ see, we position the JJ of the transmon qubit on top of a rectangular platfrom consisting of an unpatterened Si suspended membrane, as shown in Fig.~\ref{fig:JJ_topview}. The rectangular platform is tethered to the rest of the Si microchip through an acoustic metamaterial, which is designed to exhibit a microwave-frequency acoustic bandgap centered around $5.1$~GHz. Inside the bandgap, the acoustic metamaterial shields the TLS from spontaneous phonon emission into the phonon modes of the bulk materials and extends the lifetime of TLS.

The readout resonators are designed to situate $\sim 700$~MHz above the transmon's upper sweet spot frequency, with a coupling strength of $\sim 70$~MHz and a linewidth of $\sim 2$~MHz. No Purcell filters are used, yielding a Purcell limit of $\sim 10$~$\mu$s, a timescale that is on the same order of transmon's $T_1$. We believe the Purcell limit serves as the secondary contribution to the relaxation process of transmon, with the major contribution being resonant coupling to TLS at the shunt capacitor, as mentioned earlier.

For the comprehensive characterization of phonon engineering of TLS and the acoustic bandgap, two distinct microchips are designed, labelled Chip-A and Chip-B. Each chip accommodates four transmon qubits. On Chip-A, the designed upper sweet spot frequencies of the four transmon qubits span the range of $6-6.5$~GHz, strategically chosen to resolve the upper edge of the acoustic bandgap. Conversely, the four transmon qubits on Chip-B are designed to cover the frequency range $5.1-5.5$~GHz to resolve the lower edge of the acoustic bandgap.
On each chip, the four transmon qubits have identical JJs.  Adjustments to the shunt capacitance of each transmon qubit tunes its sweet spot frequency to the desired value. Between Chip-A and Chip-B, the geometry remains identical. The different transmon frequency ranges between the two chips are achieved by varying the oxidation condition during the JJ fabrication process. 

\begin{figure}
    \centering
    \includegraphics[width = 0.45\textwidth]{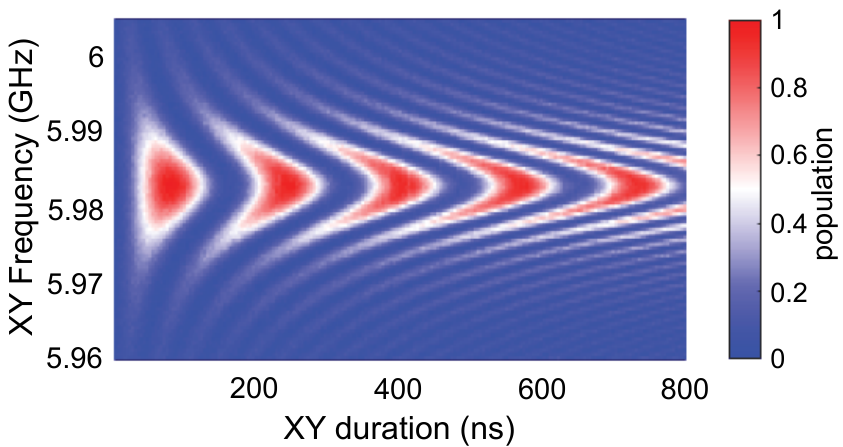}
    \caption{\textbf{Rabi chevron of a representative hybrid transmon qubit device.}  The measurements presented correspond to Q$_4$ of Chip-A, conducted during the first cool down.}
    \label{fig:Qubit_chevron}
\end{figure}

The fabricated Chip-A (Chip-B) covers upper sweet spot frequency ranges $6\textrm{--}6.5$~GHz ($5.4\textrm{--}5.8$~GHz).
Typical parameters for all transmon qubits on both chips are presented in Table~\ref{tab:qubit_params}. It's worth noting that these parameters are subject to minor changes during different cool downs. On Chip-B, Q$_4$ remains operational, indicated by the Lamb shift of the readout resonator and its susceptibility to flux tuning via the crosstalk from the Z lines of Q$_{1\textrm{--}3}$. However, Q$_4$ does not show frequency tuning through its own Z line. We suspect that potential defects in the Z line or associated wirebonds lead to an open connection. We therefore exclude Q$_4$ from Chip-B in this study. 
Parameters of the other seven transmon qubits align well with our design and simulations. Their $T_1$ times span the range $1.5\textrm{--}6.0~\mu$s, corresponding to $Q$ values between $0.5\textrm{--}2\times 10^5$, which are on par with the best SOI transmon qubits reported in literature~\cite{Keller17}. Fig.~\ref{fig:Qubit_chevron} displays a representative Rabi chevron pattern measured on Q$_4$ of Chip-A.

\begin{table}
    \centering
\begin{tabular}{c|c|c|c|c|c}
            & $\omega/2\pi$ & $\alpha$ & $\omega_{\mathrm{RR}}/2\pi$ & $g$ & $T_1$  \\
   Device   & (GHz) & (MHz) & (GHz) & (MHz)& ($\mu$s) \\
     \hline
Chip-A Q$_1$ & 6.48 & -182.5 & 7.26 & 74.0 & 4.5 \\
Chip-A Q$_2$ & 6.29 & -166.7 & 7.13 & 71.6 & 2.1 \\
Chip-A Q$_3$ & 6.11 & -159.3 & 7.00 & 66.8 & 3.0 \\
Chip-A Q$_4$ & 5.98 & -152.6 & 6.88 & 67.5 & 3.2 \\
Chip-B Q$_1$ & 5.77 & -182.1 & 6.22& 71.8 & 1.5 \\
Chip-B Q$_2$ & 5.57 & -171.3 & 6.11 & 71.0 & 6.0 \\
Chip-B Q$_3$ & 5.44 & -162.7 & 6.01 & 70.1 & 3.5
\end{tabular}
    \caption{\textbf{Transmon qubit parameters.} Chip-A is designed to resolve the upper edge of the acoustic bandgap. This chip hosts four qubits, Q$_{1\textrm{--}4}$, with upper sweet spot frequencies above the upper edge of the acoustic bandgap, simulated to be around $5.814$~GHz. As for Chip-B, designed to resolve the lower edge of the acoustic bandgap, three fully functional qubits, Q$_{1\textrm{--}3}$, have upper sweet spot frequencies above the lower edge of the acoustic bandgap, simulated to be around $4.442$~GHz. Additional qubit parameters are provided in the table, including the qubit's anharmonicity $\alpha$, the frequency of its corresponding readout resonator $\omega_\mathrm{RR}$, the coupling strength $g$ to the resonator, and the typical $T_1$ relaxation time measured at the sweet spot. We note that these values are representative, subject to slight variations after each thermal cycling.
    }
    \label{tab:qubit_params}
\end{table}

\subsubsection{Acoustic bandgap metamaterials}\label{sec:shield_design}
The acoustic metamaterial is formed from a periodic cross-shield pattern etched into the Si membrane layer. A scanning electron microscope (SEM) image showcasing the JJ, the rectangular platform, and the surrounding acoustic metamaterials is shown in Fig.~\ref{fig:JJ_topview}a. The platform region replaces the central $2\times 3$ unit cells of the cross-shield metamaterial, and is surrounded by nine periods of shielding. As shown in the SEM of the device (Fig.~\ref{fig:JJ_topview}a) and the schematic ( Fig.~\ref{fig:JJ_topview}b), the JJ leads pass through several unit cells of the acoustic metamaterial, introducing additional mass and perturbations to the band structures. Consequently, central to our design is the establishment of a large acoustic bandgap in the presence of the JJ leads. 

\begin{figure}
    \centering
    \includegraphics[width = 0.45\textwidth]{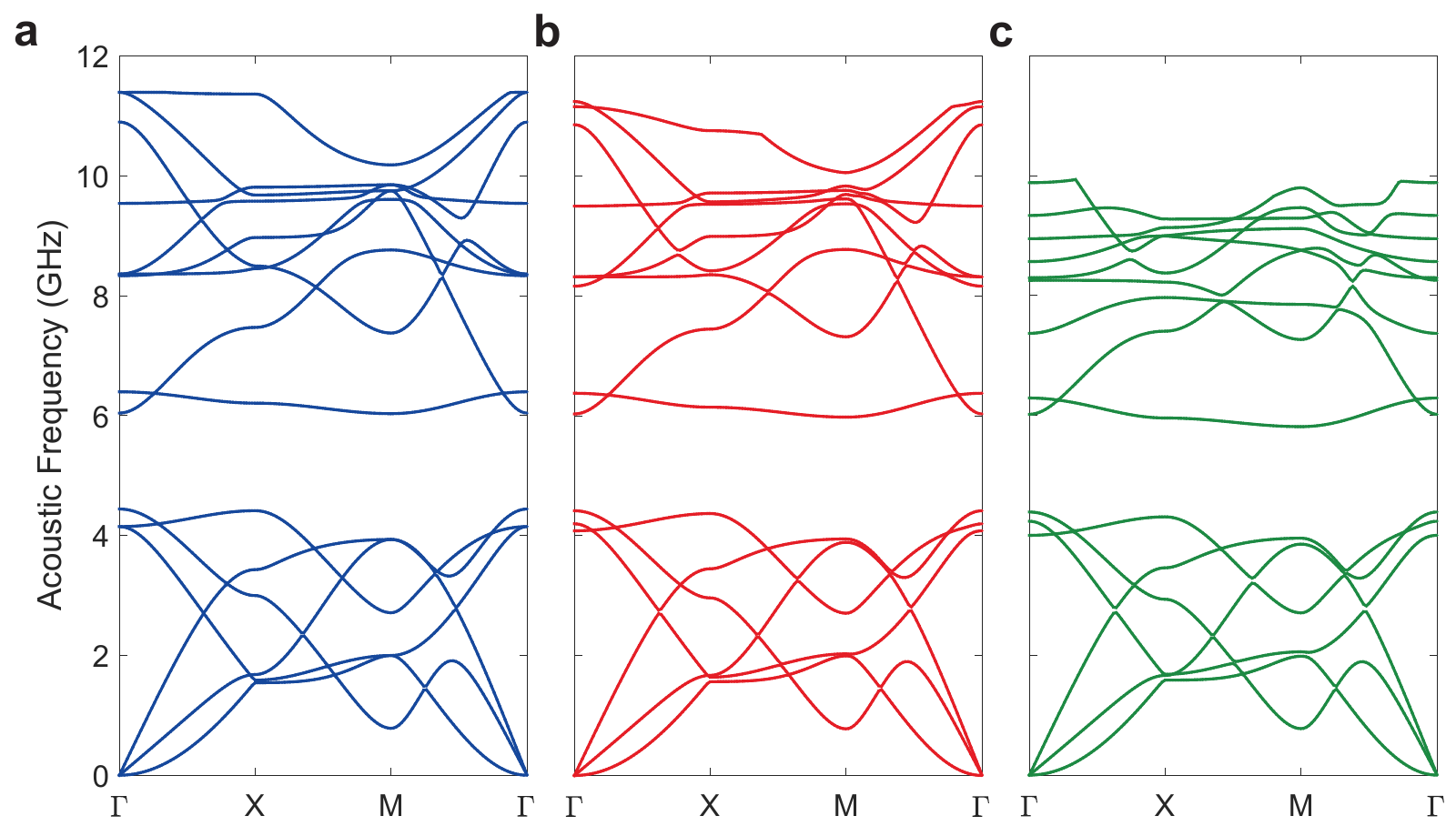}
    \caption{\textbf{Simulated acoustic band structures of the cross-shield acoustic metamaterial unit cell.} Utilizing COMSOL simulations, we present the acoustic band structures of the three types of unit cells: \textbf{a,} silicon only cross-shield, \textbf{b,} silicon cross-shield with $30$nm thick Al lead passing through, and \textbf{c,} silicon cross-shield with $50$nm thick Al lead passing through. The Si cross-shield geometry remains the same for all three cases, with Si device layer thickness of $220$~nm. The simulated bandgap frequencies are listed in Table~\ref{tab:acoustic_bandgap_sim}.}
    \label{fig:acoustic_band_sim}
\end{figure}

To account for these perturbations due to the JJ leads, we consider three different unit cell types in the COMSOL simulation of the acoustic band structure. These unit cell types are designed in accordance with the geometry of our transmon device, which include \textbf{a} silicon only (enclosed by the blue dashed box in Fig.~\ref{fig:JJ_topview}b), \textbf{b} silicon with $30$~nm thick Al leads from the JJ (enclosed by the red dashed box in Fig.~\ref{fig:JJ_topview}b), and \textbf{c} silicon with $50$~nm thick Al leads from the JJ (enclosed by the green dashed box in Fig.~\ref{fig:JJ_topview}b). Throughout all three cases, the Si geometry stays the same. The dimensions used by the COMSOL simulations are derived from the SEM image of a sister chip that is nominally identical to Chip-B. The results from the COMSOL simulations for the acoustic band structures in these three scenarios, along special paths connecting highly symmetric points in the $k$-space, are shown in Fig.~\ref{fig:acoustic_band_sim}. Additionally, the corresponding bandedge frequencies are listed in Table~\ref{tab:acoustic_bandgap_sim}. The overlap of these three simulated bandgaps yields an overall bandgap spanning $1.372$~GHz, ranging from $4.442$~GHz to $5.814$~GHz, in the presence of the JJ leads. The overall bandgap is $0.219$~GHz narrower than the Si only unit cell ($4.442\textrm{--}6.033$~GHz), due to the perturbations of Al leads. 

\begin{table}
    \centering
\begin{tabular}{c|c|c}
     Al thickness (nm) & $f_1$ (GHz) & $f_2$ (GHz) \\
     \hline
     $0$ & $\mathbf{4.442}$ & $6.033$ \\
     $30$ & $4.417$ & $5.979$ \\
     $50$ & $4.389$ & $\mathbf{5.814}$
\end{tabular}
    \caption{\textbf{Simulated acoustic bandgaps for three different unit cells.} The tabulated information outlines the acoustic bandgap frequencies for the three unit cell types, as illustrated in the blue, red, and green dashed boxes in Fig.~\ref{fig:JJ_topview}b. The corresponding band structures are plotted in Fig.~\ref{fig:acoustic_band_sim}.
    The table's columns specify the thickness (th.) of Al leads from the JJ passing through the cross-shield unit cell, as well as the respective lower and upper edges of the acoustic bandgap ($f_1$ and $f_2$). The bandgap for the entire structure, given by the frequency overlap between these three unit cell types, spans from $4.442$~GHz to $5.814$~GHz, highlighted in bold.}
    \label{tab:acoustic_bandgap_sim}
\end{table}

Additionally, we conducted COMSOL simulations to explore the effects of the widths and thicknesses of the Al leads that run through the Si cross-shield pattern. Of the two factors, thickness has the larger impact on the band structures. As the thickness of the Al lead increases, the size of the bandgap decreases from $1.591$~GHz until it vanishes at approximately $100$~nm Al thickness. 
Preserving a large acoustic bandgap therefore necessitates careful design of the JJ geometry and the process of double-angle evaporation of Al. It is critical to ensure that any cross-shield metamaterial unit cell undergoes no more than one metalization, or equivalently, avoiding the formation of parasitic junctions on the cross-shield pattern. This condition ensures that the three scenarios simulated in Fig.~\ref{fig:acoustic_band_sim} faithfully capture all the acoustic band structures encountered in our device. The elimination of parasitic junctions in the cross-shield region is achieved by a geometric argument, which will be discussed in detail in the subsequent section on device fabrication (sec.~\ref{sec:fab}). 

\subsection{Device fabrication}~\label{sec:fab}
\begin{figure}[!htbp]
    \centering
    \includegraphics[width = 0.48\textwidth]{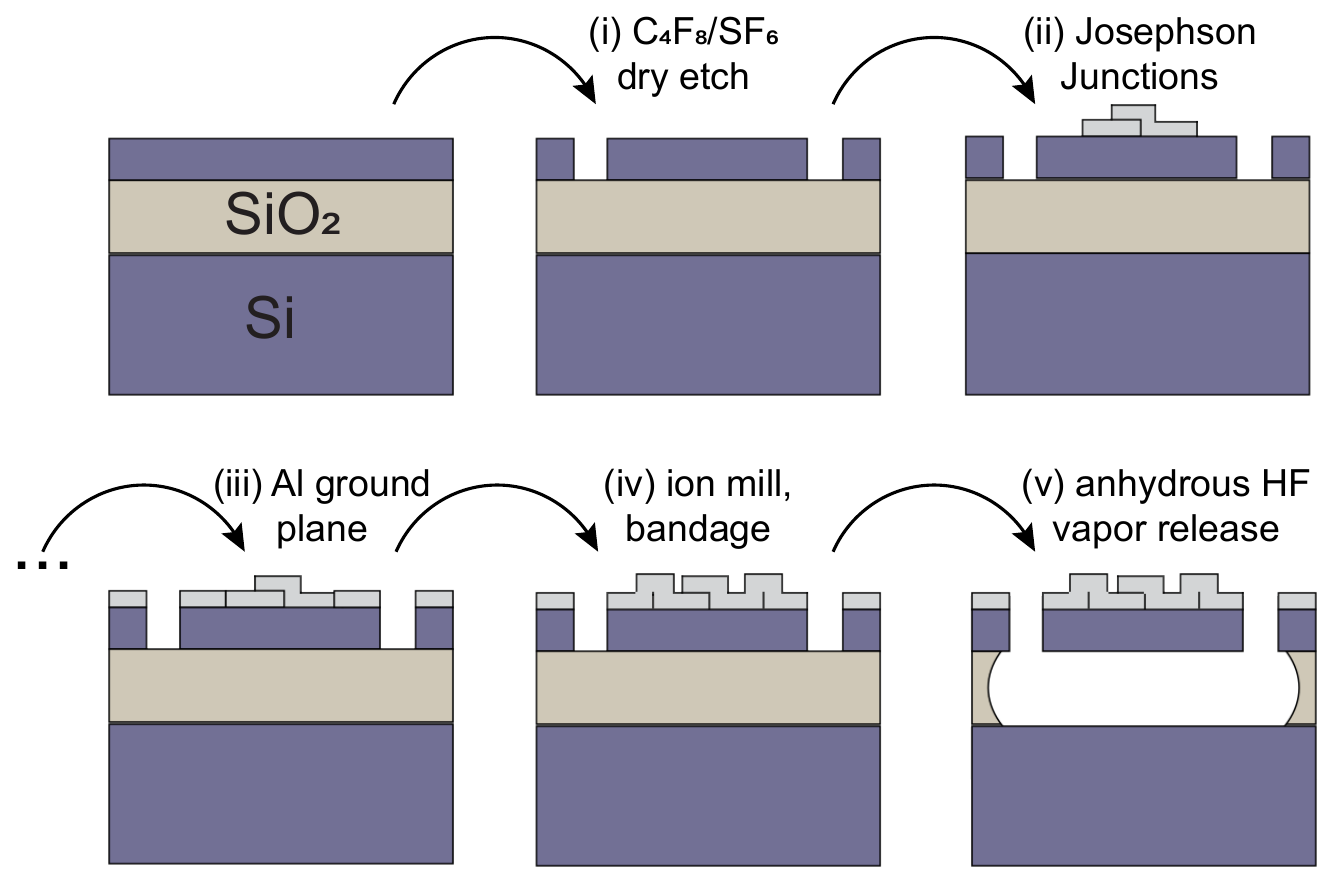}
    \caption{\textbf{Fabrication process of the hybrid device on SOI substrate.} All beam writes employ $100$~keV electron-beam lithography (Raith EBPG 5200). All metal depositions are realized by electron beam evaporation (Plassys MEB550S) and a liftoff process.}
    \label{fig:SOI_fab}
\end{figure}

Our fabrication process of the hybrid device stems from the fabrication recipe for transmon qubit on silicon-on-insulator (SOI) substrate outlined in ref.~\cite{Keller17}. Our modified process is illustrated in Fig.~\ref{fig:SOI_fab}.
We start with an SOI wafer (SEH) with the following specifications: silicon device layer, $220$~nm in thickness, resistivity $\rho\geq 5~\mathrm{k}\Omega\cdot \mathrm{cm}$; buried silicon dioxide layer, $3$~$\mu$m in thickness; and a silicon handle, $750$~$\mu$m in thickness, $\rho\geq 5~\mathrm{k}\Omega\cdot \mathrm{cm}$. First, the wafer is diced along the $\langle 100\rangle$ direction into chips of dimensions $20~\mathrm{mm}\times 10~\mathrm{mm}$. We then perform the following fabrication steps, all using $100$~keV electron-beam lithography (Raith EBPG5200) for patterning, and electron beam evaporation (Plassys MEB 550S) for metalization: 
(i) Si device layer patterning using inductively coupled plasma reactive ion etching (ICP-RIE) with C$_4$F$_8$/SF$_6$ (Oxford Plasmalab 100) to define the cross-shield acoustic metamaterials, as well as the release holes for device suspension. 
(ii) $30^\circ$ double-angle evaporation for the Manhattan-style JJ ($30$~nm/$50$~nm) using a single layer photo resist (ZEP520A). The oxidation steps are performed at $130$~mbar, for a duration of $84'$ and $102'$ for Chip-A and Chip-B, respectively. 
(iii) Al ground plane patterning by liftoff. 
(iv) Ar ion milling, bandage deposition and liftoff. 
(v) Device release in anhydrous vapor-HF (SPTS uEtch).

Our fabrication process is fine-tuned to accurately realize our design and ensure the preservation of a large acoustic bandgap. The key in our fabrication is to minimize any perturbations to the acoustic structure, in particular: 
a) Preventing any metal deposition on the vertical sidewalls of the silicon acoustic structures, 
and b) eliminating the formation of parasitic junctions on the silicon cross-shield structures during the double-angle evaporation process. 

\subsubsection{Preventing metal deposition on the Si sidewalls}
To avoid any undesired metal deposition on the vertical sidewalls of the silicon acoustic structures, we make sure 1. the JJ lead is small enough to completely locate on top of the Si structure, and 2. there is good alignment between the JJ and acoustic metamaterial patterns. For the first part, we make the width of the JJ leads narrower than the width of the Si tether, which is the narrowest part of the cross-shield acoustic metamaterial that the JJ lead runs through, given by $w_\mathrm{lead} = 45$~nm~$ < w_\mathrm{tether} = 72$~nm, as illustrated in Fig.~\ref{fig:JJ_topview}b. For the second part, we employed local markers during the e-beam lithography, which contributes to consistent alignment accuracy, resulting in small misalignment of $\lesssim 10$~nm between the acoustic metamaterial pattern and the JJ, as evidenced in Fig.~\ref{fig:JJ_topview}a. Even considering a worst case scenario with a $10$~nm misalignment, the metal deposition of the JJ leads remains confined to the top silicon surface, avoiding any undesired metal deposition onto the sides of the silicon acoustic structures. 

\subsubsection{Eliminating the formation of parasitic junctions}
\begin{figure}
    \centering
    \includegraphics[width = 0.48\textwidth]{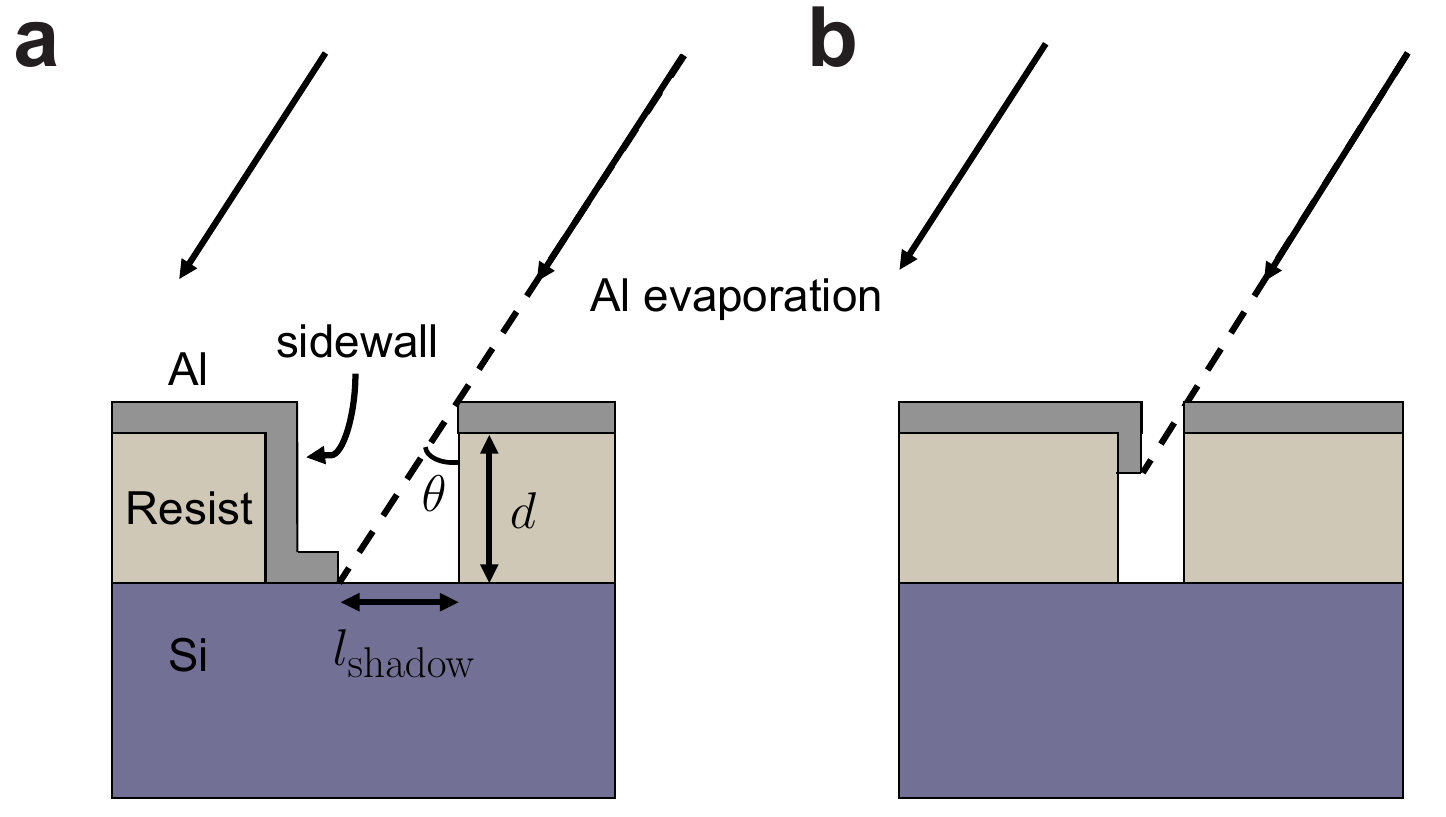}
    \caption{\textbf{Sideview of angled evaporation in a single layer resist process.} 
    \textbf{a,} From a geometric standpoint, there is a shadow area of size $l_\mathrm{shadow}=d\tan\theta$ without metalization, where $d$ is the thickness of the resist, and $\theta$ the evaporation angle from normal incidence. In addition, during angled evaporation of Al, due to the lack of an undercut in a single layer resist process, Al is deposited both on the Si substrate (excluding the shadow area) and on the sidewalls of the resist. The latter sometimes remains as free-standing vertical sidewalls post the liftoff process. Notably, these residual vertical sidewalls do not influence the transmon qubit performance in our experiment. 
    \textbf{b,} In scenarios where the feature size is smaller than the shadow size, e.g., $w_\mathrm{lead}\ll l_\mathrm{shadow}$, there is no metalization on Si. Consequently, a Manhattan-style JJ that takes advantage of the shadow could avoid the formation of parasitic junctions within target regions.}
    \label{fig:angled_evap_sideview}
\end{figure}

To eliminate the formation of parasitic junctions on the cross-shield metamaterials, we implement a geometric strategy in the Manhattan-style JJ, as illutrated in Fig.~\ref{fig:angled_evap_sideview}. During the angled evaporation process, an unmetalized `shadow' area of size $l_\mathrm{shadow}=d\tan\theta$ is created, where $d$ is the thickness of photoresist and $\theta$ the evaporation angle from normal incidence. In our process, the shadow size is approximately $l_\mathrm{shadow}\approx150$~nm. 
In the design of the Manhattan-style JJ, we enforce the condition that $w_\mathrm{lead}\ll l_\mathrm{shadow}$, where $w_\mathrm{lead} = 45$~nm is the width of our JJ leads. 
This condition guarantees that only one layer of Al is metalized on the Si structure, avoiding the formation of parasitic junctions in the Manhattan-style JJ configuration. Specifically, the fabrication of Manhattan-style JJ involves two separate Al evaporations, whose in-plane evaporation angles are perpendicular to each other. When the in-plane direction of the evaporation aligns with the length direction of the JJ lead (into the plane in Fig.~\ref{fig:angled_evap_sideview}), Al is deposited to the Si structure. When the in-plane direction of the evaporation is perpendicular to the length direction of the JJ lead, as shown in Fig.~\ref{fig:angled_evap_sideview}b, Al only deposits onto the resist, and is subsequently lifted off. Consequently, no parasitic junctions are formed during the double-angle evaporation process. This is important to the preservation of a large acoustic bandgap, as excessive Al deposition on the cross-shield acoustic metamaterial can quickly diminish the bandgap, as discussed previously.

We note that parasitic junctions still exist in our fabrication process, where we broaden up the JJ leads for the bandage. However, these parasitic junctions are strategically positioned outside the acoustic metamaterial region. As a result, the increased thickness and weight of Al in these areas do not affect the acoustic bandgap. These parasitic junctions are shorted by a bandage at the final stage of the fabrication (step v). We remark that for a single JJ qubit (fixed frequency qubit), it is indeed possible to completely eliminate the parasitic junctions, through purely geometric considerations. This is important in future work when we embed the whole merged-element transmon qubit~\cite{Zhao20,Mamin21} into the acoustic structure and remove the shunt capacitor from our design. 

\subsubsection{Single resist layer JJ process}
In conjunction with the criterion $w_\mathrm{lead}\ll l_\mathrm{shadow}$, we make an additional effort to minimize the size of the shadow area $l_\mathrm{shadow}$. 
The shadow area is inherently part of the rectangular platform on which the JJ resides, as depicted in Fig.~\ref{fig:JJ_topview}. As such, a larger shadow region requires a larger rectangular platform. A larger platform in turn supports more local acoustic modes inside the acoustic bandgap, which might potentially influence TLS performance. 

In order to suppress the number of these local acoustic modes, and avoid their potential couplings to TLS, we have developed a single resist layer JJ fabrication process, similar to that outlined in ref.~\cite{Zhang17sr}, using ZEP520A instead of the more conventional PMMA-MMA double layer resist process. This process minimizes the size of the shadow area to $l_\mathrm{shadow}\approx 150$~nm, resulting in free spectral range of the local acoustic modes within the acoustic bandgap on the order of $\sim 100$~MHz according to COMSOL simulations.

\begin{figure}
    \centering
    \includegraphics[width = 0.3\textwidth]{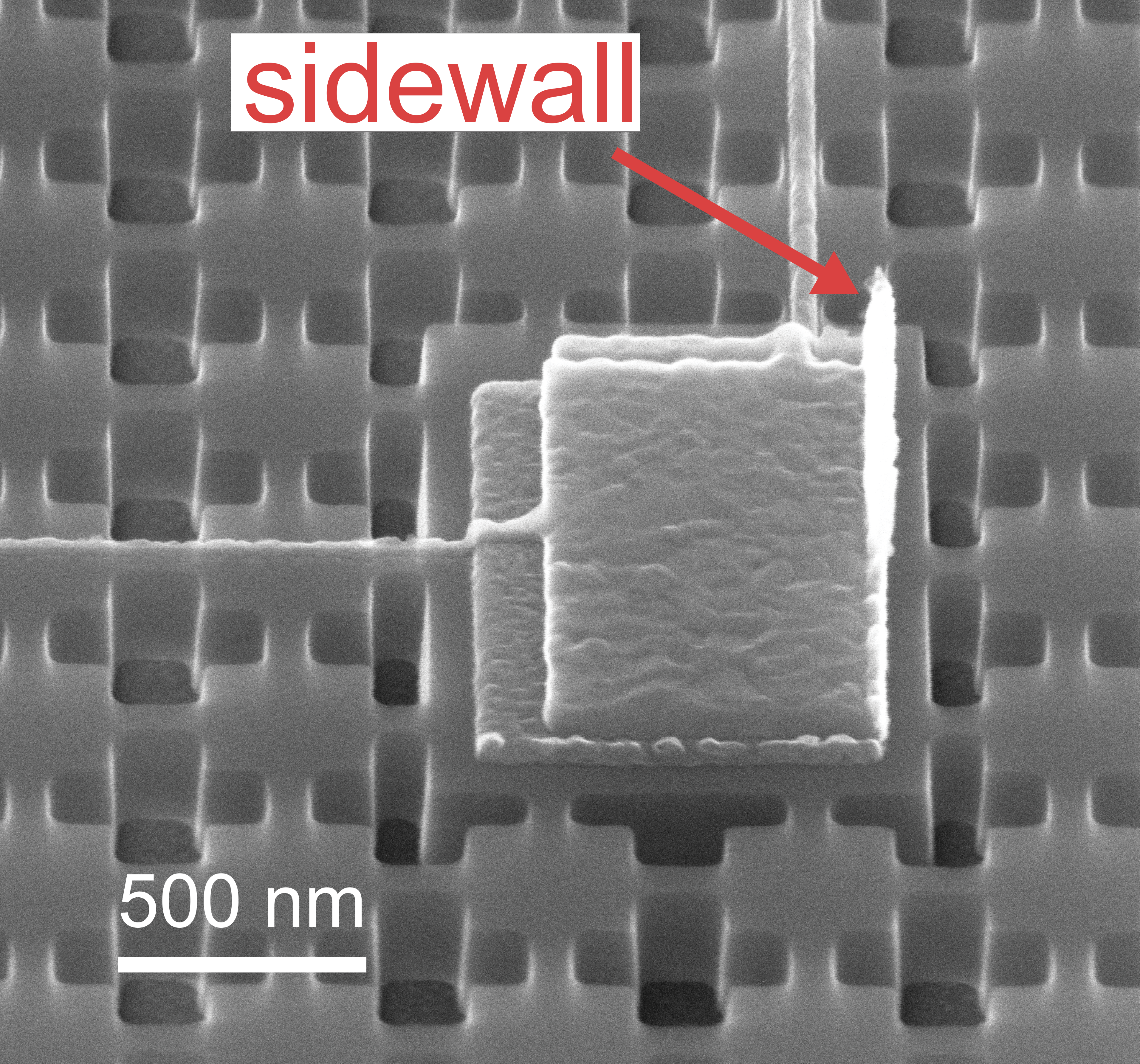}
    \caption{\textbf{Angled SEM of a Josephson junction with Al sidewall}. The free-standing Al sidewall comes from the liftoff process, which is a consequence of the lack of an undercut in conjugation with angled evaporation in our single resist JJ process.}
    \label{fig:JJ_sidewall}
\end{figure}

It is important to note, however, that we have noticed the formation of free-standing vertical Al sidewalls post the liftoff process in some of our devices, as shown in Fig.~\ref{fig:JJ_sidewall}. This phenomenon is anticipated when single layer resist is used in angled evaporation, without an undercut. In this scenario, the metal deposited on the sidewall of the resist, as indicated in Fig.~\ref{fig:angled_evap_sideview}a, might not be entirely removed through the liftoff process. However, despite the presence of these residual vertical Al sidewalls, we have not observed impacts on the performance of the transmon qubits.

\subsubsection{JJ oxidation}
\begin{figure}
    \centering
    \includegraphics[width = 0.4\textwidth]{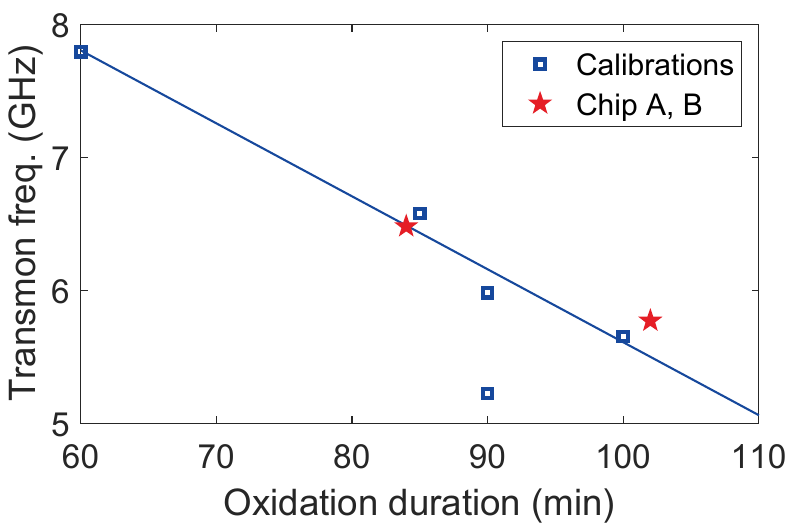}
    \caption{\textbf{Transmon frequency-dependence on the JJ oxidation time.} All the collected data points except for one 
   exhibit an empirical linear relation, which we use to inform the fabrication of Chip-A and Chip-B. The one outlier data point ($90$~min, $5.22$~GHz) represent a chip that was aged for roughly a month before measurements were taken, which possibly explains the abnormal behavior. The pentagrams represent frequencies of Q$_1$'s of Chip-A and Chip-B, which agree well with the empirical linear fit.}
    \label{fig:JJ_oxidation}
\end{figure}

The oxidation condition for the merged-element-style JJs are determined based on measurements of previous calibration chips, as shown in Fig.~\ref{fig:JJ_oxidation}. An extended oxidation duration at high static oxygen pressure grows a thicker AlO$_x$ barrier layer of the JJ, which decreases the Josephson energy $E_J$ and the transmon frequency.
Calibration data show an empirical linear dependence between the transmon frequency and the oxidation duration, to which we fit and inform our fabrication of Chip-A and Chip-B. The one outlier is a chip that aged for approximately one month prior to measurement, which explains the atypical behavior. The frequencies of fabricated Q$_1$'s of Chip-A and Chip-B, represented by pentagrams agree well with the empirical linear fit.

\subsection{Measurement setup}
\begin{figure}
    \centering
    \includegraphics[width = 0.45\textwidth]{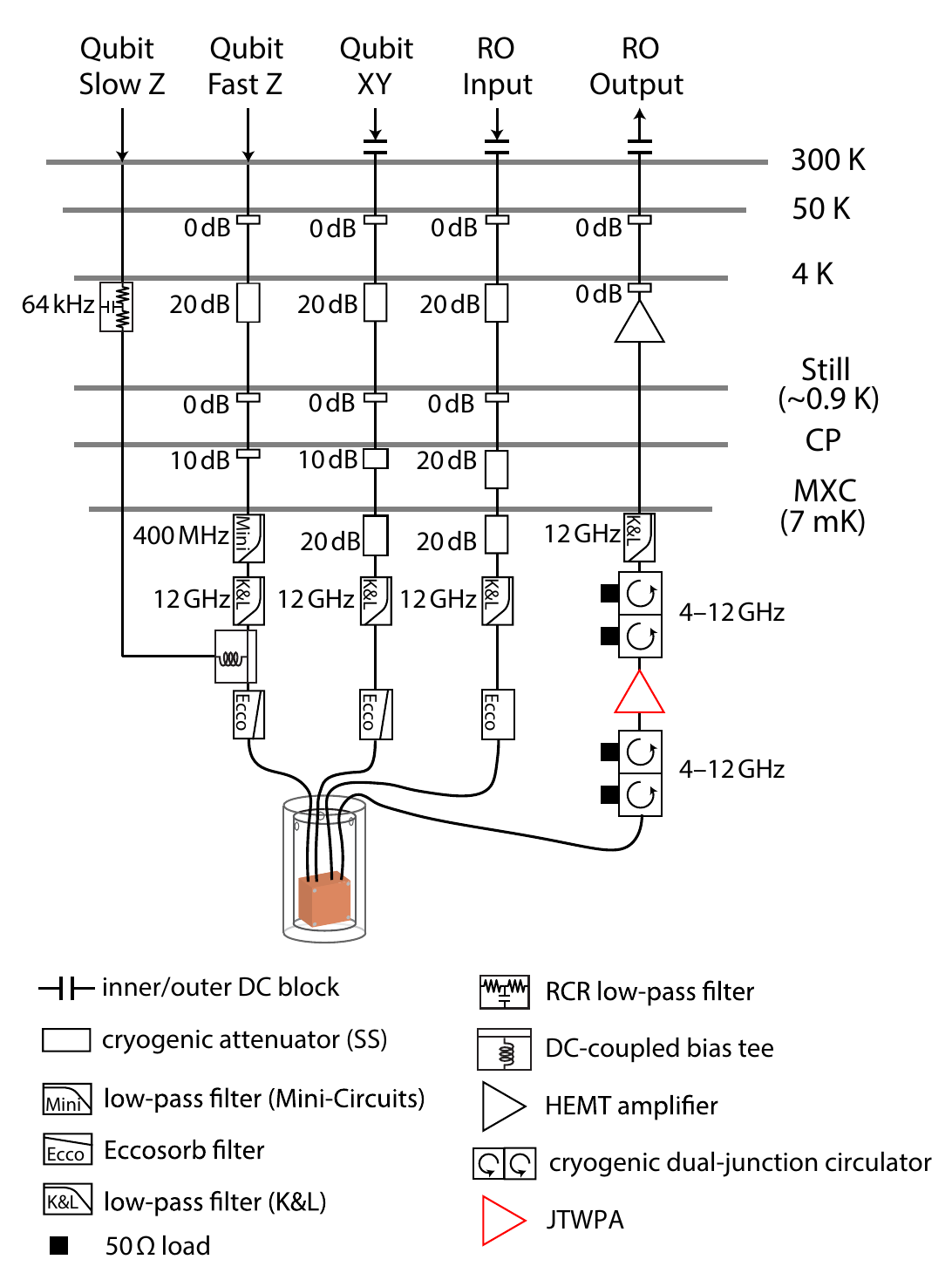}
    \caption{\textbf{Schematic of the measurement setup inside the dilution fridge.} The setup includes slow and fast Z lines for qubit frequency tuning, XY lines for qubit drive, and RO input, RO output lines for qubit dispersive readout. The values for cryogenic attenuation and filters at different temperature stages are listed in the diagram. The pump line for JTWPA is not shown in the diagram for brevity.}
    \label{fig:fridge_setup}
\end{figure}

Fig.~\ref{fig:fridge_setup} shows a schematic of the measurement setup inside the cryogen-free dilution refrigerator (Bluefors LD400), which includes standard shielding and filtering for superconducting transmon qubit experiments~\cite{Ferreira22,Zhang23}. 
The refrigerator consists of multiple temperature stages, which in descending order of temperature are $300$~K, $50$~K, $4$~K, still, cold plate (CP), and mixing chamber (MXC) flanges. The experimental sample is mounted to the MXC plate. Under standard operating conditions, the MXC plate achieves a base temperature of $7$~mK, providing the low temperature environment required for the experiments. 

The frequency control of each transmon qubit is achieved by a bias current that generates a magnetic field threading through the SQUID loop of the transmon qubit. The bias current consists of two parts: the static DC bias (slow Z) and the dynamic RF pulse (fast Z). The static DC bias is generated by a stable DC voltage source (QDevil QDAC) passing through a $2.8$~k$\Omega$ resistor at room temperature. The DC current is filtered by a resistor-capacitor-resistor low-pass filter (Aivon Therma-uD25-G) at $64$~kHz placed at the $4$~K stage. The DC bias provides a broad tuning range and high tuning precision for the transmon qubit frequency. The static DC bias is combined with a dynamic RF pulse (fast Z) through a DC-coupled bias tee (Mini-Circuits ZFBT-4R2GW+ with the capacitor shorted). The fast Z pulse is generated directly by an arbitrary waveform generator (AWG, Keysight M3202A), introducing dynamic tuning capabilities for the transmon qubit frequency. 
For the present experiment, we have not performed corrections for Z line distortions, as discussed in ref.~\cite{Rol20}. Consequently, a slight drift in the patterns of vacuum Rabi oscillations at short Z duration is observed, as seen in Fig.~2 of the main text and Fig.~\ref{fig:TLS_saturation}.

The resonant control of transmon qubit is achieved by the XY line, which couples capacitively to the transmon qubit through a coupling capacitance of approximately $\sim 80$~aF. We use a total of $50$~dB attenuation (XMA cryogenic attenuators) in the fridge XY lines, to accommodate the need of higher microwave driving power for the direct control of TLS (to be discussed in sec.~\ref{sec:TLS_drive}). The microwave signal is generated at room temperature. A pair of intermediate frequency (IF) signals from the AWG (Keysight M3202A), in conjugation with a local oscillator (LO) signal from a microwave signal generator (Rhode\&Schwarz SMB100A), undergoes IQ mixing (Marki Microwave MMIQ-0218L) and generates a single sideband microwave signal that achieves the XY control of the transmon qubit as well as TLS.

For the readout (RO) of the transmon qubits, a microwave RO input signal (generated and filtered similarly to the XY signal) is passed down to the feedline of the sample. The RO output signal from the feedline is first amplified by a JTWPA (Josephson traveling wave parametric amplifier) which is sandwiched by two sets of circulators (Low Noise Factory LNF-CICI4\_12), a HEMT (high electron mobility transistor, Low Noise Factory LNF-LNC4\_16B or LNF-LNC0.3\_14A) amplifer, a low-noise room temperature amplifier (MITEQ LNA-30-0400800-07-10P), a high pass-filter (Mini-Circuits VHF-4600+), a tunable attenuator (Vaunix Lab Brick LDA-133), and another MITEQ low-noise amplifier (MITEQ LNA-30-0400800-07-10P). The RO output signal is then downconverted at room temperature by an IQ mixer and the same LO used to generate the RO input signal. The resulting in-phase (I) and quadrature (Q) signals are filtered (Mini-Circuits VLF-160+), amplified (Mini-Circuits ZFL-500HLNB+), and digitized (Keysight M3102A) for qubit readout. In addition to the aforementioned filtering, low-pass filters (Mini-Circuits VLFX-400+, K\&L Microwave 6L250-12000/T26000) and infrared Eccosorb filters (custom made) are added at the MXC plate where appropriate. All the microwave instruments are synchronized to an external $10$~MHz reference clock from a Rubidium frequency standard (Stanford Research Systems FS725). The AWG and digitizer are both triggered by a delay generator (Stanford Research Systems DG645).

\subsection{Phonon density of states}
\begin{figure}
    \centering
    \includegraphics[width = 0.45\textwidth]{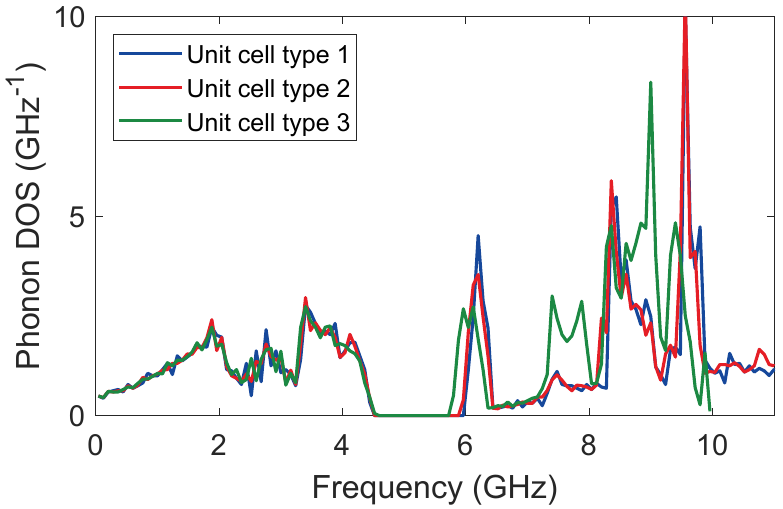}
    \caption{\textbf{Simulated phonon density of states for the three unit cell types.} Colors in the plot correspond to the three types of unit cells as illustrated in the blue, red, and green dashed boxes in Fig.~\ref{fig:JJ_topview}b.}
    \label{fig:phonon_DOS}
\end{figure}

In this section we describe the process of finding the phonon density of states (DOS) based on the COMSOL simulated band structures. To achieve this, we expand upon the simulations presented in Fig.~\ref{fig:acoustic_band_sim}, which focus on special paths connecting points of high degrees of symmetry. These simulations are efficient in finding the bandgap frequencies. However, they do not represent the entire band structures, and consequently, the phonon DOS. To extract the phonon DOS, we leverage symmetries in our structure and uniformly sample one quarter of the first Brilloin zone in the two-dimensional $k$-space, given by $k_x, k_y\in[0,\pi/a]$, using $N$ steps for the $k_x, k_y$ values. Here, $a$ denotes the length of the square unit cell. We then count the total number of $k$ states in the first Brilloin zone, accounting for symmetries. The results are then grouped into frequency bins of $80$~MHz interval based on the frequencies of the eigenstates, and normalized by a factor of $1/(2N-2)^2$ to arrive at the phonon DOS. The resulting phonon DOS for all three unit cell types is shown in Fig.~\ref{fig:phonon_DOS}, and a zoom-in view is displayed in Fig.~3e in the main text.

\subsection{Thermal cycling of the device}
In our experiment, we employ a method known as thermal cycling to generate new distributions of TLS on the same devices. This method has been shown to be effective in ref.~\cite{Shalibo10} when the MXC plate temperature rises above $\sim 20$~K. 
In this study, we perform thermal cycling of the fridge to room temperature, then back down to the base temperature, to ensure the absence of correlation between the two sets of TLS characterized during different cool-down cycles. 

\section{Discussions}
In this section we provide analysis and discussions regarding the phonon engineering of TLS. We start by presenting a comprehensive list of parameters for all 56 TLS characterized and analyzing their distributions. Next, we derive the frequencies of the average acoustic bandgap, shared across all devices, driven by the TLS data. This is followed by an examination of individual devices, where we identify the distinctive acoustic bandgaps corresponding to each of the seven fabricated devices. The results extend and complement the data presented in the main text. Additionally, this analysis unveils disorders in the frequencies of individual device bandgaps, providing an explanation for some of the outlier data points mentioned in the main text. We then provide an explanation for the significant variations observed in the TLS $T_1$ relaxation times, based on the confined geometry of the device, thereby addressing the remaining outlier data points mentioned in the main text. Following this, we present experiments and data that corroborates the anharmonicity of TLS. Our experimental findings suggest that TLS is highly anharmonic. Intriguingly, our data also implies a three-mode coupling involving TLS, the transmon qubit, and an additional TLS. Furthermore, we present data and analysis on the temperature-dependent relaxation of both the transmon qubit and TLS. This motivates a detailed discussion on possible relaxation channels for the TLS, offering a comprehensive view on the temperature-dependent TLS relaxation. Finally, we showcase direct XY control for TLS, which has been used to characterize both the energy relaxation $T_1$ and dephasing $T_2^*$ of TLS. The result raises intriguing questions regarding the interactions between quasiparticles and TLS.

\subsection{TLS parameters}
\begin{figure}
    \centering
    \includegraphics[width = 0.45\textwidth]{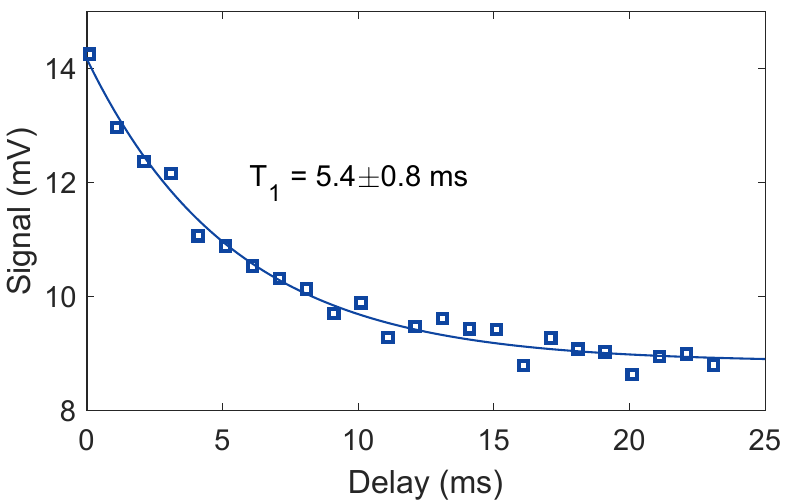}
    \caption{\textbf{$T_1$ relaxation curve of TLS22.} This measurement represents the longest $T_1$ relaxation time of all the 56 TLS characterized. Markers represent experimental data, and the solid line is a simple exponential fit, given by $A\exp(-t/T_1)+B$. Note the `delay' represented on the x-axis is in units of `millisecond', as opposed to `microsecond' used in other plots, to accommodate the long relaxation time.}
    \label{fig:5ms_T1}
\end{figure}

A complete list of TLS frequencies, their respective coupling strengths $g$ to the transmon qubit, and $T_1$ relaxation times, measured on Chip-A and Chip-B, is provided in Table~\ref{tab:TLS_chipA} and Table~\ref{tab:TLS_chipB}, respectively. This dataset that includes 56 distinct TLS has been acquired across seven transmon devices. The TLS characterized span frequencies from $3.7421$~GHz to $6.3935$~GHz, and their $T_1$ values range from $0.25 \pm 0.02$~$\mu$s to $5400\pm 800$~$\mu$s (Fig.~\ref{fig:5ms_T1}). The TLS frequencies and coupling strengths $g$ are extracted through fitting the avoided crossings in the microwave spectroscopy of transmon qubits to the transmon-TLS interaction model

\begin{equation}
\begin{split}
&\ham = \frac{\omega_q}{2} \hat{\sigma}^z_{q} + \frac{\omega_{\mathrm{TLS}}}{2} \hat{\sigma}^z_{\mathrm{TLS}} + \hat{H}_{\mathrm{int}},\\
&\ham_{\mathrm{int}} = g(\hat{\sigma}^+_q\hat{\sigma}^-_{\mathrm{TLS}} + \hat{\sigma}^-_q\hat{\sigma}^+_{\mathrm{TLS}}),
\end{split}
\end{equation}

\noindent where $\omega$ denote their frequencies, $\hat{\sigma}^z, \hat{\sigma}^\pm$ are the Pauli operators.

We remark that certain TLS $T_1$'s measurements are conducted using a strong microwave pulse that directly drives the TLS to its excited-state. These particular TLS $T_1$ values are distinctly marked by $\dagger$ in both Table~\ref{tab:TLS_chipA} and Table~\ref{tab:TLS_chipB}). We note that the $T_1$ values obtained by this method appear comparatively shorter than those measured using SWAP with the transmon qubit. Moreover, TLS relaxation curves measured by this method can sometimes deviate from a simple exponential decay.
When such deviations are evident, we report the relaxation values derived from fitting to a double exponential model (Eq.~\ref{eq:double_exp}). We attribute both phenomenon to interactions with the quasiparticles (QP) induced by the strong microwave pulse~\cite{Wang14,Vool14}, which will be discussed later in sec.~\ref{sec:TLS_de_relax}.

We make a final comment on the TLS parameters regarding TLS35. TLS35 exhibits $T_1 = 261 \pm 21$~$\mu$s, close to the Purcell limit from the transmon qubit at $T_{1,\mathrm{Purcell}} = 210$~$\mu$s (Q$_1$ of Chip-B). In light of this proximity to the Purcell limit, and the likelihood that the measured TLS35 $T_1$ value falls short of its intrinsic $T_1$, TLS35 is excluded from all the median and mean $T_1$ statistics. Among the remaining 55 TLS characterized, their Purcell limits from the transmon are considerably higher than their measured relaxation times. As such, the $T_1$'s of these 55 TLS are all used in the calculation of the median and mean values reported in this study.

\begin{table}[!htbp]
    \centering
    \begin{tabular}{c|c|c|c|c}
    TLS index & freq. (GHz) & g (MHz) & $T_1~(\mu$s$)$ & host qubit \\
    \hline
    1 & $6.3935$ & $9.4$ & $1.9 \pm 0.5^\dagger$ & Q$_1$ CD1 \\
    2 & $5.8818$ & $7.4$ & $7.1 \pm 0.6^\dagger$ & Q$_1$ CD1 \\
    3 & $5.2063$ & $21.1$ & $283 \pm 51^\dagger$ & Q$_1$ CD1 \\
    4 & $5.0730$ & $22.3$ & $1611 \pm 188^\dagger$ & Q$_1$ CD1 \\
    5 & $5.8996$ & $7.2$ & $0.49 \pm 0.02^\dagger$ & Q$_1$ CD1 \\
    6 & $5.7980$ & $20.3$ & $35 \pm 5^\dagger$ & Q$_1$ CD1 \\
    7 & $6.1647$ & $3.7^\$$ & $199 \pm 27^\dagger$ & Q$_1$ CD1 \\
    8 & $5.4359$ & $19.6$ & $948 \pm 223^\dagger$ & Q$_1$ CD1 \\
    9 & $6.1819$ & $3.4^\$$ & $2.7 \pm 0.3$ & Q$_1$ CD2 \\
    10 & $6.0677$ & $26.6$ & $1.87 \pm 0.16$ & Q$_1$ CD2 \\
    11 & $5.9024$ & $16.2$ & $1.90 \pm 0.12$ & Q$_1$ CD2 \\
    12 & $5.7953$ & $6.9$ & $215 \pm 15$ & Q$_1$ CD2 \\
    13 & $5.6563$ & $21.7$ & $1116 \pm 203$ & Q$_1$ CD2 \\
    14 & $6.2740$ & $4.5^\$$ & $7.2 \pm 1.6$ & Q$_2$ CD1 \\
    15 & $5.6891$ & $4.8^\$$ & $2726 \pm 1026^\dagger$ & Q$_2$ CD1 \\
    16 & $5.6534$ & $9.7$ & $544 \pm 131^\dagger$ & Q$_2$ CD1 \\
    17 & $4.9745$ & $15.3$ & $25 \pm 3^\dagger$ & Q$_2$ CD1 \\
    18 & $6.0877$ & $30.2$ & $13.2 \pm 3.9$ & Q$_3$ CD1 \\
    19 & $5.9581$ & $24.9$ & $10.7 \pm 1.5$ & Q$_3$ CD1 \\
    20 & $5.7359$ & $3.8^\$$ & $3.6 \pm 0.3$ & Q$_3$ CD1 \\
    21 & $5.4867$ & $10.2$ & $135 \pm 85^\dagger$ & Q$_3$ CD1 \\
    22 & $4.8196$ & $4.2^\$$ & $5424 \pm 830$ & Q$_3$ CD1 \\
    23 & $4.6952$ & $8.1$ & $90 \pm 38$ & Q$_3$ CD1 \\
    24 & $5.2905$ & $9.4$ & $524 \pm 74$ & Q$_3$ CD2 \\
    25 & $4.6925$ & $6.9$ & $571 \pm 77$ & Q$_3$ CD2 \\
    26 & $4.5098$ & $15.7$ & $5.6 \pm 0.8$ & Q$_3$ CD2 \\
    27 & $4.4604$ & - & $3.8 \pm 0.7$ & Q$_3$ CD2 \\
    28 & $5.4069$ & - & $451 \pm 82^\dagger$ & Q$_4$ CD1 \\
    29 & $5.4097$ & - & $831 \pm 185^\dagger$ & Q$_4$ CD1 \\
    30 & $5.2404$ & $7.3$ & $178 \pm 97^\dagger$ & Q$_4$ CD1 \\
    31 & $5.1759$ & $47.7$ & $177 \pm 10$ & Q$_4$ CD1 \\
    32 & $5.8521$ & $11.3$ & $18 \pm 2$ & Q$_4$ CD2 \\
    33 & $4.9567$ & $10.4$ & $893 \pm 289^\dagger$ & Q$_4$ CD2 \\
    34 & $4.3428$ & $6.4$ & $4.5 \pm 0.5^\dagger$ & Q$_4$ CD2
    \end{tabular}
    \caption{\textbf{List of TLS parameters measured on Chip-A.} The provided list compiles the parameters obtained from characterizing 34 TLS using the four transmon qubit devices on Chip-A. The parameters listed include TLS frequency, interaction strength $g$ with the transmon qubit, as well as their $T_1$ relaxation times, with one standard deviation uncertainty quoted. The hosting transmon qubit device for the TLS and the specific cool-down cycle when these TLS were characterized, are specified at the end of the list.
    $\dagger$: Measurements conducted using the direct TLS control method (sec.~\ref{sec:TLS_drive}), which likely results in shorter TLS $T_1$ measurements due to microwave generated quasiparticles. $\$$ or -: Signify cases where the coupling strength $g$ could not be extracted with high confidence. This might arise from small $g$ values, or overlapping TLS avoided crossings, as identified in the SWAP spectroscopy. CD1/CD2: Indicate the cool-down cycle during which the TLS were measured.}
    \label{tab:TLS_chipA} 
\end{table}

\begin{table}[!htbp]
    \centering
    \begin{tabular}{c|c|c|c|c}
    TLS index & freq. (GHz) & g (MHz) & $T_1~(\mu$s$)$ & host qubit \\
    \hline
    35 & $5.6481$ & $11.7$ & $261 \pm 21$ & Q$_1$ CD1 \\
    36 & $4.9813$ & $9.2$ & $1106 \pm 458^\dagger$ & Q$_1$ CD1 \\
    37 & $4.7006$ & $11.3$ & $85 \pm 8^\dagger$ & Q$_1$ CD1 \\
    38 & $4.4365$ & $28.4$ & $0.25 \pm 0.02^\dagger$ & Q$_1$ CD1 \\
    39 & $5.2866$ & $4.8^\$$ & $257 \pm 46^\dagger$ & Q$_1$ CD1 \\
    40 & $4.8542$ & - & $255 \pm 24$ & Q$_1$ CD2 \\
    41 & $4.7279$ & $19.6$ & $287 \pm 33$ & Q$_1$ CD2 \\
    42 & $4.5474$ & $11.7$ & $478 \pm 39$ & Q$_1$ CD2 \\
    43 & $4.3888$ & $7.5$ & $2.6 \pm 0.5$ & Q$_1$ CD2 \\
    44 & $4.2304$ & - & $1.4 \pm 0.1$ & Q$_1$ CD2 \\
    45 & $3.6385$ & $30.2$ & $3.2 \pm 0.3$ & Q$_1$ CD2 \\
    46 & $5.2956$ & $10.9$ & $474 \pm 298$ & Q$_2$ CD1 \\
    47 & $4.4225$ & $12.1$ & $4.0 \pm 0.3$ & Q$_2$ CD1 \\
    48 & $4.0957$ & $23.2$ & $4.3 \pm 0.6$ & Q$_2$ CD1 \\
    49 & $4.8908$ & $3.9$ & $506 \pm 91$ & Q$_2$ CD2 \\
    50 & $4.3205$ & $27.4$ & $0.71 \pm 0.02$ & Q$_2$ CD2 \\
    51 & $4.0277$ & $10.3$ & $102 \pm 11$ & Q$_2$ CD2 \\
    52 & $5.1151$ & $3.0$ & $652 \pm 103$ & Q$_3$ CD2 \\
    53 & $4.9870$ & $9.2$ & $866 \pm 116$ & Q$_3$ CD2 \\
    54 & $4.3282$ & $9.2$ & $4.6 \pm 1.6$ & Q$_3$ CD2 \\
    55 & $3.7421$ & - & $11.5 \pm 2.8$ & Q$_3$ CD2 \\
    56 & $3.7567$ & - & $22.1 \pm 1.0$ & Q$_3$ CD2
    \end{tabular}
    \caption{\textbf{List of TLS parameters measured on Chip-B.} The provided list compiles the parameters obtained from characterizing 22 TLS using the three transmon qubit devices on Chip-B. The parameters listed include TLS frequency, interaction strength $g$ with the transmon qubit, as well as their $T_1$ relaxation times, with one standard deviation uncertainty quoted. The hosting transmon qubit device for the TLS and the specific cool-down cycle when these TLS were characterized, are specified at the end of the list.
    $\dagger$: Measurements conducted using the direct TLS control method (sec.~\ref{sec:TLS_drive}), which likely results in shorter TLS $T_1$ measurements due to microwave generated quasiparticles. $\$$ or -: Signify cases where the coupling strength $g$ could not be extracted with high confidence. This might arise from small $g$ values, or overlapping TLS avoided crossings, as identified in the SWAP spectroscopy. CD1/CD2: Indicate the cool-down cycle during which the TLS were measured.}
    \label{tab:TLS_chipB} 
\end{table}

\subsection{TLS distributions}\label{sec:TLS_dist}
In this section, we analyze three distributions of the TLS parameters: 1. the distribution of TLS $T_1$, 2. the distribution of their coupling strengths $g$ to the transmon qubit, and 3. the distribution of TLS $T_1$ against $g$.

\subsubsection{TLS $T_1$ distribution}
To begin, we look at the distribution of TLS $T_1$ from all 55 TLS, in supplementary to Fig.~3c in the main text. 
In Fig.~\ref{fig:TLS_vs_T1}, we present the cumulative distribution of TLS $T_1$ values for family A (blue squares) and family B (red triangles), respectively. To characterize this distribution, we employ three commonly used models: the normal distribution, the exponential distribution, and the log-normal distribution. These models are given by their cumulative distribution functions (CDF),
\begin{equation}
    \begin{split}
        &\mathrm{CDF}_\mathrm{norm}(x) = \frac{1}{2}[1+\erf(\frac{x-\mu}{\sqrt{2}\sigma})],\\
        &\mathrm{CDF}_\mathrm{exp}(x) = 1-\exp(-\lambda x),\\
        &\mathrm{CDF}_\mathrm{logn}(x) = \frac{1}{2}[1+\erf(\frac{\ln{x} - \mu}{\sqrt{2}\sigma})].
    \end{split}
\end{equation}

\noindent Based on the fittings using the three models in Fig.~\ref{fig:TLS_vs_T1}, represented by the solid lines, we identify the log-normal distribution as the best representation for our data. 
The resulting parameters yield distinct median $T_1$ values of $4.1\pm 0.2~\mu$s for family A and $414\pm 17~\mu$s for family B. These fitted values are consistent with those outlined in Table~\ref{tab:TLS_stats_by_device} based on the frequencies of the acoustic bandgap, which will be discussed later.

\begin{figure}
    \centering
    \includegraphics[width = 0.45\textwidth]{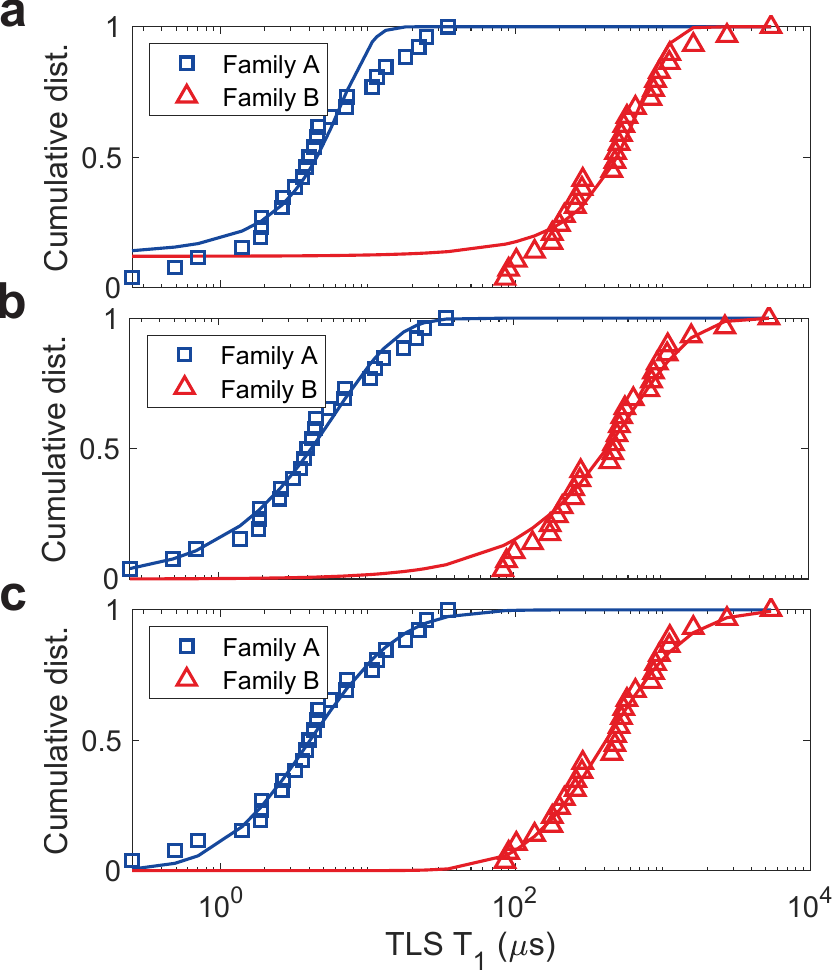}
    \caption{Cumulative distributions of TLS $T_1$ values for family A (blue squares) and family B (red triangles). The distributions are fitted to three commonly used models: \textbf{a,} the normal distribution, \textbf{b,} the exponential distribution, and \textbf{c,} the log-normal distribution.
    The solid lines in each subfigure correspond to the fits using the respective models. The log-normal distribution in \textbf{c} emerges as the best overall fit to the data. This distribution model yields median $T_1$ values of $4.1\pm 0.2~\mu$s for Family A and $414\pm 17~\mu$s for Family B.}
    \label{fig:TLS_vs_T1}
\end{figure}

\subsubsection{TLS coupling strength distribution}
\begin{figure}
    \centering
    \includegraphics[width = 0.45\textwidth]{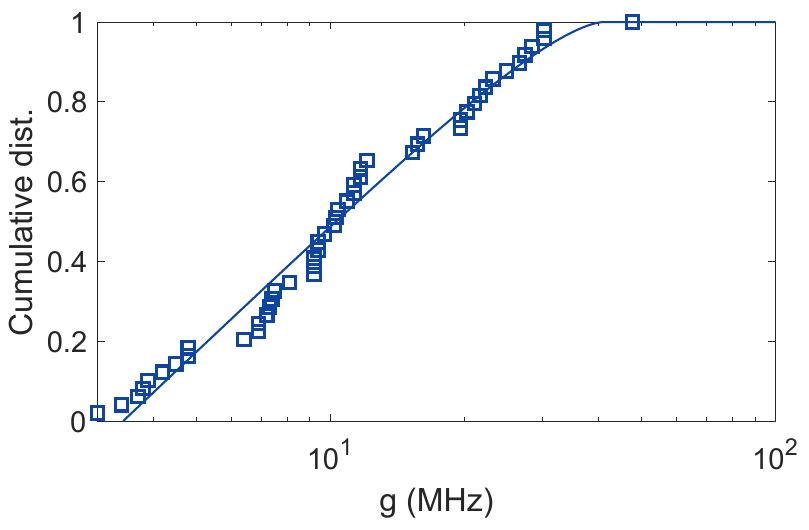}
    \caption{\textbf{TLS cumulative distribution over coupling strength $g$}. Blue square markers represent experimental data, and blue solid line is fitting to the STM prediction of Eq.~\ref{eq:TLS_density}, which yields $\sigma = 0.6~\mathrm{GHz}^{-1} \mathrm{\mu m}^{-2}$. }
    \label{fig:TLS_vs_g}
\end{figure}

Next, we show the distribution of TLS coupling strengths $g$ to the transmon qubit in Fig.~\ref{fig:TLS_vs_g}. According to the standard tunneling model (STM), this distribution is a reflection of the electric dipoles of TLS, which has a density of~\cite{Martinis05}

\begin{equation}\label{eq:TLS_density}
    d^2N/dE dg = \sigma A \sqrt{1-g^2/g_{\mathrm{max}}^2}/g,
\end{equation} 

\noindent where $E, A, \sigma$ are the energy of TLS, the area of the JJ, and the TLS density, respectively. The measured TLS distribution over coupling strength $g$ overall aligns well with STM predictions. To determine the TLS density $\sigma$, we normalize the fitted parameter by the total size of JJ in the transmon qubit, which is approximately $1.66~\mu$m$^2$, and the collective frequency span of the seven transmons in our search for these TLS, which amounts to $22$~GHz. This results in a TLS density of $\sigma = 0.6~\mathrm{GHz}^{-1} \mathrm{\mu m}^{-2}$, in agreement with literature~\cite{Martinis05,Lisenfeld19,Mamin21}. 

\subsubsection{TLS $T_1$ distribution over coupling strength}
In addition, the STM ascribes TLS relaxation to spontaneous phonon emission~\cite{Phillips72,Anderson72} via the interaction between TLS' elastic dipole and the acoustic environment. In this context, the TLS' elastic dipole is proportional to its electric dipole, governed by $\propto \Delta_0/E\propto \vec{d}$. Here $\Delta_0$ is the tunneling energy, $E$ the eigenenergy, and $\vec{d}$ the electric dipole of the TLS. As discussed above, the coupling strength $g$ reflects the electric dipole of TLS. Consequently, a power-law dependence of $1/T_1\propto g^\alpha$ is expected, and has indeed been observed for TLS located inside the Josephson Junctions of a phase qubit~\cite{Shalibo10}. 

\begin{figure}
    \centering
    \includegraphics[width = 0.45\textwidth]{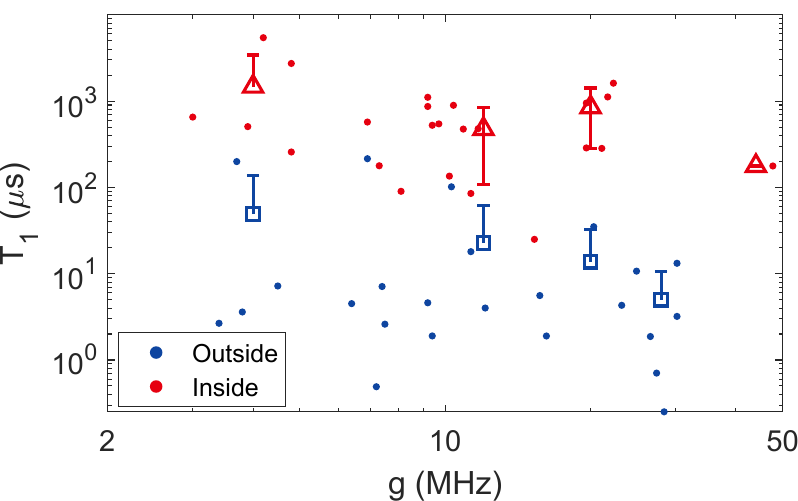}
    \caption{\textbf{Scatter plot of TLS $T_1$ vs coupling strength $g$.} Blue and red filled circles represent the $T_1$ values for TLS outside and inside the average acoustic bandgap (avg. BG), respectively. The corresponding mean $T_1$ values (binned every $8$~MHz by coupling strength $g$) are depicted in blue open squares and red open triangles. Errorbars denote one standard deviation. Missing lower part of errorbars marks standard deviation larger or equal to the mean value.}
    \label{fig:T1_vs_g}
\end{figure}

We remark that this power-law dependence is not unique to the spontaneous phonon emission process. Based on the STM, any relaxation process mediated through either the electric or elastic dipole of TLS would yield a power-law dependence. Therefore, it could also apply to our device, where the spontaneous phonon emission has been suppressed. In Fig.~\ref{fig:T1_vs_g} we show the TLS $T_1$ distribution against their coupling strengths $g$ to the transmon qubit. Here, the blue and red filled circles respectively represent TLS located outside and inside the average acoustic bandgap. At first glance, our data does not readily exhibit a clear power-law dependence for TLS either within or outside the average acoustic bandgap, partly due to the wide spread of the $T_1$ data points that obscures any underlying correlations. 

To address this, we follow the method in ref.~\cite{Shalibo10}, and group the data into bins based on their $g$ values. We then compute the mean and standard deviation in each bin, which are represented by the open markers and their errorbars, respectively, with corresponding colors in Fig.~\ref{fig:T1_vs_g}. These data points reveal a trend of negative correlation between the mean $T_1$ values and the coupling strength $g$, in alignment with expectations from the STM. However, this trend does not convincingly conform to a power-law dependence. We attribute this deviation to the relatively limited size of our available dataset. Additionally, the deviation could arise from the extreme ways in which the acoustic bandgap metamaterial structures the acoustic environment. This influence can even extend to frequencies outside the acoustic bandgap. In such cases, the substantial alteration in the acoustic DOS, rather than the susceptibility to the acoustic environment, may prevail in determining the TLS $T_1$ distribution. 

\subsection{Identification of the average acoustic bandgap}\label{sec:average_bandgap}
As described in the main text, we select a $T_1$ cutoff between $35~\mu$s and $85~\mu$s to categorize all TLS into two groups: family A, characterized by shorter TLS $T_1$, and family B, characterized by longer TLS $T_1$. Remarkably, we observe a strong correlation between this categorization based solely on $T_1$ values and the frequency distribution of TLS within the two families. This correlation motivates us to identify an average acoustic bandgap across all seven transmon devices, using the following cost function, 
\begin{equation}\label{eq:cost_fun}
    \mathcal{C}(f_1,f_2) = \log[1 - F_A(f_1,f_2)\times F_B(f_1,f_2)],
\end{equation}
where the frequency band is specified between $f_1$ and $f_2$. $F_A(f_1,f_2)$ denotes the fraction of TLS in family A whose frequencies lie outside this defined frequency band, while $F_B(f_1,f_2)$ represents the fraction of TLS in family B that fall within this frequency band. 

\begin{figure}
    \centering
    \includegraphics[width = 0.45\textwidth]{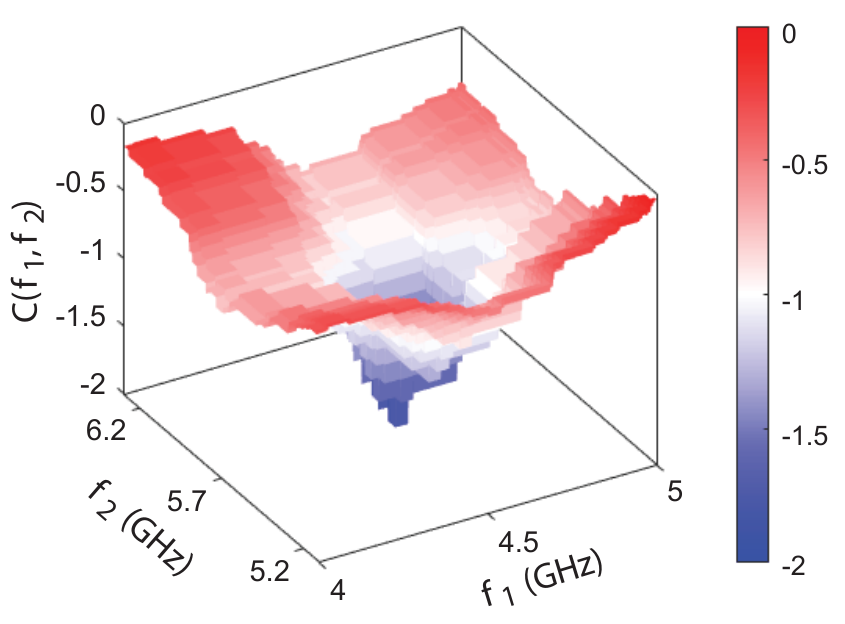}
    \caption{\textbf{Landscape of the cost function $\mathcal{C}(f_1,f_2)$.} $\mathcal{C}_\mathrm{min} = -1.98$ identifies the average acoustic bandgap across all seven transmon devices, with $f_{1,\mathrm{avg.bg}}\in [4.510, 4.547]$~GHz for the lower bandedge and $f_{2,\mathrm{avg.bg}}\in [5.690, 5.735]$~GHz for the upper bandedge.}
    \label{fig:cost_function}
\end{figure}

The landscape of the cost function $\mathcal{C}(f_1,f_2)$ is provided in Fig.~\ref{fig:cost_function} as a function of the lower bandedge frequency $f_1$ and upper bandedge frequency $f_2$. 
The minimum in the landscape yields $\mathcal{C}_\mathrm{min} = -1.98$, which identifies the average acoustic bandgap present across all seven transmon devices. This average bandgap is characterized by 
\begin{equation}
    \begin{split}
        f_{1,\mathrm{avg.bg}}&\in [4.510, 4.547]~\mathrm{GHz}, \\
        f_{2,\mathrm{avg.bg}}&\in [5.690, 5.735]~\mathrm{GHz}.
    \end{split}
\end{equation}

\noindent This average bandgap, in turn, yields a median TLS $T_1$ of $M_{\mathrm{out, 2D}}(T_1) = 4.4$~$\mu$s outside the bandgap and $M_{\mathrm{in, 2D}}(T_1) = 506$~$\mu$s inside the bandgap. Our preference for using median over mean is justified by the large skewness of $1.8$ and $3.3$ for the TLS $T_1$ distributions of the two families.

It's important to note that the average bandgap, shared across different fabricated devices and chips, represents a lower-bound estimate, due to fabrication disorder on individual devices, which will be discussed shortly in sec.~\ref{sec:single_device}. Despite this, the average acoustic bandgap still boasts a width exceeding $1$~GHz. Furthermore, it exhibits a remarkable similarity to the COMSOL simulated bandgap, differing by merely $\lesssim100$~MHz, as shown in Table~\ref{tab:TLS_stats_by_device}. This high degree of agreement underscores the reproducibility and robustness of the overall fabrication process for the acoustic bandgap metamaterial.

Lastly, we emphasize that the determination of the frequencies of the average acoustic bandgap does not depend on any a priori knowledge of the existence of an acoustic forbidden band. Instead, these bandgap frequencies arise naturally from the TLS data itself.

\subsection{Acoustically-shielded TLS on individual devices}\label{sec:single_device}
\begin{figure}
    \centering
    \includegraphics[width = 0.45\textwidth]{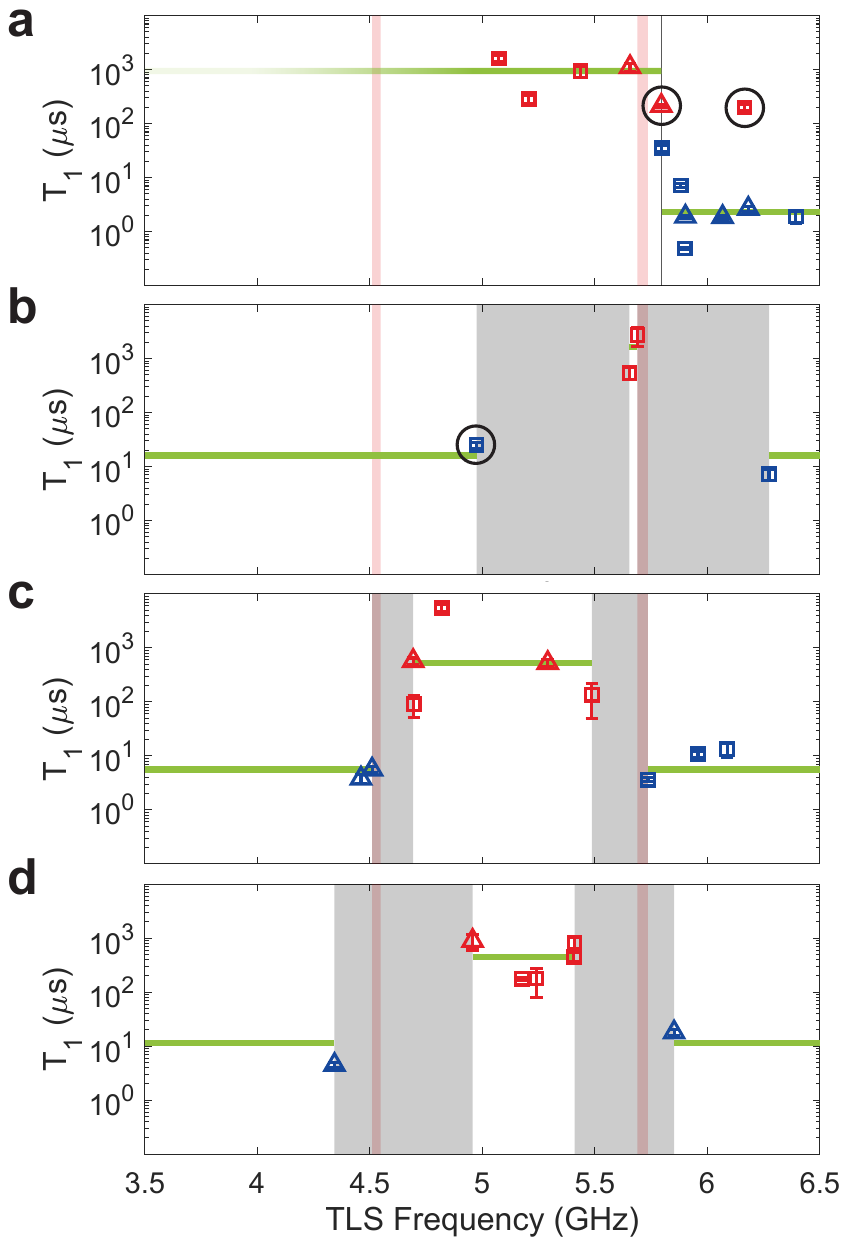}
    \caption{\textbf{TLS $T_1$ measured on individual transmon devices on Chip-A.} \textbf{a,} Chip-A Q$_1$, \textbf{b,} Chip-A Q$_2$, \textbf{c,} Chip-A Q$_3$, and \textbf{d,} Chip-A Q$_4$. Blue and red markers denote TLS belonging to family A and family B, respectively. Additionally, square and triangular markers differentiate between TLS characterized during the first and second cool-down cycles. The gray shading represents the frequencies of the bandedges determined using the cost function for each individual device, while the pink shading corresponds to the average bandgap, serving as a reference.
    Solid green lines are guides to the eye, illustrating median TLS $T_1$ values both inside and outside the acoustic bandgap. Outlier TLS, classified using the average acoustic bandgaps, are marked by black circles.
    Data in \textbf{a, c, d} shows no apparent correlation between the TLS measured during different cool-downs. In the cases of \textbf{a} and \textbf{b}, upward shifts in bandgap frequencies are observed, which are attributed to fabrication disorder.}
    \label{fig:TLS_T1_single_device_A}
\end{figure}

\begin{figure}
    \centering
    \includegraphics[width = 0.45\textwidth]{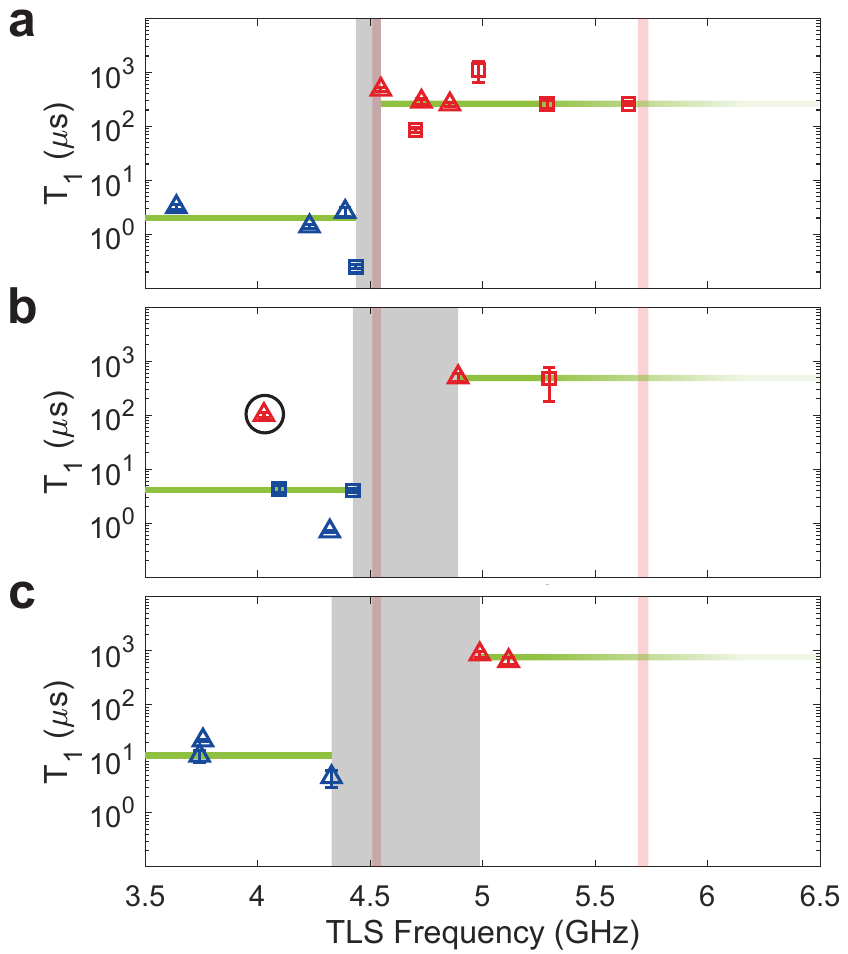}
    \caption{\textbf{TLS $T_1$ measured on individual transmon devices on Chip-B.} \textbf{a,} Chip-B Q$_1$, \textbf{b,} Chip-B Q$_2$, and \textbf{c,} Chip-B Q$_3$. Blue and red markers denote TLS belonging to family A and family B, respectively. Additionally, square and triangular markers differentiate between TLS characterized during the first and second cool-down cycles. The gray shading represents the frequencies of the bandedges determined using the cost function for each individual device, while the pink shading corresponds to the average bandgap, serving as a reference.
    Solid green lines are guides to the eye, illustrating median TLS $T_1$ values both inside and outside the acoustic bandgap. Outlier TLS, classified using the average acoustic bandgaps, are marked by black circles.
    Data in \textbf{a, b} shows no apparent correlation between the TLS measured during different cool-downs.}
    \label{fig:TLS_T1_single_device_B}
\end{figure}

Using the full set of TLS data collected from all seven fabricated transmon devices, we present additional details complementing the information in Fig.~3 from the main text, and demonstrate the robust TLS $T_1$ enhancement on all devices from the acoustic bandgap. In Fig.~\ref{fig:TLS_T1_single_device_A} and Fig.~\ref{fig:TLS_T1_single_device_B}, we present the TLS $T_1$ relaxation data measured on individual transmon devices, for Chip-A and Chip-B, respectively. These plots reveal the existence of two families of TLS, based on their frequencies and $T_1$ times, for each device. We use the same method for identifying the average acoustic bandgap to analyze the bandgaps of these individual devices. For each transmon device, we search for the frequency range of the acoustic bandgap $[f_1, f_2]$ that minimizes the cost function $\mathcal{C}(f_1,f_2)$. This analysis yields the frequencies of either one or both of the bandedges, depending on the frequency ranges and total number of TLS characterized on the particular transmon device. The determined frequencies of the bandedges $f_1$ and $f_2$ are depicted using gray shading in Fig.~\ref{fig:TLS_T1_single_device_A} and Fig.~\ref{fig:TLS_T1_single_device_B}. For reference, we also plot the average bandgap frequencies $f_{1,\mathrm{avg.bg}}$ and $f_{2,\mathrm{avg.bg}}$ determined above in sec.~\ref{sec:average_bandgap}, using pink shading. The overlap between the individual device bandgaps and the average bandgap highlights the robustness of the fabrication process of acoustic bandgap metamaterial. These experimentally identified frequencies of the bandedges are listed in Table~\ref{tab:TLS_stats_by_device}, along with the frequency range of the bandgap given by COMSOL simulations. The table also includes the median and mean TLS $T_1$ values both inside and outside the corresponding bandgaps. 

Upon comparing the experimentally identified bandgaps across all seven devices, we observe disorder in the bandgap frequencies, which is most pronounced in Chip-A Q$_2$, as illustrated in Fig.~\ref{fig:TLS_T1_single_device_A}b. In this case, we identify a bandgap that is up-shifted in frequency, which likely stems from fabrication disorder in the acoustic metamaterials. We remark that the upward shift in the bandgap frequencies for Chip-A Q$_2$ results in the TLS in family A, circled out in black in Fig.~\ref{fig:TLS_T1_single_device_A}b, appearing as an outlier when using the average acoustic bandgap for analysis. However, when we apply the acoustic bandgap specific to this individual device, the TLS falls outside the bandgap, aligning with our expectation for family B. Similarly, we claim that in Fig.~\ref{fig:TLS_T1_single_device_A}a, the TLS circled out on the left side is also misclassified as an outlier when using the average acoustic bandgap. When using the acoustic bandgap of Chip-A Q$_1$, the frequency of this TLS (in family B) actually lies within the acoustic bandgap. 

Regarding the remaining two outliers, marked by black circles in Fig.~\ref{fig:TLS_T1_single_device_A}a and Fig.~\ref{fig:TLS_T1_single_device_B}b, their frequencies locating outside the acoustic bandgap cannot be accounted for by the shift in individual device bandgap frequencies. It's worth noting that both of these outlier TLS belong to family B and exhibit long $T_1$ values, but reside outside the acoustic bandgap. This particular phenomenon is related to the confined geometry of our device, and will be explained in sec.~\ref{sec:variations}.

\begin{table*}
    \centering
    \begin{tabular}{c|c|c|c|c}
    & $f_1$ (GHz) & $f_2$ (GHz) & Median in/out ($\mu$s) & Mean in/out ($\mu$s)\\
    \hline
    Chip-A Q$_1$ & - & 5.796--5.798 & 948/2.3 & 835/31.2 \\
    Chip-A Q$_2$ & 4.975--5.653 & 5.690--6.274 & 1635/16.1 & 1635/16.1 \\
    Chip-A Q$_3$ & 4.510--4.692 & 5.487--5.735 & 524/5.6 & 1349/7.4 \\
    Chip-A Q$_4$ & 4.343--4.956 & 5.410--5.852 & 451/11.2 & 506/11.2 \\
    Chip-B Q$_1$ & 4.437--4.547 & - & 476/4.0 & 496/14.2 \\
    Chip-B Q$_2$ & 4.423--4.891 & - & 490/4.2 & 490/27.6 \\
    Chip-B Q$_3$ & 4.329--4.987 & - & 759/11.5 & 759/12.7 \\
    avg. BG & 4.510--4.547 & 5.690--5.735 & 506/4.4 & 796/24.6 \\
    sim. BG & 4.442 & 5.814 & 462/4.3 & 680/18.5
    \end{tabular}
    \caption{\textbf{Acoustic bandgap and TLS $T_1$ enhancement on individual devices.} Here we outline the experimentally identified bandgap frequencies $[f_1, f_2]$ for individual transmon devices. These bandgaps are determined by minimizing the cost function in Eq.~\ref{eq:cost_fun}. Both the median and mean $T_1$ inside and outside the resultant bandgap are listed to demonstrate the robust two-orders-of-magnitude enhancement consistently observed across all devices. Additionally, the average bandgap determined using data from all TLS (avg. BG) is provided for reference. This average bandgap is compared to the prediction from COMSOL simulation (sim. BG), revealing a small difference of less than $100$~MHz in their bandedge frequencies, which highlights the robustness of the fabrication process for the acoustic metamaterials.}
    \label{tab:TLS_stats_by_device}
\end{table*}

\subsection{Variations in the relaxation time of TLS}\label{sec:variations}
The best fit to the log-normal distribution in sec.~\ref{sec:TLS_dist} suggests significant variations in TLS $T_1$ exceeding an order of magnitude. 
This phenomenon has been predicted and experimentally observed in various systems with confined geometries~\cite{Behunin16}, including opto-mechanical cavities (OMC)~\cite{MacCabe20} and nanomechanical resonators~\cite{Wollack21}. 
For example, numerical modeling in ref.~\cite{MacCabe20} uncovered significant variations in the relaxation rate of both acoustic modes and TLS defects, spanning approximately two orders of magnitude both inside and outside the acoustic bandgap. We argue that this common characteristic in systems with confined geometries is also responsible for the large $T_1$ variations observed in our devices.

When a system is confined to a small scale, its thermal bath responsible for system relaxation is often composed of mesoscopic or even microscopic modes. As a result, these modes possess discrete frequencies in the frequency domain. Depending on the frequency of the system with respect to these discrete frequencies, the relaxation to the thermal bath modes can be dominantly resonant, or off-resonant, which significantly changes the relaxation rate.
This interplay between the system's frequency and the discrete nature of the thermal bath modes contributes to the large variations observed in the system's (in this case, TLS) lifetime in confined geometries.

Let's delve into more details, and first consider the scenario where the TLS frequency lies outside the acoustic bandgap. In this case, the geometric limitations of the device results in discrete phonon modes across the frequency spectrum, with the free spectral range characteristic of the effective size of the system. 
As a consequence, TLS decays through both resonant and relaxation processes into neighboring discrete acoustic modes. The overall relaxation rate therefore heavily depends on the precise frequency configuration of both the TLS and the acoustic modes as well as their interaction strength. If the TLS frequency closely aligns with the resonance frequency of an acoustic mode, and the coupling between them is strong, the TLS will exhibit a fast relaxation rate. On the other hand, when the TLS frequency falls between neighboring acoustic modes, and their couplings are weak, the TLS will have slow relaxation rate. This results in large variations in the $T_1$ relaxation times of TLS outside the bandgap, which also indicates the presence of long-lived TLS outside the bandgap.
These long-lived TLS outside the acoustic bandgap are predicted numerically in ref.~\cite{MacCabe20}, and have been observed experimentally in TLS7 and TLS51 of our device, which account for the outermost two outliers in Fig.~3d in the main text and in Fig.~\ref{fig:TLS_T1_single_device_A}a, Fig.~\ref{fig:TLS_T1_single_device_B}b.

The same argument applies to scenarios within the acoustic bandgap. Here, the TLS and the bulk/local phonon modes of the device exhibit larger frequency detunings on average. Consequently, TLS experience an even weaker relaxation from the interactions with acoustic modes. On average, this leads to prolonged TLS lifetimes inside the bandgap, but still with large variations in their $T_1$'s. 

In summary, the observed variations in TLS relaxation rate, as predicted by numerical modeling in ref.~\cite{MacCabe20} and subsequently verified experimentally in this study, stem from the discrete nature of the thermal bath modes that govern the relaxation process. This discreteness in the thermal bath modes represents a common characteristic intrinsic to devices with a confined geometry.

\subsection{Anharmonicity of TLS}\label{sec:anharmonicity}
\subsubsection{Two-excitation SWAP spectroscopy}
Let us now address a long-standing debate regarding whether coherent TLS are harmonic oscillator modes~\cite{Lupascu09,Bushev10,Cole10,Muller19}. This question is of particular relevance in this study, because there potentially exist high-$Q$ local acoustic modes, which may mimic TLS-like behaviors observed in our experiments. 
We resolve this concern by demonstrating that individual TLS observed in our experiments become saturated with a single quanta of excitation, revealing their anharmonic nature.

\begin{figure}[!htbp]
    \centering
    \includegraphics[width = 0.45\textwidth]{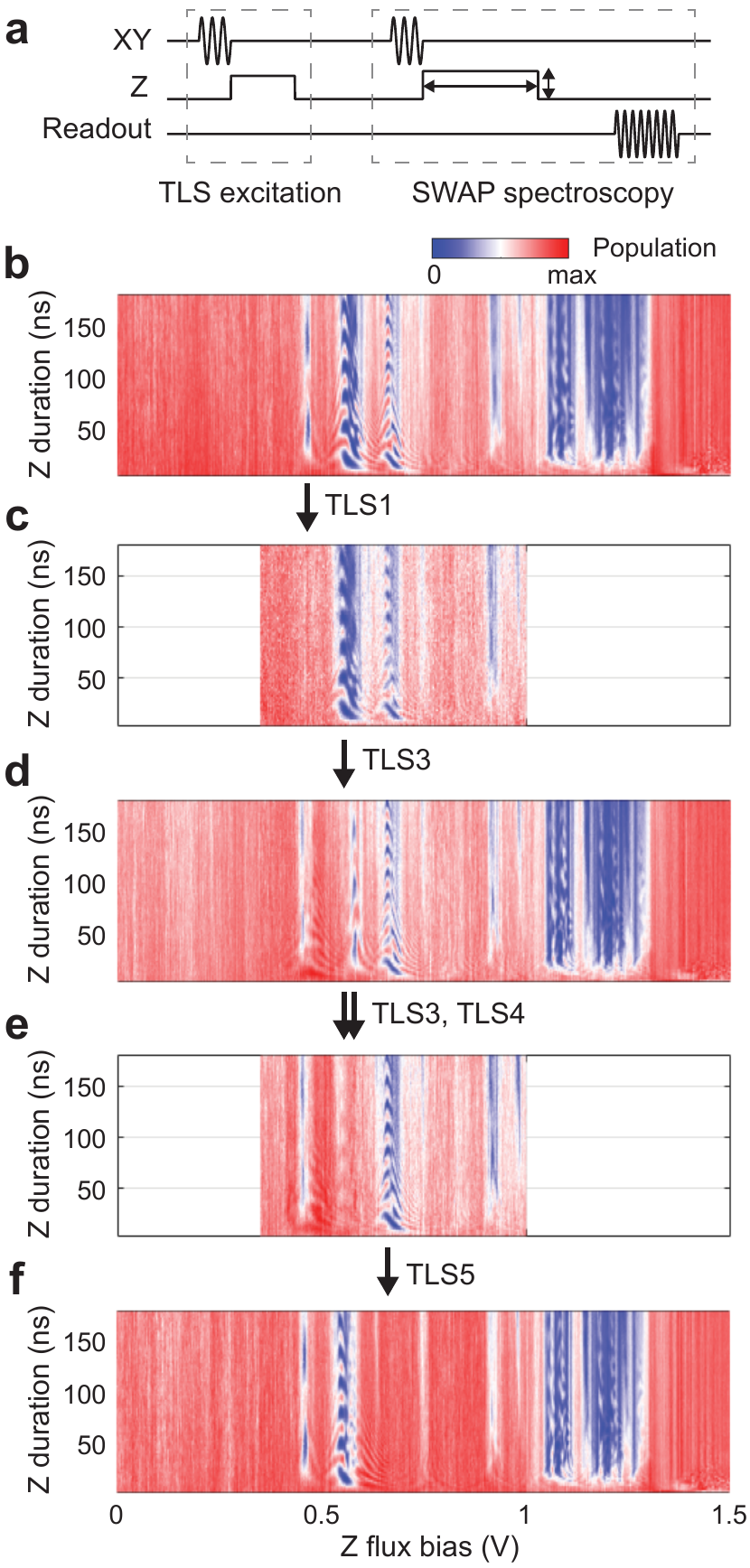}
    \caption{\textbf{SWAP spectroscopy of two quanta of excitations.} \textbf{a,} Experimental sequence for the two excitation SWAP spectroscopy experiment, involving the initial excitation of the target TLS, followed by a regular SWAP spectroscopy that exchanges a second excitation from the transmon qubit. \textbf{b,} Reference SWAP spectroscopy when all TLS are in their ground states $\vert 0\rangle$. \textbf{c,} SWAP spectroscopy with TLS1 in the excited-state $\vert 1\rangle$. \textbf{d,} SWAP spectroscopy with TLS3 in excited-state $\vert 1\rangle$. \textbf{e,} SWAP spectroscopy with both TLS3, TLS4 in excited-state $\vert 1\rangle$. \textbf{f,} SWAP spectroscopy with TLS5 is in excited-state $\vert 1\rangle$. Notably, TLS3 and TLS5 have stronger coupling $g\sim 20$~MHz, making it easier to resolve the presence of vacuum Rabi oscillations at lower frequencies. As a result, we selected TLS3 and TLS5 for a larger range scan of up to $1.5$~V flux bias.}
    \label{fig:TLS_saturation}
\end{figure}

The experimental sequence is illustrated in Fig.~\ref{fig:TLS_saturation}a. We initialize the TLS of interest in its excited-state, $\vert 1\rangle$, and subsequently attempt to transfer a second quanta of excitation from the transmon qubit to this same TLS through SWAP spectroscopy.
If TLS represents harmonic modes, it interacts with the transmon qubit via the interaction Hamiltonian
\begin{equation}\label{eq:int_ham}
\mathcal{H}_{int} = g(\sigma^+ a + \sigma^- a^\dagger),
\end{equation}
where $\sigma^{+(-)}$ is the raising (lowering) operator for the transmon qubit, and $a~(a^\dagger)$ the annihilation (creation) operator for the harmonic mode associated with the TLS. In this context, the TLS harmonic mode would be able to absorb additional excitations at the same frequency (i.e. same flux bias of the transmon qubit in the SWAP spectroscopy) and climb up the Fock state ladder, according to the interaction Hamiltonian in Eq.~\ref{eq:int_ham}.
However, our experimental results, as shown in Fig.~\ref{fig:TLS_saturation}c--f for TLS1, TLS3, TLS4, and TLS5, reveal the absence of vacuum Rabi oscillations between the transmon qubit and excited-state TLS at the TLS frequency. The absence of vacuum Rabi oscillations unambiguously demonstrates that each TLS is fully saturated by a single quanta of excitation. 
This result establishes the anharmonic nature of all four characterized TLS. Furthermore, it validates that the TLS-like behavior we have characterized does not emerge from high-$Q$ local acoustic modes supported by the rectangular platform region of our device.

We note that, to enhance the clarity of the TLS4 patterns, particularly in light of the overlapping TLS3 patterns, an additional step is taken in the experiment concerning TLS4 (Fig.~\ref{fig:TLS_saturation}e). In this experiment, we have prepared both TLS3 and TLS4 in their excited-states through sequential SWAP operations with the excited-state transmon qubit.   
Faint fringes from TLS3 are still visible in the obtained results, due to the $T_1$ relaxation of TLS3 back to its ground state while the preparation of TLS4 in the excited-state is in progress.

Furthermore, we conducted an extensive search across a wide frequency range for the potential $\vert1\rangle\leftrightarrow\vert2\rangle$ transitions of these four TLS. We compare the resulting SWAP spectroscopy of two excitations (Fig.~\ref{fig:TLS_saturation}c--f) to that of a single excitation (Fig.~\ref{fig:TLS_saturation}b). If the transition dipole moments of the $\vert1\rangle\leftrightarrow\vert2\rangle$ transition and the $\vert0\rangle\leftrightarrow\vert1\rangle$ transition are comparable, and the second excited-state of the TLS possesses a decent coherence time ($\gtrsim$ few hundred nanoseconds), we would anticipate the emergence of additional vacuum Rabi oscillations between the transmon and the second excitation of the TLS when they are on resonance. In particular, TLS3 and TLS5 exhibit a stronger coupling to the transmon qubit, $g\sim 20$~MHz, than the other two TLS. Given the larger value of $g$, it becomes easier to discern the presence of vacuum Rabi oscillations, particularly for low-frequency transitions, as the frequency resolution of the transmon, when functioning as a spectrometer, is limited by the amplitude resolution of the Z flux bias. This resolution tends to degrade as the transmon is tuned towards lower frequencies and becomes more flux sensitive. 

In light of this, we selected TLS3 and TLS5 to conduct a broader frequency scan, extending up to $1.5$~V of the flux bias. Notably, our scans up to a flux bias of $1.5$~V did not reveal any additional vacuum Rabi oscillations. If TLS were to have a third state, these measurements would place bounds to the anharmonicity of the TLS, as listed in Table \ref{tab:TLS_anharmonicity_bound}, where bound1 and bound2 denote the bounds for positive and negative anharmonicity, respectively, that satisfy $\alpha>\mathrm{bound1}>0$ or $\alpha<\mathrm{bound2}<0$. The data from all four TLS collectively indicate a conservative bound for the anharmonicity $\alpha$, given by $\alpha > 0.41$~GHz or $\alpha < -1.3$~GHz, for positive and negative anharmonicity, respectively. Furthermore, the absence of emergent vacuum Rabi oscillations in the extensive scan range ($0\textrm{--}1.5$~V, corresponding to transmon frequencies of approximately $6.48\textrm{--}1.47$~GHz) for TLS3 and TLS5 suggests that either TLS is highly anharmonic ($\alpha > 0.59$~GHz or $\alpha < -4.1$~GHz), or TLS has only two levels. It's important to note that all the data presented here were acquired on Chip-A Q$_1$ during CD2. While these TLS are designated TLS9--13 in Table~\ref{tab:TLS_chipA}, in this context, we refer to them as TLS1--5 to maintain consistency with the main text. 

We would like to conclude this section by addressing a noticeable difference between the SWAP spectroscopy presented here (Fig.~\ref{fig:TLS_saturation}) and in Fig.~2d of the main text. This slight difference is attributed to the frequency shifts of TLS during the $200$~mK thermal cycling in the temperature-dependent relaxation measurements, to be discussed in sec.~\ref{sec:Temp_dependence}. The SWAP spectroscopy of two excitations was performed approximately two months after the initial single-excitation SWAP spectroscopy on the same qubit (Fig.~2 in the main text). In the meantime, thermal cycling up to $200$~mK (sec.~\ref{sec:Temp_dependence}) was carried out and lasted for over one month, inducing frequency shifts in all the TLS. We compare the TLS frequencies pre- and post-thermal cycling using the SWAP spectroscopy data (microwave spectroscopy was not taken post-thermal cycling). The comparison reveals that TLS1 frequency shifted by approximately $\gtrsim -100$~MHz, TLS3 by $\gtrsim -15$~MHz, TLS4 by $\gtrsim 40$~MHz, and TLS5 by $\gtrsim -10$~MHz. We note that TLS2 frequency drifted beyond the range of our scan, rendering us capable of only measuring the two-excitation SWAP spectroscopy for TLS1, TLS3, TLS4, and TLS5.

For comparison, the data in Fig.~\ref{fig:TLS_saturation} were taken over the span of one month, and we did not observe noticeable frequency drifts of TLS1, 3, 4, 5 through SWAP spectroscopy. This suggests that the large frequency shifts were more likely provoked by elevated temperatures during the $200$~mK thermal cycling, rather than being a sole consequence of the long time gap between measurements. 
Furthermore, when we tracked individual TLS frequencies for up to $90$~hours at the $7$~mK base temperature of the fridge (data not shown), we measured TLS frequency drifts in the range of a few MHz, with the largest TLS frequency jump $\lesssim 2$~MHz. The observed TLS frequency jumps are significantly smaller than the frequency shifts of TLS1, 3, 4, 5, as described above. Notably, these few MHz frequency drifts are more than $10\times$ smaller than reported in ref.~\cite{Klimov18}, which is worth further investigation.

\begin{table}
    \centering
\begin{tabular}{l|c|c}
     &  bound1 (GHz) & bound2 (GHz)  \\
     \hline
    TLS1 & 0.41 & -1.6 \\
    TLS3 & 0.59 & -4.4 \\
    TLS4 & 0.65 & -1.3 \\
    TLS5 & 0.87 & -4.1
\end{tabular}
    \caption{\textbf{Bounds on TLS anharmonicity from SWAP spectroscopy of two excitations.} The values bound1 and bound2 are determined by the experiments shown in Fig.~\ref{fig:TLS_saturation}, assuming positive and negative anharmonicity, that satisfy $\alpha>\mathrm{bound1}>0$ and $\alpha<\mathrm{bound2}<0$, respectively.}
    \label{tab:TLS_anharmonicity_bound}
\end{table}

\subsubsection{TLS-transmon-TLS three-mode coupling}
\begin{figure}
    \centering
    \includegraphics[width = 0.45\textwidth]{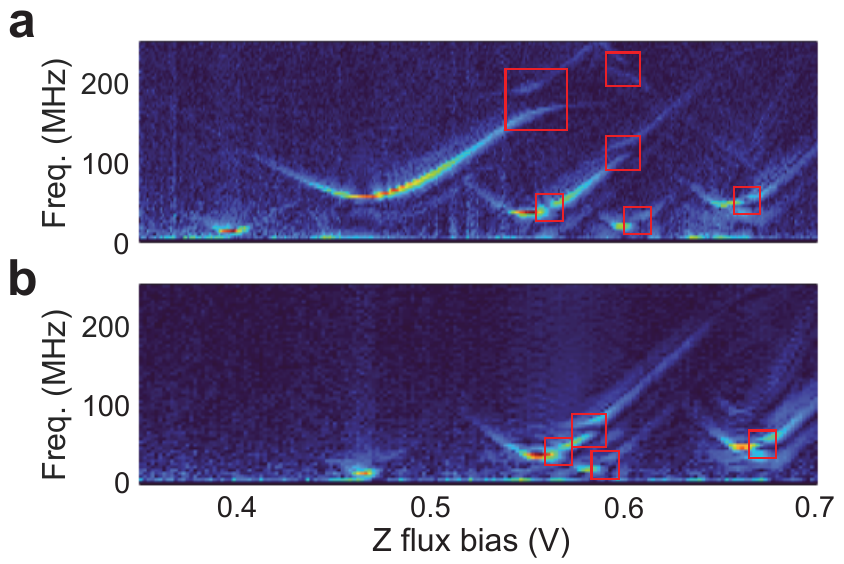}
    \caption{\textbf{Fourier transform of SWAP spectroscopy:} \textbf{a,} of Fig.~2d of main text, and \textbf{b,} of Fig.~\ref{fig:TLS_saturation}b. The two datasets are taken from the same transmon device before and after a $200$~mK thermal cycling. Red boxes highlight gaps in the vacuum Rabi frequencies, a result of TLS-transmon-TLS three-mode couplings.}
    \label{fig:TLS_SWAP_FFT}
\end{figure}

In the SWAP spectroscopy of both Fig.~2 of the main text and Fig.~\ref{fig:TLS_saturation}, we have observed deviations from ideal chevron patterns, most noticeable for TLS3 and TLS5. These deviations are attributed to TLS-transmon-TLS three-mode couplings. This becomes clear when examining the Fourier transform of the single-excitation SWAP spectroscopy data, which reveals the frequencies of the vacuum Rabi oscillations, as shown in Fig.~\ref{fig:TLS_SWAP_FFT}. Within the single-excitation manifold, the Hamiltonian governing the TLS-transmon coupled system can be expressed as follows:

\begin{equation}\label{eq:transmon-TLS}
\ham_{\mathrm{TLS}-q} =
\begin{pmatrix}
\Delta/2 & g\\
g & -\Delta/2
\end{pmatrix},
\end{equation}

\noindent where $\Delta = \omega_\mathrm{TLS} - \omega_q$ is the detuning between the transmon and the TLS, and $g$ their coupling strength. Diagonalizing the Hamiltonian yields two eigenstates, featuring a frequency gap of $\sqrt{\Delta^2 + 4g^2}$. This particular frequency corresponds to the vacuum Rabi oscillations between the transmon and the TLS, which is captured by our single-excitation SWAP spectroscopy, shown in Fig.~\ref{fig:TLS_SWAP_FFT}.

Now, we introduce the interaction of the transmon qubit with a second TLS. We expand the single-excitation manifold Hamiltonian in Eq.~\ref{eq:transmon-TLS} to account for TLS-transmon-TLS three-mode coupling as follows

\begin{equation}
    \ham_{\mathrm{TLS}-q-\mathrm{TLS}} = 
    \begin{pmatrix}
        \Delta_1 & g_1 & 0\\
        g_1& 0 & g_2\\
        0 & g_2 & \Delta_2
    \end{pmatrix},
\end{equation}

\noindent where $\Delta_1$ ($\Delta_2$) and $g_1$ ($g_2$) are the detuning and coupling strength between the first (second) TLS and the transmon qubit, respectively. Upon hybridizing the transmon qubit and the second TLS, the lower-right section of the Hamiltonian is block diagonalized, yielding

\begin{equation}\label{eq:TLS-transmon-TLS_block_diag}
    \ham_{\mathrm{TLS}-q-\mathrm{TLS}}' = 
    \begin{pmatrix}
        \Delta_1 & g_1 & 0\\
        g_1& \omega_{h2}^- & 0\\
        0 & 0 & \omega_{h2}^+
    \end{pmatrix},
\end{equation}

\noindent where $\omega_{h2}^\pm =\frac{1}{2}(\Delta_2 \pm\sqrt{\Delta_2^2+4g_2^2})$ are the eigenfrequencies of the hybridized transmon-TLS2 states. It's worth noting that the states associated with eigenfrequencies $\omega_{h2}^-$ and $\omega_{h2}^+$ correspond to the bright and dark states, respectively. Further diagonalization of the Hamiltonian yields,

\begin{equation}
\ham_{\mathrm{TLS}-q-\mathrm{TLS}}'' = 
    \begin{pmatrix}
        \omega_{h1}^+ & 0 & 0\\
        0& \omega_{h1}^- & 0\\
        0 & 0 & \omega_{h2}^+
    \end{pmatrix},
\end{equation}

\noindent where $\omega_{h1}^\pm = \frac{1}{2}(\Delta_1 + \omega_{h2}^- \pm \sqrt{(\Delta_1 - \omega_{h2}^-)^2 + 4g_1^2})$.

In the case where $\Delta_2\gg g_2$, the state with eigenfrequency $\omega_{h1}^-\approx 0$ is transmon like. The interaction between TLS1 and the transmon is unaffected by the presence of TLS2, bringing us back to the case of Eq.~\ref{eq:transmon-TLS}, with a single Rabi frequency of $\omega_{h1}^+ - \omega_{h1}^-\approx \sqrt{\Delta_1^2 + 4 g_1^2}$. 
However, when $\Delta_2\sim g_2$, the transmon and TLS2 become strongly hybridized, perturbing the TLS1-transmon interaction. Let's consider the SWAP spectroscopy experiment, where the excited-state transmon is flux-tuned close to the resonance frequency of TLS2. Due to the strong hybridization between TLS2 and the transmon, this experimental sequence effectively prepares a superposition of the states associated with eigenfrequencies $\omega_{h1}^-$ and $\omega_{h2}^+$. Measurement of the transmon then reveals two Rabi frequencies for TLS1, one at $\omega_{h1}^+ - \omega_{h1}^-$ and another one at $\omega_{h1}^+ - \omega_{h2}^+$. This splits the original TLS1-transmon Rabi frequency curve of $\sqrt{\Delta_1^2+4g_1^2}$, opening up a gap. The gap size is determined at $\Delta_2 = 0$, given by
\begin{equation}\label{eq:gap_size}
\begin{split}
&(\omega_{h1}^+ - \omega_{h1}^-) - (\omega_{h1}^+ - \omega_{h2}^+)\\ 
=& \frac{1}{2}[\sqrt{(\Delta_1+g_2)^2+4g_1^2} - (\Delta_1 - 3 g_2)]\\
\approx& 2g_2 + \frac{g_1^2}{\Delta_1+g_2},
\end{split}
\end{equation}
where the last line in Eq.~\ref{eq:gap_size} takes the approximation $(\Delta_1+g_2)^2\gg 4g_1^2$.

We have highlighted such gaps in red boxes in Fig.~\ref{fig:TLS_SWAP_FFT}.  
The flux biases at which these gaps occur correspond to resonance conditions with a second TLS. In our experiment, the flux biases coincide with the deviations from ideal chevron patterns in the time domain. Therefore it is likely that the imperfect patterns observed in the SWAP spectroscopy arise from the three-mode coupling of TLS-transmon-TLS, where a second TLS is involved. Importantly, it is worth noting, that rather than causing these spectral anomalies, the three-mode coupling could facilitate useful operations such as TLS-TLS entanglement~\cite{Grabovskij11}.

\subsection{Dependence of relaxation on temperature}\label{sec:Temp_dependence}
\subsubsection{Temperature-dependent relaxation of transmon}\label{sec:transmon_T1_model}
Before delving into possible TLS relaxation channels and presenting data on the temperature-dependent relaxation of TLS, we first visit the relaxation mechanisms for the transmon qubit. The understandings of these mechanisms are well-established.
We employ a widely adopted model that accounts for relaxations from TLS/dielectric loss, quasiparticles (QP), and other contributions. Following ref.~\cite{Gao08,Catelani11,Crowley23}, the $Q$-factor of the transmon is determined by,

\begin{equation}\label{eq:Q_tot}
    \frac{1}{Q} = \frac{1}{Q_\mathrm{TLS}} + \frac{1}{Q_\mathrm{QP}} +\frac{1}{Q_\mathrm{other}},
\end{equation}

\noindent where

\begin{equation}\label{eq:Q_TLS}
    Q_\mathrm{TLS}(\bar{n}, T) = Q_{\mathrm{TLS}, 0} \frac{\sqrt{1+(\frac{\bar{n}^{\beta_2}}{D T^{\beta_1}})\tanh(\frac{\hbar\omega}{2k_B T})}}{\tanh(\frac{\hbar\omega}{2k_B T})},
\end{equation}

\noindent and

\begin{equation}\label{eq:Q_QP}
    Q_\mathrm{QP} = Q_{\mathrm{QP},0}\frac{e^{\Delta_0/k_B T}}{\sinh(\frac{\hbar\omega}{2k_B T})K_0(\frac{\hbar\omega}{2k_B T})}.
\end{equation}

\noindent $\omega$ is the angular frequency of the transmon qubit; $T$ the temperature; $\bar{n}$ the effective excited-state population; $Q_{\mathrm{TLS},0},~Q_{\mathrm{QP},0}$ the inverse of linear absorption due to TLS and quasiparticles; $D, \beta_1, \beta_2$ are parameters that characterize TLS saturation; $\Delta_0$ the superconducting gap; $K_0$ the zeroth order modified Bessel function of the second kind.

This model gives a quantitative good fit, as shown in Fig.~\ref{fig:TLS5_vs_temp}a in the gray dash-dotted line, that characterizes three distinct regimes as we increase the MXC plate temperature of the fridge: between $7$--$20$~mK, transmon $Q$ factor remains temperature-independent, described by $Q_\mathrm{other}$; between $20$--$130$~mK transmon $Q$ increases slightly due to the saturation of TLS by thermal phonons; and above $130$~mK, transmon $Q$ decreases due to interactions with the thermally-activated QPs. We remark that the observed plateau in the measured $Q$-factor, that corresponds to the contribution of $Q_\mathrm{other}$, is not necessarily interpreted as a truly temperature-independent loss channel. This plateau could potentially arise from a discrepancy between the true temperature of the transmon device and the temperature measured on the MXC plate, especially at the lowest MXC plate temperatures. At MXC plate temperatures below $20$~mK, it is common for the true temperature of the transmon device to plateau at a higher temperature, typically around $\sim 50$~mK due to limited thermalization with the MXC plate~\cite{Lucas23}. When the MXC plate temperature is applied in Eq.~\ref{eq:Q_tot} instead of the true device temperature, the effects from the plateaued transmon device temperature can manifest as a temperature-independent residual loss term $Q_\mathrm{other}$. This phenomenon has been observed in various systems at milliKelvin temperatures, where the true device temperatures are estimated to plateau to a level around $50$~mK~\cite{Vepsalainen20}.

\subsubsection{Temperature-dependent relaxation of TLS}\label{sec:TLS_T1_model}
\begin{figure}[!htbp]
    \centering
    \includegraphics[width = 0.45\textwidth]{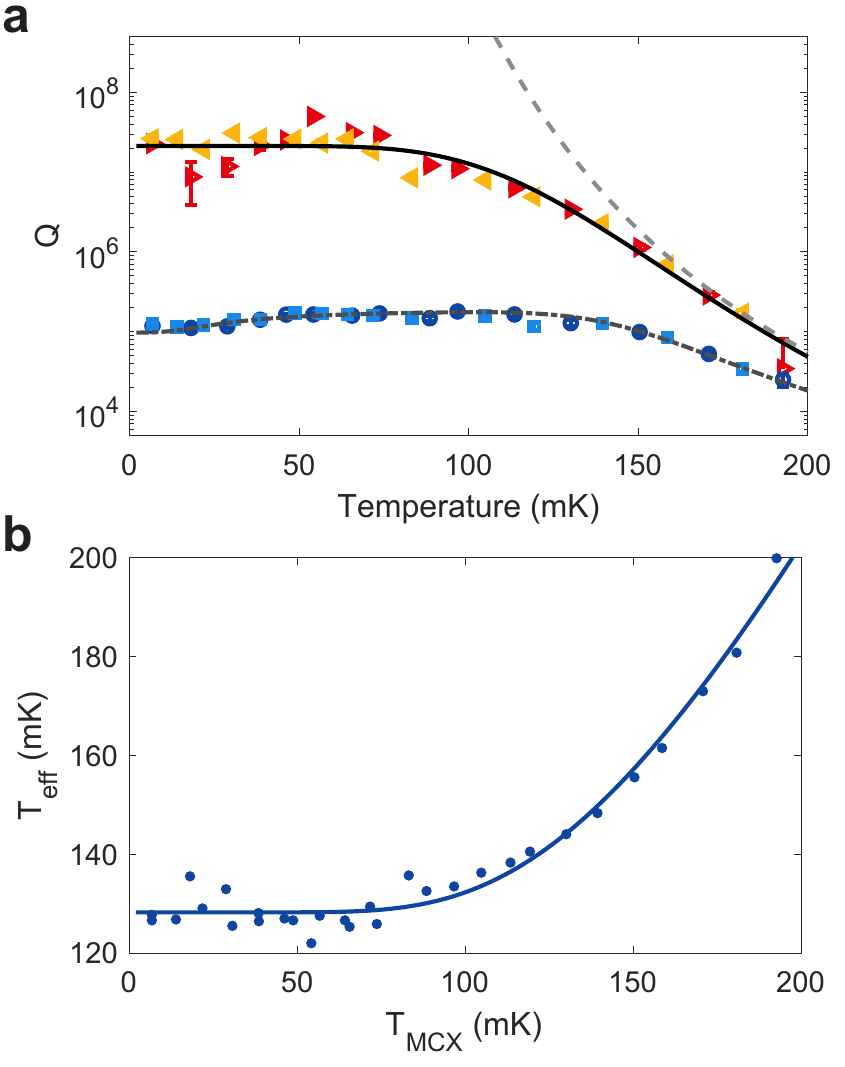}
    \caption{\textbf{Investigating relaxation channels of TLS.} 
    \textbf{a,} Plot of the $Q$-factor of both TLS5 and the transmon qubit as a function of the mixing plate temperature. 
    The red (dark blue) markers denote measurements of the TLS (transmon qubit) during device warm-up (WU), and the yellow (light blue) markers during the device cooldown (CD). The gray dashed line corresponds to a phenomenological model of QP damping of TLS using the mixing plate temperature.
    The black solid line represents a correction to the gray dashed line, when using the effective temperature from \textbf{b}. The gray dash-dotted line is a fit to the transmon curve with a model including thermal saturation of weakly coupled TLS and damping from thermally-activated QPs, using the same effective temperature as TLS.
    \textbf{b,} Plot of the effective temperature against the mixing plate temperature. The effective temperature is deduced by assuming a single TLS energy relaxation channel of QPs. The empirical fit assumes the functional form $T_\mathrm{eff} = A\sqrt{1+B\tanh(C/T_\mathrm{MXC})}/\tanh(C/T_\mathrm{MXC})$.
    }
    \label{fig:TLS5_vs_temp}
\end{figure}

Now, we elaborate on the case study of TLS5 regarding its temperature-dependent relaxation, presented in Fig.~4b of the main text, as a prelude to the rich TLS physics enabled by the significantly extended TLS lifetime. In this experiment, we investigate the thermal bath by monitoring TLS5 as we warm up (WU) the mixing plate of the fridge from the base temperature of $7$~mK to $193$~mK, and cool down (CD) back to the base temperature, shown in Fig.~\ref{fig:TLS5_vs_temp}a. No hysteresis is observed between the WU (red markers) and CD (yellow markers) paths. The quality factor $Q$ of the TLS saturates to $Q\sim 2.5\times10^7$ at temperatures below $\sim 75$~mK, then drops by three orders of magnitude to $Q<5\times10^4$ at $193$~mK. Transmon $Q$ is superimposed for reference, which agrees with a widely adopted model that considers effects from resonant TLS and quasiparticles~\cite{Catelani11,Crowley23,Gao08,SOM}, shown by the gray dash-dotted line, as discussed in the previous section, Eq.~\ref{eq:Q_tot}--\ref{eq:Q_QP}. The TLS and transmon curves have inconsistent trends, suggesting that the temperature-dependent TLS relaxation is not dominated by the Purcell limit of the transmon. 

For temperatures above $150$~mK, the drop in TLS $Q$-factor seems to follow that of the transmon qubit, which is dominated by thermally-activated QPs. Here we assumes a phenomenological QP loss-model given by:

\begin{equation}\label{eq:TLS_relaxation}
Q_\mathrm{TLS}(T)^{-1} = \Gamma_\mathrm{qp} \frac{\sinh(\frac{\hbar\omega}{2k_\mathrm{B} T})K_0(\frac{\hbar\omega}{2k_\mathrm{B} T})}{e^{\Delta_0/k_\mathrm{B} T}},
\end{equation}

\noindent where $T$ is temperature, $\omega$ the TLS transition frequency, $k_\mathrm{B}$ the Boltzmann constant, $\Delta_{0} = 1.764\times 1.2$~K the superconducting gap of Al, and $K_{0}$ the 0-th order modified Bessel function of the second kind. This model describes quasiparticles (QP) in thermal equilibrium, which tunnel through the JJ and interact with the electric dipole of the TLS. The functional form resembles the thermal equilibrium QPs interacting with SC qubits~\cite{Catelani11}, as will be discussed in Sec.~\ref{sec:TLS_QP_relax}. This loss channel could result in the many orders-of-magnitude $Q$-factor change within a small temperature range of $200$~mK. 

We fix the superconducting gap of Al $\Delta_0$, which leaves only one free parameter, $\Gamma_\mathrm{qp}$, in the model. Using the mixing plate temperature for the model, Eq.~\ref{eq:TLS_relaxation} yields the gray dashed line in Fig.~\ref{fig:TLS5_vs_temp}a. As can be seen, predictions of this simple model diverges from experimental data at temperatures below $150$~mK. However, recent studies~\cite{Vepsalainen20,Gustavsson16,Serniak18} show an excess of non-equilibrium (ne) QP population, which corresponds equivalently to thermal equilibrium QP population at approximately $150$~mK. Comparing the measured $Q$-factor to the gray dashed line, we map each mixing plate temperature $T_\mathrm{MXC}$ to an effective temperature $T_\mathrm{eff}$, shown in Fig.~\ref{fig:TLS5_vs_temp}b. $T_\mathrm{eff}$ is empirically fitted, using $T_\mathrm{eff}(T_\mathrm{MXC}) = A\sqrt{1+B\tanh(C/T_\mathrm{MXC})}/\tanh(C/T_\mathrm{MXC})$, represented by the blue solid line in Fig.~\ref{fig:TLS5_vs_temp}b. Using the effective temperature $T_\mathrm{eff}(T_\mathrm{MXC})$ in Eq.~\ref{eq:TLS_relaxation}, yields the black solid line in Fig.~\ref{fig:TLS5_vs_temp}a, in qualitative agreement with measured data. The transmon fit using $T_{\mathrm{eff}}$ remains largely unaffected. If QPs were to explain the plateauing behavior in TLS $Q$-factor, this analysis predicts a QP saturation temperature of approximately $130$~mK. This is consistent with recent studies~\cite{Vepsalainen20,Gustavsson16,Serniak18} of ne QP population in Al superconducting circuits, which infer a QP population with effective temperature of $120$-$150$~mK.

We emphasize that TLS-QP coupling is simply one possible explanation for the observed temperature-dependent TLS relaxation behavior. This behavior deviates from predictions of the standard tunneling model of TLS~\cite{Behunin16}, and reveals previously unexplored TLS physics that requires further investigation. Other possible mechanisms that could contribute to the plateauing behavior in the measured TLS $Q$ includes temperature-independent channels such as TLS coupling to heavily damped grain-boundary motion in the polycrystalline Al layers~\cite{Ke47,Wollack21}. In either case, this temperature-dependent TLS $Q$ behavior deviates from predictions of the standard tunneling model of TLS~\cite{Behunin16}. It reveals previously unexplored TLS physics that requires further investigation.

Regarding the temperature-dependent measurements of TLS, it is important to note that at elevated temperatures, we observed not only a decrease in signal-to-noise ratio, but also fluctuations in the $T_1$ relaxation curve of the TLS when we repeat the measurements. These fluctuations are likely attributed to reconfigurations of the relaxation bath. To mitigate this issue, we averaged over many $T_1$ relaxation curves in our experiment. Subsequently, all the long averaged relaxation curves were fitted to a stretched exponential model, given by 
\begin{equation}
    p(t) = A\exp[-(t/T_1)^n] + B.
\end{equation}
This averaging and fitting approach was applied to TLS5 data in Fig.~\ref{fig:TLS5_vs_temp}. The distribution of fitted exponent yields $n_{\mathrm{TLS5}} = 0.77 \pm 0.14$.

\subsection{Possible relaxation channels for TLS}\label{sec:TLS_relaxation_channels}
Guided by the measurements of temperature-dependent relaxation for both TLS and the transmon, we will now discuss in more detail the possible relaxation channels for the TLS. In this context, we will outline prevalent relaxation mechanisms expected in our system, evaluate their alignment with the experimental data, and where applicable, suggest experiments or simulations for further investigation. In the subsequent sections, we will first discuss three potential origins of the temperature-independent loss of TLS at the lowest temperatures. Subsequently, our focus shifts towards possible contributions to the thermally-activated TLS relaxation, exploring relaxation mechanisms due to TLS, phonons, and QPs.

\subsubsection{Temperature-independent loss channels}\label{sec:temp_indep_channels}
First, as discussed above in Sec.~\ref{sec:TLS_T1_model}, a possible channel comes from the ne QPs interacting with TLS, causing energy relaxation of TLS. These ne QPs are thought to originate from high-energy incidents, such as cosmic $\mu$ rays and background $\gamma$ ray radiation, which would hit the microchip. These energetic events break Cooper-pairs and excite high-energy phonons that could propagate through the entire chip. Along the way, these high-energy phonons would dissipate energy, leading to the generation of ne QPs~\cite{Vepsalainen20,Wilen21,McEwen22,Thorbeck22}. This results in ne QPs of population much higher than expected for equilibrium QPs at the temperature of the microchip, resulting in an effective QP temperature at approximately $150$~mK~\cite{Vepsalainen20,Gustavsson16,Serniak18}. Unfortunately it is challenging to filter out these ne QPs. Unlike their equilibrium counterparts, the population of ne QPs arising from high-energy events, and correspondingly their contribution to TLS relaxation, remains independent of temperature. 

Similarly, at the lowest temperatures, the true device temperature could be higher than the MXC plate temperature, due to insufficient thermalization between the device and the MXC plate, as discussed for the transmon device in Sec.~\ref{sec:transmon_T1_model}. The same argument can be extended to TLS as well. In this case, TLS might experience an even higher temperature plateau than the transmon device. This is primarily attributed to their poorer thermalization, due to factors such as their smaller size, longer lifetime, and a suppressed thermal bath (of phonons). The saturation in temperature then manifest as a temperature-independent residual loss channel. Finally, there is a temperature-independent mechanical loss channel stemming from the viscous behavior of grain-boundaries in the polycrystalline aluminum~\cite{Ke47,Wollack21}. Although we have focused on the QP loss-model (Eq.~\ref{eq:TLS_relaxation}), and the associated higher effective temperature due to the temperature-independent population of ne QPs to explain our observation in Fig.~\ref{fig:TLS5_vs_temp}, the other two temperature-independent loss channels discussed here, and potentially other relaxation channels that we have not considered, should still be considered possibilities.

\subsubsection{Temperature-dependent loss channels: other TLS}
Turning our attention to the temperature-dependent loss pathways, we first explore resonant interactions between a given TLS and other nearby TLS, all with their frequencies lying inside the acoustic bandgap. In sec.~\ref{sec:TLS_dist}, we have obtained a TLS density of $\sigma = 0.6~\mathrm{GHz}^{-1} \mathrm{\mu m}^{-2}$ in the AlO$_x$ barrier layer. Extending this TLS density estimation to the Al and Si surfaces, we expect on average a total of $120$ TLS per GHz for a $10~\mu$m$\times 10~\mu$m region. This accounts for $60$ TLS on both the top and bottom side of the $220$~nm Si device layer. The chosen area of $10~\mu$m$\times 10~\mu$m approximates the size of the acoustic metamaterials and the enclosed JJ region. We make the assumption that TLS outside the JJ maintain comparable properties to those within the JJ due to the presence of the same acoustic bandgap metamaterial. This assumption gives the TLS relaxation time of $T_1\sim500~\mu$s, and the coherence time of $T_2^*\sim 1~\mu$s in this extended region. Importantly, the spectral linewidths of these TLS are dominated by dephasing, resulting in a linewidth of $\sim 1$~MHz. Given the presence of $120$ such TLS over a $1$~GHz span, the average detuning between neighboring TLS frequencies is $\sim 8$~MHz, which significantly exceeds the linewidth. As a result, it is reasonable to infer that resonant interactions among TLS are unlikely to contribute significantly to TLS relaxation.

Continuing our exploration of TLS-TLS interactions, we extend our discussions to off-resonant interactions between TLS. Our preliminary assessment leads us to conclude that off-resonant TLS-TLS interactions do not account for the plateau of TLS $Q$-factor at the lowest temperatures. As will be discussed in sec.~\ref{sec:TLS_phonon_int}, the temperature scaling of the relaxation rate arises from the frequency scaling of the thermal bath modes in the relaxation rate expression. This frequency scaling, in turn, comes from the DOS of the bath modes as well as the frequency-dependence of the system-bath interactions (the transition matrix element). Previous studies of TLS have shown that TLS DOS is either independent of frequency, or has a weak frequency-dependence $\sim\omega^\mu$, with $\mu\simeq 0.3$~\cite{Muller19,Burnett14,Behunin16,Zolfagharkhani05}. The interaction between TLS is dipolar, determined only by their dipole magnitude, orientation, and relative position, which are all frequency-independent. Consequently, the relaxation rate due to off-resonant TLS will have at most a weak power-law temperature scaling of $T^\mu$. This mismatches the strong temperature scaling we have observed for $T\gtrsim50$~mK.

The weak temperature scaling from a bath of off-resonant TLS, however, necessitates an evaluation of their potential contribution to the plateau of TLS $Q$-factor at the lowest temperatures. Given a TLS density of $\sigma = 0.6~\mathrm{GHz}^{-1} \mathrm{\mu m}^{-2}$, the average distance between TLS is $1/2\sqrt{\sigma} = 0.65~\mu$m, corresponding to a TLS-TLS coupling strength of $g = 5$~kHz, when the two dipoles are aligned. The average detuning between TLS with adjacent frequencies is $\Delta \sim 250$~MHz. The linewidth for the bath TLS is assumed around $\Gamma_{2,\mathrm{bath}} \sim 1$~MHz. Using similar approximations that will be discussed next in sec.~\ref{sec:TLS_phonon_int}, we reach at the relaxation rate attributed to the interaction with a singular off-resonant bath TLS
\begin{equation}
    \Gamma_{1,\mathrm{TLS}} \approx \frac{g^2\Gamma_{2,\mathrm{bath}}}{\Delta^2 + \Gamma_{2,\mathrm{bath}}^2} = 0.4~\mathrm{mHz}.
\end{equation}

\noindent This contribution is seven orders of magnitude weaker than the measured relaxation rate of approximately $\sim 2$~kHz. Further numerical simulation corroborates that with the TLS density of $\sigma = 0.6~\mathrm{GHz}^{-1} \mathrm{\mu m}^{-2}$ and a linewidth of $\Gamma_{2,\mathrm{bath}} \sim 1$~MHz, interactions between TLS contribute negligibly to the $Q$-factor plateau at the lowest temperatures for the central TLS.

\subsubsection{Temperature-dependent loss channels: Phonons outside the acoustic bandgap}\label{sec:TLS_phonon_int}
In this section, we explore TLS relaxation due to its interaction with phonons. We consider a single phonon process, in which the TLS relaxes by emitting a single phonon. For this discussion, we draw upon the key results and notations from ref.~\cite{MacCabe20}, where a more detailed analysis is provided. The relevant Hamiltonian between the TLS and a stress wave associated with the phonon mode $s$ is
\begin{equation}\label{eq:TLS-phonon}
    \hat{\ham}_{\mathrm{TLS}-s} = \frac{\omega_\mathrm{TLS}}{2}\hat{\sigma}_z + \omega_s (\hat{b}_s^\dagger\hat{b}_s + \frac{1}{2}) + (g_{t,s} \hat{\sigma}_x + g_{l,s} \hat{\sigma}_z) (\hat{b}_s + \hat{b}_s^\dagger),
\end{equation}
where $\omega_\mathrm{TLS}$ and $\omega_s$ are the frequencies of the TLS and the phonon mode $s$, $g_{t,s}$, $g_{l,s}$ their transverse and longitudinal coupling, $\hat\sigma$ the Pauli operator for TLS, and $\hat{b}^\dagger (\hat{b})$ the creation (annihilation) operator for phonon mode $s$. 

In the case of resonant decay from TLS into the phonon bath, the $g_t\hat{\sigma}_x$ term in Eq.~\ref{eq:TLS-phonon} dominates. In this context, neglecting pure dephasing of the phonons, we arrive at the relaxation rate of TLS induced by phonon mode $s$, approximated by
\begin{equation}\label{eq:TLS_pure_dephasing}
    (\delta\Gamma_{1,\mathrm{TLS}})_s \approx
    \frac{g_{t,s}^2 \gamma_s(2n_s+1)}{(\omega_\mathrm{TLS}-\omega_s)^2 + (\gamma_{s}/2)^2}.
\end{equation}

\noindent Considering a phonon bath in thermal equilibrium, characterized by the Bose-Einstein distribution, where $2n_s+1 = \coth[\hbar\omega_s/2k_B T]$, Eq.~\ref{eq:TLS_pure_dephasing} leads to~\cite{MacCabe20}
\begin{equation}\label{eq:TLS_T1_expr_sum}
    (\delta\Gamma_{1,\mathrm{TLS}})_{ph} \approx \sum_s
    [\frac{g_{t,s}^2 \gamma_s}{(\omega_\mathrm{TLS}-\omega_s)^2 + (\gamma_{s}/2)^2}] \coth[\hbar\omega_s/2k_B T].
\end{equation}

\noindent In the limit of a continuum phonon bath, the summation in Eq.~\ref{eq:TLS_T1_expr_sum} is replaced by an integral. The primary contribution to the integral comes from the integration range $\omega_s\in[\omega_\mathrm{TLS}-\gamma_s/2,\omega_\mathrm{TLS}+\gamma_s/2]$, yielding
\begin{equation}\label{eq:TLS_T1_expr_integral}
    (\delta\Gamma_{1,\mathrm{TLS}})_{ph,cont.} \approx 4\rho_{ph}[\omega_\mathrm{TLS}]
    g_{t,s}^2\coth[\hbar\omega_s/2k_B T],
\end{equation}
where $\rho_{ph}[\omega]$ is the phonon DOS at frequency $\omega$. For Debye model phonons in D-dimensions, $\rho_{ph}[\omega]\propto \omega^{\mathrm{D}-1}$, and $g_{t,s}\propto \sqrt{\omega_s}$ from the vacuum strain field amplitude. This leads to $(\delta\Gamma_{1,\mathrm{TLS}})_{ph,cont.}\propto \omega_\mathrm{TLS}^\mathrm{D}$. We will show in the following that the frequency scaling yields the same scaling for temperature.

We now consider the reverse process---a phonon decays into the TLS bath. In particular, we skip the discussion of resonant decay (dominated by the $g_t\hat{\sigma}_x$ term), and instead focus on the off-resonant `relaxation' process, which is dominated by the $g_l\hat{\sigma}_z$ term in Eq.~\ref{eq:TLS-phonon}. The $\hat{\sigma}_z$ interaction shifts the frequencies of phonons, displacing them away from thermal equilibrium. Through a higher-order process between the TLS and the phonon modes, the TLS draws energy from the phonon modes, giving rise to the `relaxation' process of phonons. 
Integrating the contributions from all TLS, and assuming a $T_1$ limited $T_2$ for the phonon mode $s$, yield~\cite{MacCabe20}
\begin{equation}\label{eq:phonon_relaxation}
    (\delta\gamma_s)_\mathrm{rel} 
    \approx \sum_\mathrm{TLS} (\frac{2g_{l,s}^2}{\omega_s})(\frac{\hbar\Gamma_{1,\mathrm{TLS}}}{k_B T})\sech^2[\hbar\omega_\mathrm{TLS}/2k_B T].
\end{equation}

Assuming the energy damping of TLS in the TLS bath is dominated by resonant decay into phonon modes, $\Gamma_{1,\mathrm{TLS}} = (\delta\Gamma_{1,\mathrm{TLS}})_{ph,cont.}$ , we can plug in Eq.~\ref{eq:TLS_T1_expr_integral} into Eq.~\ref{eq:phonon_relaxation}, which leads to the scaling
\begin{equation}\label{eq:T_scaling}
\begin{split}
    &(\delta\gamma_m)_\mathrm{rel} \propto \sum_\mathrm{TLS} \frac{\omega_\mathrm{TLS}^{\mathrm{D}}}{k_B T}\coth[\frac{\hbar\omega_\mathrm{TLS}}{2k_B T}] \sech^2[\frac{\hbar\omega_\mathrm{TLS}}{2k_B T}]\\
    \approx & \int_0^\infty\frac{\omega_\mathrm{TLS}^{\mathrm{D}}}{k_B T} \coth[\frac{\hbar\omega_\mathrm{TLS}}{2k_B T}] \sech^2[\frac{\hbar\omega_\mathrm{TLS}}{2k_B T}] \rho_\mathrm{TLS} d\omega_\mathrm{TLS}\\
    = & 2\rho_\mathrm{TLS} (\frac{k_B T}{\hbar})^\mathrm{D} \int_0^{\infty} x^{\mathrm{D}} \csch[x]dx.
\end{split}
\end{equation}
In the last line, we used the identity $\sech^2[\hbar\omega/2k_B T]\coth[\hbar\omega/2k_B T]=2\csch[\hbar\omega/k_B T]$, and made the assumption of a frequency-independent TLS density $\rho_\mathrm{TLS}$. As promised earlier, the frequency scaling $\propto \omega^\mathrm{D}$ in Eq.~\ref{eq:TLS_T1_expr_integral} leads to an equivalent temperature scaling $\propto T^\mathrm{D}$ in Eq.~\ref{eq:T_scaling}, which directly reflects the dimension D of the system. More generally, the temperature scaling of off-resonant relaxation processes is a result of integrating over the frequency-dependent terms in the relaxation rate. This relation has been observed across a wide variety of examples, such as TLS-phonon interaction~\cite{MacCabe20}, two-phonon Orbach-like process~\cite{Orbach61}, spin-phonon relaxation~\cite{Cambria21}, and three-phonon scattering~\cite{Srivastava90}. 

Taking the example of the TLS-phonon interaction, the frequency scaling of phonon relaxation discussed above involves two contributions, an $\omega^{\mathrm{D}-1}$ term from the phonon DOS, and an $\omega$ term from the square of TLS-phonon coupling strength. We argue that for the long-lived TLS, one expects a similar scaling for the single phonon, off-resonant `relaxation' process. 
In our case, the effective system dimension for thermally activated phonons is $\mathrm{D}=2$ at lower temperatures, and it increases to $\mathrm{D}>2$ at higher temperatures, when high frequency phonons that see a semi-3D DOS start to be thermally populated. The predicted $T^2$ dependence, however, does not align with our experimental data in Fig.~\ref{fig:TLS5_vs_temp}. Specifically, the $T^2$ dependence is too strong to account for the almost temperature-independent relaxation at $T\lesssim 50$~mK, and too weak to explain the observed rapid decrease in $T_1$ at $T\gtrsim 50$~mK.

Additionally, numerical modelling of the damping of TLS within the acoustic bandgap due to quasi-phonon modes for a similar acoustic structure was performed in Ref.~\cite{MacCabe20}, indicating that TLS damping of this nature would limit $T_1$ to $100$~ms or more, two-orders-of-magnitude greater than measured here. We also don't observe evidence of an increasing trend of the low-temperature $T_1$ of TLS as their frequencies move deeper into the acoustic bandgap, which one would expect if this was the source of the $T_1$ limit.

\subsubsection{Temperature-dependent loss channels: Quasiparticles}\label{sec:TLS_QP_relax}
For the discussions of quasiparticles (QP), we first make the distinction between the contributions of QPs in thermal equilibrium (eq), and those in non-equilibrium (ne). The ne QPs are generated by bursts of high-energy events~\cite{Vepsalainen20}. Notably, these high-energy events, and correspondingly the distribution of ne QP, are temperature-independent, and are already discussed in Sec.~\ref{sec:temp_indep_channels}. The temperature-dependent contribution comes exclusively from the eq QPs. For the eq part, we follow the treatment of ref.~\cite{Catelani11,Catelani11b}, and list the key steps here for readers' convenience.

The QP interacts with the qubit by tunneling through the Josephson Junction, resulting in the system Hamiltonian
\begin{equation}
    \ham = \ham_\phi + \ham_\mathrm{qp} + \ham_T.
\end{equation}
The first term is the SC qubit Hamiltonian, which for transmon is
\begin{equation}
    \ham_\phi = 4E_C\hat{n}^2 - E_J \cos\hat{\phi}.
\end{equation}
The second term is the BCS Hamiltonian for the QPs,
\begin{equation}
    \ham_\mathrm{qp} = \sum_{j=L,R}\ham_\mathrm{qp}^j,~\ham_\mathrm{qp}^j = \sum_{n,\sigma}\epsilon_n^j\alpha_{n\sigma}^{j\dagger}\alpha_{n\sigma}^j,
\end{equation}
where $\alpha_{n\sigma}^j (\alpha_{n\sigma}^{j\dagger})$ are the annihilation (creation) operators for quasiparticle in the lead $j$ that has spin $\sigma = \uparrow,\downarrow$, and energy $\epsilon_n$ in the single-particle energy level $n$.
The last term describes QP tunneling through the junction, and under simplifications relevant to superconducting circuits,
\begin{equation}
    \ham_T = i\tilde{t}\sum_{n,m,\sigma}\sin\frac{\hat{\phi}}{2}\alpha_{n\sigma}^{L\dagger}\alpha_{m\sigma}^R + \mathrm{h.c.},
\end{equation}
where $\tilde{t}$ is the electron tunneling amplitude.

When calculating the QP impact on the transmon qubit, Fermi's golden rule gives
\begin{equation}
\begin{split}
    \Gamma_{i\rightarrow f} = & 2\pi \sum_{\{\lambda_\mathrm{qp}\}} \vert \langle f,\{\lambda_\mathrm{qp}\}\vert\ham_T\vert i, \{\eta_\mathrm{qp}\}\rangle\vert^2 \\
    & \times \delta(E_{\lambda,qp} - E_{\eta, qp} - \omega_{if}),
\end{split}
\end{equation}
where $E_{\eta,qp} (E_{\lambda,qp})$ is the energy of the QP in its initial (final) state $\{\eta_\mathrm{qp}\}~(\{\lambda_\mathrm{qp}\})$, $\omega_{if}$ is the energy difference of the qubit in the initial and final state. An average over the initial quasiparticle state following its distribution is taken, which is inexplicit in the equation. 

In the low energy regime, it has been shown that the qubit dynamics and QP kinetics are separable~\cite{Catelani11b}, yielding
\begin{equation}\label{eq:QP_current_spectral_density}
    \Gamma_{i\rightarrow f} = 2\pi \vert \langle f \vert \sin\frac{\hat{\phi}}{2} \vert i\rangle\vert^2 S_\mathrm{qp}(\omega_{if}).
\end{equation}
This is an established model describing the QP interaction with superconducting qubit. Eq.~\ref{eq:QP_current_spectral_density} takes the form of Fermi's golden rule, where $S_\mathrm{qp}(\omega_{if})$ represents the current spectral density of quasiparticles tunneling through the JJ, that interacts with the phase degrees of freedom of the transmon qubit through the matrix element of $\langle f \vert \sin\frac{\hat{\phi}}{2} \vert i\rangle$.

Within a similar context, we consider the interaction between QPs and TLS inside the JJ, in a regime where the TLS dynamics and QP kinematics are separable. The matrix element coupling the electric dipole of the TLS to the electric current formed by QPs tunneling through the junction depends on the details of the microscopic configuration. Here we skip discussions on those details and denote it as $A_{if}$. The current spectral density of QPs tunneling through the junction $S_\mathrm{qp}(\omega)$ stays the same as in the previous discussion,
\begin{equation}\label{eq:S_qp}
\begin{split}
    S_\mathrm{qp}(\omega) = & \frac{16E_J}{\pi}\int_0^\infty dx \frac{1}{\sqrt{x}\sqrt{x+\omega/\Delta}}f_E[(1+x)\Delta]\\
    &\times \{1-f_E[(1+x)\Delta +\omega]\}),
\end{split}
\end{equation}
where $f_E$ is the distribution function, $\Delta$ the gap parameter.

In thermal equilibrium (assuming Boltzmann distribution), and at low temperatures $T\ll \Delta$, Eq.~\ref{eq:S_qp} undergoes further simplification,
\begin{equation}
    S_\mathrm{qp}^{eq}(\omega)=\frac{16E_J}{\pi}e^{-\Delta/T}e^{\omega/2T}K_0(\frac{\vert\omega\vert}{2T}).
\end{equation}
Using the relation
\begin{equation}
    S_\mathrm{qp}^{eq}(-\omega)/S_\mathrm{qp}^{eq}(\omega) = e^{-\omega/T},
\end{equation}
we find the total TLS relaxation rate due to QPs
\begin{equation}\label{eq:TLS_T1_qp}
\begin{split}
    \Gamma_\mathrm{qp} &= 2\pi\vert A_{if}\vert^2 [S_\mathrm{qp}^{eq}(\omega) - S_\mathrm{qp}^{eq}(-\omega)]\\
    &= 2\pi\vert \tilde{A}_{if}\vert^2 e^{-\Delta/T}\sinh(\frac{\omega}{2T})K_0(\frac{\vert\omega\vert}{2T}).
\end{split}
\end{equation}

\noindent It's worth noting that the functional form of this expression is the same as that of the transmon-QP relaxation in Eq.~\ref{eq:Q_QP}. The matrix element term $\tilde{A}_{if}$ describes the coupling between eq QP current and the electric dipole of TLS, which depends on the details of the microscopic configuration, that is beyond the scope of this work.

The consideration of TLS-QP interaction naturally arises for TLS located inside the JJ. In this scenario, the current of QPs tunneling through the JJ interacts with the electric dipole of TLS. Eq.~\ref{eq:TLS_T1_qp} qualitatively describes the temperature-dependent TLS relaxation well at high temperatures $T\gtrsim 150$~mK, as shown in Fig.~\ref{fig:TLS5_vs_temp}. We therefore attribute eq QPs as one of the relevant TLS relaxation channels. 

In our experiment, the prominence of the QP contribution emerges from our selection of TLS located inside the JJ. This specific configuration results in a strong coupling between the TLS and QPs tunneling through the JJ, due to the close proximity. We note that the QP current spectral density, and correspondingly the QP-TLS interaction, is expected to vary significantly across different regions of the device. An example is the substrate-air interface, where QPs are altogether absent. Looking ahead to future experiments, one could explore the identification of individually addressable and controllable TLS located at circuit interfaces, and measure their temperature-dependent behavior to test our model of TLS-QP interaction.

\begin{figure}
    \centering
    \includegraphics[width = 0.45\textwidth]{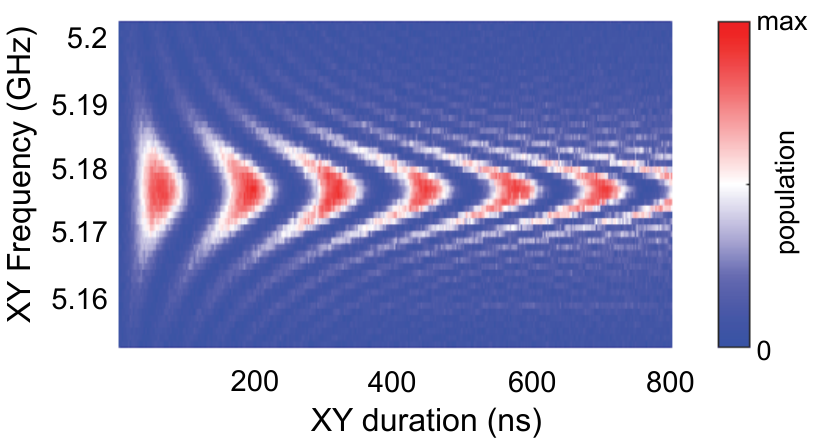}
    \caption{\textbf{Rabi chevron of TLS31.}  Measurements were taken on Q$_4$ of Chip-A during CD1. The interaction strength and detuning between the TLS and transmon qubit are $g=47.7$~MHz, $\Delta=1.7$~GHz. The XY driving power in this experiment is $\sim 17$~dB stronger than typically used for controlling the transmon qubit.}
    \label{fig:TLS_chevron}
\end{figure}

\subsection{Direct control of TLS}
\subsubsection{TLS pulses}\label{sec:TLS_drive}
In this section, we describe the technique of directly controlling a TLS. This is achieved by sending a strong microwave pulse resonating with the TLS down the XY line of the transmon qubit. The direct control of TLS is possible due to the mutual interaction and the resulting hybridization between the transmon qubit ($q$) and TLS, governed by the Hamiltonian 
\begin{equation}\label{eq:int_ham}
\begin{split}
&\ham = \frac{\omega_q}{2} \hat{\sigma}^z_{q} + \frac{\omega_{\mathrm{TLS}}}{2} \hat{\sigma}^z_{\mathrm{TLS}} + \hat{H}_{\mathrm{int}},\\
&\ham_{\mathrm{int}} = g(\hat{\sigma}^+_q\hat{\sigma}^-_{\mathrm{TLS}} + \hat{\sigma}^-_q\hat{\sigma}^+_{\mathrm{TLS}}),
\end{split}
\end{equation}
where $\omega$ denote their frequencies, $\hat{\sigma}^z, \hat{\sigma}^\pm$ are the Pauli operators.

\noindent The interaction term $\ham_\mathrm{int}$ hybridizes the transmon and the TLS. The hybridization, even at the presence of a detuning between the transmon and the TLS, gives the TLS-like eigenstate a little transmon character, that enhances TLS' coupling with the transmon's XY line. This technique has been used in the context of directly controlling TLS in a phase qubit~\cite{Lisenfeld10}, in the cross-resonance gate between two coupled superconducting qubits~\cite{Paraoanu06,Rigetti10}, as well as in accelerating nuclear spin gates in quantum registers in diamond~\cite{Chen15}. 

An example is shown in Fig.~\ref{fig:TLS_chevron}. Here, a Rabi pulse of varying duration and microwave frequency drives the state of the TLS, which is subsequently read out through the transmon qubit. The resultant Rabi chevron pattern, notably of the TLS under direct control, demonstrates the feasibility and precision in controlling TLS using this technique.
In this specific experiment, the interaction strength and detuning between the TLS and the transmon qubit are $g=47.7$~MHz, $\Delta=1.7$~GHz. The driving power of the XY pulse is approximately $\sim 17$~dB stronger than what is typically used for controlling the transmon qubit.

\subsubsection{TLS relaxation time with TLS pulses}

\begin{figure}
    \centering
    \includegraphics[width = 0.45\textwidth]{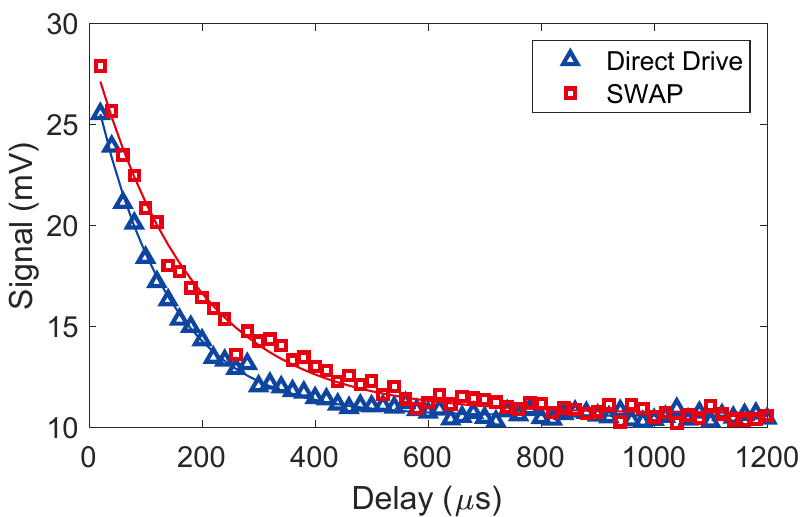}
    \caption{\textbf{$T_1$ relaxation curves of TLS31}. We contrast $T_1$ measurements obtained using two different methods: preparing the initial TLS excited-state with a direct TLS $\pi$ pulse (blue triangles) and through swapping the excitation from the transmon qubit (red squares). Exponential fits (solid lines) yield $T_1 = 129 \pm 5$~$\mu$s and $T_1=177 \pm 10$~$\mu$s for the direct drive and the SWAP methods, respectively.}
    \label{fig:T1_direct_vs_SWAP}
\end{figure}

The technique of direct TLS control, as discussed above, allows us to calibrate pulses for the TLS and prepare the TLS in its excited-state independently from the transmon qubit. However, it's important to note that the readout process still involves the transmon qubit. In Fig.~\ref{fig:T1_direct_vs_SWAP}, we present the $T_1$ energy relaxation curve of TLS31, showcasing a comparison between the two methods: preparing the initial TLS excited-state with a TLS $\pi$ pulse (direct drive, blue triangles) and through swapping the excitation from the transmon qubit (SWAP, red squares). 

The two distinct methods of preparing the TLS in its excited-state yield different relaxation curves. Notably, the direct drive method gives a shorter $T_1 = 129 \pm 5$~$\mu$s, whereas the SWAP method gives a longer $T_1=177 \pm 10$~$\mu$s. This discrepancy can be attributed to the high power of the microwave pulse required for the direct drive, which is approximately $\sim 17$~dB stronger than that used for the transmon qubit. The high power microwave can generate QPs~\cite{Wang14,Vool14} which subsequently accelerate the relaxation of TLS through the mechanism discussed in sec.~\ref{sec:TLS_QP_relax}. Following this observation from the first cool-down (CD1), we have been using the SWAP method exclusively for the preparation of TLS excited-states in the following TLS $T_1$ measurements. 

Throughout the measurements, particularly during CD1, a notable portion of TLS were characterized by the direct drive method, as indicated by the $\dagger$ symbol in Table~\ref{tab:TLS_chipA} and Table~\ref{tab:TLS_chipB}. We argue that for TLS with resonant frequency lying outside the acoustic bandgap, the relaxation rate induced by the QPs originating from high-power microwave pulses, is likely minor when compared to other relaxation mechanisms (e.g. spontaneous phonon emissions). Revisiting the case of TLS31, we compute the relaxation rate due to microwave induced QPs $\tau_\mathrm{QP}^{-1} = 1/129~\mu\mathrm{s} - 1/177~\mu\mathrm{s} = 476^{-1}~\mu\mathrm{s}^{-1}$. This value is two orders of magnitude smaller than the average relaxation rate of $4^{-1}~\mu\mathrm{s}^{-1}$ for TLS outside the bandgap, contributing negligibly to their relaxation.
While for TLS inside the bandgap, these QPs generated by the high-power microwave pulses could potentially lower the measured $T_1$ notably, as shown in Fig.~\ref{fig:T1_direct_vs_SWAP}. As a result, the reported ratio between the $T_1$ values of TLS frequencies inside vs outside the acoustic bandgap represents an \emph{underestimation} of the impact from acoustic metamaterials.

\begin{figure}
    \centering
    \includegraphics[width = 0.45\textwidth]{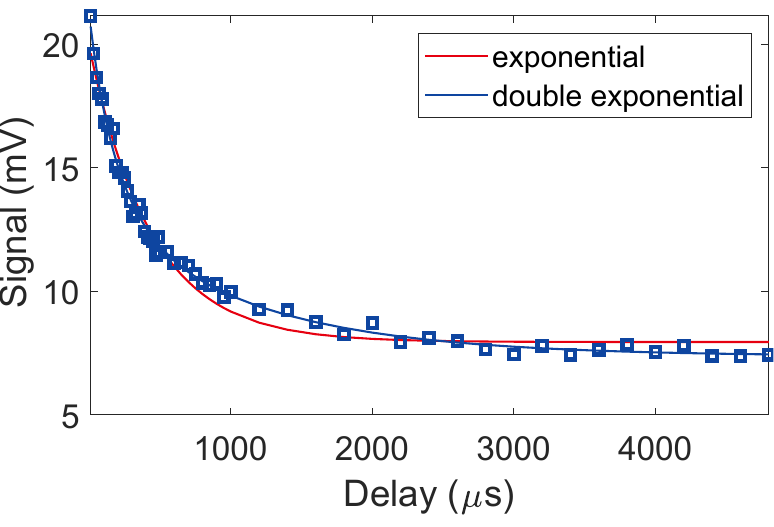}
    \caption{\textbf{$T_1$ relaxation curve of TLS36.} We analyze the shape of the relaxation curve when the initial excited-state of TLS is prepared by direct XY drive. Fits using a simple exponential decay model (red solid line) and double exponential model (blue solid line) are superimposed. The data clearly demonstrates deviation from the simple exponential decay.}
    \label{fig:double_exp_decay}
\end{figure}

\subsubsection{TLS relaxation curve with TLS pulses}\label{sec:TLS_de_relax}
Another phenomenon emerges when measuring TLS relaxation using the direct drive method. In certain cases, we have observed deviations of the TLS $T_1$ relaxation curve from a simple exponential decay. An illustrative example from TLS36 is presented in Fig.~\ref{fig:double_exp_decay}. To analyze the data, we fit it with a simple exponential decay curve, 
\begin{equation}
    p(t) = A\exp(-t/T_1) + B,    
\end{equation}
indicated by the red solid line, as well as a double exponential decay curve~\cite{Pop14,Gustavsson16},
\begin{equation}\label{eq:double_exp}
    p(t) = Ae^{\langle n_\mathrm{qp}\rangle(\exp[-t/T_{1,\mathrm{qp}}]-1)} e^{-t/T_1} + B,
\end{equation}
indicated by the blue solid line. 
Eq.~\ref{eq:double_exp} was introduced in refs.~\cite{Pop14,Gustavsson16} for superconducting qubits, to disentangle the relaxation rate induced by quasiparticles from other relaxation channels. Here, $\langle n_\mathrm{qp}\rangle$ is the average quasiparticle population, $T_{1,\mathrm{qp}}$ is the relaxation time due to one quasiparticle, and $T_1$ is the relaxation time from other decay channels. 

In experiments where the TLS relaxation curve no longer adheres to a simple exponential decay, as observed in Fig.~\ref{fig:double_exp_decay}, the double exponential fit is adopted. Consequently, we report $T_1$ from Eq.~\ref{eq:double_exp} in these cases. As previously discussed in sec.~\ref{sec:TLS_QP_relax}, we propose that the relaxation of TLS inside the JJ, induced by interaction with QPs, follows the same functional form as the interaction between the transmon qubit and QPs. This similarity justifies the application of Eq.~\ref{eq:double_exp} for characterizing TLS $T_1$ relaxation with an explicit contribution from QP. We note that the direct TLS drive method does not always lead to double exponential decays. A case of simple exponential decay under direct TLS drive can be found in Fig.~\ref{fig:T1_direct_vs_SWAP}. In these cases, we fit the measured TLS relaxation curve to a simple exponential decay model, and report the corresponding underestimate of fitted $T_1$ values.

\bibliography{Biblio}